\documentclass[english,12pt]{article}[']
\usepackage{graphicx,floatrow}
\usepackage{times}
\usepackage{cite}
\usepackage{afterpage}
\usepackage[unicode=true]{hyperref}
\usepackage[page,header]{appendix}
\usepackage{color}
\usepackage[T1]{fontenc}
\usepackage[latin9]{inputenc}
\usepackage{amsmath}
\usepackage{amssymb}
\usepackage{babel}
\usepackage{lipsum}
\usepackage{titlesec}
\usepackage{etoc}
\usepackage{enumerate}

\usepackage{ulem} 

\newcommand{\lapprox}{%
\mathrel{%
\setbox0=\hbox{$<$}\raise0.6ex\copy0\kern-\wd0\lower0.65ex\hbox{$\sim$}}}

\def\lambdabar{\lambda\kern-1ex\raise0.65ex\hbox{-}}
\usepackage[margin=1in]{geometry}
\newcommand{\be}{\begin{eqnarray}}
\newcommand{\ee}{\end{eqnarray}}
\newcommand{\bc}{\begin{center}}
\newcommand{\ec}{\end{center}}
\newcommand{\beq}{\begin{eqnarray}}
\newcommand{\eea}{\end{eqnarray}}

\titlespacing*{\subsubsection}{0pt}{0ex}{-2ex}
\usepackage{xspace}
\usepackage[percent]{overpic}

\def\Kp    {\ensuremath{K^+}\xspace}
\def\Km    {\ensuremath{K^-}\xspace}
\def\Kpm   {\ensuremath{K^\pm}\xspace}

\def\KS    {\ensuremath{K^0_{\scriptscriptstyle S}}\xspace} 
\def\piz   {\ensuremath{\pi^0}\xspace}
\def\pip   {\ensuremath{\pi^+}\xspace}
\def\pim   {\ensuremath{\pi^-}\xspace}

\def\pimp  {\ensuremath{\pi^\mp}\xspace}
\def\etac  {\ensuremath{\eta_c}\xspace}

\newcommand{\mev}{\ensuremath{\mathrm{\,Me\kern -0.1em V}}\xspace}

\usepackage{relsize}
\def\babar{\mbox{\slshape B\kern-0.1em{\smaller A}\kern-0.1em
    B\kern-0.1em{\smaller A\kern-0.2em R}}}
\begin{document}
\hfill{\bf Date:} {\today}
\begin{center}
{\large Proposal for JLab PAC48}
\end{center}
\begin{center}
\Large{Strange Hadron Spectroscopy with Secondary $K_L$ Beam in Hall~D}
\end{center}


\begin{center}
\underline{\large Experimental Support}:\\
Shankar~Adhikari$^{43}$,
Moskov~Amaryan~(\textcolor{red}{Contact Person}, \textcolor{blue}{Spokesperson})$^{43}$,
Arshak~Asaturyan$^{1}$,
Alexander~Austregesilo$^{49}$,
Marouen~Baalouch$^{8}$,
Mikhail~Bashkanov~(\textcolor{blue}{Spokesperson})$^{63}$,
Vitaly~Baturin$^{43}$,
Vladimir~Berdnikov$^{11,35}$,
Olga~Cortes~Becerra$^{19}$,
Timothy~Black$^{60}$,
Werner~Boeglin$^{13}$,
William~Briscoe$^{19}$,
William~Brooks$^{54}$,
Volker~Burkert$^{49}$,
Eugene~Chudakov$^{49}$,
Geraint~Clash$^{63}$,
Philip~Cole$^{32}$,
Volker~Crede$^{14}$,
Donal~Day$^{61}$,
Pavel~Degtyarenko$^{49}$,
Alexandre~Deur$^{49}$,
Sean~Dobbs~(\textcolor{blue}{Spokesperson})$^{14}$,
Gail~Dodge$^{43}$,
Anatoly~Dolgolenko$^{26}$,
Simon~Eidelman$^{6,41}$,
Hovanes~Egiyan~(\textcolor{red}{JLab Contact Person})$^{49}$,
Denis~Epifanov$^{6,41}$,
Paul~Eugenio$^{14}$,
Stuart~Fegan$^{63}$,
Alessandra~Filippi$^{25}$,
Sergey~Furletov$^{49}$,
Liping~Gan$^{60}$,
Franco~Garibaldi$^{24}$,
Ashot~Gasparian$^{39}$,
Gagik~Gavalian$^{49}$,
Derek~Glazier$^{18}$,
Colin~Gleason$^{22}$,
Vladimir~Goryachev$^{26}$,
Lei~Guo$^{14}$,
David~Hamilton$^{11}$,
Avetik~Hayrapetyan$^{17}$,
Garth~Huber$^{53}$,
Andrew~Hurley$^{56}$,
Charles~Hyde$^{43}$,
Isabella~Illari$^{19}$,
David~Ireland$^{18}$,
Igal~Jaegle$^{49}$,
Kyungseon~Joo$^{57}$,
Vanik~Kakoyan$^{1}$,
Grzegorz~Kalicy$^{11}$,
Mahmoud~Kamel$^{13}$,
Christopher~Keith$^{49}$,
Chan~Wook~Kim$^{19}$,
Eberhard~Klemp$^{5}$,
Geoffrey~Krafft$^{49}$,
Sebastian~Kuhn$^{43}$,
Sergey~Kuleshov$^{2}$,
Alexander~Laptev$^{33}$,
Ilya~Larin$^{26,59}$,
David~Lawrence$^{49}$,
Daniel~Lersch$^{14}$,
Wenliang~Li$^{56}$,
Kevin~Luckas$^{28}$,
Valery~Lyubovitskij$^{50,51,52,54}$,	
David~Mack$^{49}$,
Michael~McCaughan$^{49}$,
Mark~Manley$^{30}$,
Hrachya~Marukyan$^{1}$,
Vladimir~Matveev$^{26}$,
Mihai Mocanu$^{63}$,
Viktor~Mokeev$^{49}$,
Curtis~Meyer$^{9}$,
Bryan~McKinnon$^{18}$,
Frank~Nerling$^{15,16}$,
Matthew Nicol$^{63}$,
Gabriel~Niculescu$^{27}$,
Alexander~Ostrovidov$^{14}$,
Zisis~Papandreou$^{53}$,
KiJun~Park$^{49}$,
Eugene~Pasyuk$^{49}$,
Peter Pauli$^{18}$,
Lubomir~Pentchev$^{49}$,
William~Phelps$^{10}$,
John~Price$^{7}$,
J\"org~Reinhold$^{13}$,
James~Ritman~(\textcolor{blue}{Spokesperson})$^{28,68}$,
Dimitri~Romanov$^{26}$,
Carlos~Salgado$^{40}$,
Todd~Satogata$^{49}$,
Susan~Schadmand$^{28}$,
Amy~Schertz$^{56}$,
Axel~Schmidt$^{19}$,
Daniel~Sober$^{11}$,
Alexander~Somov$^{49}$,
Sergei~Somov$^{35}$,
Justin~Stevens~(\textcolor{blue}{Spokesperson})$^{56}$, 
Igor~Strakovsky~(\textcolor{blue}{Spokesperson})$^{19}$,
Victor~Tarasov$^{26}$,
Simon~Taylor$^{49}$,
Annika~Thiel$^{5}$,
Guido~Maria~Urciuoli$^{24}$,
Holly~Szumila-Vance$^{19}$,
Daniel~Watts$^{63}$,
Lawrence~Weinstein$^{43}$,
Timothy~Whitlatch$^{49}$,
Nilanga~Wickramaarachchi$^{43}$,
Bogdan~Wojtsekhowski$^{49}$,
Nicholas~Zachariou$^{63}$,
Jonathan~Zarling$^{53}$,
Jixie~Zhang$^{61}$
\vspace{0.1in}\\
\underline{\large Theoretical Support}:\\
Alexey~Anisovich$^{5,44}$,
Alexei~Bazavov$^{38}$,	
Rene~Bellwied$^{21}$,
Veronique~Bernard$^{42}$,
Gilberto~Colangelo$^{3}$,
Ale\v{s}~~Ciepl\'{y}$^{46}$,
Michael~D\"oring$^{19}$,
Ali~Eskanderian$^{19}$,
Jose~Goity$^{20,49}$,
Helmut~Haberzettl$^{19}$,
Mirza~Had\v{z}imehmedovi\'{c}$^{55}$,
Robert~Jaffe$^{36}$,
Boris~Kopeliovich$^{54}$,
Heinrich~Leutwyler$^{3}$,
Maxim~Mai$^{19}$,
Terry Mart$^{65}$,
Maxim~Matveev$^{44}$,
Ulf-G.~Mei{\ss}ner$^{5,29}$,		
Colin~Morningstar$^{9}$,
Bachir~Moussallam$^{42}$,
Kanzo~Nakayama$^{58}$,
Wolfgang~Ochs$^{37}$,
Youngseok~Oh$^{31}$,
Rifat~Omerovic$^{55}$,
Hedim~Osmanovi\'{c}$^{55}$,		
Eulogio~Oset$^{62}$,
Antimo Palano$^{64}$,
Jose~Pel\'aez$^{34}$,
Alessandro~Pilloni$^{66,67}$,
Maxim Polyakov$^{48}$,
David~Richards$^{49}$,
Arkaitz~Rodas$^{49,56}$,
Dan-Olof~Riska$^{12}$,	
Jacobo~Ruiz~de~Elvira$^{3}$,
Hui-Young~Ryu$^{45}$,
Elena~Santopinto$^{23}$,
Andrey~Sarantsev$^{5,44}$,
Jugoslav~Stahov$^{55}$,			
Alfred~\v{S}varc$^{47}$,
Adam~Szczepaniak$^{22,49}$,
Ronald~Workman$^{19}$,
Bing-Song~Zou$^{4}$
\end{center}
\vspace{0.3in}
$^{1}$ A.~I.~Alikhanian National Science Laboratory (Yerevan Physics Institute (YerPhi)) \\
$^{2}$ Departamento de Ciencias Fisicas, Universidad Andres Bello, Sazie 2212, Piso 7, Santiago, Chile \\
$^{3}$ University of Bern, CH-3012 Bern, Switzerland \\
$^{4}$ Institute of Theoretical Physics, CAS, Beijing 100190, People's Republic of China \\
$^{5}$ Helmholtz-Institut f\"ur Strahlen- und Kernphysik, Universit\"at Bonn, Bonn 53115, Germany \\
$^{6}$ Budker Institute of Nuclear Physics SB RAS, Novosibirsk 630090, Russia \\
$^{7}$ California State University, Dominguez Hills, Carson, CA 90747, USA \\
$^{8}$ C.E.A. L'energie Atomique Et Aux Energies Alternatives, 91190 Saclay, France \\
$^{9}$ Carnegie Mellon University (CMU), Pittsburgh, PA 15213, USA \\
$^{10}$ Christopher Newport University (CNU), Newport News, VA 23606, USA \\
$^{11}$ The Catholic University of America (CUA), Washington, DC 20064,  USA \\
$^{12}$ Finnish Society of Science and Letters, Helsinki 00130, Finland \\
$^{13}$ Florida International University (FIU), Miami, FL 33199, USA \\
$^{14}$ Florida State University (FSU), Tallahassee, FL 32306, USA \\
$^{15}$ G\"othe University Frankfurt, Frankfurt 60323, Germany \\
$^{16}$ GSI Helmholtzzentrum f\"ur Schwerionenforschung GmbH, Darmstadt 64291, Germany \\
$^{17}$ Justus Liebig-University of Gie{\ss}en, Gie{\ss}en 35392, Germany \\
$^{18}$ University of Glasgow, Glasgow G12 8QQ, UK \\
$^{19}$ The George Washington University (GW), Washington, DC 20052, USA\\
$^{20}$ Hampton University, Hampton, VA 23668, USA \\
$^{21}$ University of Houston, Houston, TX 77204, USA \\
$^{22}$ Indiana University (IU), Bloomington, IN 47403, USA \\
$^{23}$ I.N.F.N. Sezione di Genova, Genova 16146, Italy \\
$^{24}$ I.N.F.N. Sezione di Roma, Roma 00185, Italy \\
$^{25}$ I.N.F.N. Sezione di Torino, Torino 10125, Italy \\
$^{26}$ National Rsearch Centre ``Kurchatov Institute", Institute for Theoretical and Experimental Physics (ITEP), Moscow 117218, Russia \\
$^{27}$ James Madison University (JMU), Harrisonburg, VA 22807, USA \\
$^{28}$ Institute f\"ur Kernphysik \& J\"ulich Center f\"ur Hadron Physics, J\"ulich 52425, Germany \\
$^{29}$ Institut f\"ur Advanced Simulation, Institut f\"ur Kernphysik and J\"ulich Center f\"ur Hadron Physics, J\"ulich 52425, Germany \\
$^{30}$ Kent State University (KSU), Kent, OH 44242, USA \\
$^{31}$ Kyungpook National University, Daegu 41566, Republic of Korea \\
$^{32}$ Lamar University, Beaumont, TX 77710,  USA \\
$^{33}$ Los Alamos National Laboratory (LANL), Los Alamos, NM 87545, USA \\
$^{34}$ Universidad Complutense de Madrid, 28040 Madrid, Spain \\
$^{35}$ National Research Nuclear University Moscow Engineering Physics Institute (MEPhI), Moscow 115409, Russia \\
$^{36}$ Massachusetts Institute of Technology (MIT), Cambridge, MA 02139, USA \\
$^{37}$ Max-Planck-Institut f\"ur Physik, M\"unchen D-80805, Germany \\
$^{38}$ Michigan State University (MSU), East Lansing, MI 48824, USA \\
$^{39}$ North Carolina A\&T State University (N.C.A\&T), Greensboro, NC 27411, USA \\
$^{40}$ Norfolk State University (NSU), Norfolk, VA 23504, USA \\
$^{41}$ Novosibirsk State University, Novosibirsk 630090, Russia \\
$^{42}$ Universite Paris-Sud 11, 91400 Orsay, France \\
$^{43}$ Old Dominion University (ODU), Norfolk, VA 23529, USA \\
$^{44}$ National Research Centre ``Kurchatov Institute", Petersburg  Nuclear Physics Institute (PNPI), Gatchina 188300, Russia \\
$^{45}$ Pusan National University, Busan 46241, Republic of Korea \\
$^{46}$ Nuclear Physics Institute, Rez 250 68, Czech Republic \\
$^{47}$ Rudjer Bo\v{s}kovi\'{c} Institute, Zagreb 10002, Croatia \\
$^{48}$ Institut f\"ur Theoretische Physik II - Ruhr-Universit\"at, D-44780 Bochum, Germany \\
$^{49}$ Thomas Jefferson National Accelerator Facility (JLab), Newport News, VA 23606, USA \\
$^{50}$ Tomsk State University, Tomsk 634050, Russia \\
$^{51}$ Tomsk Polytechnic University, Tomsk 634050, Russia \\
$^{52}$ Institut f\"ur Theoretische Physics, T\"ubingen Universit\"at, T\"ubingen 72076, Germany \\
$^{53}$ University of Regina (UR), Regina, SA S4S 0A2, Canada \\
$^{54}$ Universidad T\'ecnica Federico Santa Mar\'ia (UTFSM), Casilla 110-V Valpara\'iso, Chile \\
$^{55}$ University of Tuzla, Tuzla 75000, Bosnia and Herzegovina \\
$^{56}$ College of William and Mary (W\&M), Williamsburg, VA 23185, USA \\
$^{57}$ University of Connecticut (UConn), Storrs, CO 06269, USA \\
$^{58}$ University of Georgia (UGA), Athens, GA 30602, USA \\
$^{59}$ University of Massachusetts (UMASS Amherst), Amherst, MA 01003, USA \\
$^{60}$ University of North Carolina at Wilmington (UNCW), Wilmington, NC  28403, USA \\
$^{61}$ University of Virginia (UVa), Charlottesville, VA 22904, USA \\
$^{62}$ Centro Mixto Universidad de Valencia-CSIC, 46071 Valencia, Spain \\
$^{63}$ University of York (UoY), Heslington, York YO10 5DD, UK \\
$^{64}$ INFN Sezione di Bari, 70125 Bari, Italy \\ 
$^{65}$ FMIPA, Universitas Indonesia, Depok 16424, Indonesia \\
$^{66}$ European Centre for Theoretical Studies in Nuclear Physics and related Areas (ECT$^\ast$) and Fondazione Bruno Kessler, Villazzano (Trento), I-38123, Italy\\
$^{67}$ INFN Sezione di Genova, Genova, I-16146, Italy\\
$^{68}$	Institut f\"ur Experimentalphysik I - Ruhr-Universit\"at, Bochum 44780, Germany \\
\noindent

\hspace{0.1in}
\begin{center}
\large{(The \textsc{KLF} Collaboration)}
\end{center}

\begin{figure}[htpb]
\centering
{
    \includegraphics[width=0.2\textwidth,keepaspectratio]{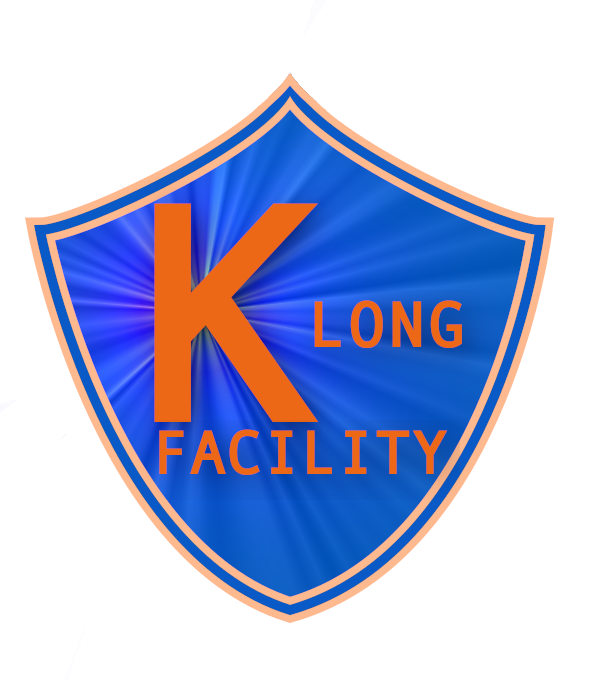} }
\end{figure}

\newpage
\begin{center}{\bf Abstract}\end{center}

We propose to create a secondary beam of neutral kaons in Hall~D at Jefferson 
Lab to be used with the GlueX experimental setup for strange hadron 
spectroscopy. The superior CEBAF electron beam will enable a flux on the order 
of $1\times 10^4~K_L/sec$, which exceeds the flux of that previously attained 
at SLAC by three orders of magnitude. It will allow a broad range of 
measurements that will correspondingly improve the statistics of earlier data 
obtained on a hydrogen target likewise by three orders of magnitude. The use 
of a deuteron target will provide first measurements ever with neutral kaons 
on neutrons.

The experiment will measure both differential cross sections and self-analyzed 
polarizations of the produced $\Lambda$, $\Sigma$, $\Xi$, and $\Omega$ hyperons 
using the GlueX detector at the Jefferson Lab Hall~D. The measurements will 
span CM $\cos\theta$ from $-0.95$ to 0.95 in the range W = 1490~MeV to 2500~MeV.
The new data will significantly constrain the partial wave analyses and reduce 
model-dependent uncertainties in the extraction of the properties and pole 
positions of the strange hyperon resonances, and establish the orbitally 
excited multiplets in the spectra of the $\Xi$ and $\Omega$ hyperons. Comparison 
with the corresponding multiplets in the spectra of the charm and bottom 
hyperons will provide insight into he accuracy of QCD-based calculations over 
a large range of masses.

The proposed facility will have a defining impact in the strange meson sector 
through measurements of the final state $K\pi$ system up to 2~GeV
invariant mass. This will allow the determination of pole positions and widths 
of all relevant $K^\ast(K\pi)$ $S$-,$P$-,$D$-,$F$-, and $G$-wave resonances, 
settle the question of the existence or nonexistence of scalar meson 
$\kappa/K_0^\ast(700)$ and improve the constrains on their pole parameters. 
Subsequently improving our knowledge of the low-lying scalar nonet in general.


\newpage
\tableofcontents


\newpage
\pagenumbering{arabic}
\setcounter{page}{1}
\section{Preamble}
\label{sec:Exec}

This proposal is an update to the conditionally approved K-Long Facility 
in Jefferson Lab Hall~D proposal [C2--19--001] and in particular intended 
to address several specific questions raised by the PAC47 in 2019.   
Detailed simulations of several hyperon production channels and their 
inclusion into a global Partial Wave Analysis have been studied in order 
to determine the full systematic uncertainties associated with the experiment 
to extract the hyperon pole parameters. One important conclusion of these 
studies is that measurements with a neutral kaon beam are essential in order 
to resolve ambiguities in charged kaon beam measurements of hyperon resonance 
parameters.  

We have also developed a method for measuring the $K\pi$ final system without 
reconstructing the recoil nucleon, allowing access to $|t|$ values for strange 
meson spectroscopy all the way down to threshold, and have extended these 
studies to $K\pi$ production off the $\Delta^{++}$, allowing for the 
separation of the different isospin amplitudes.  Conceptual designs for the 
components of the new $K_L$ beamline are progressing well, and we have done 
detailed simulations of the neutron flux on the GlueX spectrometer, and how 
this could contribute to the reconstruction of $K_L$ and neutron-induced 
reactions.  These studies will be described in detail.

The KLF project proposes to establish a secondary K$_L$ beamline at JLab 
Hall~D for scattering experiments on both proton and neutron targets in 
order to determine the differential cross sections and the self-polarization 
of strange hyperons with the GlueX detector. These data will allow for precise 
PWA for the determination of all resonances up to 2500~MeV in the spectra of 
the $\Lambda$, $\Sigma$, $\Xi$, and $\Omega$ hyperons, the knowledge of which 
is very poor compared to the nucleon. The firm establishment of the lowest 
hyperon multiplets will allow for tests of models of hyperon structure, and 
comparison to future Lattice QCD calculations. Together with the progress 
made in understanding the spectrum of baryons containing charm and beauty 
quarks by experiments such as LHCb and Belle~II, these hyperon measurements 
will provide new insight into the implications of QCD over a wide range of 
mass scales.  In addition, this facility provides a unique environment to 
study strange meson spectroscopy through the $K\pi$ interaction, to locate 
the pole positions in $S$-, $P$-,$D$-,$F$- and $G$-waves, particularly for 
the low-lying $S$-wave strange scalar meson $\kappa/K_0^\ast(700)$. 

The $K_L$ beam will be generated by directing a high energy, high 
intensity photon beam onto a Be-target upstream of the GlueX detector. The 
flux of the $K_L$ beam will be $\sim$1$\times 10^4~K_L$/sec on a liquid 
hydrogen/deuterium cryogenic target within the GlueX detector, which has a 
large acceptance with coverage of both charged and neutral particles. This 
flux will allow statistics in the case of the hydrogen target to exceed that 
of earlier experiments by almost three orders of magnitude. The main 
components of the experimental setup related to the $K_L$ beamline are the 
Compact Photon Source, the Kaon Production Target, the sweeping magnet, and 
the Kaon Flux Monitor.

At the first stage,  the KLF program will focus on two-body and quasi-two-body 
reactions: elastic $K_Lp\to K_Sp$ and charge-exchange $K_Lp\to K^+n$ reactions, 
then on two-body reactions producing $S = -1$ ($S = -2$) hyperons as 
$K_Lp\to\pi^+\Lambda$, $K_Lp\to \pi^+\Sigma^0(\pi^0\Sigma^+)$, and $K_Lp\to 
K^+\Xi^0$, as well as three body $K_Lp\to K^+K^+\Omega^-$.  The differential 
cross section and self-polarization measurements of these reactions will be 
used as PWA inputs to determine the resonance pole parameters of the states 
described above.

The best fit will determine the partial wave amplitudes and the 
resonance pole positions, residues and Breit-Wigner (BW) parameters. These 
will provide a benchmark for results of forthcoming Lattice QCD calculations 
and elucidate the structure of the hyperons.

Based on the technical requirements and minimum construction 
timelines, we expect that experiments with the K$_L$ beam in Hall~D could 
begin in 3 years, consistent with the completion of the currently approved 
GlueX physics program. This time scale would allow the KLF program to begin 
prior to competitive and complementary experiments at upcoming international 
facilities.

\newpage
\section{Scope of the Proposal} 
\label{sec:Scope}

The nature of QCD confinement continues to provide a challenge to our understanding of soft QCD. Experimental investigation of the baryon spectrum provides the obvious avenue to understand QCD in this region, since the location and properties of the excited states depend on the confining interaction and the relevant hadronic degrees of freedom.

Through analyses of decades worth of data, from both hadronic and electromagnetic scattering experiments, numerous baryon resonances have been observed, and many of which have had their masses, widths, and quantum numbers fully determined. There are 109 baryons in the PDG2018 listings, but only 58 of them with $4^\ast$ or $3^\ast$ quality~\cite{Tanabashi:2018oca}. Many more states are predicted by quark models (QMs). For example, in the case of $SU(6)_F\times O(3)$ symmetry, 434 resonances would be required, if all partly revealed multiplets were completed (three 70-plets and four 56-plets).

The light and strange quarks can be arranged in six baryonic families, $N^\ast$, $\Delta^\ast$, $\Lambda^\ast$, $\Sigma^\ast$, $\Xi^\ast$, and $\Omega^\ast$. The possible number of members in a family is not arbitrary~\cite{Nefkens:1998xz}. Under the $SU(3)_F$ symmetry these are the octet: $N^\ast$, $\Lambda^\ast$, and $\Sigma^\ast$, and the decuplet: $\Delta^\ast$, $\Sigma^\ast$, $\Xi^\ast$, and $\Omega^\ast$. The number of experimentally identified resonances in each baryon family in PDG2018 summary tables is 17 $N^\ast$, 24 $\Delta^\ast$, 14 $\Lambda^\ast$, 12 $\Sigma^\ast$, 7 $\Xi^\ast$, and 2 $\Omega^\ast$. Constituent 
QMs, for instance, predict the existence of no fewer than 64 
$N^\ast$ and 22 $\Delta^\ast$ states with masses less than 3~GeV. The ``missing-states" problem~\cite{Koniuk:1979vw} is obvious from these numbers. To complete $SU(3)_F$ multiplets, one needs no fewer than 17 $\Lambda^\ast$s, 41 $\Sigma^\ast$s, 41 $\Xi^\ast$s, and 24 $\Omega^\ast$s.

If these ``missing resonances" exist, they have either eluded 
detection or have produced only weak signals in the existing data sets. The search for those resonances provides a most natural motivation for future measurements at Jefferson Lab.  As stated in the \textit{2015 Long Range Plan for Nuclear Science}~\cite{Geesaman:2015fha}: \textit{The new capabilities of the 12-GeV era facilitate a detailed study of baryons containing two and three strange quarks. Knowledge of the spectrum of these states will further enhance our understanding of the manifestation of QCD in the three-quark arena.}

The JLab 12~GeV energy upgrade, with the new Hall~D, provides an ideal tool for extensive studies of both non-strange and, specifically, strange baryon resonances~\cite{AlGhoul:2017nbp,
AlekSejevs:2013mkl}. Our plan is to take advantage of the existing high-quality photon beamline and the experimental area in the Hall~D complex at Jefferson Lab to deliver a beam of $K_L$ particles onto a 
LH$_2$/LD$_2$ target within the GlueX detector. The recently 
constructed GlueX detector is a large-acceptance spectrometer with good coverage for both charged and neutral particles that can be adapted to this purpose. Obviously, a $K_L$ beam facility with good momentum resolution is crucial for providing the data needed to identify and characterize the properties of hyperon resonances. 

The masses and widths of the lowest $\Lambda$ and $\Sigma$ baryons were determined mainly with kaon beam experiments in the 1970s~\cite{Tanabashi:2018oca}. The first determinations of the pole position in the complex-energy plane for a hyperon, for instance, for the $\Lambda(1520)3/2^-$, have been made only 
recently~\cite{Qiang:2010ve}. An intense $K_L$ beam would open a new window of opportunity, not only to locate ``missing resonances", but also to establish their properties by studying different decay channels systematically.

A recent white paper, dedicated to the physics with meson beams and endorsed by a broad physics community, \underline{summarized} unresolved issues in hadron physics, and outlined the vast opportunities and advances that only become possible with a 
``secondary beam facility"~\cite{Briscoe:2015qia}.  The Hall~D K-long Facility (KLF) measurements will allow studies of very poorly known multiplets of $\Lambda^\ast$, $\Sigma^\ast$, $\Xi^\ast$, and even $\Omega^\ast$ hyperons with unprecedented statistical precision.  These measurements also have the potential to observe dozens of predicted (but heretofore unobserved) states and to establish the quantum numbers of already observed hyperon resonances listed in 
PDG2018~\cite{Tanabashi:2018oca}. Interesting puzzles exist for PDG-listed excited hyperons that do not fit into any of the low-lying excited multiplets, and these need to be further revisited and investigated. Excited $\Xi$s, for instance, are very poorly known. Establishing and discovering new states is important, in particular, for determination of the multiplet structure of excited baryons.

Additionally, the proposed facility will also have great impact in the strange meson sector by measurements of the final-state $K\pi$ system from threshold up to 2~GeV in invariant mass to establish and improve on pole positions and widths of all $K^{\ast}(K\pi)$ $P$-wave states and the $S$-wave scalar meson $\kappa$ or $K_0^\ast(700)$. In particular, the $\kappa/K_0^\ast(700)$ meson has been under discussion for decades and still remains to be unequivocally confirmed with corresponding quantum numbers by detailed phase-shift analysis with high statistics data~\cite{kappa,DescotesGenon:2006uk}. 

We have organized four Workshops: \textit{Physics with Neutral Kaon Beam at JLab} (KL2016) (February 2016)~\cite{KL2016}, \textit{Excited Hyperons in QCD Thermodynamics at Freeze-Out} (YSTAR2016) (November 2016)~\cite{YSTAR2016}, \textit{New Opportunities with High-Intensity Photon Sources} (HIPS2017) (February 2017)~\cite{HIPS2017}, and \textit{Pion-Kaon Interactions} (PKI2018) (February 2018)~\cite{PKI2018}.  They were dedicated to the physics of hyperons produced by the neutral kaon beam.  The KL2016 Workshop~\cite{Albrow:2016ibs} followed our LoI--12--15--001~\cite{LoI} to help address the comments made by PAC43 and to prepare the full proposal for PAC45~\cite{Amaryan:2017ldw}. The proposed KLF program is complementary, for instance, to the CLAS12 baryon spectroscopy experiments~\cite{VSP,KY} and would operate at Hall~D for several years.  The YSTAR2016 Workshop~\cite{Alba:2017cbr} was a successor to the recent KL2016 Workshop and considered the influence of possible ``missing" hyperon resonances on QCD thermodynamics, on freeze-out in heavy ion collisions and in the early universe, and in spectroscopy. Then, the HIPS2017 Workshop~\cite{Horn:2017vkz} aimed at producing an optimized photon source concept with potential increase of scientific output at Jefferson Lab, and at refining the science for hadron physics experiments benefitting from such a high-intensity photon source.  Finally, the PKI2018 Workshop is dedicated to the physics of strange mesons produced by the neutral kaon beam~\cite{pki}. 

The proposal is organized in the following manner.   
The motivations for hyperon and kaon spectroscopy are reviewed 
in Sec.~\ref{sec:highlights}, with summaries of the high-impact measurements possible with the KLF given in Sec.~\ref{sec:hyperon_spectroscopy}  for hyperon spectroscopy, and 
in Sec.~\ref{sec:kaon_spectroscopy} for kaon spectroscopy.  
In Sec.~\ref{sec:simulations}, we give results of detailed simulation  studies of key reaction channels. Full detector simulations, including background studies, of several hyperon production channels are described in Sec.~\ref{sec:hyperon_sims}, along with PWA studies showing the impact of these measurements on the hyperon spectrum. 
In Sec.~\ref{sec:kaon_spectroscopyA}, we describe the current status of our studies of the $K\pi$ system, including reactions with missing protons, and how they will impact our knowledge of the resonance parameters of the $\kappa/K_0^\ast(700)$.  
In Sec.~\ref{sec:neutron_sims}, we give additional details of our studies of neutron-induced reactions.  Our proposed K$_L$ beam facility is described in Sec.~\ref{sec:beam} including: K$_L$ production and properties of the secondary K$_L$ beam, measurements of K$_L$ flux, and 
cryogenic target. Further details of the organizational status and 
planning are given in Sec.~\ref{sec:project_planning} while summary and a 
beam time request is given in Sec.~\ref{sec:summary}. The Appendices contain many technical details of our proposal.

\newpage
\section{Physics Highlights}
\label{sec:highlights}

\subsection{Hyperon Spectroscopy}
\label{sec:hyperon_spectroscopy}
\begin{figure}[ht]
    \centering
{    \includegraphics[width=0.8\textwidth]{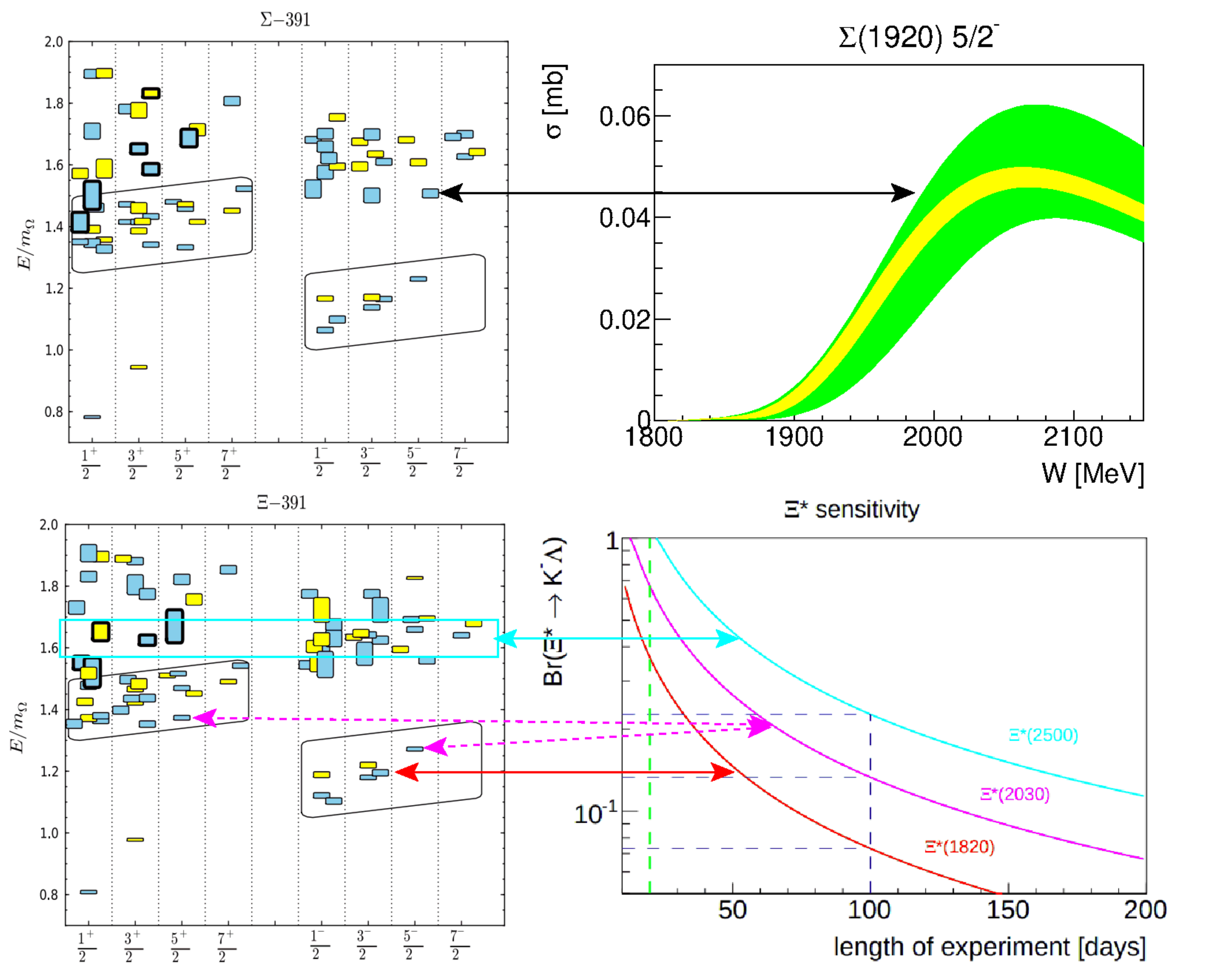} }

\centerline{\parbox{0.80\textwidth}{
 \caption[] {\protect\small 
 Example of comparison between expected KLF measurements (right) and Lattice QCD 
	predictions for the hyperon spectrum~\protect\cite{Edwards:2012fx} (left), 
	see Section~\protect\ref{sec:hyperon_sims}, Appendix~\protect\ref{sec:A3} and 
	Ref.~\protect\cite{HypSim} of the text for details.}
    \label{fig:money} } }
\end{figure}

Lattice QCD and quark model calculations predict a rich spectrum of $\Lambda^\ast$ 
and $\Sigma^\ast$ states, many of which have large widths that can be studied at KLF 
but are not easily accessible in other reactions. As a typical example in 
Fig.~\ref{fig:money} we demonstrate a complete PWA extraction of a fairly low lying 
but already broad $\Sigma^\ast$ resonance in the reaction $K_Lp\rightarrow\Xi^0 K^+$ 
assuming 20 (100) days of running shown in green (yellow), see 
Sec.~\ref{sec:hyperon_sims} for details. A clean discrimination of broad excited 
states on top of many overlapping resonances with various different quantum numbers 
is a key feature of the KLF, unmatched by comparable experiments. The precision of 
KLF data clearly allows for the identification of these excited states in a mass 
range not accessible with previous measurements, and determination of their quantum 
numbers and pole positions, which can be then  compared with calculations from Lattice 
QCD~\protect\cite{Edwards:2012fx}.

An additional power of  KLF data is the ability to distinguish between different isospin 
exchanges, which helps to identify new hyperon resonances and resolves 
ambiguities of fits 
to existing data.  As discussed in Sec.~\ref{sec:sigma_pwa}, we consider model of $\Sigma$ 
hyperons which describes existing $\pi\Lambda/\pi\Sigma$ data well even with the addition 
of three $\Sigma^*$ hyperons based on quark model expectations.  However, the additional 
$\Sigma^*$ lead to drastic differences in the expected differential cross sections and 
polarization observables, up to two orders of magnitude depending on if these expected, 
but unidentified states are included or not.  This example shows the essential nature of 
$K_L$ beam data in the global PWA analysis of hyperons.

The spectrum of $\Xi$ hyperons also clearly has significant discovery potential with 
implications for heavy quark symmetry and relationships to mass splittings 
in the charm and 
beauty hyperons.  The bottom panel of Fig.~\ref{fig:money} demonstrates the branching 
ratio sensitivity for several excited $\Xi^\ast\rightarrow\Lambda K$ as a function of 
running days for the experiment (see Appendix~\ref{sec:A3} and Ref.\cite{HypSim} for 
details). These states are expected to be narrow and can be directly compared to the 
spectrum predicted by Lattice QCD~\protect\cite{Edwards:2012fx}.

\textbf{In summary}, this proposal aims to address many of the key open questions in 
strange hyperon spectroscopy. The detailed studies described in the following sections 
demonstrate the ability to determine the pole positions and decay modes of many resonances 
using rigorous PWA methods, beyond naive ``bump hunting" assignments listed by in the 
Particle Data Group~\cite{Tanabashi:2018oca}. With these ability to perform these PWAs, 
the KLF experiment will settle the spectrum of $\Lambda^\ast, \Sigma^\ast, \Xi^\ast$ and 
even $\Omega^\ast$ spectra in the mass range up to 2500~MeV. More details about the 
theoretical and experimental techniques to be employed in KLF experiments can be found 
in following sections and supplementary materials.
\newpage
\subsection{Kaon Spectroscopy}
\label{sec:kaon_spectroscopy}

One of the most controversial states in light meson spectroscopy is the elusive $\kappa/K^{\ast}(700)$. 
This broad resonance ``still needs confirmation'' according to  the PDG2018~\cite{Tanabashi:2018oca}. 
The reason for the importance of this state is twofold. First of all, the unambiguous determination of 
its existence would complete the lightest scalar meson nonet, together with the already observed 
$a_0(980)$, $f_0(980)$, and $\sigma/f_0(500)$ mesons.  Secondly, the precise determination of the 
resonance parameters of the $\kappa/K^\ast(700)$ is necessary to distinguish between different models 
of its internal structure, which also will provide insight into its lighter cousin, the $\sigma/f_0(500)$.  
In addition to this state, there are several other strange-quark resonances belonging to heavier nonets 
which decay to $K\pi$ with sizeable branching ratios. Unfortunately most of the resonance parameters of 
these states have not been extracted with high accuracy due statistical limitations or uncontrolled 
systematic effects due to the simple models used to describe their lineshapes.
\begin{figure}[ht]
    \centering
{    \includegraphics[width=0.8\textwidth]{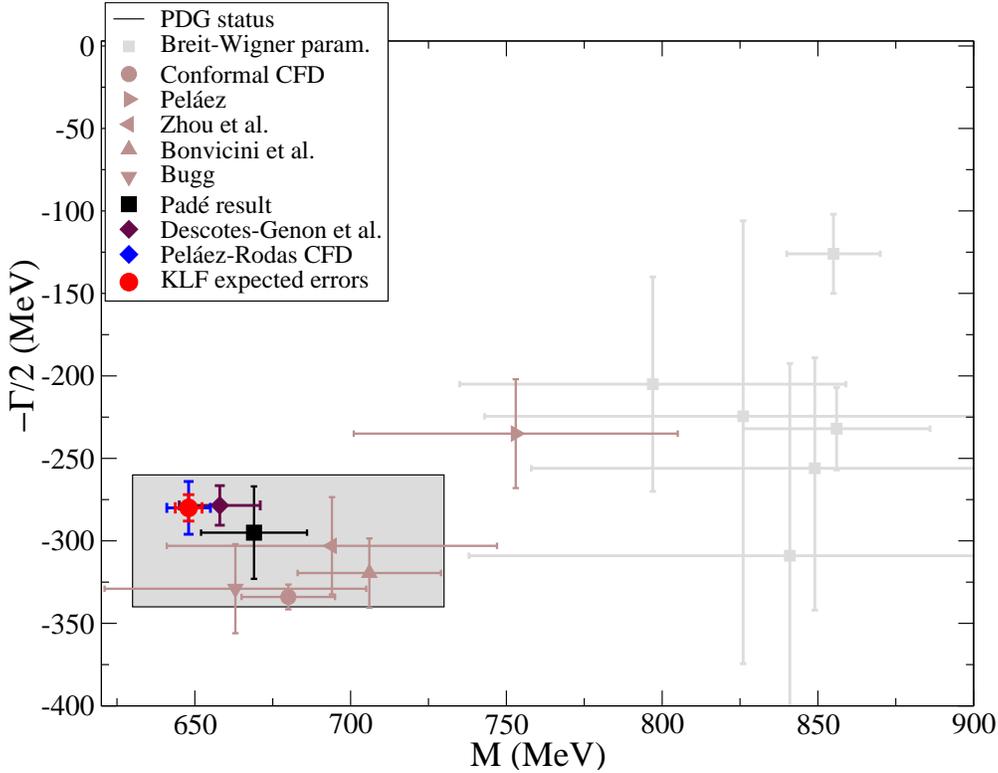} }

 \centerline{\parbox{0.80\textwidth}{
 \caption[] {\protect\small 
 Expected precision on the $\kappa/K_0^\ast(700)$ pole parameters for 100~days of running time. The 
	uncertainties of KLF prediction  are presented in a red color within the blue error bars 
	obtained without KLF data. The shadowed rectangle stands for PDG2018 uncertainties. (see 
	Section~\protect\ref{sec:kaon_spectroscopyA} and Appendix~\protect\ref{sec:A4} for details).  
 }
    \label{fig:money_pik} } }
\end{figure}

The best way to unravel these states and  improve the current knowledge on them is to study the 
$I = 1/2$ partial waves of $K\pi$ scattering, particularly in the elastic region. In order to have 
access to this scattering process we have to perform an analysis of production experiments like 
$KN\to K\pi N$ or $KN\to K\pi\Delta$, where the total transferred momentum to the final state 
baryon $t$ (Mandelstam variable) is small. In this region, the interaction between the kaon and 
nucleon on the initial state can be approximated by a one pion exchange. Provided the systematic 
effects of this approximation are under control, the $K \pi$ scattering 
can be determined with high 
precision from the aforementioned production experiments. In the past, charged kaon beams were 
used for this purpose~\cite{Estabrooks:1977xe,Aston:1987ir,Cho:1970fb,Bakker:1970wg,Linglin:1973ci,
Jongejans:1973pn}.

In order to directly measure $S$-wave phase-shift for isospin $I = 1/2$, as opposed to the  sum of 
$I = 1/2$ plus $I = 3/2$ which was measured at LASS, in this proposal we simulated data in $K_Lp 
\to K^-\pi^0\Delta^{++}$ and $K_L p \to K_L\pi^-\Delta{++}$ with different linear combinations of 
Clebsch-Gordan coefficients for $I = 1/2$ and $I = 3/2$.  Measuring both of these channels allows 
us to isolate the contribution of the isospin $I = 1/2$ amplitude, as discussed in more detail in 
Sec.~\ref{sec:kaon_spectroscopyA}.

Measurements of $K\pi$ scattering in the $S-$wave will allow the $K_L$-Facility to contribute to 
our understanding of the elusive $\kappa$ meson, as described in Section~\ref{sec:kaon_spectroscopyA} 
and Appendix~\ref{sec:A4}.  Figure~\ref{fig:money_pik} shows the expected improvement in the 
determination of the mass and width of the $\kappa$ that can be reached with 100~days of running 
time. A phase-shift analysis of this scattering data will allow for a decisive determination of its 
properties by  using above mentioned reactions with recoiling $\Delta^{++}$ in final state.

\newpage
\section{Simulations of Physics Reactions}
\label{sec:simulations}

Several simulations on various channels were performed to obtain an insight on the 
expected results and the beam time requirements for precision measurements. The 
simulations results that follow are based on a 100~days of a beam time with a $1\times 
10^4$~$K_L$/sec impinged on a 40~cm long target.  Generated events assuming standard 
beam/target conditions listed in Tables~\ref{tab:klbm} and \ref{tab:kltg} are processed 
through a full Geant3-based Monte Carlo (MC) simulation of the GlueX 
detector~\cite{Ghoul:2015ifw}. Below we provide a summary of studies performed on 
various channels for Hyperon spectroscopy (more details can be found in Ref.~\cite{HypSim}). 

\subsection{Hyperon Production Reactions}
\label{sec:hyperon_sims}
\begin{figure}[h!]
\centering
  \includegraphics[width=1\textwidth,keepaspectratio]{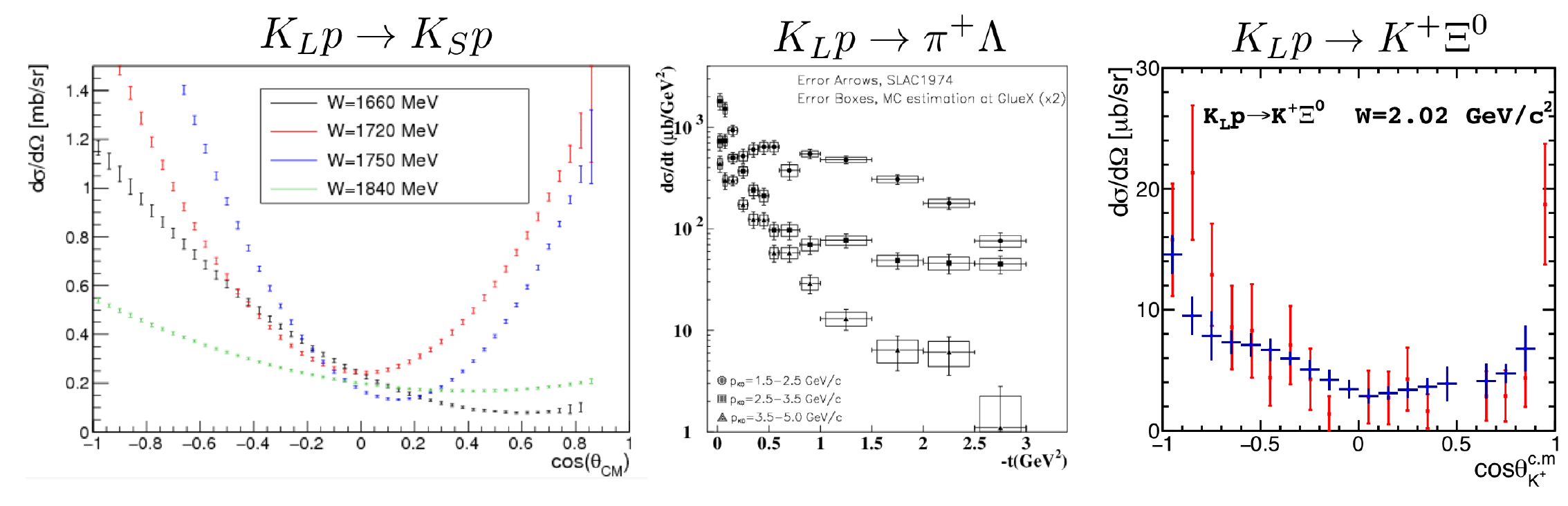} 
  \centerline{\parbox{0.80\textwidth}{
    \caption[] {\protect\small
    Differential cross sections for reactions (1), (2), and (3) exclusively 
	reconstructed illustrating the statistical uncertainties after 
	100~days of running. 
    \underline{Left}: Cross sections of $K_Lp\to K_Sp$
    binned in various $W$ bins as a function of the center of mass angle 
	(one can compare to existing data from 
	Ref.~\protect\cite{Capiluppi:1982fj}.
    \underline{Middle}: Differential cross section for various 
    kaon momentum bins as a function of $t$ for $K_Lp\to\pi^+\Lambda$. The 
	boxed errors indicate the statistical uncertainty increased by a 
	factor of 2 after 100~days of running as compared to existing 
	uncertainties~\protect\cite{Yamartino:1974sm}.  
    \underline{Right}: Cross section of $K_Lp\to K^+\Xi^0$ reconstructed 
	exclusively for a specific $W$ bin as a function of the kaon 
	production angle. The uncertainties after 100~days of running are 
	shown with blue error bars and are compared to the existing results 
	(red error bars)~\protect\cite{Sharov:2011xq}. }  
	\label{fig:HadSpeXsec}
    }} 
\end{figure}
For the case of hyperon spectroscopy, a number of channels that are key
to studying hyperon resonances were simulated and studied in detail.
Specifically, a summary of the simulation results on the following 
two-body reactions:
\begin{eqnarray}
	K_Lp\rightarrow K_Sp , \\
    K_Lp\rightarrow\pi^+\Lambda , \\
    K_Lp\rightarrow K^+\Xi^0 , \\
    K_Ld\rightarrow K^+\Xi^-p_{spectator} , \\
    K_Lp\rightarrow K^+n .
\end{eqnarray}
are presented here (Fig.~\ref{fig:HadSpeXsec}). An analog to the $N\pi$ 
reactions for the $N^\ast$ spectra is the $\pi^+\Lambda$ (and $\pi^+\Sigma^0$, 
not shown here) and can provide crucial information on the excited spectrum of 
hyperons. The study of cascade data allows us to place stringent constraints 
on dynamical coupled-channel models and identify resonances that do not couple 
strongly to the $\pi\Lambda$ but decay preferably to a $K\Xi$ channel; 
analogous to $N^\ast$ resonances that do not couple strongly to $\pi N$ but 
are cleanly seen in $K\Lambda$ and $K\Sigma$ channels. In addition, cascade 
data on proton and neutron targets provide us with long-sought information 
on missing excited $\Xi$ states, which can be easily identified and isolated 
using missing-mass and invariant-mass techniques, and the possibility to 
measure the quantum numbers of the already established $\Xi$(1690) and 
$\Xi$(1820) from a double-moment analysis. Finally, non-resonant 
contributions that could interfere with hyperon production amplitudes and 
distort hyperon signals can also be studied. This is because due to strangeness 
conservation, formation of intermediate resonances is forbidden in the $K_Lp\to 
K^+n$ reaction.  Therefore, a detailed study of this reaction provides a clean 
and controlled way to study and eliminate nonresonant contributions.
\begin{figure}[ht]
\centering
  \includegraphics[width=0.9\textwidth,keepaspectratio]{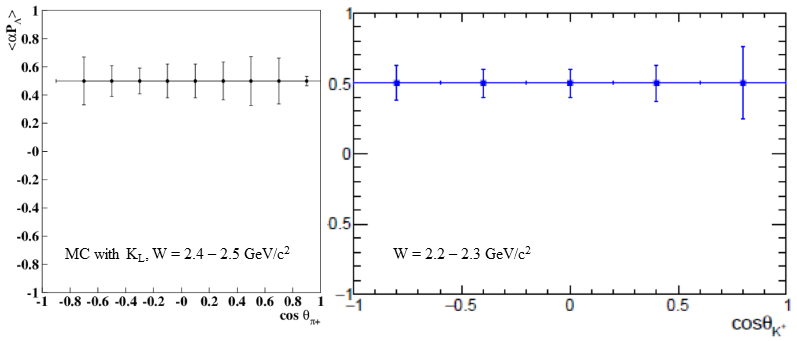}
\centerline{\parbox{0.80\textwidth}{
    \caption[] {\protect\small
    Two-fold differential results on the induced polarization of
    $\Lambda$ (left) and $\Xi^0$ (right) from $K_Lp\to\pi^+\Lambda$ and 
	$K_Lp\to K^+\Xi^0$ after 100~days of running. }
	\label{fig:LambdaXiPl} } }
\end{figure}

The statistical uncertainties obtained after 100~days of running for the 
differential cross sections of reactions $K_Lp\to K_S p$, $K_Lp\to\pi^+
\Lambda$, and $K_Lp\to K^+\Xi^0$ are shown in Fig.~\ref{fig:LambdaXiPl}. 
The proposed experiment will allow us to reconstruct about 2.7M $K_Sp$ 
events in the $K_S\to\pi^+\pi^-$ channel, allowing an unprecedented and 
detailed investigation of the cross section ambiguities evident in existing 
results.  Furthermore, the proposed experiment will significantly improve 
measurements of $K_Lp\to\pi^+\Lambda$ and $K_Lp\to K^+\Xi^0$ as illustrated 
by the uncertainties in Fig.~\ref{fig:LambdaXiPl}. 
\begin{figure}[ht]

\centering
  \includegraphics[width=0.44\textwidth,keepaspectratio]{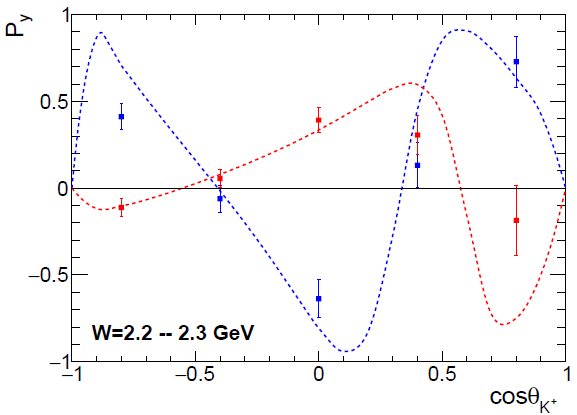}
  \includegraphics[width=0.54\textwidth,keepaspectratio]{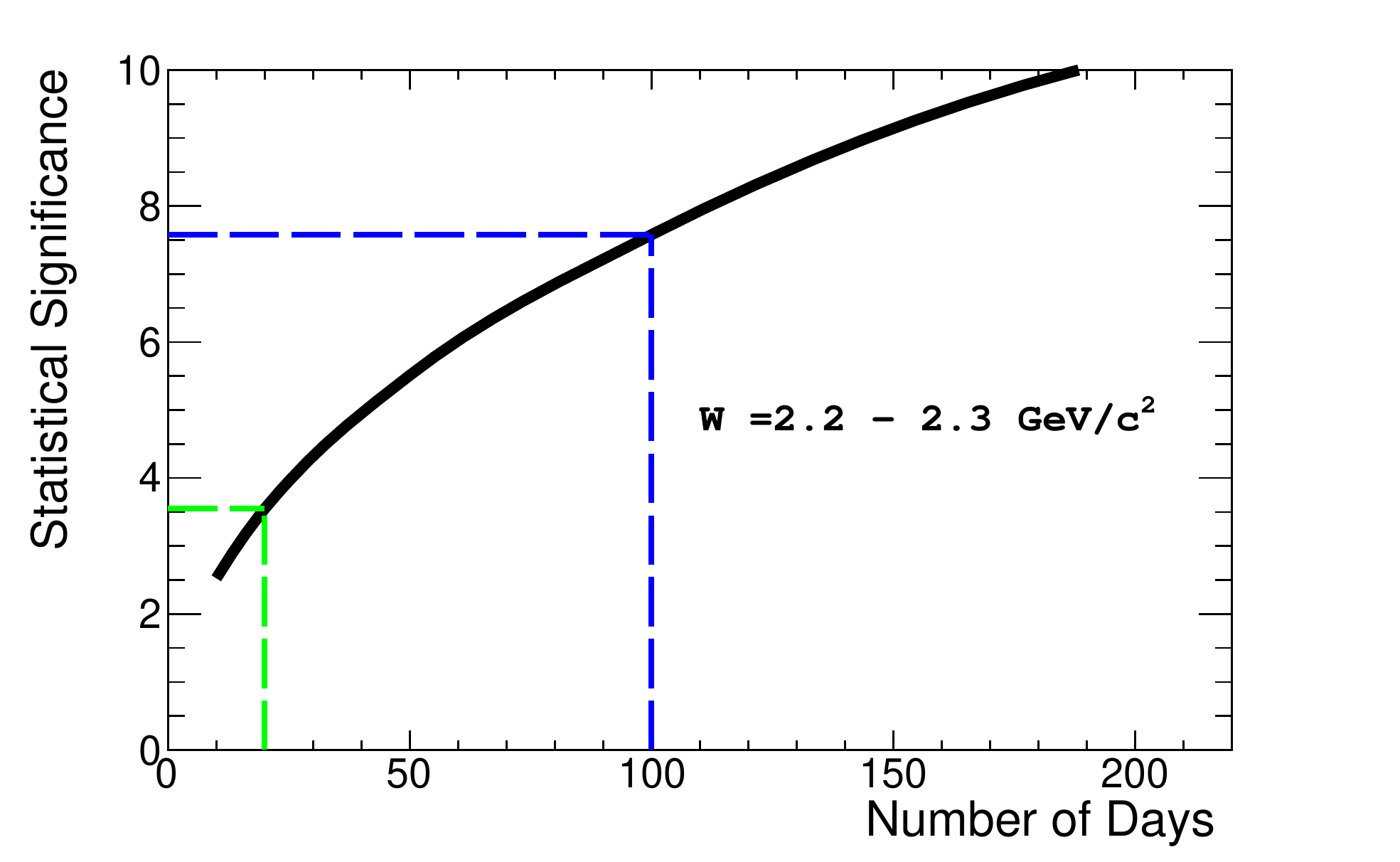} 
  \centerline{\parbox{0.80\textwidth}{
    \caption[] {\protect\small
    \underline{Left}: Estimated statistical uncertainties of the induced 
	polarization of the $\Xi^-$ in a $K_Ln\rightarrow K^+\Xi^-$ reaction 
	as a function of CM\ $\cos\theta_{K^+}$ (two-fold differential). The 
	curves show the theoretical predictions based on two solutions as 
	described in Appendix~\protect\ref{sec:Maxim}. 
    \underline{Right}: Expected statistical significance, in units of 
	$\sigma$s, to distinguish two models as a function of the running 
	time. Two benchmark cases of 20 and 100~days are highlighted by the 
	dashed green and blue curves, respectively.}   
	\label{fig:KlKpXiminusPY} } }
\end{figure}
\begin{figure}[ht]
\centering
   \includegraphics[width=0.48\textwidth,keepaspectratio]{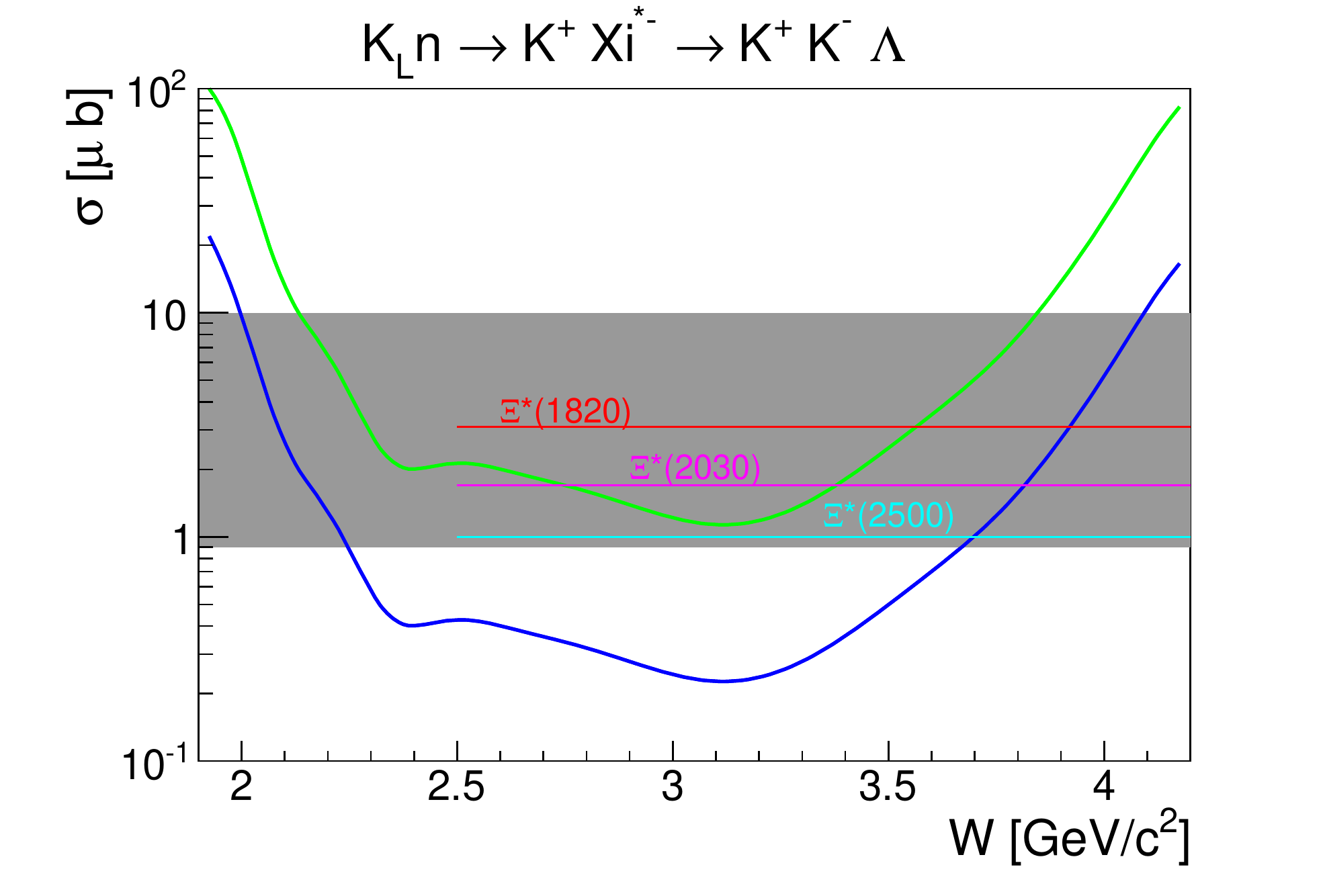}
   \includegraphics[width=0.48\textwidth,keepaspectratio]{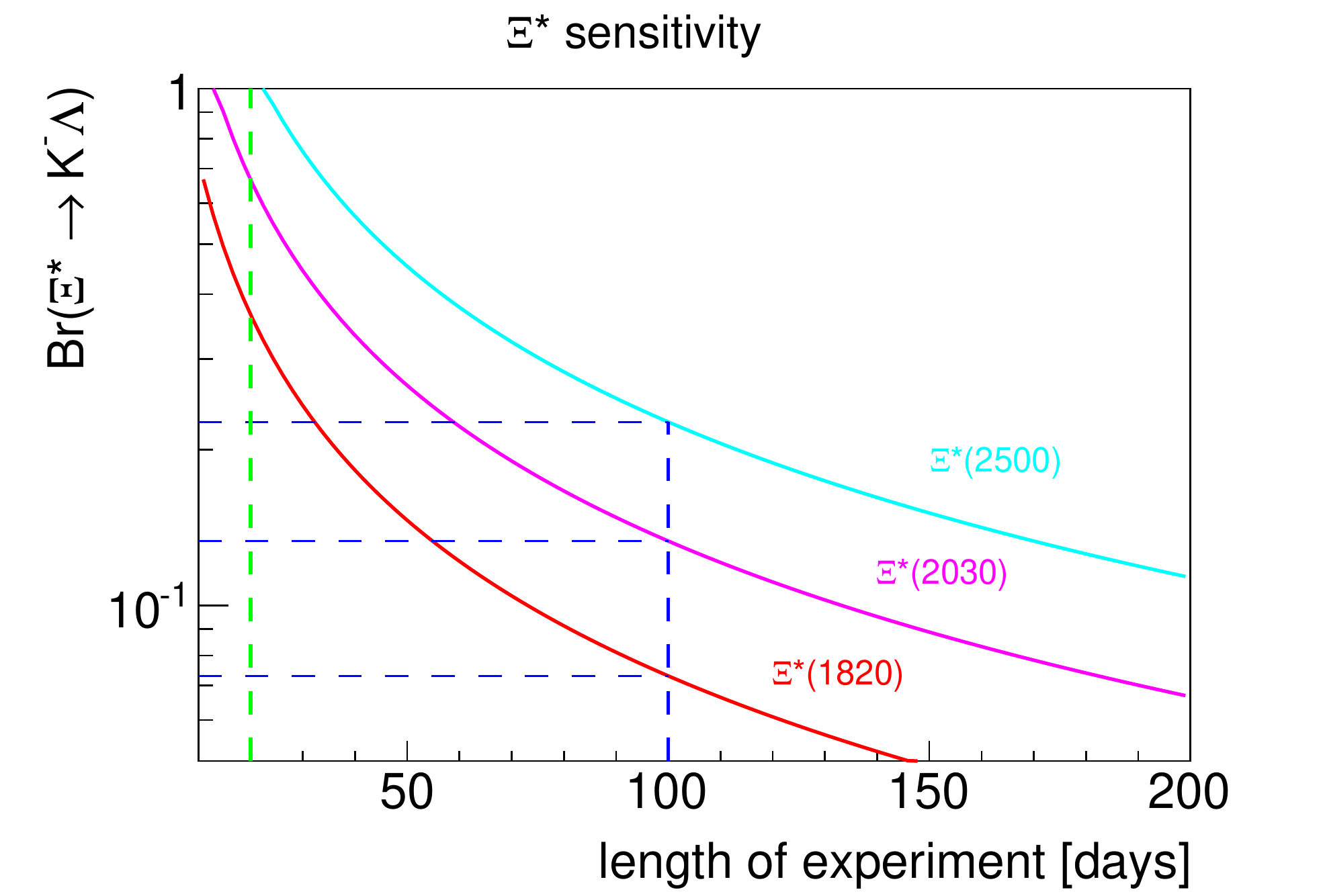} 
   \centerline{\parbox{0.80\textwidth}{
    \caption[] {\protect\small
   \underline{Left}: $\Xi^\ast$ discovery potential achievable at KLF during the 
	100 (blue) and 20 (green) day experiment, under assumption of 10~\% statistical  
	accuracy and $Br(\Xi^\ast\to\bar K\Lambda) = 1$. The gray band corresponds to 
	typical $\Xi^\ast$ cross sections and horizontal lines are few examples of BNL 
	cross sections from Ref.~\protect\cite{Jenkins:1983pm}.
   \underline{Right}: Estimation of lowest measurable 
   $\Xi^\ast\to\bar K\Lambda$ branching fraction at KLF as a function of experiment 
	duration at W$\sim$3.1$\pm$0.025~GeV. Two benchmark cases of 100 (20)~days 
	are highlighted by dashed blue (green) curves.\label{fig:Xi_star1} } } } 
\end{figure}

Utilizing the large self-analyticity of hyperons through their parity violating weak 
decay, one can easily determine the hyperon induced polarization by studying the 
angular distribution of the hyperon decay products. Such measurements place further 
stringent constraints on the underlying dynamics and are shown to be an invaluable 
tool in identifying PWA amplitudes. Figure~\ref{fig:LambdaXiPl} shows the two-fold 
differential induced polarisation and the statistical uncertainties obtained after 
100~days of running for $K_Lp\to\pi^+\Lambda$ (left) and $K_Lp\to K^+\Xi^0$ (right). 
The proposed experiment will increase significantly the kinematic coverage of available 
results and provide a statistical improvement by at least a factor of $2$ on existing 
measurements.

The so-called coupled-channel Chiral Unitary approaches (UChPT) implement unitarity 
exactly via a re-summation of a chiral potential to a certain chiral order. They 
successfully describe all available anti-kaon-nucleon scattering data. In the most 
advanced formulation, such a UChPT approach relies on a chiral amplitude for meson-baryon 
scattering up to next-to-leading chiral order. The unitarity constraint is imposed via 
the Bethe-Salpeter equation either in the full off-shell formulation~\cite{Mai:2012dt,
Bruns:2010sv} or in the so-called on-shell approximation~\cite{Mai:2014xna,Ikeda:2012au}. 
It was found there that various models, which typically have many free parameters, 
adjusted to the same experimental data, predict very different behavior of the 
scattering amplitude on and off the real energy-axis. This ambiguity can be traced 
back to the fact that the experimental data used to fix the parameters of the models 
is rather old and imprecise. In particular, there are missing good quality data 
filtering the isoscalar and isovector parts of the antikaon-nucleon amplitude. In 
this respect, the proposed measurement of the $K^{0}_{L}p$ reactions to the $\pi 
\Lambda$, $\pi \Sigma$ and $K \Xi$ states will provide valuable constraints on the 
isovector sector of the $\bar{K}N$ interactions.

The $K_L$ beam can be scattered on a neutron target, while measuring the strangeness 
$S = -1$ final meson-baryon states. In such a setup, the proposed experiment will 
become a new and very strongly desired source of experimental data to pinpoint the 
properties of the antikaon-nucleon scattering amplitude. To make this statement more 
quantitative, we compare predictions of both solutions of the model from 
Ref.~\cite{Mai:2014xna}. These solutions agree with all scattering, threshold as well 
as the photoproduction data for the $\Sigma\pi$ line shapes by the CLAS 
Collaboration~\cite{Moriya:2013eb}.

An analysis of generated polarization data from the reaction $K_Ln\to K^+\Xi^-$ using 
models described in Appendix~\ref{sec:Maxim} indicate than with 100~days of beamtime 
the expected statistical uncertainties allow a clear identification between the two 
available models, which give very different predictions. The left panel in 
Fig.~\ref{fig:KlKpXiminusPY} illustrates the expected two-fold differential results on 
the induced cascade polarization ($\Xi^-$) after 100~days of running using generated 
data from the two model predictions.  The expected statistical significance for the 
model separation at the same W-bin as a function of experiment duration is shown in the 
right panel. The right panel indicates that a 100~days experiment would reach a decisive 
level of 7.6~$\sigma$ separation power, compared to only a 3.5~$\sigma$ separation after 
20~days.

The spectrum of excited cascades is barely known and practically nothing is known about 
their quantum numbers. Detailed studies utilizing generated data were performed to 
investigate the discovery potential achievable at KLF including measurements on the 
neutron. The left panel of Fig.~\ref{fig:Xi_star1} shows the production cross sections 
that will be measurable at KLF with 100~days (blue) and 20~days (green) of running 
considering a 10~\% statistical uncertainty and a branching ratio~\footnote{From BNL 
measurements given in Ref.~\cite{Jenkins:1983pm} the $\Xi^{\ast -}$ production cross 
section should be on the order of $1-10~\mu b$ and the higher $\Xi^\ast$ mass the lower 
the cross section, from $3.7\mu b$ for the $\Xi^\ast(1820)$ to $1\mu b$ for the 
$\Xi^\ast(2500)$.} $Br(\Xi^\ast\to \bar K\Lambda) = 1$. The right panel shows the 
lowest measurable $\Xi^\ast\to\bar K\Lambda$ branching fraction at KLF as a function 
of experiment duration indicating the lowest measurable branching ratios at 20 and 
100~days. 100~days of running allows us to study the several $\Xi^\ast$ states even 
with somewhat suppressed $\Xi^\ast\rightarrow \bar K \Lambda$ decay of heavy 
$\Xi^\ast$'s~\footnote{The $\Xi^\ast\to\bar K\Lambda$ is ``dominant" for many $\Xi^\ast$ 
states according to PDG2018~\cite{Tanabashi:2018oca}}.  A $W$-variation of the $\Xi^\ast$ 
production cross-section provide and important information on $\Xi^\ast\to\bar 
K\Lambda^\ast$ and $\Xi^\ast\to\bar K \Sigma^\ast$ couplings as an inverse process 
allowing   further insight into $\Xi^\ast$ internal structure.

Finally, the nonresonant reaction $K_Lp\rightarrow K^+n$, can be studied in a clean 
and controlled way and one can use this to identify nonresonant contributions to the 
hyperon production amplitudes. In 100~days of a beamtime, we expect to detect $\sim60$M 
events significantly improving the statistical significance of existing 
measurements~\cite{Cline:1970db,Baillon:1977ja} and provide precision measurements in  
the energy range 2 < $W$ < 3.5~GeV where there are no data on this reaction at all. 

\subsubsection{Expectations for $\Lambda^{\ast}$ and $\Sigma^{\ast}$ Spectroscopy via 
$K^+ \Xi$ PWA:}
The observation of $\Lambda^{\ast}$ and $\Sigma^{\ast}$ hyperons at KLF will require 
a coupled-channel PWA using the measured differential cross sections and recoil 
observables, which have been simulated as described in Appendix~\ref{sec:A3}.  
The resonance poles in the complex energy plane will be used to confirm previously 
observed states and identify new $\Lambda^\ast$ and $\Sigma^\ast$ resonances in the 
hyperon spectrum.

The existing $K_Lp$ database is so poor that PWAs of individual $K_Lp$-induced 
reactions may not be possible based on currently available data (Appendix~\ref{sec:A3}). 
In particular, there are no $K_Lp\rightarrow K^+\Xi^0$ polarization data available and 
there is only one energy for the $K_Lp\rightarrow\pi^+\Lambda$ reaction with both 
$d\sigma/d\Omega$ and polarization data.  Our proposal does not consider the use of a 
polarized target at this stage and, for that reason, we will be able to measure 
polarization data for recoil observables only.  Overall, one certainly cannot perform 
a reliable PWA for reactions in which only $d\sigma/d\Omega$ data are available.  The 
existing $K_Ln$ database is nonexistent.
\begin{table}[ht]

\centering \protect\caption{Comparison of masses and widths (in MeV
	units) of simulated values with the results of 
	PDG2018~\protect\cite{Tanabashi:2018oca} and 
	LQCD~\protect\cite{Edwards:2012fx} predictions.}
\vspace{2mm}
{%
\resizebox{16cm}{!}{
\begin{tabular}{|c|c|c|c|c|}
\hline
Resonance           & 20~days: M, $\Gamma$                 & 100~days: M, $\Gamma$                & PDG2018: M, $\Gamma$   & LQCD: M \\
\hline
$\Sigma(1920)5/2^-$ & 1977$\pm$21$\pm$25 327$\pm$25$\pm$25 & 1923$\pm$10$\pm$10 321$\pm$10$\pm$10 &                        & 2027 \\
                    &                                      &                                      &                        & 2487 \\
                    &                                      &                                      &                        & 2659 \\
                    &                                      &                                      &                        & 2781 \\ 
\hline
$\Sigma(2030)7/2^+$ & 1981$\pm$30$\pm$30 350$\pm$80        & 1930$\pm$20$\pm$30 400$\pm$40        & 2030$\pm$10 180$\pm$30 & 2686 \\
                    &                                      &                                      &                        & 2709 \\
                    &                                      &                                      &                        & 2793 \\
                    &                                      &                                      &                        & 2806 \\
\hline
\end{tabular}}} \label{tab:res}
\end{table}

To estimate the impact that new $K_L$ measurements will have on fits, we have carried 
out a study $K_Lp\to K^+\Xi$ reaction. Using the recent BnGa 
solution~\cite{Sarantsev:2017edu,Sarantsev:2019jau}, we generated pseudo-data for the 
$K_Lp\rightarrow K^+\Xi$ reaction, which were subsequently passed through the 
Geant-simulated detector setup and analysed the same way as usual data. The 
$d\sigma/d\Omega$ and recoil polarization $P$ with associated statistical errors were 
extracted for the PWA within BnGa framework. The pseudo-data were generated for our 
worse case of statistics for a $K_Lp\rightarrow K^+\Xi^0$ binning of 20~MeV in CM\ 
energy $W$ and $\theta$ = 5 (10) 175$^\circ$ for $d\sigma/d\Omega$ and $\cos\theta$ = 
$-0.8$ (0.4) 0.8 for $P$.  A series of global fits were obtained using settings 
associated with 20 and 100~days of running time within BnGa framework.

The simulated solution was obtained from the fit of the $K^-p\to K\Xi$ data using the 
K-matrix fit which included all $3$ and $4$ star $\Lambda$ and 
$\Sigma$-hyperons~\cite{Tanabashi:2018oca} and an additional $\Sigma$ state needed to 
fit the data. The quantum numbers of this state were found to be $5/2^-$ and its mass 
is located just above 1900~MeV. This fit also showed a notable deviation for the pole 
position of the $\Sigma(2030)7/2+$ state from the PDG2018 value (Table~\ref{tab:res}). 
The data from the $K_Lp$ collision where only $\Sigma$ hyperons are produced can provide 
a crucial information which can confirm or reject this solution.

The data simulated assuming 20 and 100~days of the data taken were fitted as a sum of 
the Breit-Wigner (BW) states. Although 20 day data set allowed us to reproduce the 
quantum numbers of the all contributions from the simulated solution the properties 
of the $\Sigma(1920)5/2^-$ and $\Sigma(2030)7/2^+$ hyperons were defined with large 
uncertainties. The mass of the $\Sigma(1920)5/2^-$ state was found to be notably higher 
than that used in the simulation (Table~\ref{tab:res}). The fit of the 100-day data sets 
solve all the above-mentioned problems. The masses and the widths of the simulated states 
were fond to be much closer to those simulated in the solution and with much smaller 
errors. The fit of the 100~day data sets allowed us to define with much larger accuracy 
the couplings of the fitted states into $K\Xi$ channel.

\begin{figure}[ht]
\centering
{
    \includegraphics[width=0.48\textwidth,keepaspectratio]{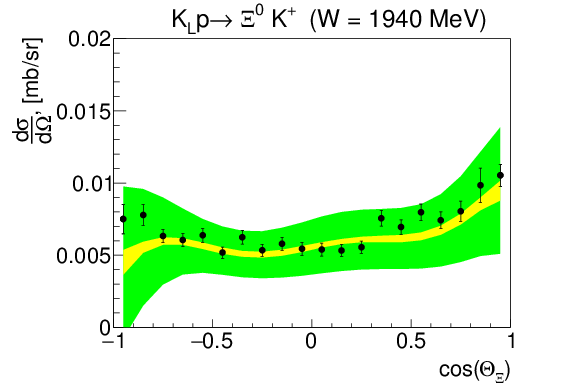} }
{
    \includegraphics[width=0.48\textwidth,keepaspectratio]{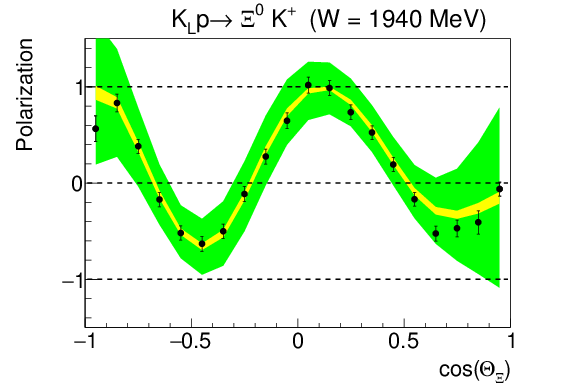} }

    \centerline{\parbox{0.80\textwidth}{
    \caption[] {\protect\small
    Two examples (W = 1940~MeV) showing the impact of the proposed data on the BnGa 
	solutions. The green (yellow) hatched band indicates the present uncertainties 
	for 20 (100)~days of running time. The solid black points corresponds to the 
	quasi-data expected for 100~days of experiment. $d\sigma/d\Omega$ on the left 
	and $P$ on the right panels.  \label{fig:BoGa_data} } } }

\end{figure}

\begin{figure}[ht]
\centering
{
   
\includegraphics[width=0.48\textwidth,keepaspectratio]{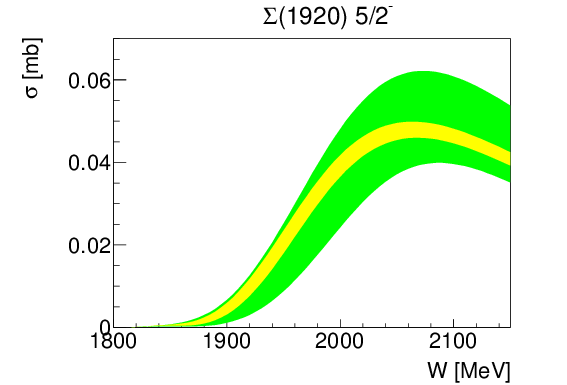} 
\includegraphics[width=0.48\textwidth,keepaspectratio]{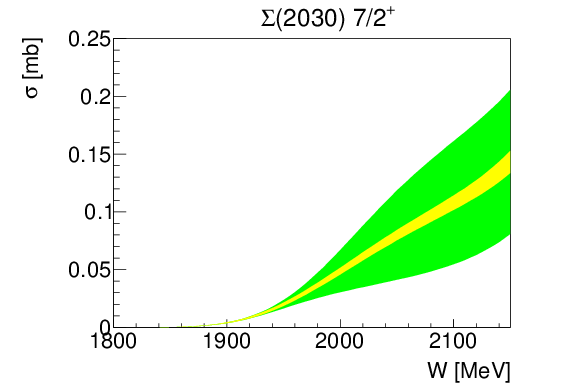} 
}

    \centerline{\parbox{0.80\textwidth}{
    \caption[] {\protect\small
    Resulting effect of 20 (green) and 100 (yellow) days of running time on two 
	$\Sigma^\ast$ resonances within BnGa PWA solution. $\Sigma^\ast(1920)5/2^-$ 
	(left) and $\Sigma^\ast(2030)7/2^+$ (right).} \label{fig:BoGa_fit} } }
\end{figure}

An example of the fit results for $d\sigma/d\Omega$ and $P$ for 20 and 100~days settings 
together with 100~days pseudo-data are shown on Fig.~\ref{fig:BoGa_data}. To illustrate 
the effect for 20 (100)~days data taking on resonance parameters, we plot the total 
cross section for two prominent $\Sigma^\ast$ states: $\Sigma(1920)5/2^-$ and 
$\Sigma(2030)7/2^+$, see Fig.~\ref{fig:BoGa_fit}.

The new measurements, specifically the 100~days of running time case will significantly 
reduce the uncertainties of the observables. The total angular resolution will therefore 
be greatly improved, which will enhance the possibility of determining the number of 
amplitudes that are involved. With this greater understanding of these observables, 
effects of higher-spin resonances can be investigated.

From the PWA of $K_Lp\rightarrow K^+\Xi$, we extracted Breit-Wigner (BW) resonance 
parameters for two example states in the $\Sigma^\ast$ spectrum, $\Sigma(1920)5/2^-$ 
and $\Sigma(2030)7/2^+$.  The mass and width parameters are shown in Table~\ref{tab:res} 
under the 20- and 100-day scenarios.  The precision of the resonance parameters 
significantly deteriorates in a 20-day scenario. With 100~days of running time, we could 
reach a precision level comparable to modern results of the SAID $\pi$N 
PWA~\cite{Alekseev:2014pzu}.

\textbf{To summarize}: With 100~days of running time, we can provide a reliable solution 
for all the resonances having elastic branching ratios larger than 4~\%, at least up to 
$l=4$.  With 20 days of beamtime, we could only carry out simple ``bump-hunting" - an 
identification of well-defined and well-separated resonances with regular shapes. All 
irregular cases (e.g., molecular states with skewed shapes and complex 
energy-dependent-widths, threshold-effects, multiple interferences, etc.) and all the 
exotic states that are predicted to populate the hyperon spectrum will require 
high-precision polarization observables on the order of 0.1 or better to be identified. 
From our $K^+\Xi$ PWA study, we can infer that the precision of resonance parameters 
extracted from PWA of KLF data for the higher-mass $\Lambda^\ast$ and $\Sigma^\ast$ 
states we propose to measure will deteriorate without sufficient running time.  The 
spectrum of excited $\Lambda^\ast$ and $\Sigma^\ast$ states is expected to be densely 
populated with typical mass differences of about 100~MeV for states with the same quantum 
numbers~\cite{Tanabashi:2018oca}.  Therefore to disentangle the spectrum of observed 
hyperon states, we require sufficient precision for the extracted mass and width 
parameters, provided by the proposed 100~days of running time.  With this 100~days of 
running time we could reach a precision level in the hyperon spectrum comparable to 
modern results of the SAID $\pi$N PWA~\cite{Alekseev:2014pzu}.

\subsubsection{Expectations for $\Sigma^\ast$ Spectroscopy via a $\pi^+\Lambda$ and 
	$\pi\Sigma$ PWA:} 
\label{sec:sigma_pwa}
Another effect of the measurement of $K_L$-induced reactions on our knowledge of the 
hyperon spectrum was not fully understood until recently. $\Sigma$-baryons are both 
members of octet and decuplet of baryons, hence for each discovered $N^\ast$ or 
$\Delta^\ast$ resonance we should find a complimentary $\Sigma$ baryon. Having 17 
$N^\ast$ states and 10 $\Delta$ states we should expect no less than 27 $\Sigma$'s 
instead of the 12 currently known~\cite{Tanabashi:2018oca}.

It was long believed that large data sets accumulated with charged kaon beams in the 
simplest $\pi\Lambda/\pi\Sigma$ production reactions would be sufficient to identify 
all missing hyperon states, but these reactions themselves were found to not provide 
the sensitivity necessary to extract the missing states. However, the effect of 
neutral kaon beams was overlooked in these evaluations.
\begin{figure}[ht]
\vspace{-10mm}
\centering
{
   \includegraphics[width=0.95\textwidth,keepaspectratio]{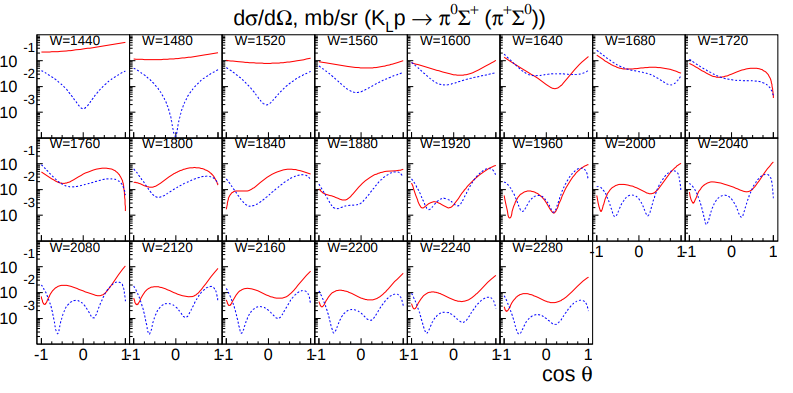} }

    \centerline{\parbox{0.80\textwidth}{
    \caption[] {\protect\small
    Resulting effect of $K_Lp$ data on three new $\Sigma^\ast$ resonances within BnGa 
	PWA solution on differential observables in $K_Lp\to \pi\Sigma$ 
	reactions~\protect\cite{Sarantsev:2020}, note the semi-log scale.}
    \label{fig:BoGa_Sigmafit} } }
\end{figure}
\begin{figure}[htb!]
\centering
{
   \includegraphics[width=0.7\textwidth,keepaspectratio]{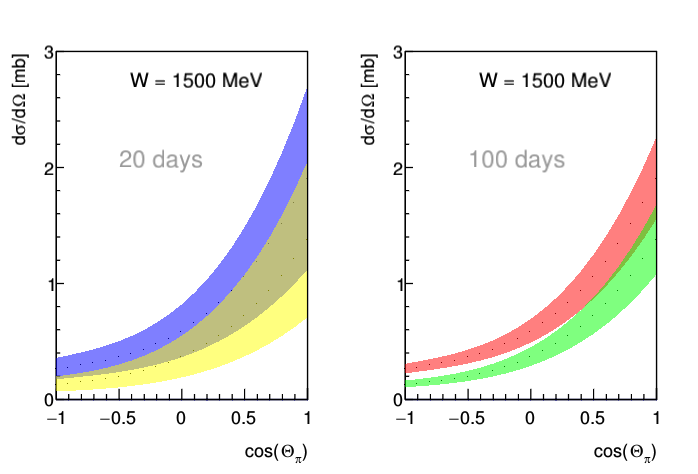} }

    \centerline{\parbox{0.80\textwidth}{
    \caption[] {\protect\small
    Resulting effect of $K_Lp$ data on three new $\Sigma^\ast$ resonances within BnGa 
	PWA solution on the differential cross section for the $K_Lp\to\pi\Lambda$ 
	reaction~\protect\cite{Sarantsev:2020} for the 20~days (left) and 100~days 
	(right) beamtime.}
    \label{fig:BoGa_Lambdafit} } }
\end{figure}

Let's consider the possible $\Sigma$-production channels with various kaon beams and 
nucleon target. The dynamics for the $K^-p\to\Sigma\pi$ reaction can be described with 
two isospin amplitudes $A_1$ and $A_0$, see Eq.~(\ref{eq:SigmaAmps}).
\begin{eqnarray}
|A(\Sigma^+\pi^-)|^2 & = & |\frac{1}{\sqrt{2}}A_1 + \frac{1}{\sqrt{3}}A_0|^2 = \frac{1}{6}(3|A_1|^2 + 2|A_0|^2 + 2\sqrt{6}Re(A_1^\ast \cdot A_0)) \nonumber \\
|A(\Sigma^-\pi^+)|^2 & = & |-\frac{1}{\sqrt{2}}A_1 + \frac{1}{\sqrt{3}}A_0|^2 = \frac{1}{6}(3|A_1|^2 + 2|A_0|^2 - 2\sqrt{6}Re(A_1^\ast \cdot A_0)) \nonumber \\
|A(\Sigma^+\pi^0)|^2 & = & |A(\Sigma^0\pi^+)|^2 = \frac{1}{2}|A_1|^2 .
\label{eq:SigmaAmps}
\end{eqnarray}

The most abundant $K^-$ reactions on proton contain both $A_1$ ($\Sigma^*$- channel), 
$A_0$ ($\Lambda^\ast$-channel) and their interference. The $K^0$ reactions on neutron 
contain the same amplitudes but with the different sign in the interference term, 
which is extremely helpful for separation of interference effects. Finally, the $K^0$ 
reactions on proton contain only the $A_1$ amplitude, with no ambiguities, 
Eq.~(\ref{eq:SigmaAmps}).

To quantify the effect of including $K^0$-induced $\pi\Sigma$ data on the  possible 
$\Sigma^\ast$ resonances we performed a dedicated study, where we fit existing 
$\pi\Lambda/\pi\Sigma$ data under two assumptions. In the first case we assumed 
known $\Sigma$ resonances only, in the second one we have added three extra $\Sigma$ 
states expected from baryon multiplet correspondence. Both models reproduce existing 
data very well. The marginal improvement of the fit results when including the additional 
$\Sigma$ states can only be seen from the $\chi^2/dof$ and can hardly be differentiated 
from experimental plots. However the predictions we get from these two fits differs 
drastically in $K^0$-induced reaction, as is seen in Fig.~\ref{fig:BoGa_Sigmafit}.

The predictions of these two fits differs by nearly two orders of magnitude even in 
the simplest differential cross sections distributions. Both models also show very 
different behaviour in polarization observables.

A similar, but somewhat smaller effect is expected in $\Lambda \pi$ channel, see 
Fig.~\ref{fig:BoGa_Lambdafit} (same models as in Fig.~\ref{fig:BoGa_Sigmafit} or 
Ref.~\protect\cite{Sarantsev:2020}). The difference in sensitivity is expected to 
originate from the difference in coupling constants of $\Sigma^\ast\to \Sigma\pi$ 
or $\Sigma^\ast\to \Lambda\pi$ decay branches.

Our study shows that the missing $\Sigma$ states not seen in the simplest possible 
$\pi\Lambda/\pi\Sigma$ reactions originate not from a low sensitivity of theses 
reactions to the various $\Sigma^\ast$-excited states, but rather it shows the 
inability of theoretical extraction of these states in the absence of $K^0$ data.
\subsection{$K\pi$ Spectroscopy}
\label{sec:kaon_spectroscopyA}
The simplest hadronic reaction that involves strange quarks is $K\pi$ scattering,  therefore 
its experimental study plays a crucial role  for our understanding of QCD in the non-perturbative 
domain. Theoretically  the $K\pi$  scattering amplitude can be calculated based on Chiral 
Perturbation Theory  at one  loop~\cite{Bernard:1990kw,Bernard:1990kx} and at two 
loops~\cite{Bijnens:2004bu}. There are also LQCD calculations of $K\pi$ scattering from the first 
principles treatment of QCD~\cite{Miao:2004gy,Beane:2006gj,Nagata:2008wk,Fu:2011wc,Lang:2012sv,
Sasaki:2013vxa,Wilson:2014cna,Helmes:2018nug,Wilson:2019wfr,Guruswamy:2020uif}.

The $K\pi$ scattering has two possible isospin channels, $I = 1/2$ and $I = 3/2$. For $S$-wave 
scattering, both are significant below 2~GeV, whereas for $P$-wave scattering, $I = 3/2$ is 
almost negligible. Below 1~GeV, the $P$-wave is basically a narrow elastic wave peaking at 
892~MeV, interpreted as the $K^\ast(892)$ resonance. A second $P$-wave resonance, the 
$K^\ast_1(1410)$, exists above 1~GeV, although its properties are less precisely known.  The 
$I = 3/2$ $S$-wave is elastic and repulsive up to 1.7~GeV and contains no known resonances. The 
$P$-wave $I = 3/2$ has been measured in Ref.~\cite{Estabrooks:1977xe} and is also repulsive but 
very small. However, the $I = 1/2$ $S$-wave has a peaking broad resonance above 1350~MeV, 
interpreted as $K^\ast_{0}(1430)$. 

Most of the experimental studies of the $K\pi$ system have been performed by experiments 
scattering a kaon beam off of nucleons, where the single pion exchange process at low $|t|\leq 
0.2$~GeV$^2$ provides access to the $K\pi$ scattering dynamics.  At KLF there are a total of 9 
possible production processes (listed in Eq.~\eqref{eq:prodreactions}) to study the $K\pi$ system 
that could be accessed through a $K_L$ beam, proton target and recoiling proton, neutron or $\Delta^{++}$.
\begin{align}
        K_L p&\to K_L\pi^0 p, \nonumber \\
        K_L p&\to K^{\pm}\pi^{\mp} p, \nonumber \\
        K_L p&\to K_{(L,S)}\pi^+ n, \nonumber \\
        K_L p&\to K^+\pi^0 n, \nonumber \\
        K_L p&\to K^-\pi^0 \Delta^{++}, \nonumber \\
        K_L p&\to K_{(L,S)}\pi^-\Delta^{++}. 
        \label{eq:prodreactions}
\end{align}

These reactions are proportional to different combinations of the $I = 1/2$ and $I = 3/2$ $K\pi$ 
scattering amplitudes, as explained in Appendix~\ref{sec:A4}. Isospin separation requires studying 
multiple reactions with linearly independent isospin combinations, which in this case requires 
measuring a reaction with and without a $K_L$ in the final state.  We have selected three 
complementary reactions to simulate and demonstrate the precision of the proposed measurements 
which are summarized in this section and described in detail in Ref.~\cite{Amaryan:2020zz}.  

The first reaction, $K_L p\to K^{\pm}\pi^{\mp} p$, provides access to the $K\pi$ scattering 
amplitudes via neutral pion exchange.  Simulations of this reaction have demonstrated that it can 
be measured by a) not detecting the proton, providing acceptance down to $t_{min}$ or b) detecting 
the complete final state, providing excellent resolution $W$ and $|t|$ resolution but limiting the 
acceptance to $|t|> 0.01$~GeV$^2$.

The remaining two simulated reactions proceed via charge exchange resulting in a $\Delta^{++}$ recoil: 
$K_L p\to K^-\pi^0 \Delta^{++}$ and $K_L p\to K_{L}\pi^-\Delta^{++}$.  The charge exchange process has 
the advantage that it has been studied extensively by the LASS Collaboration and Estabrooks \textit{et 
al.} and is known to be dominated by single pion exchange at low $|t|$.  The energy available in the 
$\Delta^{++}$ decay also reduces the minimum accessible $|t|$ when the proton is reconstructed.  However, 
kinematically $t_{min}$ does increase significantly with $M_{K\pi}$ as pointed out in 
Ref.~\cite{Estabrooks:1977xe}. 

The extraction of the $K\pi$ scattering amplitude requires isolating the pion exchange contribution 
from other possible production mechanisms.  The pion exchange contribution is expected to dominate 
at low $|t|$, and the simulation studies demonstrate the acceptance in this critical regime.  However, 
significant theoretical modeling of the $|t|$ dependence of the production model will be required to 
extract the $K\pi$ scattering amplitudes.  Systematic uncertainties due to this theoretical modeling 
are very difficult to estimate and thus were not taken into account in previous measurements.

While the simulations described in this proposal have focused on the proton target data, similar 
reactions will be analyzed with the 100 days of deuterium target running.  This data will complement 
the proton target data and provide essential systematic checks on the extraction of the charged pion 
exchange contribution.  In addition, the opportunity to study both proton and neutron target reactions 
at low $t$-Mandelstam will provide unprecedented access to all four isospin partners of $K_0^\ast(700)$ 
in a single experiment at KLF.

\subsubsection{Impact on $P$-Wave Phase-Shift Study:}
\begin{figure}[ht]
\centering
\includegraphics[height=6.2cm]{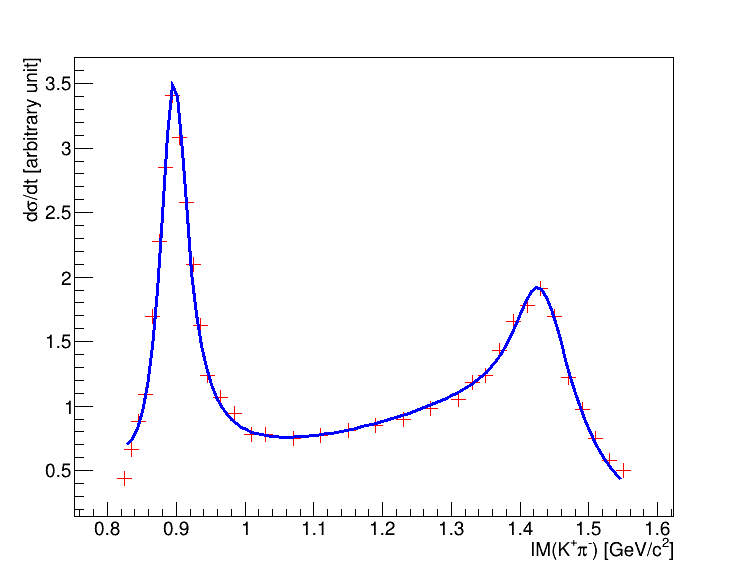}
\includegraphics[height=6.2cm]{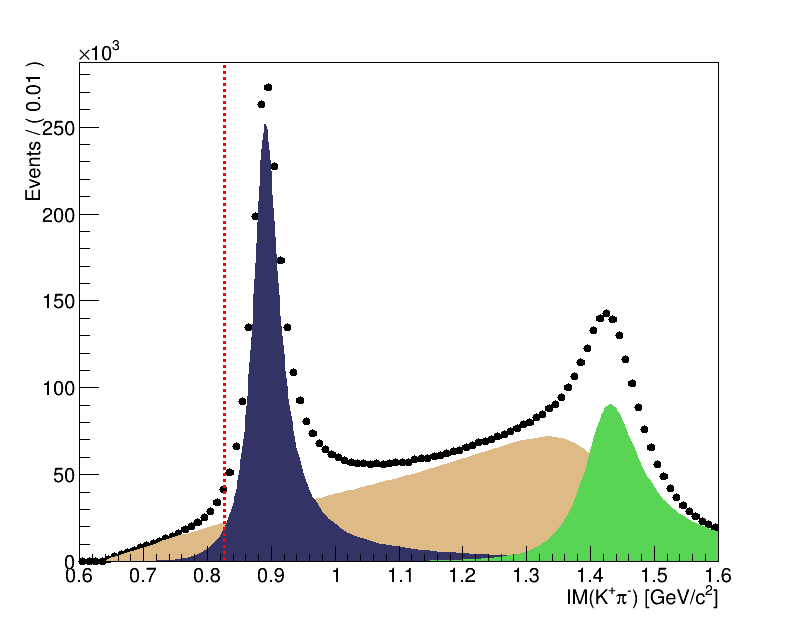}

        \centerline{\parbox{0.70\textwidth}{
        \caption[] {\protect\small 
        \underline{Left}: Cross section of $K^-p\to K^+\pi^-n$ as a function of the invariant mass 
	from LASS results~\cite{Aston:1987ir}. The blue line is the fit to the cross section using 
	composite model containing two RBWs, spin-1 and spin-2, and $S$-wave LASS parameterization.
        \underline{Right}: Expected distribution of the $K^+\pi^-$ invariant mass below 1.6~GeV from 
	KLF after 100~days of running. The dark blue function represents the $K^+\pi^-$ $P$-wave, 
	light brown the $S$-wave and green the $D$-wave. The dashed line represents the threshold 
	of $K\pi$ invariant mass in LASS results~\protect\cite{Aston:1987ir}.} 
        \label{fig:totinv} } }
\end{figure}

The pion exchange in the hadro-production mechanism of $K^{\ast 0}(892)$ occurs mostly at low $-t$, 
therefore allowing access to the scattering amplitude of $K^0\pi^0\to K^+\pi^-$, as illustrated in 
Fig.~\ref{fig:totinv}. Using the resolutions and efficiencies from our simulations, we can estimate 
the improvement that can be made on the scattering amplitude analysis of $K\pi\to K\pi$. The range 
of $-t$ that will be used in this comparison will be [0.14, 0.2]~GeV$^2$ to ensure that the $t$ 
efficiency is uniform. The efficiency of this $t$ range selection is $\epsilon_{\pi} = 17.85~\%$. 
The expected number of events in this case is $2\times 10^6$ based on 100 days of  running time.

The precise and robust determination of the $P$-wave elastic phase-shift is crucial for many process 
involving strangeness, particularly in that it is the dominant contribution to the $K\pi$ form factor. 
There exist several different alternative studies to the production experiments like the extraction of 
the vector form factor $f_{\pm}(t)$ by Ref.~\cite{Bernard:2013jxa}, where $t$ is the four-momentum 
transfer. The vector form factor, at the optical point $f_+(0)$, has an impact on the measurement of 
the CKM matrix element $V_{us}$~\cite{Boito:2010me,Bernard:2013jxa}, where the precision on this 
measurement plays an important role on probing the physics beyond the Standard Model. The 
phenomenological studies~\cite{Boito:2010me,Bernard:2013jxa} analyzed the $K\pi$ $P$-wave phase-shift 
produced by the Belle Experiment~\cite{Epifanov:2007rf} using the decay $\tau\to K\pi\nu_{\tau}$ and 
LASS data~\cite{Aston:1987ir} using the scattering reaction $K^-p\to K^+\pi^-n$. A 
major concern from these studies is that as explained in~\cite{Bernard:2013jxa} the phase-shift 
extraction by the LASS experiment~\cite{Aston:1987ir} could be affected by 
systematic effect.  Furthermore, there is a clear tension regarding the parameters of the $K^\ast(892)$ 
between the modern $\tau$ decay experiments and the older LASS experiments. 

A new, high-statistics measurement of this amplitude at the KLF could resolve these systematic effects 
and provide a new precision measurement of the  $K^\ast(892)$ resonance parameters. We can evaluate the 
improvement that can be performed by KLF in these type of studies. A comparison between the projected 
KLF results and the LASS results~\cite{Aston:1987ir}, which perform a similar scattering reaction study 
with a charged kaon beam, is presented in Fig.~\ref{fig:totinv} (left). The resulting model from the 
fit to LASS amplitude is used to simulate the distribution of the invariant mass of $K^+\pi^-$ in a  
$P$-wave with KLF.

\subsubsection{$S$-wave and $D$-wave Production in $K_Lp\to K^+\pi^-p$:} 
The $K\pi$ scattering $S$-wave has two possible isospin channels. Of those the isospin $I = 1/2$ 
$S$-wave contains two resonances, $\kappa$ and $K_0^\ast(1430)$, both of which are not well defined.  
The isospin $I = 3/2$ $S$-wave is a repulsive interaction, and thus no resonances are seen with just 
a smooth linearly decreasing phase-shift instead. So far, the available data used to study the dynamics 
of the $S$-wave for both isospins are SLAC data~\cite{Estabrooks:1977xe,Aston:1987ir}. The $K\pi$ 
$P$-wave and $D$-wave are better understood. The former has one elastic vector-meson resonance 
$K^\ast(892)$ which dominates the description of the partial wave, and two inelastic resonances with 
small branching ratios to $K\pi$ below 2~GeV, whereas the latter has one inelastic tensor meson 
resonance, the $K^\ast_2(1430)$. The simulation of the reaction $K_Lp\to K^{\ast0}(892)p\to K^+\pi^-p$ 
in KLF can be used  to estimate the total production rate of the different $K\pi$ waves. 
Fig.~\ref{fig:totinv} (left)  shows the fit to the cross section of LASS results. After 100~days of KLF 
running, we expect $3.5\times 10^6$ events for $S$-wave production and $1.2\times 10^6$ event in the 
$D$-wave. The total 100~days production statistics for the $K^+\pi^-$ system is expected to be $\approx 
7\times 10^6$ events for the $S$, $P$, and $D$-waves combined. This production includes isospin $I = 
1/2$ and $I = 3/2$ and represent about 50 times the dataset collected by LASS 
experiment~\cite{Aston:1987ir}.  Figure~\ref{fig:totinv} (right) shows the expected $K^+\pi^-$ invariant 
mass distribution produced by the reaction $K_Lp\to K^{\ast0}(892)p\to K^+\pi^-p$ in KLF. 

It is important to note that especially in the region below 0.75~GeV we expect this new data to provide 
a dramatic improvement in these scattering amplitude determinations not only due to the very high 
statistics, but also because LASS did not provide any data below 0.75~GeV.  These data are therefore 
very relevant for the extraction of the scalar-meson scattering lengths that will test the predictions 
and convergence of SU(3) Chiral Perturbation Theory. Moreover, the rigorous variable for analytic 
continuation to the complex plane is $s$, and due to the large width of the $\kappa$, the real part 
of the $\kappa$ pole position in the $s$-plane, ${\rm Re}(M_\kappa-i\Gamma/2)^2\simeq 
(M_\kappa^2-\Gamma^2/4)\simeq 0.39\,$GeV$^2$, is much closer to the threshold $s_{th}\simeq 
0.40$~GeV$^2$ 
than to its nominal mass $M_\kappa^2\simeq 0.465$~GeV$^2$, which makes the determination of the pole 
especially sensitive to the threshold region.  The KLF low-mass results are therefore of even greater 
relevance.

\subsubsection{Kappa Investigation:}
\begin{table*}[ht]

    \centering \protect\caption{Illustrative values of $\kappa/K_0^\ast(700)$-pole determinations from 
	models (Lines 2 to 7). Line 8 is a model independent prediction from a dispersive analysis 
	without using $S$-wave data below 1~GeV. We also compare in the last two lines the model 
	independent extraction using present data versus the extraction using the expected KLF data.}
\vspace{2mm}
\resizebox{16cm}{!}{
\begin{tabular}{|c|c|c|}
\hline
Reference & Pole Position (MeV)                 & Comment  \\
          & $\sqrt{s_\kappa}\equiv M-i\Gamma/2$ &  \\
\hline
Bonvicini~\protect\cite{Bonvicini:2008jw}   & 706.0$\pm$24.6-i 319.4$\pm$22.4
  & $T$-matrix pole model from CLEO \\
Bugg~\protect\cite{Bugg:2003kj}             & 663$\pm$42-i 342$\pm$60
  & Model with LO Chiral symmetry   \\
Pel\'aez~\protect\cite{Pelaez:2004xp}       & 753$\pm$52-i 235$\pm$33
  & Unitarized ChPT up to NLO       \\
Conformal CFD~\protect\cite{Pelaez:2016tgi} & 680$\pm$15-i 334$\pm$8
  & Conformal parameterization from dispersive fit   \\
Pad\'e~\protect\cite{Pelaez:2016klv}        & 670$\pm$18-i 295$\pm$28
  & Analytic local extraction from dispersive fit  \\
Zhou \textit{et al.}~\protect\cite{Zhou:2006wm} & 694$\pm$53-i 303$\pm$30
  & partial-wave dispersion relation. Cutoff on left cut.   \\
Descotes-Genon \textit{et al.}~\protect\cite{DescotesGenon:2006uk} & 658$\pm$13-i 279$\pm$12
  & Roy-Steiner prediction. No S-wave data used below 1~GeV. \\
Pelaez-Rodas HDR~\protect\cite{pki,Pelaez:2020uiw,inprep}& 648$\pm$7-i 280$\pm$16
  & Roy-Steiner analysis of scattering data   \\
\hline
\textbf{KLF expected errors} & \textbf{648$\pm$4-i 280$\pm$8}   &
\textbf{As previous line but with KLF expected errors}   \\
\hline
\end{tabular}}
\label{tab:bunchCharge}
\end{table*}

\begin{figure}[ht]
\centering
{
\includegraphics[width=0.45\textwidth,keepaspectratio]{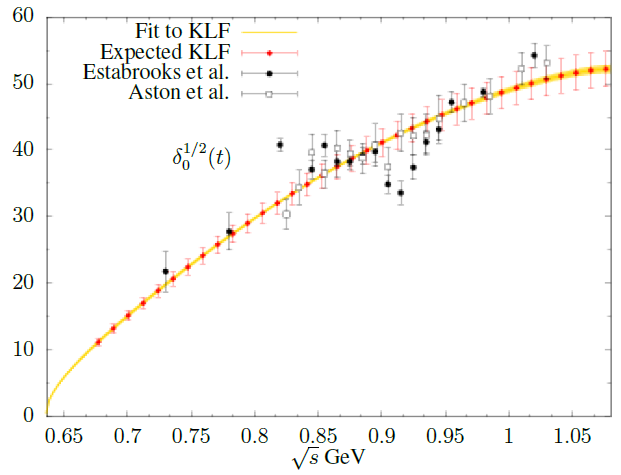} 
}{
\includegraphics[width=0.45\textwidth,keepaspectratio]{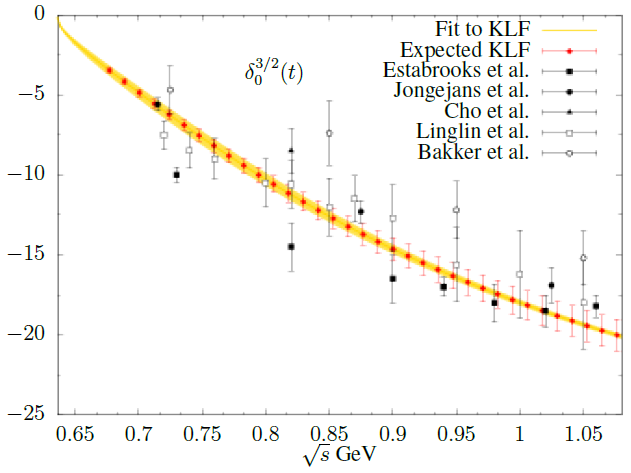} 
}

    \caption{
    \underline{Left}: The $S$-wave phase-shift for isospin $I = 1/2$ and 
    \underline{Right}: isospin $I = 3/2$ (in degrees) as a function of $\sqrt{s}$, invariant mass of 
	$K\pi$ system (see text for details). The yellow band corresponds to the uncertainty of the 
	fit to the world data without KLF data, whereas the red diamonds are KLF expected measurements 
	with statistical and experimental systematic uncertainties added in  quadrature. The projected 
	KLF data correspond to 100~days of running.}
\label{fig:swave}
\end{figure}

The $\kappa$ or $K_0^\ast(700)$ meson is a $0^+$ resonance with strangeness. The pole of this resonance 
is found in the $K\pi$ $S$-wave with isospin $I = 1/2$. In case of neutral kaon scattering off the proton 
producing a $K\pi$ system with neutral or charged exchange, the $S$-wave final state is composed of the 
two isospin components $I = 1/2$ and $I = 3/2$. Therefore if possible, a separate determination of the 
two components to study the pole of  the $\kappa$ meson would be very important. It should be noted 
that this separation was not performed in the last LASS experiment~\cite{Aston:1987ir}. Actually, the 
existing $I = 3/2$ data are older than the LASS experiment and are of much lower precision, which 
produces a large source of uncertainty that contaminates the extraction of the $I = 1/2$ amplitude 
and thus the $\kappa/K_0^\ast(700)$ or other strange resonance poles. Simulations of the two reactions 
$K_Lp\to K^-\pi^0\Delta^{++}$ and $K_Lp\to K_L\pi^-\Delta^{++}$ for 100~days of running provide estimates 
of the precision that will be obtained for the isospin separate phase-shifts, as presented in 
Appendix~\ref{sec:A4} and an Analysis Note~\cite{Amaryan:2020zz}, and shown in Fig~\ref{fig:swave}.  
\textbf{Note:} these phase shifts assume that the pion exchange contribution to the scattering amplitude 
will be obtained through a model dependent fit to the $|t|$ dependence of the amplitudes as performed by 
the LASS Collaboration~\cite{Aston:1987ir} and Estabrooks \textit{et 
al.}~\cite{Estabrooks:1977xe}, however theoretical uncertainties due to this 
extraction were not included in these studies.

In Figure~\ref{fig:swave}, we present $S$-wave phase-shift for an isospin $I = 1/2$ and $I = 3/2$, 
which includes all world data on production experiments. The yellow band corresponds to the 
uncertainties of the fit to  previous data. The red points with error bars are projected KLF data 
for 100~days of running  with statistical and experimental systematic uncertainties added in quadrature.

\begin{figure}[ht]
\centering
\floatbox[{\capbeside\thisfloatsetup{capbesideposition={right,center},capbesidewidth=7cm}}]{figure}[\FBwidth]
{

        \caption{Present situation of the determinations of the
        $\kappa/K_0^\ast(700)$ pole.  The figure is from Refs.~\protect\cite{pki,Pelaez:2020uiw,inprep} 
	but we have added as a red point with uncertainties, the simulation of the pole position that 
	would be obtained by means of a Roy-Steiner analysis by using simulated data from KLF experiment 
	after 100~days of run, note that the blue error bars are for parameters obtained without KLF data 
	and the tiny red error bars within the blue ones are   obtained with KLF data. These calculations 
	also include estimates of systematic effects. Note that the other points are either
        predictions~\protect\cite{DescotesGenon:2006uk} or illustrative models that may have additional 
	systematic uncertainties due to their model dependence, like BW determinations. Shadowed rectangle 
	stands for PDG2018 uncertainties.}
        \label{kes}

}
    {\includegraphics[width=0.55\textwidth,keepaspectratio,trim=0 0 0 0, clip]{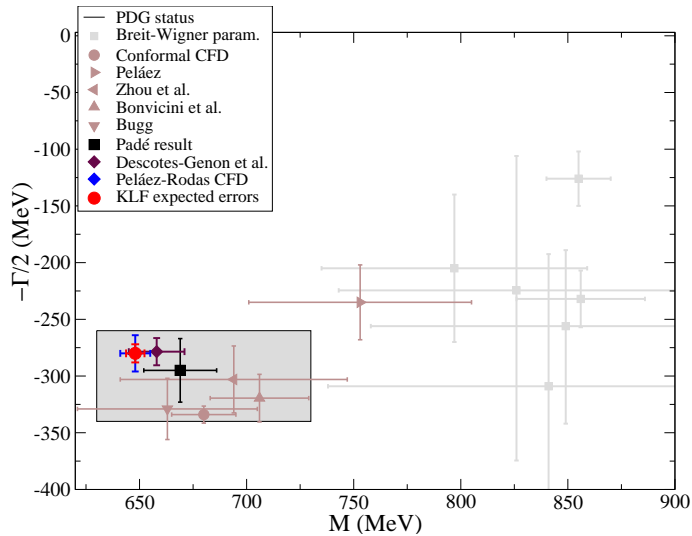}}
\end{figure}

There are many {\it models} describing the $\kappa/K_0^\ast(700)$ and its associated pole (see the 
PDG2018~\cite{Tanabashi:2018oca} for an exhaustive compilation). For illustration we show some 
representative results in Fig.~\ref{kes}. Note that many of them still use BW parameterizations, 
which unfortunately are not applicable in this case  because they violate chiral symmetry and do 
not have the left and circular cuts that are numerically relevant for precise determinations of the 
$\kappa/K_0^\ast(700)$ pole. 

The other analyses we list in Table~\ref{tab:bunchCharge} are: a model of a T-matrix 
pole~\cite{Bonvicini:2008jw} and more sophisticated models including some implementation of 
chiral symmetry~\cite{Bugg:2003kj,Pelaez:2004xp}, but still with some model dependence that is 
not included in their uncertainties. We also show a dispersive evaluation~\cite{Zhou:2006wm}, where 
the difficult left and circular cut contributions have been approximated with some assumptions (like 
a cut-off), but with very conservative systematic uncertainties. In addition, we show two extractions 
of the pole, one exploiting the analyticity in the whole complex plane by means of a conformal 
expansion~\cite{Pelaez:2016tgi} and another one using  Pad\'e approximants to extract the pole 
parameters from local information of the amplitude near the pole without assuming a specific 
parameterization~\cite{Pelaez:2016klv}. Both of them use as input a fit to data constrained with 
Forward Dispersion Relations and their uncertainties include an estimate of systematic effects. 
Other determinations in the literature, not shown here, are usually based on models and often quote 
uncertainties that do not include systematic effects.

Finally, as already commented, the most rigorous determination of the $\kappa/K_0^\ast(700)$ pole 
with a realistic estimate of both statistical and systematic uncertainties, can be made by means of 
Roy-Steiner Dispersion Relations. There is actually such an estimate of the pole~\cite{DescotesGenon:2006uk}, 
although it does not use data on the scalar wave below 1~GeV. Actually, the scalar partial waves in that 
region are obtained as solutions of the Roy-Steiner equations with input from other waves and higher 
energies. In this sense, the $\kappa$ pole and the whole low-energy region in 
Ref.~\cite{DescotesGenon:2006uk} are a prediction, not a determination from data.

Thus, in order to estimate the effect of the proposed KLF experiment, we have recalculated the pole 
obtained by using a Roy-Steiner analysis either using all the existing data~\cite{pki,Pelaez:2020uiw,inprep} 
or with the LASS data rescaled to the expected accuracy of the KLF experiment.  In the first case, without 
KLF, Pel\'aez \textit{et al.}~\cite{Pelaez:2020uiw,inprep} finds $M_\kappa\simeq(648\pm7)$~MeV and 
$\Gamma_\kappa = (580\pm32)$~MeV, whereas by using the expected KLF data the uncertainties are divided 
by slightly more than a factor of two for the mass, so that we find: $M_\kappa\simeq(648\pm4)$~MeV; and 
by more than a factor of three for the width, finding: $\Gamma_\kappa = (580\pm16)$~MeV. With this 
assumed precision for the KLF measurement, a significant improvement on the $\kappa/K_0^\ast(700)$ 
search can be performed, especially by improving the elastic region of the $K\pi$ invariant mass. 
Fig.~\ref{kes} shows as pole positions in the complex plane, the different determinations of the 
$\kappa$ mass and width, that we have just described, including the determination with the expected 
amplitude and phase-space that will be produced by KLF. The expected result for the kappa pole is 
$\sqrt{s_\kappa}\equiv M-i\Gamma/2$ = (648$\pm$4~-~i 280$\pm$8)~MeV (the error coming from $\pi K$ 
scattering is less than 1~MeV, the rest comes basically from the high energy Regge input and the 
crossed channel $\pi\pi \to K \bar K$ input to the dispersive integrals).

\textbf{To summarize}: The KLF experiment will provide the first measurement of the $K\pi$ scattering 
process in more than 30~years, utilizing modern experimental techniques which will provide unprecedented 
statistical precision and access to much lower $K\pi$ masses than previous measurements.  Of particular 
interest is the expected impact on the $I = 1/2$ channel where it will help to lower the tension between 
phenomenological dispersive determinations of scattering lengths from data versus those from Chiral 
Perturbation Theory and lattice QCD.  Finally, it will significantly reduce the uncertainty in the mass 
and width determination of the controversial $\kappa$ or $K_0^\ast(700)$. 
\subsection{Neutron-induced Reactions}
\label{sec:neutron_sims}

Since all charged particles from the KPT will be wiped out from the beam by the sweeping magnet 
and the photon flux is also expected to be low, the only background which requires special 
considerations is due to neutron-induced reactions. The neutron flux on the cryogenic target 
is expected to be moderate, $6.6\times 10^5~n/sec$~\cite{Strakovsky:2020wtw},  and moreover the 
neutron flux drops exponentially with energy (see Fig.~\ref{fig:yield} (left) for details), so 
generally the high-energy neutron flux is tiny. We have studied possible neutron-induced background 
thoroughly and found all neutron-related reaction to contribute at the level of $10^{-3}$ or lower. 
The highest ($10^{-3}$ level) neutron background contribution is expected for inclusive reactions, 
while for exclusive reactions the neutron induced reactions can be completely eliminated due to 
their very different event topology and kinematics.

There are several neutron energy ranges which requires separate treatments:\\ 
1) $E_n > 1.6$~GeV, strangeness production threshold ($np\to K^+n\Lambda$ reaction is allowed), 
$\sim 1\%$ of the neutron flux; \\
2) $1.6 > E_n > 0.3$~GeV, single pion production threshold ($np\to\pi^+nn$ reaction is allowed; 
$np\to pn$ reaction has sufficient energy to be measured), $\sim 5\%$ of the neutron flux; \\
3) $E_n < 0.3~GeV$, $\sim 94\%$ of the neutron flux - do not contribute at all.

We have identified three major sources of physical background: $np\to K^+n\Lambda$ (the next most 
probable reaction with strangeness, $np\to K^+n\Sigma$, has an order of magnitude smaller 
cross-section) and $np\to\pi^+nn$ with $\pi^+$ misidentifed as $K^+$.
\begin{figure}[ht]
\centering
{
    \includegraphics[width=0.32\textwidth,keepaspectratio]{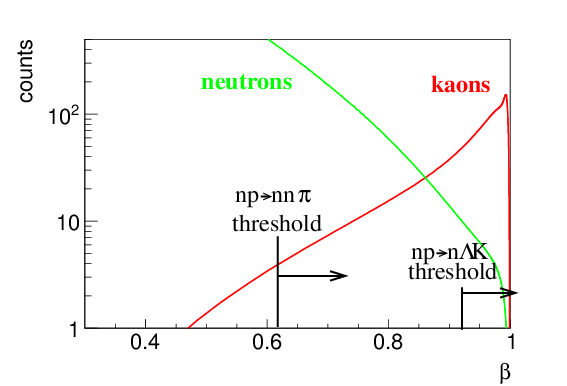} }
    {
   \includegraphics[width=0.32\textwidth,keepaspectratio]{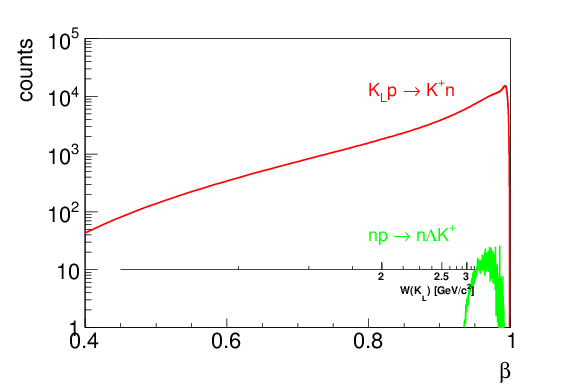} }
   {
   \includegraphics[width=0.32\textwidth,keepaspectratio]{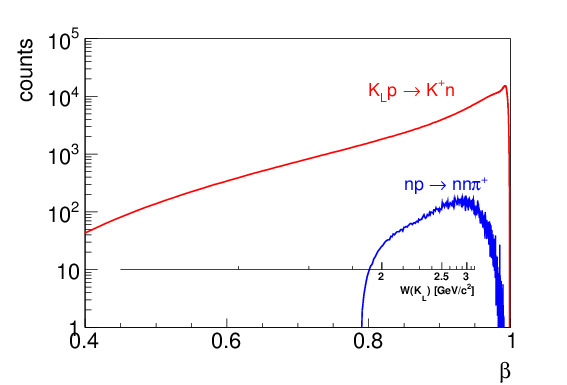} }

    \centerline{\parbox{0.70\textwidth}{
    \caption[] {\protect\small
    \underline{Left}: Neutron and $K_L$ fluxes as a function of velocity $\beta$. 
    \underline{Middle}: $K^+$ flux as a function of projectile 	velocity $\beta$ for neutron-induced 
	(green) and kaon-induced (red) reactions.
	\underline{Right}: Same but for $np\to\pi^+nn$ (blue) and $K_Lp\to K^+n$ (red) reactions. 
	Pion misidentification efficiency for the neutron-induced reaction is extracted from the 
	full MC Geant simulation. Secondary axis showed corresponding $W$ value under assumption of 
	$K_Lp\to K^+n$ reaction.}
    \label{fig:NeutFluxBeta} } }
\end{figure}

As mentioned above, neutron flux drops exponentially with energy and generally the high-energy 
neutron flux is small, but nonvanishing. If neutrons and $K_L$s have the same velocity, they 
cannot be separated by time-of-flight. However if $K_L$ and neutrons have the same velocities 
they have very different energies, hence very different reaction kinematics. Moreover neutron-induced 
reactions always have extra particle in a final state compare to $K_L$-induced reactions (2-body final 
state for $K_Lp\to K^+n/K_Lp\to K^+\Xi$ vs 3-body $np\to K^+\Lambda n$). 

Neutron-induced reactions have high cross sections, which is why one needs to consider them as a 
possible source of background. In Fig.~\ref{fig:NeutFluxBeta} (left), one can see a comparison of 
kaon and neutron fluxes, similar to Fig.~\ref{fig:yield} (left) but in terms of $\beta$. Particles 
with the same $\beta$ cannot be separated by time-of-flight. At $\beta = 0.86$ neutron and kaon 
fluxes become equal. This velocity corresponds to a neutron momentum of p$_n$ = 1.6~GeV/$c$ and 
kaon momentum of p$_K$ = 0.8~GeV/$c$.

To evaluate the amount of background, we need to fold this flux with production cross section and 
reconstruction efficiency. The cross-section for the $np\rightarrow K^+\Lambda n$ reaction can be 
extracted using data from Ref.~\cite{Valdau:2011jg} and isospin factors. The cross-section for 
another background reaction, $np\to\pi^+nn$, can be taken from Ref.~\cite{Adlarson:2017kou} 
($np\to\pi^-pp$) using isospin coefficients as well. The result of folding of these cross-sections 
with the flux and detector reconstruction efficiency can be seen on Fig.~\ref{fig:NeutFluxBeta} 
(middle and right).


As one can see, both reactions contribute at a very low level, Fig.~\ref{fig:NeutFluxBeta}, and in 
a very narrow range of energies. One can further suppress this type of backgrounds by exclusivity 
cuts. Altogether one can easily suppress these types of background below $10^{-4}$. 

One can make a more general statement: due to their high production threshold the $np\to K^+ X$ 
reactions cannot serve as a background channel to any of kaon-induced hyperon production channels, 
neither for the s-channel $\Sigma$-production nor for the associated $\Xi^\ast$-production.
The only inclusive reaction planned to be measured at KLF so far is $K_Lp\to K^+n$. Here the 
$np\to\pi^+nn$ reaction can play a role of a possible background at a sub-percent level. However it 
does not contribute to any higher multiplicity reactions due to exclusivity conditions. Indeed, to 
mimic the simplest $K_Lp\to \pi^+\Lambda$ reaction by the neutron induced process, one should have:
\begin{enumerate}[(i)]
\setlength{\itemsep}{0pt}
    \item $np\to\pi^+nn$ with pion rescattering within detector volume to adjust pion momentum-angle 
	relation of a 3-body process to two-body $\Lambda\pi$ \\
    \item have secondary rescattering of ejected neutron on supporting structures with the reaction 
	$nn\to\pi^-pn$ to produce $p\pi^-$ pair \\
    \item have both $p$ and $\pi^-$ rescatter to tune their angle-energy to mimic $\Lambda$.
\end{enumerate}
A combination of such low probability events make neutron induced background, absolutely impossible 
for the reactions we have considered so far in our proposal. 

Neutron induced events can become a sizable background only to a very rare kaon-induced reactions 
with tiny cross-sections where the final state would be reconstructed inclusively/semi-exclusively. 
Neither of such reactions are considered in this proposal so far or planned to be investigated at 
the first stage of analysis. 

\subsubsection{Useful Neutron-induced Reactions:} 
\begin{figure}[ht]
\centering
{
    \includegraphics[width=0.42\textwidth,keepaspectratio,angle=90]{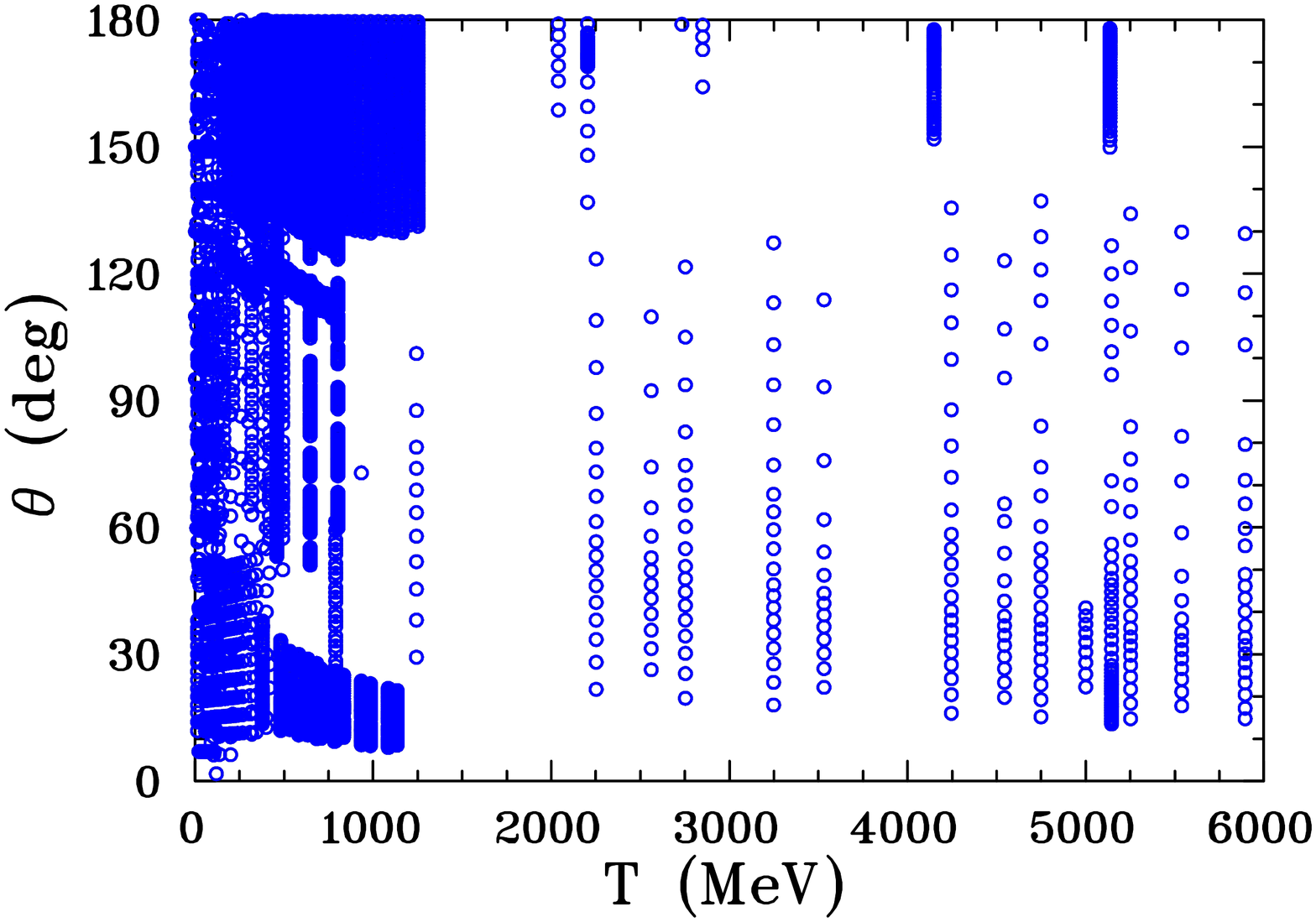} }
    {
    \includegraphics[width=0.38\textwidth,keepaspectratio]{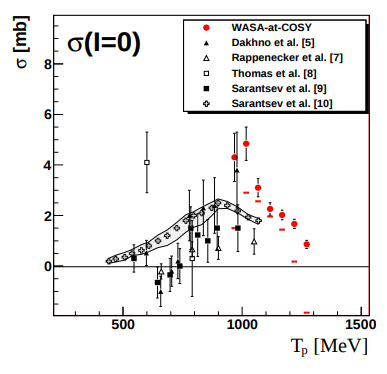} }

    \centerline{\parbox{0.70\textwidth}{
    \caption[] {\protect\small
    \underline{Left}: Differential cross section data available for $np\to np$ reactions as a 
	function of neutron kinetic energy~\protect\cite{SAIDweb}.
    \underline{Right}: Total cross section as a function of incident nucleon energy for the 
	isoscalar single pion production from Ref.~\protect\cite{Adlarson:2017kou}. } 
    \label{fig:npElastic} } }
\end{figure}

Nucleon-nucleon interactions is a central piece of nuclear physics studies since the discovery of 
nuclei. Nuclear forces can be decomposed into an isospin $I = 1$ part accessible by proton-proton 
scattering which has been studied with reasonable accuracy in a wide range of energies, and an 
isospin $I = 0$ part which requires a neutron beam or target. In  Fig.~\ref{fig:npElastic} (left), 
one can see the currently available data for the $I = 0$ channel. In a view of the upcoming 
Electron-Ion Collider (EIC)~\cite{Accardi:2012,USNAC:2018} which enables a strong program in 
nucleon short-range correlations (SRC)~\cite{Duer:2018sby,Schmookler:2019nvf}, where 
$np$-correlations comprise 90\% of all events, our pure knowledge of nuclear potential and 
$np$-elastic scattering at energies as low as 1~GeV is truly deplorable.

Neutrons are massive particles with lower velocities compare to $K_L$'s, so excellent 
time-of-flight resolution available at KLF would allow for even more precise momentum 
measurements for the neutron beams. A very high $np$-elastic cross section and a good 
detection efficiency of a GlueX setup for a single proton track events, make this channel 
a golden by-product of a KLF program.

Keeping in mind that even the reaction with the highest cross-section, $np$-elastic scattering, 
is so poorly known one can immediately guess that all other less abundant reactions have fewer 
measurements. The next best examples here is a single pion production in an isoscalar channel. 
As can be seen from Fig.~\ref{fig:npElastic} (right), the data exist only for very narrow range 
of energies and agreement between various data sets is extremely pure. This is particularly 
unfortunate since the HADES Collaboration recently showed particularly clean non-interfering 
$N^\ast$'s and $\Delta^\ast$'s excitations in high energy $NN$-collisions~\cite{Agakishiev:2014}. 
This topic might be a second strong by-product of a KLF program. 

\textbf{To summarize}: One can expect more than 100M neutron induced single pion production events 
to be detected within 100~days of KLF beamtime, and an  even larger number for $np$-elastic 
scattering events. Several other reactions with two- and three-pion production, and $\eta$-production, 
are also feasible.

\section{Proposed KL Beam Facility}
\label{sec:beam}

\begin{figure}[ht]
\centering
{
    \includegraphics[width=0.6\textwidth,keepaspectratio]{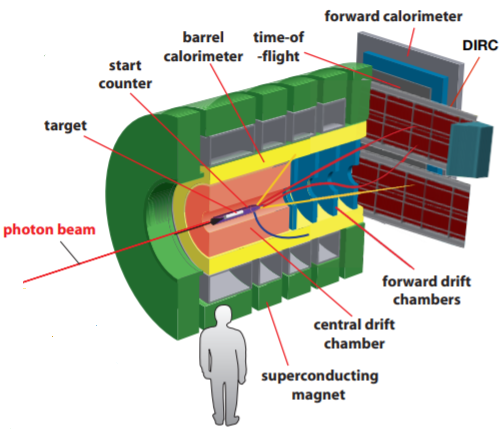}
}

\centerline{\parbox{0.80\textwidth}{
 \caption[] {\protect\small The GlueX spectrometer.}
        \label{fig:GlueX} } }
\end{figure}

The GlueX spectrometer in Hall~D at Jefferson Lab, shown in Figure~\ref{fig:GlueX}, 
is a powerful tool employed by the GlueX Collaboration to investigate a wide range 
of topics in meson and baryon spectroscopy and structure, particularly the search 
for mesons with excited gluonic content, using the recently upgraded 12~GeV electron 
beam of CEBAF accelerator. The spectrometer is carefully designed~\cite{Adhikari:2020cvz} 
to measure the complete electromagnetic response for nucleons and nuclei across the 
kinematic plane: elastic, resonance, quasi-elastic, and deep inelastic reactions with 
almost 4$\pi$ acceptance for all final state particles.

\subsection{Beamline Delivery for Secondary $K_L$ Beam}
\label{sec:beamT}

The proposed secondary $K_L$ experiment requires time-of-flight 
measurements, which in turn requires substantially lower bunch 
repetition rates in CEBAF than the nominal 249.5~MHz or 499~MHz.  
Gun laser hardware lends itself to powers of two reductions in 
repetition rates, so this proposal includes beam delivery at either the 
32nd (15.59~MHz) or 64th (7.80~MHz) harmonic of the nominal 499~MHz.

An average dump power limit of 30~kW to 60~kW for 12~GeV electrons in 
the CPS translates to average beam currents of 2.5~$\mu$A to 
5.0~$\mu$A.  Combining these beam requirements leads to individual 
bunch charges shown in Table~\ref{TAB:bunchCharge}.
\begin{table}[ht]

\centering \protect\caption{CEBAF injector bunch currents and 
	repetition rates for Secondary $K_L$ experiment.}
\vspace{2mm}

\begin{tabular}{|c|c|c|c|c|}
\hline
Current  &Repetition& Harmonic   & Bunch  & Equivalent \\ 
         &   Rate   & of 499~MHz & Charge & 499~MHz Current \\
($\mu$A) &  (MHz)   &            &  (pC)  & ($\mu$A) \\
\hline
     2.5 & 15.59    & 32nd       & 0.16   &  80  \\
     2.5 &  7.80    & 64th       & 0.32   & 160  \\
     5.0 & 15.59    & 32nd       & 0.32   & 160  \\
     5.0 &  7.80    & 64th       & 0.64   & 320  \\
\hline 
\end{tabular}
\label{TAB:bunchCharge}
\end{table}

Operations with 0.16~pC to 0.32~pC bunch charge has been demonstrated 
but shown to be challenging in the 12~GeV era. Injector setup time of 
up to a week is required to limit bunch tails that cause beam trips 
and background, and intervention on the order of every few days is 
currently required to maintain a reasonable accelerator availability. 
The G0 experiment~\cite{Androic:2011rha} ran 1.6~pC/bunch, but only at 
3~GeV with the 6~GeV machine and in a dedicated configuration that 
required substantial interception to trim beam tails~\cite{REF:Joe}.

These concerns may be mitigated somewhat by completion of the injector 
upgrade program, including operations of a 200~keV gun, in the 2021 
timeframe. The HV gun was installed in summer 2018, and some initial 
tests with modest gun voltages occurred in FY19, but the full injector 
upgrade for potential high bunch charge availability will not be 
available for study until FY21~\cite{REF:Reza}.

Low frequency, high power amplifier use has been attempted at CEBAF 
in recent years, resulting in substantial damage and high amplifier 
failure rates even near 30~MHz because of high peak power required as 
repetition rate is lowered. The low bunch repetition rate with high 
bunch charge therefore also requires considerable investment.

With the existing and planned gun confirmation, laser development is 
required to achieve any of the planned bunch repetition rates. This 
requires construction of a pulse picker that would pass a sub-harmonic 
of the 249.5~MHz system (for example, 15.6~MHz) to avoid major impact 
to the existing 249.5/499~MHz laser systems. Amplification is then 
required before doubling to the proper wavelength to achieve useful 
power, even for 10~$\mu$A beam. Additional power amplification is 
necessary for the higher beam currents required here.

To  build up a beamline delivery system for the secondary $K_L$ beam 
requires the pulse picking system and the laser amplifier.  The lead 
time on amplifiers can be long so ideally a year of advance funding 
would be necessary to design, build and demonstrate the system 
performance~\cite{REF:Joe}.

The G0 experiment~\cite{Androic:2011rha} used a commercial 
Ti:Al$_2$SO$_3$ laser with a very long ($\sim$5~m) optical cavity that 
was very difficult to keep on and locked to the accelerator RF. This 
solution is not practical for the 12~GeV era. 

\subsubsection{Raster for $K_L$ Beam}
\label{sec:raster}

KLF requires a large electron beam incident on the CPS to minimize power 
density, and convergence of a projected photon beam on the target and 
collimator $\approx$65~m downstream of the CPS.

One option to address these requirements is installation of a beam raster, 
similar to rasters used in other CEBAF high-current halls. However, a raster 
intrinsically enlarges the beam divergence to achieve a larger beam size. 
This is inconsistent with the convergence requirements of the projected 
photon beam. We have instead considered adjustments to the existing Hall~D 
final focus optics to maximize the beam size at the CPS while maintaining 
required projected photon beam convergence.

Reference~\cite{Todd:2020} evaluates the existing Hall~D final focus optics 
and quadrupole apertures for three conditions: 95\% full width horizontal 
beam sizes of 1.0~cm, 1.4~cm, and 1.7~cm. A 95\% full width vertical beam 
size of $\mathcal{O}$(1~cm) is expected at the CPS; it cannot be smaller 
than this to maintain reasonable projected photon beam convergence.

The 95\% full width horizontal 1.0~cm beam size case is quite similar to 
existing optics for GlueX. Under these conditions, KLF should expect 
similar beam size stability to that observed during GlueX-II operations. 

The 1.4~cm case requires more aggressive focusing that results in a maximum 
beam size in the existing final focus quadrupoles that is $\approx$65\% of 
the existing aperture. At these beam sizes chromatic and nonlinear effects 
start contributing substantially to beam quality. It may be feasible to run 
KLF with this beam size at the CPS face, but beam size stability and 
sensitivity of tune may be problematic.

The 1.7~cm case requires substantially more aggressive focusing. The 
maximum beam size in the existing final focus quadrupoles in this 
condition would be at least 75\% of the existing aperture. Here chromatic 
conditions and sensitivity of tune to energy fluctuations starts to dominate, 
and there is very little room for orbit and beam size variation.

For 95\% full width horizontal beam sizes on the CPS dump face above 1.5~cm, 
new final focus quadrupoles would likely be required with larger apertures 
of 20--30~mm radius compared to the existing radii of 16~mm.
\subsection{$K_L$ Beam Overview}
\begin{figure}[ht]
\centering
{
    \includegraphics[width=0.93\textwidth,keepaspectratio]{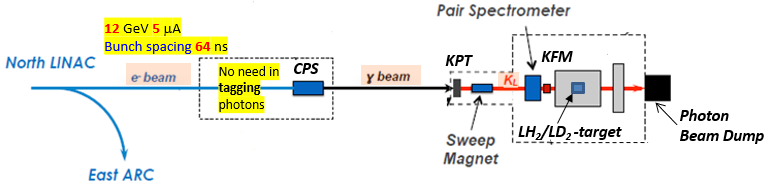} 
}

\centerline{\parbox{0.80\textwidth}{
    \caption[] {\protect\small 
    Schematic view of Hall~D beam line with the production chain  
    $e\rightarrow\gamma\rightarrow K_L$. The main components are the CPS, KPT, 
	sweep magnet, and KFM (see text for details). We do not need in pair 
	spectrometer~\protect\cite{Barbosa:2015bga}. Beam goes from left to right.} 
    \label{fig:beam} } }
\end{figure}
We propose to create a secondary beam of neutral kaons at Hall~D at Jefferson Lab to be 
used with the GlueX experimental setup for strange hadron spectroscopy. The superior CEBAF 
electron beam will enable a flux on the order of $1\times 10^4 K_L/sec$, which exceeds the 
kaon flux previously attained at SLAC~\cite{Yamartino:1974sm} by three orders of magnitude. 
Using a deuterium target in addition to the standard liquid hydrogen target will provide 
the first measurements ever with neutral kaons interacting with neutrons. The ability of 
the GlueX spectrometer to measure reaction fragments over wide ranges in polar $\theta$ 
and azimuthal $\phi$ angles with good coverage for both charged and neutral particles (see, 
for instance, Refs.~\cite{Adhikari:2019gfa,Ali:2019lzf,AlGhoul:2017nbp}), together with the 
K$_L$ energy information from the K$_L$ time-of-flight, provides an ideal environment for 
these measurements. 

A schematic view of the Hall~D beam line showing the production chain $e\to\gamma\to K_L$ 
is given in Fig.~\ref{fig:beam}. Tables~\ref{tab:klbm} and \ref{tab:kltg} summarize beam 
properties and dimensions of targets, respectively.
\begin{table}[ht]

\centering \protect\caption{Expected electron/photon/kaon beam
        conditions at the $K_L$ experiment.}
\vspace{2mm}
{%
\begin{tabular}{|l|r|}
\hline
Property                                                & Value \\
\hline
Electron beam current                        ($\mu$A)   & $5$ \\
Electron flux at CPS                         ($s^{-1}$) & $3.1\times10^{13}$ \\
Photon flux at Be-target $E_\gamma>1500$~MeV ($s^{-1}$) & $4.7\times 10^{12}$ \\
$K_L$ beam flux at cryogenic target          ($s^{-1}$) & $1\times 10^4$ \\
$K_L$ beam $\sigma_p/p$ @ 1~GeV/$c$          (\%)       & $\sim$1.5 \\
$K_L$ beam $\sigma_p/p$ @ 2~GeV/$c$          (\%)       & $\sim$5 \\
$K_L$ beam nonuniformity                     (\%)       & $<2$ \\
$K_L$ beam divergence                        ($^\circ$) & $<0.15$ \\
$K^0/\overline{K^0}$ ratio at Be-target          & 2:1 \\
Background neutron flux at cryogenic target ($s^{-1}$)  & $6.6\times 10^5$ \\
Background $\gamma$ flux at cryogenic target ($s^{-1}$), $E_{\gamma}>100$~MeV  & $6.5\times 10^5$ \\
\hline
\end{tabular}} \label{tab:klbm}
\end{table}
\begin{figure}[ht]
\centering
{
    \includegraphics[width=0.45\textwidth,keepaspectratio]{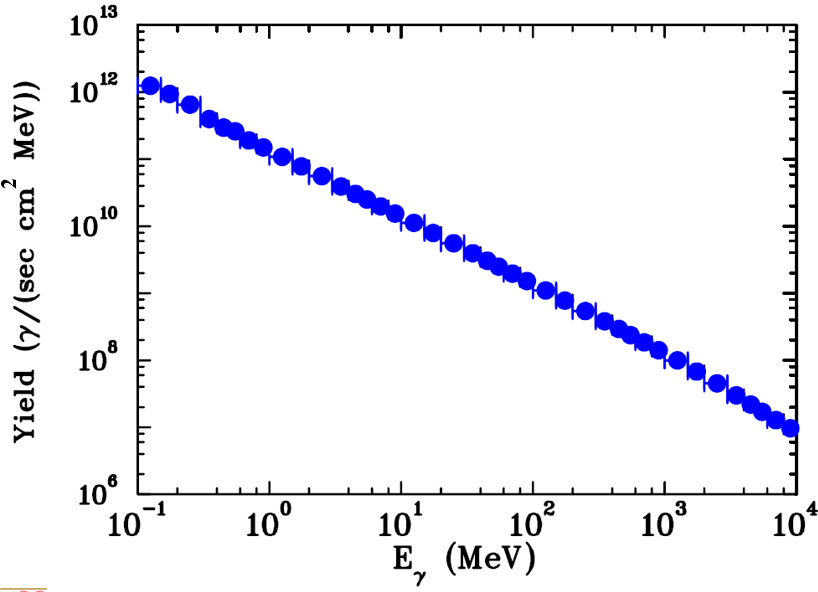}
    \includegraphics[width=0.45\textwidth,keepaspectratio]{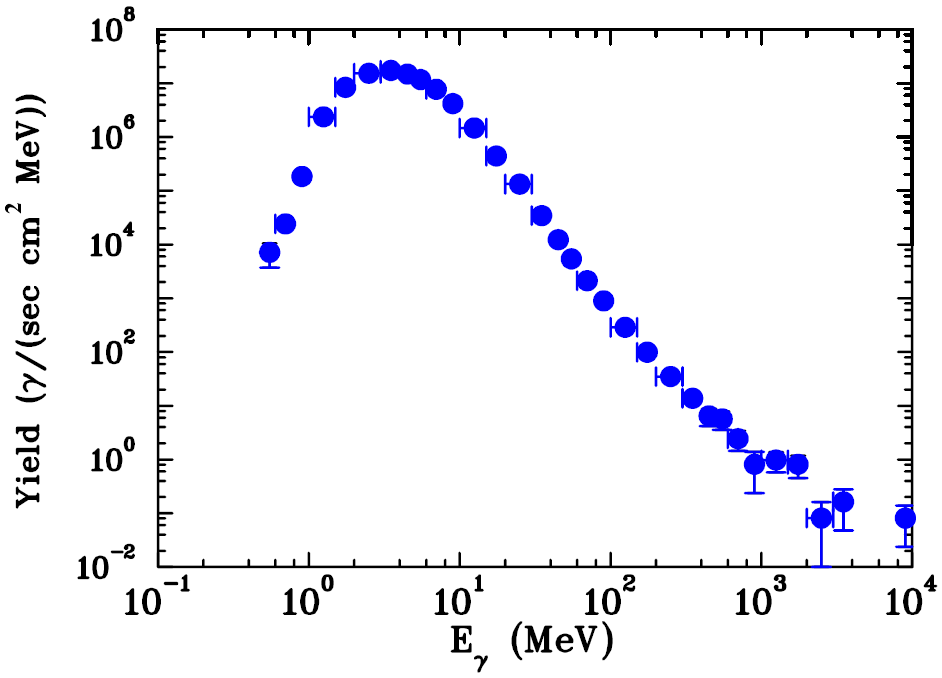}
}

\centerline{\parbox{0.8\textwidth}{
  \caption[] {\protect\small 
  Energy spectrum of bremsstrahlung photons on the face of the Be-target (left) and on the 
	face of the cryogenic target (right).
  Calculations were performed using the MCNP radiation transport code~\protect\cite{Werner:2018}.} 
  \label{fig:gam} } }
\end{figure}
\begin{table}[ht]

\centering \protect\caption{Expected targets properties at the 
        $K_L$ experiment.}
\vspace{2mm}
{%
\begin{tabular}{|l|r|}
\hline
Property                                       & Value  \\
\hline
Copper radiator in CPS              (\%R.L.)   &  10    \\
$\varnothing$Be-target              (m)        &   0.06 \\
Be-target length                    (m)        &   0.40 \\
$\varnothing$LH$_2$/LD$_2$ cryogenic target (m)&   0.06 \\
LH$_2$/LD$_2$ cryogenic target length       (m)&   0.40 \\
Photon beamline length              (m)        &  67    \\
Kaon beamline length                (m)        &  24    \\
\hline
\end{tabular}} \label{tab:kltg}
\end{table}
\begin{figure}[ht]
\centering
{
    \includegraphics[width=0.93\textwidth,keepaspectratio]{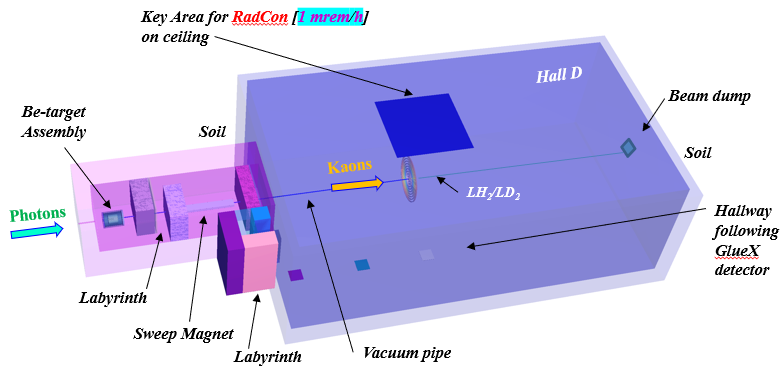}
}

\centerline{\parbox{0.80\textwidth}{
    \caption[] {\protect\small 
    Schematic view of Hall~D setting for the MCNP radiation transport 
	code~\protect\cite{Werner:2018} calculations. The model is presented as 
	semi-transparent for demonstration purposes.  Beam goes from left to right.}  
	\label{fig:hall} } }
\end{figure}

At the first stage, 12-GeV electrons ($3.1\times 10^{13}~e/sec$) will scatter in the radiator 
inside the CPS~\cite{Day:2019qdz}, generating an intense beam of untagged bremsstrahlung 
photons with intensity  ($4.7\times 10^{12}~\gamma/sec$, for E$_\gamma >$1.5~GeV) impinging 
on the face of the Be-target. The main source of K$_L$ production from the target is the 
$\phi$-meson decay, whose photoproduction threshold is E$_\gamma \sim$1.58~GeV. The full 
energy spectrum of photons on the face of the Be-target is shown in Fig.~\ref{fig:gam} (left). \\

The energy spectrum of secondary photons on the face of the cryogenic target is shown on 
Fig.~\ref{fig:gam} (right). The flux is not sufficient to provide any significant background 
in the case of $\gamma p$ or $\gamma d$ interactions in the cryogenic target.

The CPS contains a copper radiator (10\% $X_0$) that is capable of handling the power deposited 
in it by the 12-GeV, 60~kW electron beam, which will be fully absorbed inside the CPS dump. The 
CPS will be located downstream of the Hall-D tagger magnet. The Hall~D tagger magnet and detectors 
will not be used. 

At the second stage, the bremsstrahlung photons will hit the Be target (KPT)~\cite{Strakovsky:2020wtw} 
located at the beginning of the collimator alcove (Fig.~\ref{fig:hall}) in 67~m from CPS, and produce 
neutral kaons ($1\times 10^4~K_L/sec$), along with neutrons ($6.6\times 10^5~n/sec$), photons, and 
charged particles. 

The vacuum beam pipe has a $\varnothing$0.07~m and prevents neutron
rescattering in air. Finally, $K_L$ mesons will reach the LH$_2$/LD$_2$ cryogenic target located 
inside the GlueX spectrometer. The distance between the primary Be and cryogenic targets is 24~m. 
\subsection{Compact Photon Source for $K_L$}
\begin{figure}[ht]
\vspace{-7mm}
\centering
{
    \includegraphics[width=0.60\textwidth,keepaspectratio]{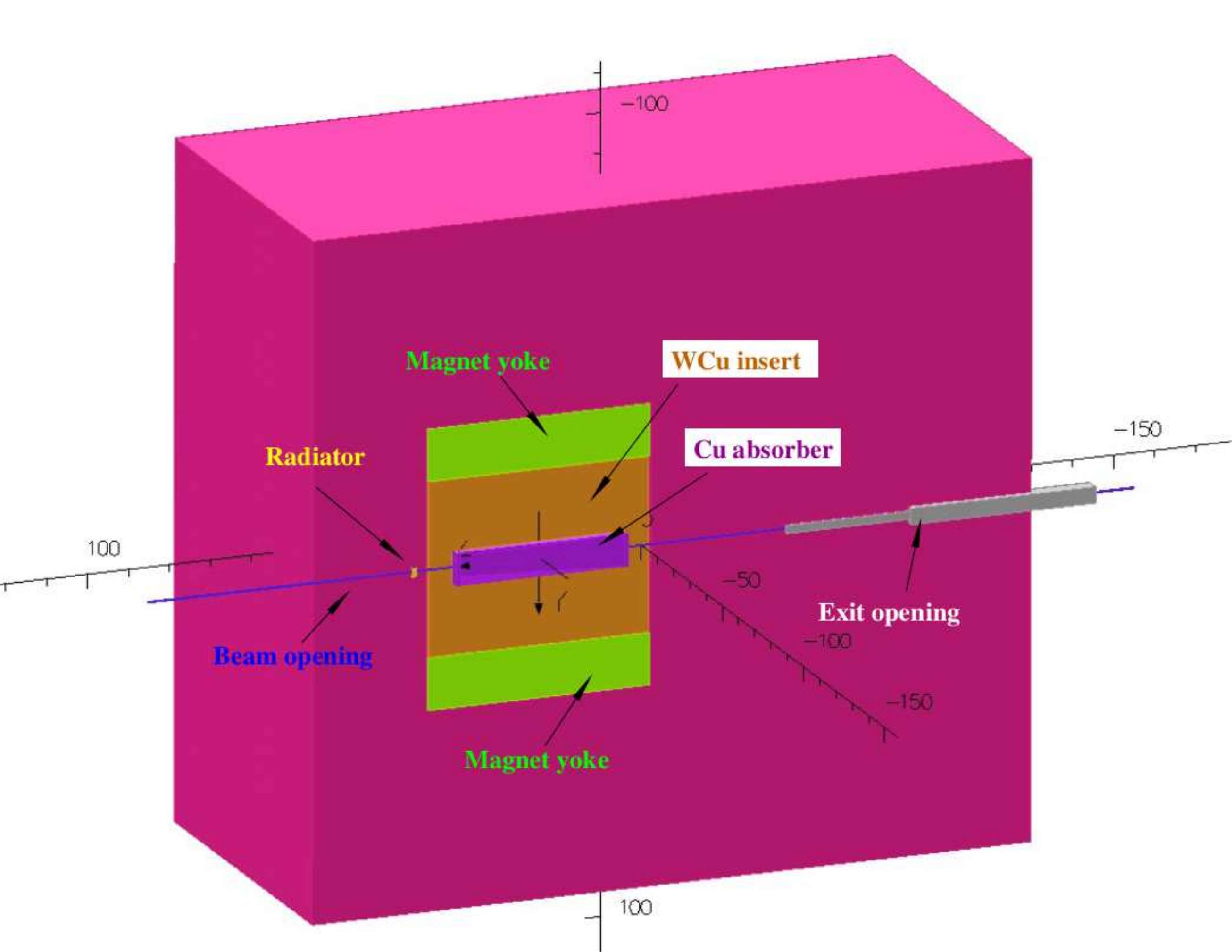}
}

\centerline{\parbox{0.70\textwidth}{
    \caption[] {\protect\small 
    A cut-out diagram view of the CPS. Deflected electrons strike a copper absorber, surrounded 
	by a W-Cu insert inside the magnet yoke. The outer rectangular region in this view is 
	the shielding.  In Hall~D, this will consist primarily of lead.   
	From Ref.~\protect\cite{Day:2019qdz}.} \label{fig:cps1} } }
\end{figure}

\begin{figure}[ht]
\centering
{
    \includegraphics[width=0.75\textwidth,keepaspectratio]{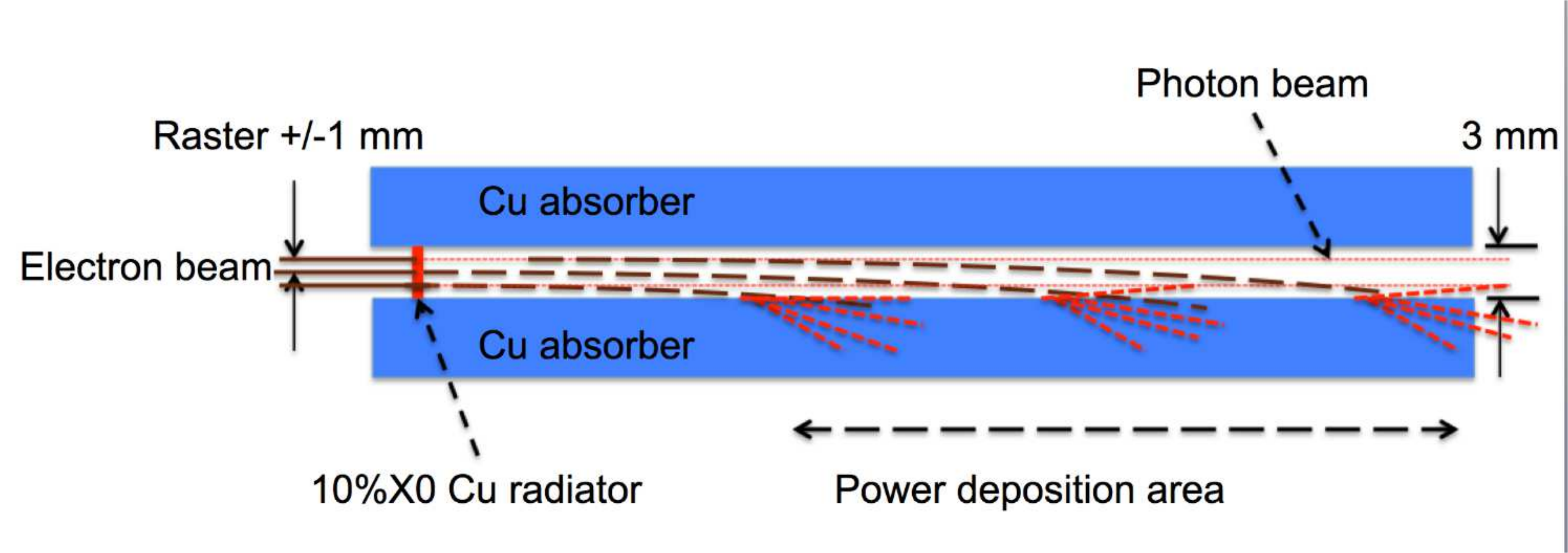}
}

\centerline{\parbox{0.70\textwidth}{
    \caption[] {\protect\small 
    The scheme of beam deflection in the magnetic field to the absorber/dump.  From 
	Ref.~\protect\cite{Day:2019qdz}. In the Hall~D design, the opening will be 1.5~cm 
	vertically and horizontally.}   \label{fig:cps2}  } }
\end{figure}

An intense photon beam is needed to produce the flux of $K_L$ mesons required for this experiment, 
with a photon flux several orders of magnitude larger than that used for current GlueX operations.  
At this very high flux, the current GlueX beam line would require substantial additional shielding 
and other modifications to protect against radiation damage.  Instead, we propose to generate the 
photon beam using a device based on the novel Compact Photon Source (CPS),  currently being developed 
for use in Hall~C,  which combines the photon beam radiator and dump in one properly shielded 
enclosure.  The conceptual design of the CPS is well advanced and has been published earlier this 
year~\cite{Day:2019qdz}.  To briefly summarize the CPS design (illustrated in Fig.~\ref{fig:cps1}): 
immediately downstream of a 10\% R.L. copper radiator lies the electron dump, which consists of a 
copper absorber, a tungsten/copper insert to provide additional shielding, and a radiation-hard 
magnet.  The magnet bends the electrons that have passed the radiator into the absorber, in which 
all of their energy is dissapated,  and only the photons continue through to pass through the 
collimated aperture (see Fig.~\ref{fig:cps2}).  This dump is surrounded by additional shielding, 
primarily lead, to reduce both the prompt photon radiation and the activation dose.  Outside of 
this main mass of shielding lies a 10~cm layer of borated polyethylene and an additional 5~cm of 
lead, the combination of which were found to dramatically decrease the radiation dose due to neutrons.

This design is well matched to the requirements of the KLF photon beam, with a only a few 
modifications required, and can easily fit in the area of the Hall~D Tagger Hall, downstream 
of the current tagging spectrometer. The KLF electron beam power is expected to be twice that 
delivered to the Hall~C CPS (60~kW compared to 30~kW), so the size of the absorber and the 
magnet must be increased to accommodate the extra beam power. Additionally, while the 
polarized target is just downstream of the CPS in the Hall~C configuration, for KLF the 
photon beam must travel 67~m between the tagger and main experimental halls and kept aimed 
at the KPT.  As described in the previous section, an electron beam of width $\sim1.5$~cm in 
both the horizontal and vertical directions is possible with the current Hall D beam line, 
with an acceptable focus on the KPT.  Finally, the design should be kept roughly within the 
weight limit of 150~t.
\begin{figure}[ht]
\centering
{
    \includegraphics[width=1\textwidth,keepaspectratio]{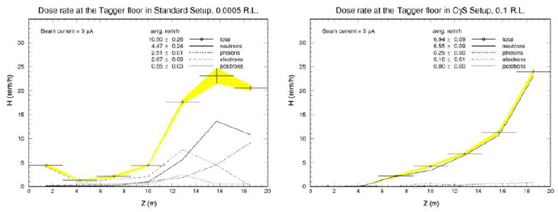}
}

\centerline{\parbox{0.70\textwidth}{
    \caption[] {\protect\small 
    A comparison of dose rate estimates in the Tagger alcove in the two conditions. 
    \underline{Left}: Nominal Hall~D configuration with an amorphous radiator of 0.05\% R.L.
    \underline{Right}: Radiator at 10\% R.L., used as part of the CPS setup. 
    This Figure is taken from Ref.~\cite{Degtyarenko:2016}.}   
    \label{fig:pavel} } }
\end{figure}

Simulations are being performed in order to optimize the power deposition and shielding distribution.  
One major difference in the requirements for the different CPS configurations is in the difference in 
radiation sensitivity in the regions downstream and upstream of the incoming electron beam.  The Hall~C 
CPS is situated just over a meter from a polarized target, and therefore has strict constraints on the 
radiation levels downstream from the CPS.  The Hall~D CPS has few beamline elements downstream before 
the end of the Tagger Hall and continuation of the beamline underground, and so has looser constraints 
on radiation levels compared to the upstream part of the Hall, where the tagged photon spectrometer, 
electronics racks, and other equipment is placed. 

Additionally, we show in Fig.~\ref{fig:pavel} a comparison of radiation dose rates in the Tagger Hall 
at $5~\mu$A beam current between a radiator of 0.05\%~R.L. in the standard GlueX beamline configuration  
and an early configuration of the CPS. Similar levels are seen for both configurations, showing more 
evidence that the CPS will deliver a level of radiation not higher than what could potentially be seen 
in standard GlueX configuration operating at the maximum beam current of $5~\mu$A.

\textbf{To summarize}: The KLF CPS concept provides the intense photon beam required to produce the 
$K_L$ beam, and is being developed in collaboration with the CPS Collaboration based on their published 
conceptual design.  The design is being optimized in parallel with calculations for the rest of the 
beamline, though initial studies show that minor modifications to the baseline design will provide 
sufficient shielding and heat dissipation for the KLF experiment.  The use of lead shielding as 
opposed to the tungsten powder/lead mix for the baseline CPS design will save on the cost.

\subsection{$K_L$ Be Production Target}
Calculations for the Kaon production Target (KPT) were performed for different shielding 
configurations to minimize the neutron and gamma prompt radiation dose rate and cost of the 
KPT~\cite{Strakovsky:2020wtw}.

The prompt background radiation condition is one of the most important parameters of the 
K$_L$ beam for the JLab KL Facility. Beryllium targets were used for K$_L$ production at 
SLAC~\cite{Brody:1969mx} and NINA~\cite{Albrow:1970pd}. We have performed comprehensive 
simulations of the neutron, gamma, and muon backgrounds and their possible influence on the 
proposed measurement. The most important and damaging background comes from neutrons. To 
estimate the neutron and gamma flux in the beam and the neutron prompt radiation dose rate 
in the experimental hall from scattered neutrons and gammas, we used the MCNP6 N-Particle 
(MCNP) radiation transport code~\cite{Werner:2018}.

For the MCNP calculations (in terms of flux [part/s/cm$^2$/MeV] or biological dose rate 
[mrem/h]), many tallies (spots were we calculated a flux or dose rate) were placed along 
the beam at the experimental hall and alcoves for neutron and gamma fluence estimation. 
Fluence-to-Effective Dose conversion factors from ICRP~116~\cite{ICRP:2010} were implemented 
to convert neutron and gamma fluence to effective dose rate. We used the material composition 
data for the radiation transport modeling from Ref.~\cite{PNNL:2006}.

The realism of MCNP simulations is based on the advanced nuclear cross section libraries 
created and maintained by several DOE National Laboratories. The physical models implemented 
in the MCNP6 code take into account bremsstrahlung photon production, photonuclear reactions, 
neutron and photons multiple scattering processes. The experimental hall, collimator alcove, 
and photon beam resulting from the copper radiator within CPS were modeled using the 
specifications from the layout presented in Figure~\ref{fig:hall}, shown as a 3D graphic 
model of the experimental setup.
\begin{figure}[ht]
\centering
{
    \includegraphics[width=0.4\textwidth,keepaspectratio]{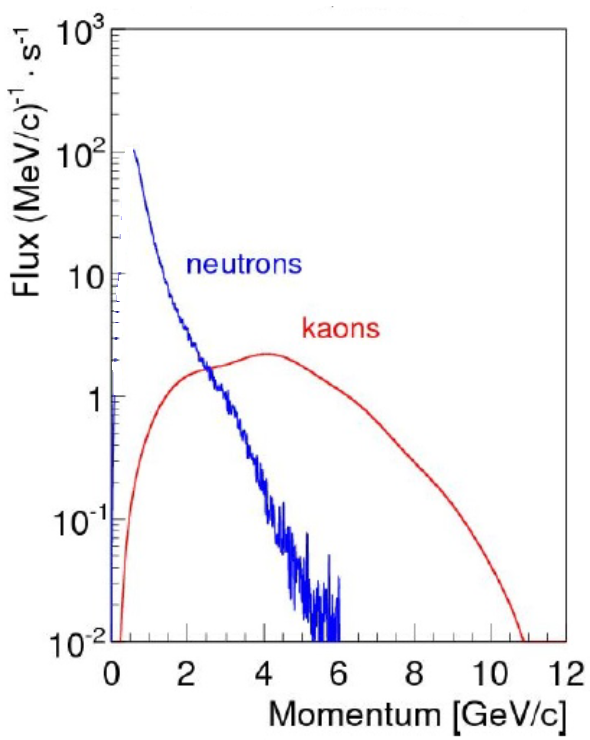}~~~
    \includegraphics[width=0.4\textwidth,keepaspectratio]{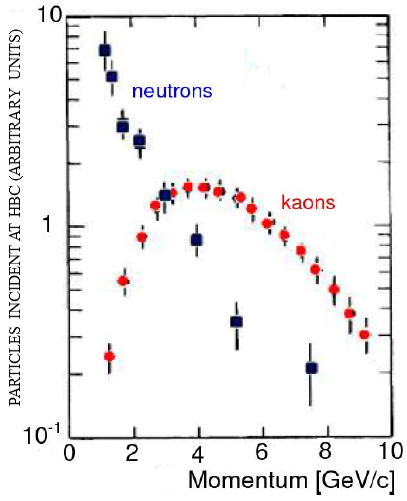}
}

\centerline{\parbox{0.70\textwidth}{
 \caption[] {\protect\small 
    The K$_L$ and neutron momentum spectra on the  
    cryogenic target.
   \underline{Left}: Rate of $K_L$ (red) and neutrons (blue) on the LH$_2$/LD$_2$ cryogenic 
	target of Hall~D as a function of their generated momenta, with a total rate of 
	$1\times10^4~K_L/sec$ and $6.6\times10^5~n/sec$, respectively. Kaon calculations 
	were performed using Pythia generator~\cite{Pythia} while neutron calculations were 
	performed using the MCNP transport code~\protect\cite{Werner:2018}.
   \underline{Right}: Experimental data from SLAC measurements using a 16~GeV/$c$ electron 
	beam were taken from Ref.~\protect\cite{Brandenburg:1972pm} (Figure~3).} 
   \label{fig:yield}  } }
\end{figure}

\subsubsection{Kaon and Neutron Flux:} 
Neutral kaon production was simulated for a photon bremsstrahlung beam produced by the 12~GeV 
electron beam in the Hall~D CPS. The main mechanism of $K_L$ production in our energy range is 
via $\phi$-meson photoproduction, which yields the same number of $K^0$ and $\bar{K^0}$. 
Calculations of the $K_L$  flux~\cite{Larin:2016} are performed using the Pythia MC 
generator~\cite{Pythia}, while the neutron flux calculations were performed using the MCNP 
radiation transport code~\cite{Werner:2018}.

The MCNP model simulates a 12~GeV 5~$\mu$A electron beam hitting the copper radiator inside 
of the CPS. Electron transport was traced in the copper radiator, vacuum beam pipe for 
bremsstrahlung photons, and Be-target. Neutrons and photons were traced in all components of 
the MCNP model. The areas outside the concrete walls of the collimator alcove and bremsstrahlung 
photon beam pipe was excluded from consideration to facilitate the calculations. Additionally, 
we ignore Pair Spectrometer (PS)~\cite{Barbosa:2015bga} and KFM magnets but took into account 
five iron-blocks around beam pipe in front of the GlueX spectrometer.

Figure~\ref{fig:yield} demonstrates that our simulations for the KLF kaon and neutron flux 
(Fig.~\ref{fig:yield} (left)) are in a reasonable agreement with the $K_L$ spectrum measured 
by SLAC at 16~GeV~\cite{Brandenburg:1972pm} (Fig.~\ref{fig:yield} (right)).

\subsubsection{Target and Plug Materials:}
The $K_L$ beam will be produced with forward emission kinematics due to the interaction of the 
photon beam with a Be-target. Beryllium is used because lighter elements have a higher 
photoproduction yield with a lower absorption of kaons, as pointed out in previous SLAC 
studies~\cite{Brandenburg:1972pm}.  These studies showed that beryllium is the optimal 
material for neutral kaon photoproduction. Another common target material is carbon, which is 
easier to handle than beryllium, however the simulations we performed show that a beryllium 
target performs significantly better than a similar target made of carbon. The Pythia~\cite{Pythia} 
simulations showed that the kaon yield from beryllium is higher than that from carbon at the same 
radiation length. The ratio of beryllium to carbon gives a factor of 1.51 for kaon yield. At the 
same time, MCNP simulations demonstrated that the beryllium target reduces the neutron yield more 
effectively than carbon. The ratio of generated particles from beryllium to the carbon appears to 
be about $\sim$~1.45 for neutrons. 
\begin{figure}[ht]
\centering
{
    \includegraphics[width=0.6\textwidth,keepaspectratio]{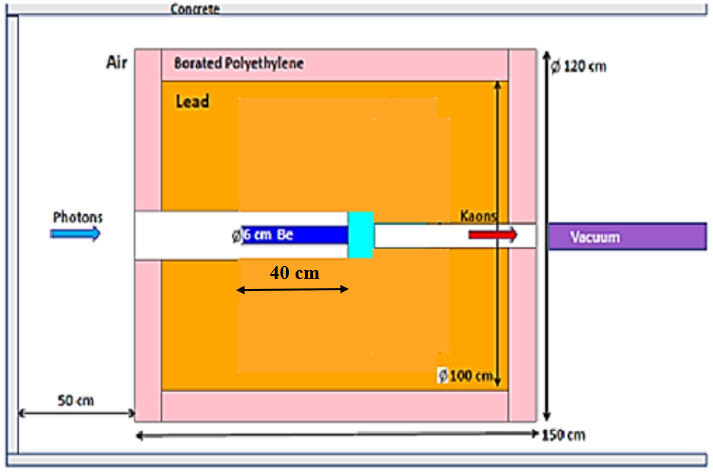}
}

\centerline{\parbox{0.70\textwidth}{
    \caption[] {\protect\small 
    Schematic view of the Be-target ($K_L$ production target) assembly. Concrete, borated polyethylene, 
	lead, tungsten, beryllium, vacuum beam pipe, and air shown by grey, pink, brown, light blue, 
	blue, violet, and white color, respectively. Beam goes from left to right.} \label{fig:be} } }
\end{figure}

A tungsten beam plug of a 10~cm thickness (30~X$_0$) and 16~cm diameter is attached to the beryllium 
target (Fig.~\ref{fig:be}) to clean up the beam and absorb induced radiation. In the same SLAC studies 
referenced above, it was shown that tungsten is the optimal material for the plug and that tungsten 
has a lower absorption factor for kaons as compared to copper. Our Pythia simulations showed that the 
ratio of tungsten to copper (20\%) gives 1.16 (1.36) at kaon momentum 1~GeV/$c$ (0.5~GeV/$c$). Our MCNP 
simulations additionally demonstrated that the tungsten plug reduces the yield of neutrons and gamma 
compared to a plug of lead or copper of the same length. The production ratio for lead (copper) to 
tungsten is 2.25 (9.29) for neutrons and 8.11 (66.8) for gammas.

It was found that increasing the plug diameter will increase the neutron background. For example, 
increasing the diameter to 24~cm from 16~cm in diameter yields an increase of neutron production by 
a factor of 2.8. This effect is due to re-scattered neutrons in the plug.  There is no effect for 
gammas.  It was also found that increasing the plug length will decrease the neutron background.  
For example, increasing the length to 15~cm from 10~cm in length gives a factor of 0.60 in neutron 
production. For gammas, the effect is more significant. However, we do not plan to increase 
the length to prevent similar losses in $K_L$ yield.

\subsubsection{Location of the Be-target Assembly:}
To reduce the effect of the neutron and gamma background coming from the beryllium target and tungsten 
plug into the experimental hall, we place the KPT upstream of the GlueX spectrometer in the collimator 
alcove (Fig.~\ref{fig:hall}). Additional shielding inside the collimator alcove is added to minimize 
the neutron and $\gamma$ background in the experimental hall and to satisfy the JLab RadCon requirement 
establishing the radiation dose rate limit in the experimental hall (1~mrem/h), roughly based on the 
requirement to limit the yearly dose accumulation at the CEBAF boundary at 10~mrem. The key area for 
the dose rate evaluation is the area of $(6\times 6)~m^2$ on ceiling of the experimental hall centered 
above the GlueX detector. The dose rate limit at that location roughly correspond to the expected dose 
rate at the CEBAF fence at the level of 1~$\mu$rem/h, both evaluated, and observed at other locations 
at CEBAF (vicinity of the high power End Stations of Halls~A and C). The vacuum beam pipe (between KPT 
and cryogenic target) prevents neutrons re-scattering in the air in the experimental hall. Directly 
downstream of the Be target there will be a sweeping magnet with a field integral of $0.8~T\cdot m$ 
to clean up the charged particle component from the beam (including muons).
\begin{figure}[ht]
\centering
{
    \includegraphics[width=0.75\textwidth,keepaspectratio]{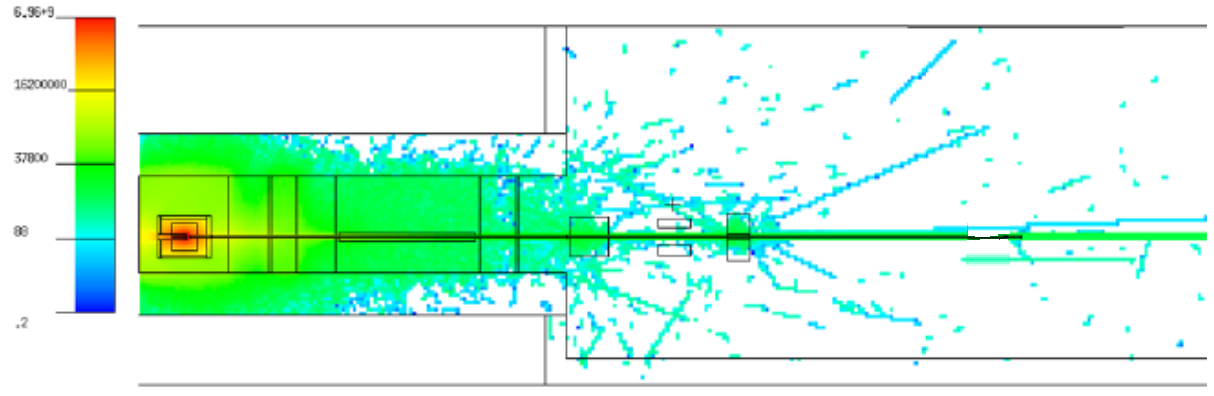}
    \includegraphics[width=0.75\textwidth,keepaspectratio]{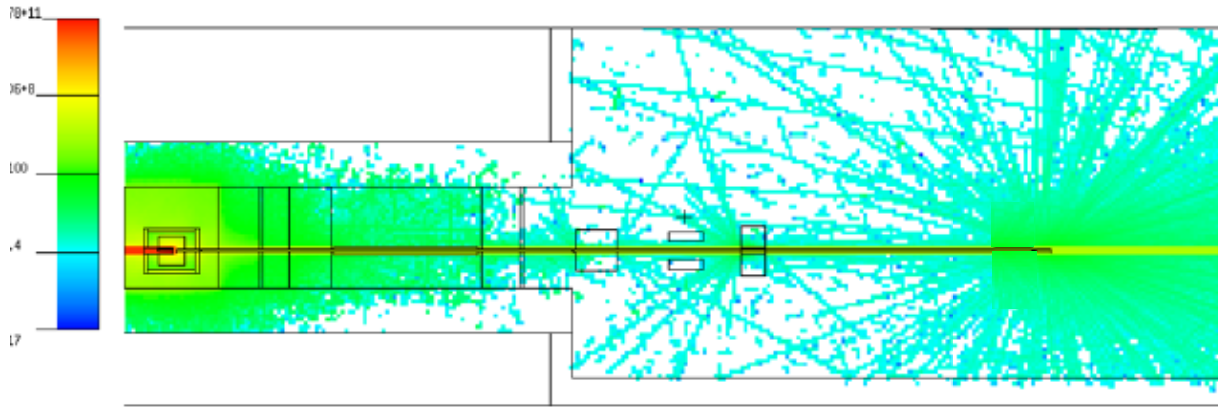}
}

\centerline{\parbox{0.80\textwidth}{
    \caption[] {\protect\small 
    Prompt neutron and gamma yield.
    \underline{Top}: (\underline{Bottom}:) Vertical cross section of the neutron (gamma) flux calculated 
	using the MCNP model. Beam goes from left to right.}   \label{fig:prompt} } }
\end{figure}

\subsubsection{Design of the Be Target Assembly:} 
A schematic view of the Be-target assembly is given in Fig.~\ref{fig:be}.  Changing from the photon to 
the kaon beam line and vice versa is expected to take about half year or less, and thus should fit well 
into the breaks of current CEBAF schedule.  The collimator alcove has enough space (with 4.52~m width) 
for Be-target assembly to remain far enough from the beam line (Fig.~\ref{fig:hall}) and to not obstruct 
GlueX operations in regular photon beam mode. The water cooling is available in experimental hall, and 
is sufficient to dissipate 6~kW of power delivered by the photon beam to Be-target and W-plug.

The conceptual design of KPT with combination of lead shielding and tungsten plug is shown on 
Fig.~\ref{fig:be}. The prompt radiation dose rate for neutrons (gammas) in the experimental hall at the 
key area for RadCon on the ceiling is (0.35$\pm$0.17)~mrem/h (0.078$\pm$0.005~mrem/h).  Replacing all 
tungsten by lead (including plug), one can get (0.61$\pm$0.25)~mrem/h ((0.527$\pm$0.006)~mrem/h). 
Finally, in the case of lead shielding and tungsten plug (Fig.~\ref{fig:be}), one can get 
(0.27$\pm$0.08)~mrem/h ((0.065$\pm$0.002)~mrem/h). The prompt neutron and gamma yield calculated by 
MCNP code is demonstrated in Fig.~\ref{fig:prompt}.

\textbf{To summarize}: The optimization of the Be-target assembly resulted  in the weight of the device 
12~t and the estimated cost of \$0.134M (note that the final total cost depends on the cost of tungsten). 
\subsection{$K_L$ Flux Monitor}
\begin{figure}[ht]
\vspace{-5mm}
\centering
{
       \includegraphics[width=0.75\textwidth,keepaspectratio]{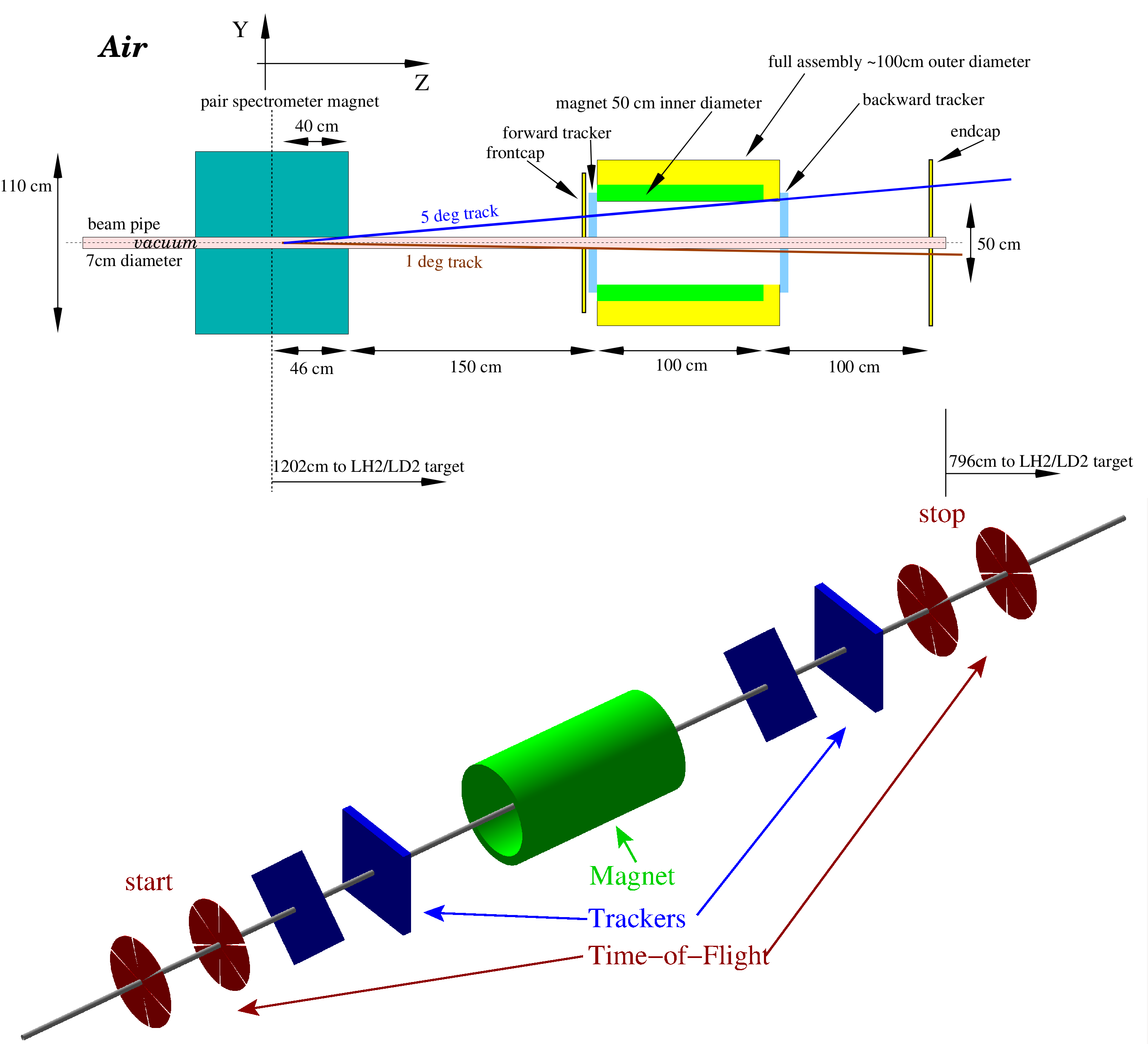} }
\centerline{\parbox{0.70\textwidth}{
    \caption[] {\protect\small 
    \underline{Top}: Schematic view of the Kaon Flux Monitor setup.  
    \underline{Bottom}: Expanded view of the KFM GEANT4 simulation.
    Beam goes from left to right.}
    \label{fig:FM_setup} } }
\end{figure}

An accurate determination of the $K_L$ beam flux is necessary to maximize the physics impact 
of the resulting data~\cite{Misha}. To reach an accuracy of <5~\% in the determination of the 
flux, we plan to build a dedicated flux monitor= which would utilize in-flight decays of the 
$K_L$. To account for various possible acceptance effects during $K_L$ beam propagation from 
the Be-target, we plan to measure the $K_L$ flux upstream of the GlueX detector, utilizing the 
Hall~D Pair Spectrometer~\cite{Barbosa:2015bga} as shielding against $K_L$ which have decayed 
further upstream. The Kaon Flux Monitor (KFM) will measure a small fraction of decayed $K_L$'s, 
concentrating on the portion decaying within a distance of 2~m downstream of the Pair Spectrometer 
magnet center (see Fig.~\ref{fig:FM_setup}). The KFM is a combination of solenoid magnetic field 
spectrometer and a time of flight detectors. The KFM consists of the following major parts: the 
front cap, forward tracker, backward tracker, endcap, and solenoidal magnet. The KFM can be further 
equipped with a plastic scintillator barrel, covering inner part of a magnet and a Flux Monitor 
Start Counter (FMSC), comprising plastic scintillator bars covering the beampipe, from the location 
of the Pair Spectrometer magnet to the KFM magnet (see Fig.~\ref{fig:FM_setup}). All our MC 
calculations have been performed for the JLab Hall~D beamline geometry. 
\begin{figure}[htb!]
\vspace{-3mm}
\centering
{
    \includegraphics[width=0.47\textwidth,keepaspectratio]{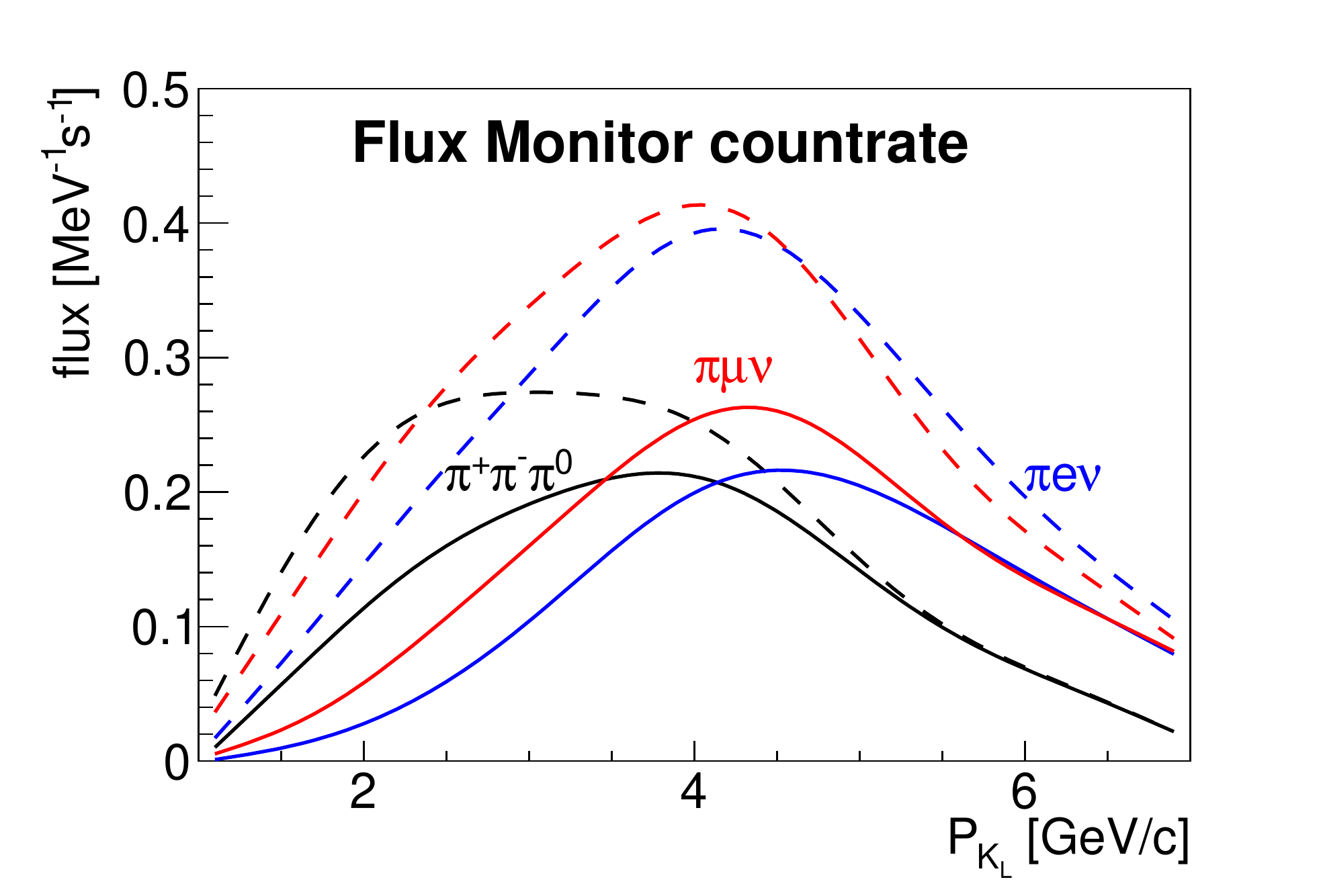}}
{
    \includegraphics[width=0.47\textwidth,keepaspectratio]{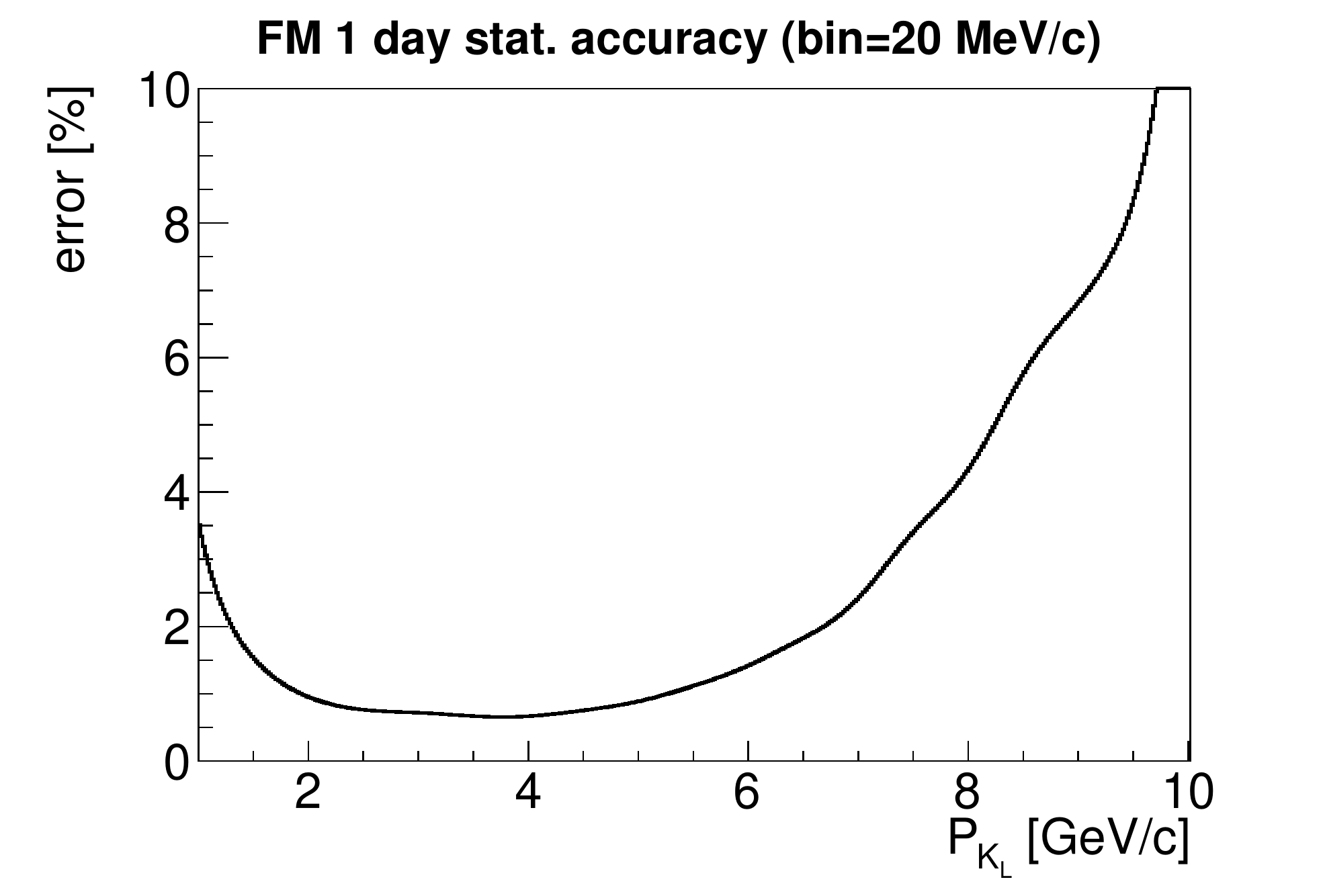} }

\centerline{\parbox{0.80\textwidth}{
    \caption[] {\protect\small 
    The Kaon Flux Monitor performance.
    \underline{Left}: Visible $K_L$ flux for various decay 
    channels within the KFM acceptance. Solid lines correspond to a system with front/end-caps 
	only. Dashed lines show the improvement one can obtain with the additional barrel part 
	extension to the FM.
    \underline{Right}: Expected statistical accuracy for 1~day FM measurement ($\pi^+\pi^-\pi^0$ 
	branch only) in 20~MeV/$c$ momentum bin.}  \label{fig:flux} } }
\end{figure}

To be measured by the KFM, both charged particles from the kaon decay need to be incident within 
the KFM acceptance. Taking into account the different branching ratios, we expect to reconstruct 
the following number of $K_L$ from various decay channels (see Fig.~\ref{fig:flux} (left)). One 
can quantify the expected rate in terms of the achievable statistical error within a one day 
measurement (see Fig.~\ref{fig:flux} (right)). For the kaon beam momenta range appropriate for 
the hyperon  program a 1~\% statistical error of the $K_L$ flux determination is achievable in 
less than a day.

An accurate flux monitoring requires determination of the kaon flux as both a function of transversal 
position within the beampipe and kaon energy. The most inner 3~cm of the transverse beam profile at 
the position of the KFM would correspond to a 6~cm profile at the cryogenic target. 
A $\varnothing$7~cm beam pipe allows sufficient margins and the clean definition of fiducial regions 
of the transverse beam profile at the KFM position. All in all we expect to measure $1\times 
10^4~K_L/sec$ in the KFM.
\newpage
\subsection{$K_L$ Beam Properties and Neutron, Gamma, and Muon Background}
\subsubsection{Neutron and Gamma Background:}
\begin{figure}[ht]
\centering
{
    \includegraphics[width=0.46\textwidth,keepaspectratio]{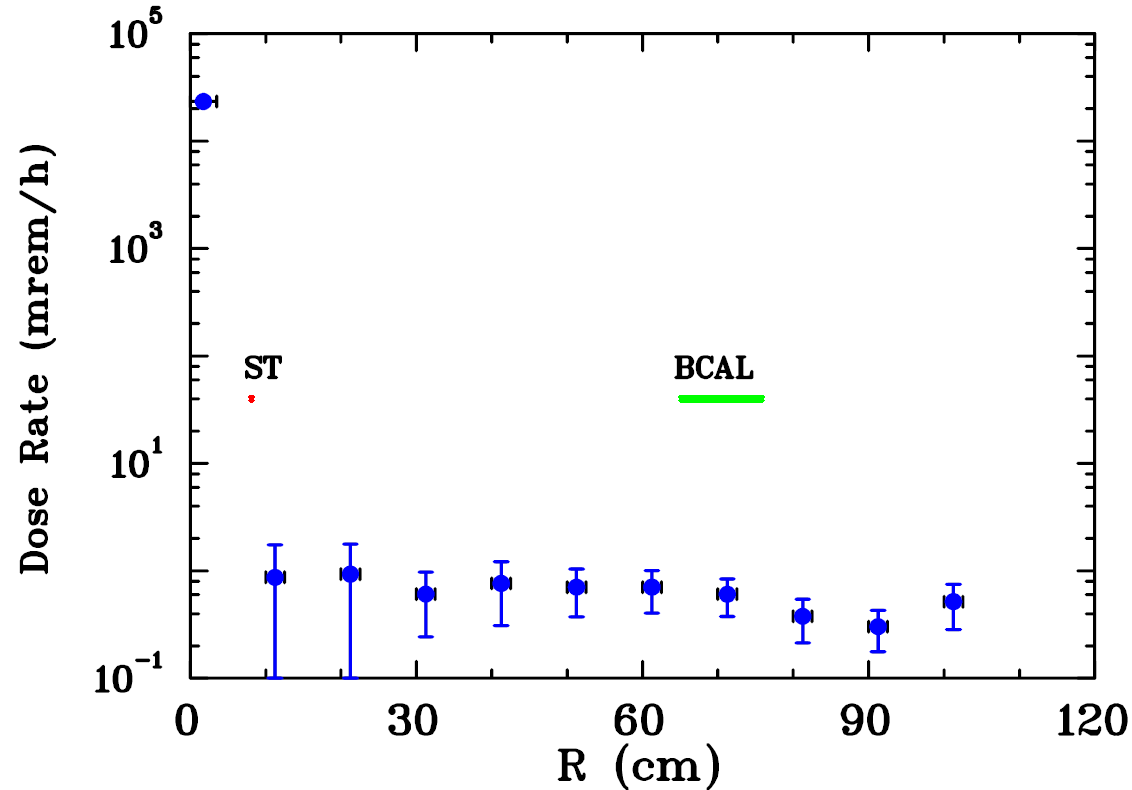}
    \includegraphics[width=0.45\textwidth,keepaspectratio]{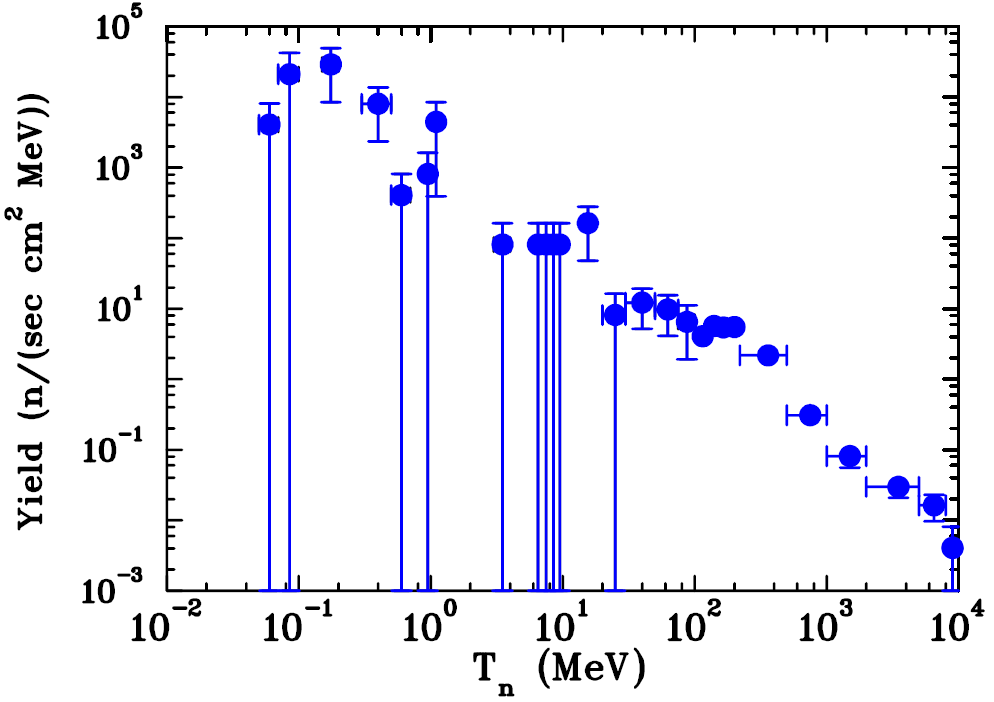}
}

\centerline{\parbox{0.80\textwidth}{
    \caption[] {\protect\small 
    \underline{Left}: Neutron prompt radiation dose rate background calculated for SiPM of SC 
	and BCAL on the face of the cryogenic target. In this case, we did not take into account 
	additional shieldings in the experimental hall.  
    \underline{Right}: Neutron energy spectrum at the beam on the face of the cryogenic target.} 
	\label{fig:rad} } }
\end{figure}

The neutron flux on the face of the $LH_2/LD_2$ cryogenic target is found to be  
$6.6\times10^5~n/sec$. This energy spectrum of this flux drops exponentially to 10~GeV 
(Fig.~\ref{fig:rad} (right)). The SiPM detectors are only sensitive to neutron energies 
above 1~MeV~\cite{Somov:2011}. The prompt neutron dose rate for the silicon photomultipliers 
(SiPMs) of the Start Counter (SC)~\cite{Qiang:2012zh,Pooser:2019rhu,pooser-thesis,
Degtiarenko:2011} and Barrel Electromagnetic Calorimeter (BCAL)~\cite{Beattie:2018xsk,
Degtiarenko:2011} is given in Fig.~\ref{fig:rad} (left). The SiPMs used in the SC and BCAL 
are expected to tolerate the calculated neutron background shown in Fig.~\ref{fig:rad}.  
Previous studies state that the dose rate of 30~mreh/h increases a dark current at SiPM by 
a factor of 5 after 75~days of running period~\cite{Somov:2011}, and the expected dose is 
well below this rate.  In the extreme worst case, it is reasonable to expect to replace the 
affected SiPMs once per year. 

The flux is additionally not sufficient to provide any significant background in the case of 
$np$ or $nd$ interactions in the cryogenic target (see Sec.~\ref{sec:neutron_sims} for details).

\subsubsection{Muon Background:}
\begin{figure}[ht]
\centering
{
    \includegraphics[width=0.45\textwidth,keepaspectratio]{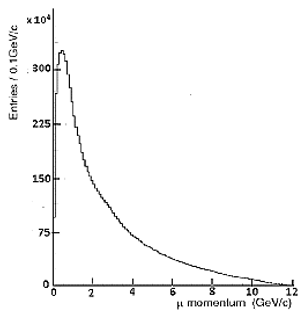}
}

\centerline{\parbox{0.80\textwidth}{
 \caption[] {\protect\small 
     Muon momentum spectrum for the Bethe-Heitler
    (see details in text).
    } \label{fig:muon} } }
\end{figure}

Following Keller~\cite{Keller:2015}, our Geant4~\cite{Allison:2016lfl} simulations included 
Bethe-Heitler muon background from KPT and photon dump at CPS, both of which contribute to 
the background at the GlueX detector and the muon dose rate outside Hall~D~\cite{Larin:2016}. 
The number of produced muons in the Be-target and W-plug are about the same, but muons originating 
in tungsten have much softer momenta.  The estimated number of produced muons is less than 
$10^7~\mu/sec$. Their momentum spectrum is shown in Fig.~\ref{fig:muon}.  Half of the muons 
will have momenta higher than 2~GeV/$c$, 10\% of them will have momenta higher than 6~GeV/$c$, 
and 1\% will have momenta above 10~GeV/$c$. Overall, the muon flux for the KLF experiment is 
tolerable to the RadCon requirement, and such muons are deflected well by the sweeping magnet 
downstream of the target.

\textbf{To summarize}: Calculations for KPT were performed for different shielding configurations 
to minimize the neutron and gamma prompt radiation dose rate and the cost of KPT. Our studies have 
shown that the KPT will produce a high K$_L$ flux of the order of $1\times 10^4~K_L/sec$ and the 
neutron and gamma fluxes and prompt dose rates for the KLF experiment are below the JLab RadCon 
requirement establishing the radiation dose rate limits in the experimental hall. 
\subsection{$K_L$ Momentum Determination and Beam Resolution}
The mean lifetime of the $K_L$ is 51.16~nsec ($c\tau = 15.3$~m) whereas the mean lifetime 
of the $K^-$ is 12.38~nsec ($c\tau = 
3.7$~m)~\cite{Tanabashi:2018oca}.  For this reason, it is much easier to perform measurements 
of $K_Lp$ scattering at low beam energies compared with $K^-p$ scattering.
\begin{figure}[ht]
\centering
{
    \includegraphics[width=0.25\textwidth,keepaspectratio]{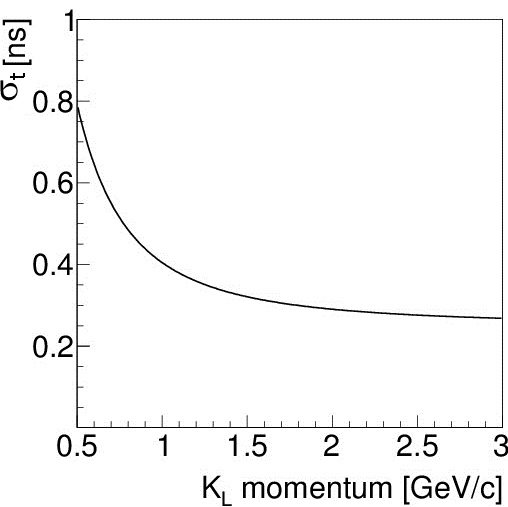} } 
{
    \includegraphics[width=0.26\textwidth,keepaspectratio]{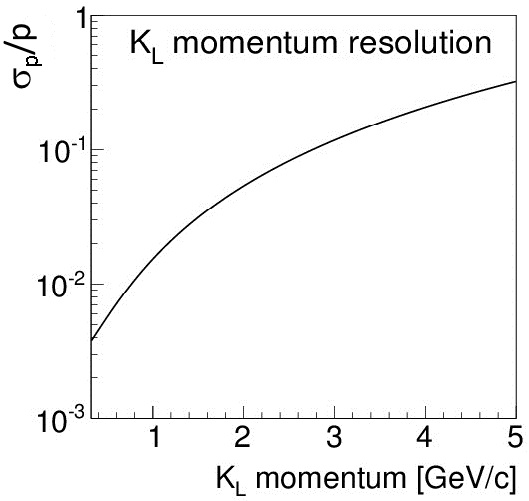} }
{
    \includegraphics[width=0.36\textwidth,keepaspectratio]{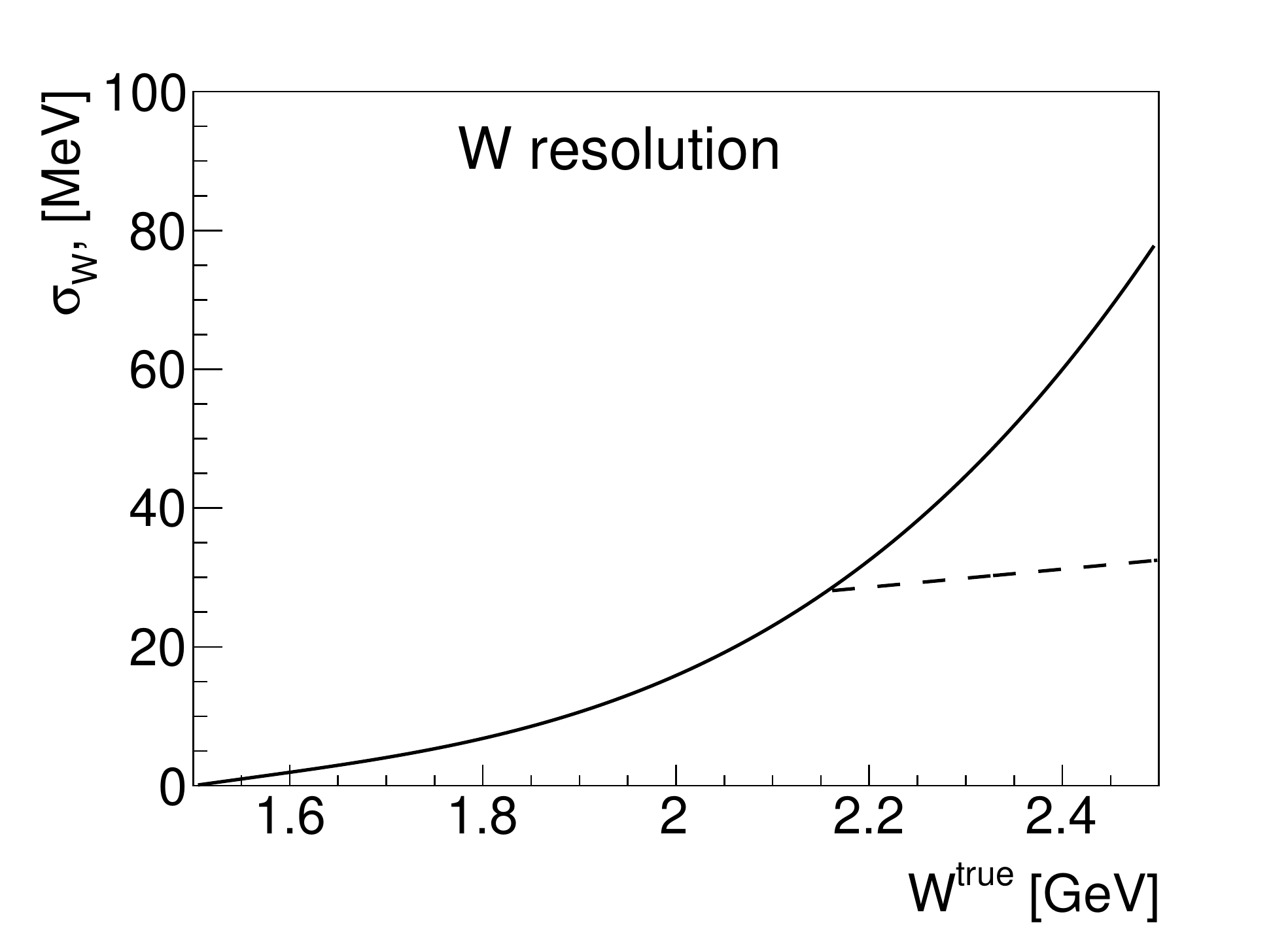} 
}
\centerline{\parbox{0.70\textwidth}{
    \caption[] {\protect\small 
    \underline{Left}: Time resolution ($\sigma_t$) for $K_L$ beam as a function of $K_L$-momentum.
    \underline{Middle}: Momentum resolution ($\sigma_p/p$) as a function of momentum (note, log scale).
    \underline{Right}: Energy resolution ($\sigma_W$) as a 
    function of energy. The dashed line shows approximate $W$ resolution from reconstruction of the 
	final-state particles.} \label{fig:mom} } }
\end{figure}

The momentum of a K$_L$ beam will be measured using time-of-flight (TOF) - the time between the 
accelerator bunch (RF signal from CEBAF) and the reaction in the LH$_2$/LD$_2$ target as detected 
by the GlueX spectrometer. Thus the TOF resolution is a quadratic sum of accelerator time and GlueX 
spectrometer time resolutions. Since the accelerator signal has a very good time resolution on the 
order of few picoseconds, the TOF resolution will be defined mainly by the GlueX detector. In our 
calculations, we used currently achieved Start Counter (SC) time resolution of 250~psec to show the 
time and beam momentum resolution vs. kaon momentum (Fig.~\ref{fig:mom}).  All hyperon production 
reactions have very similar TOF and final state reconstructed $W$-resolution.

To get precise TOF information, the electron beam needs to have a narrow bunch time structure. As 
discussed in Sec.~\ref{sec:beamT}, the electron beam can be delivered with predetermined repetition 
rate. For the $K_L$ experiment, the 64~nsec bunch spacing structure is an optimal choice. It allows 
no cross-bunch overlap for the full range of kaon beam momentum from p > 320~MeV/$c$. 

The uncertainty in a neutral kaon production position at lower momenta
(p < 0.5~GeV/$c$) affects timing resolution caused by the TOF difference between the photon and kaon 
time traversing the Be~target, however, as $\Delta p/p = \gamma^2\Delta t/t$ momentum resolution is 
below 1~\% at lower momenta. The TOF resolution is flat for momenta higher than 1~GeV/$c$. The momentum 
resolution decreases with momentum: for 1~GeV/$c$ it is $\sim$1.5~\% and for 2~GeV/$c$ it is $\sim$5~\%.  
For fully reconstructed final states $W$ can be reconstructed directly, providing a better resolution 
in the region where the TOF method deteriorates, W > 2.2~GeV (see dashed curve in Fig.~\ref{fig:mom} 
(right)).

The $K_L$ beam momentum and time resolution are governed by the time resolution provided by the GlueX 
detector from the reconstruction of charged particles produced in the LH$_2$/LD$_2$ target.  There are 
three detector systems that can provide precision timing information for reconstructed charged particles 
in GlueX: the SC~\cite{Pooser:2019rhu}, Barrel Calorimeter (BCAL)~\cite{Beattie:2018xsk}, and the TOF 
detectors.  In the current studies, we rely on the SC for beam momentum determination.

The GlueX SC is a cylindrical plastic scintillator detector surrounding the LH$_2$/LD$_2$ target, with 
3~mm thick scintillator bars and a tapered nose region that bends toward the beamline at the downstream 
end.  The scintillation light from each of the 30 scintillator bars is detected by an array of four 
$3\times 3~{\mathrm{mm^2}}$ Hamamatsu S10931-050P surface mount silicon photomultipliers 
(SiPMs)~\cite{pooser-thesis}.  The time resolution of the SC was determined to be 250~psec during the 
2016 and 2017 GlueX run periods and thus provided adequate separation of the 250~MHz photon beam bunch 
structure delivered to Hall~D during that time.  This performance was achieved using the recommended 
operating gain and bias voltages supplied by Hamamatsu to provide both the FADC 250 analog signals and 
precision F1TDC discriminator signals used in the GlueX reconstruction.

\textbf{To summarize}: The simulation studies in this proposal have assumed a time resolution of 250~psec, 
which is adequate for the proposed physics program. With the current detector, the overall K$_L$-momentum 
resolution will be determined by utilizing the timing information from the SC, BCAL, and TOF detectors and 
will probably overshoot a very conservative 250~psec specification.  Finally, we are exploring potential 
upgrades to improve the SC time resolution significantly; however, such improvements would not influence 
much on the resonance parameters extracted by the PWA, hence they have low priority for the proposed 
hyperon spectroscopy program. 
\subsection{LH$_2$/LD$_2$ Cryogenic Target for KL Beam at Hall~D} 
\begin{figure}[ht]
\centering
{
    \includegraphics[width=0.8\textwidth,keepaspectratio]{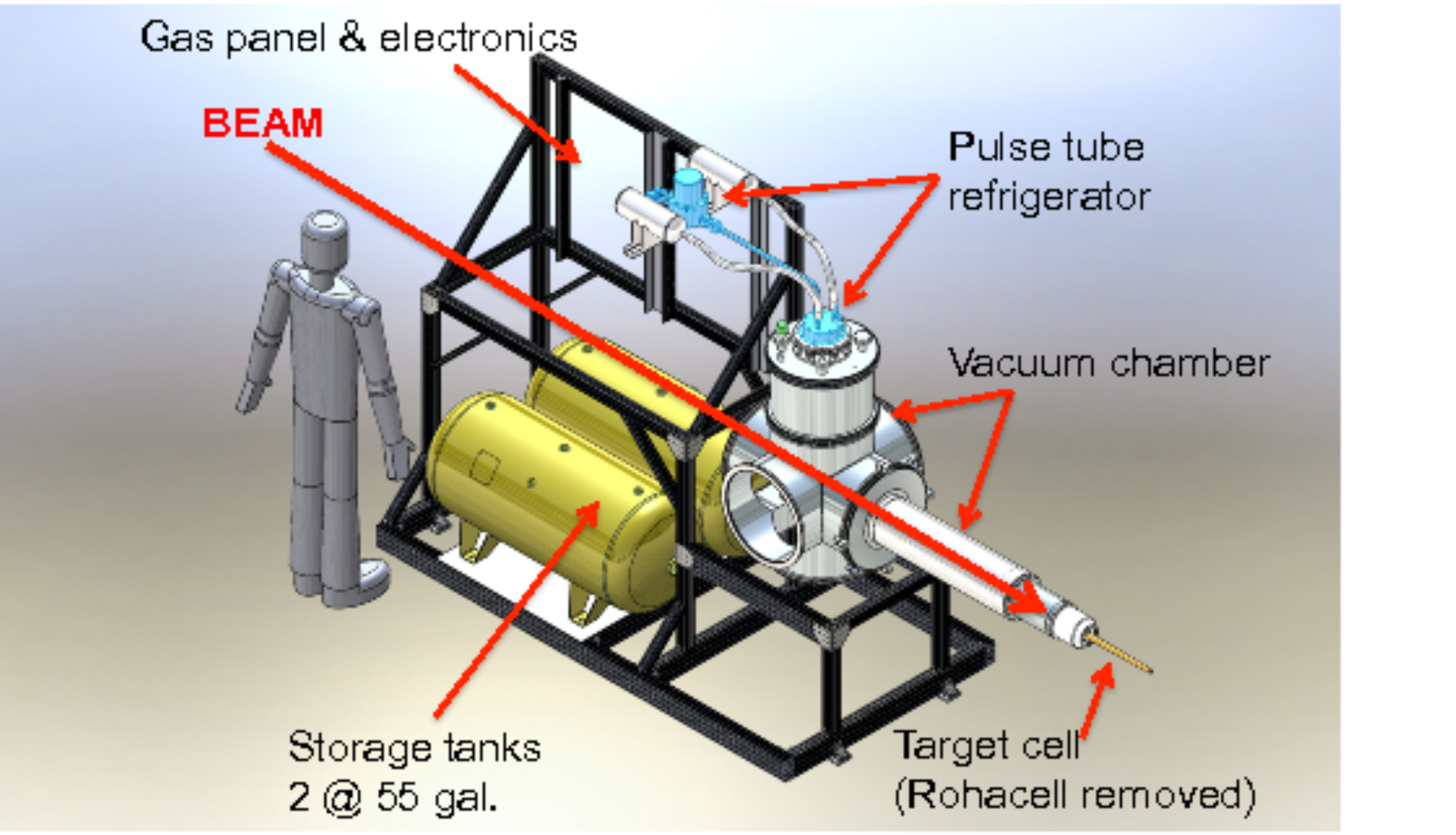} }

{
    \includegraphics[width=0.45\textwidth,keepaspectratio]{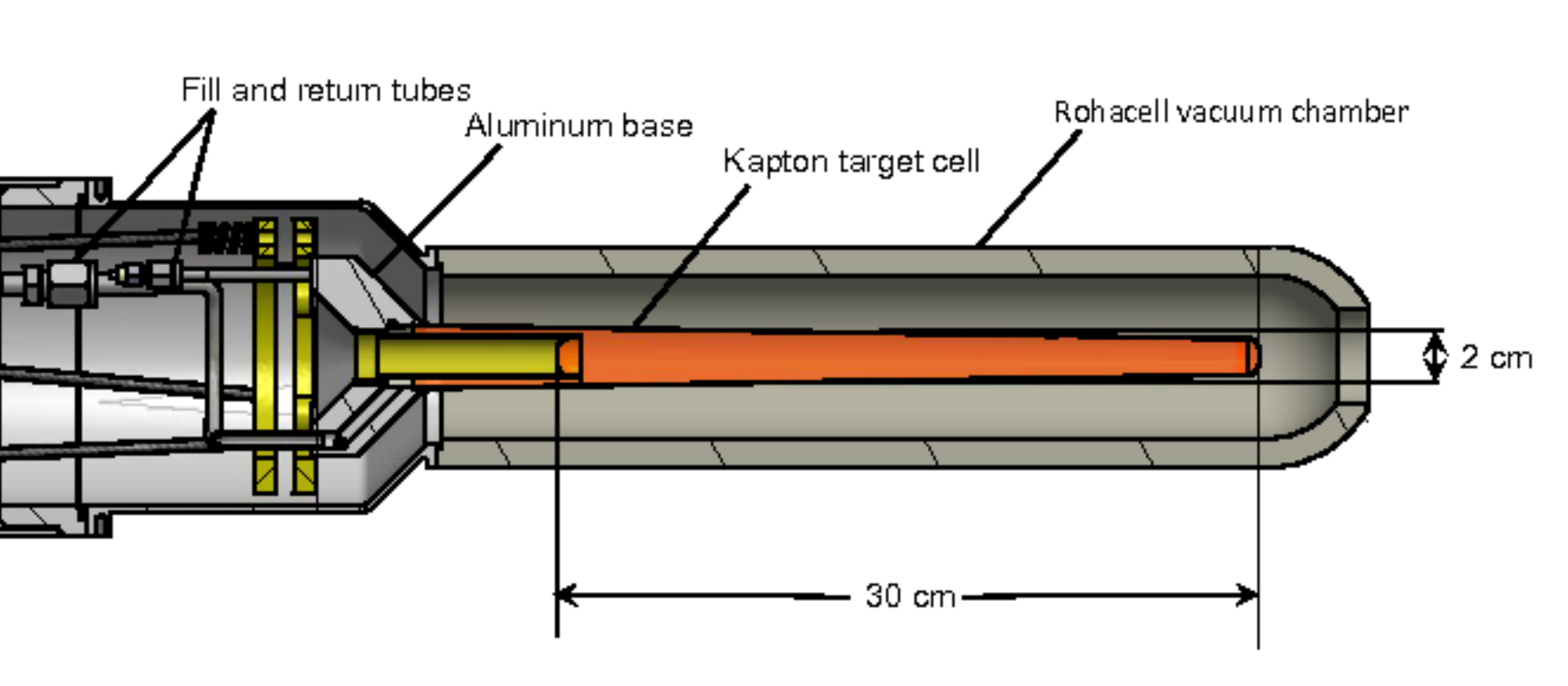} }
{
    \includegraphics[width=0.45\textwidth,keepaspectratio]{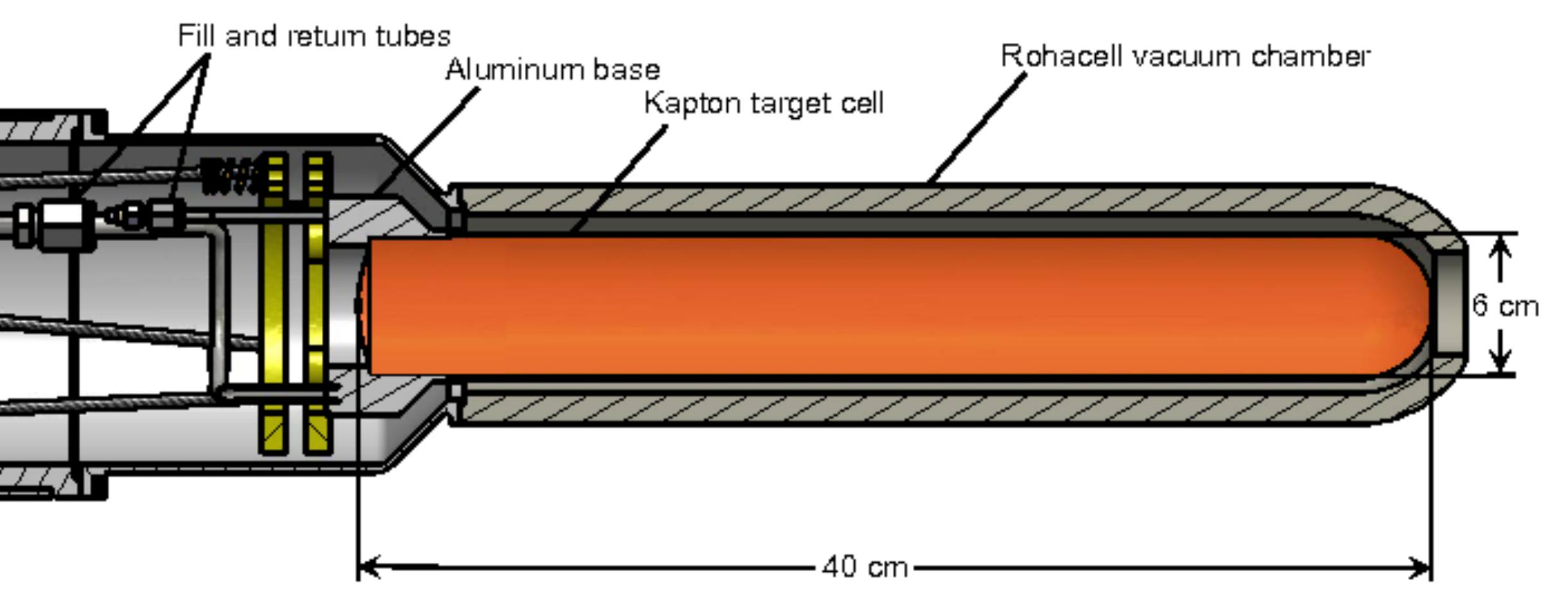} }

  \centerline{\parbox{0.80\textwidth}{
    \caption[] {\protect\small
    \underline{Top}: The GlueX liquid hydrogen target.
    \underline{Bottom left}: Kapton target cell for the GlueX 
    LH$_2$/LD$_2$ cryogenic target.
    \underline{Bottom right}: Conceptual design for a larger target cell for the proposed $K_L$ 
	beam at Hall~D experiment.}  \label{fig:lh2a} } }
\end{figure}

The proposed experiment will utilize the existing GlueX liquid hydrogen cryogenic target 
(Fig.~\ref{fig:lh2a}) modified to accept a larger diameter target cell~\cite{Keith:2016rfh}. 
The GlueX target is comprised of a kapton cell containing liquid hydrogen at a temperature and 
pressure of about 20~K and 19~psia, respectively. The 100~ml cell is filled through a pair of 
1.5~m long stainless steel tubes (fill and return) connected to a small container where hydrogen 
gas is condensed from two room-temperature storage tanks. This condenser is cooled by a pulse 
tube refrigerator with a base temperature of 3~K and cooling power of about 20~W at 20~K. A 
100~W temperature controller regulates the condenser at 18~K.

The entire target assembly is contained within an ``L"-shaped
stainless steel and aluminum vacuum chamber with a Rohacell extension 
surrounding the target cell. The SC for the GlueX experiment fits 
snugly over this extension. The vacuum chamber, along with the 
hydrogen storage tanks, gas handling system, and control electronics, 
is mounted on a custom-built beamline cart for easy insertion into the 
Hall~D solenoid. A compact I/O system monitors and controls the 
performance of the target, while hardware interlocks on the target 
temperature and pressure and on the chamber vacuum ensure the system's 
safety and integrity. The target can be cooled from room temperature 
and filled with liquid hydrogen in about 5~hours. For empty target
runs, the liquid can be boiled from the cell in about 20 minutes (the 
cell remains filled with cold hydrogen gas), and then refilled with 
liquid in about 40 minutes.

The GlueX cell (Fig.~\ref{fig:lh2a} (bottom left)) is closely modeled 
on those utilized at Hall~B for more than a decade and is a horizontal, tapered cylinder 
about 0.38~m long with a mean diameter of 0.02~m. The cell walls are 130~$\mu$m kapton 
glued to an aluminum base. A $\varnothing$0.02~m reentrant beam window defines the length 
of LH$_2$/LD$_2$ in the beam to be about 0.30~m. Both entrance and exit windows on the 
cell are 75~$\mu$m kapton. In normal operation, the cell, the condenser, and the pipes 
between them are all filled with liquid hydrogen. In this manner, the liquid can be 
subcooled a few degrees below the vapor pressure curve, greatly suppressing bubble 
formation in the cell. In total, about 0.4~liter of LH$_2$ is condensed from the storage 
tanks, and the system is engineered to recover this quantity of hydrogen safely back into 
the tanks during a sudden loss of insulating vacuum, with a maximum allowed cell pressure 
of 49~psia~\cite{Meekins}.

A conceptual design for the neutral kaon beam target is also shown in Fig.~\ref{fig:lh2a} 
(bottom right). The proposed target cell has a $\varnothing$0.06~m and a 0.40~m length from 
entrance to exit windows, corresponding to a volume of about 1.1~liter, which will require 
filling the existing tanks on the target cart to about 50~psia. The Collaboration will work 
with the JLab Target Group to investigate alternative materials and construction techniques 
to increase the strength of the cell.  As an example, the LH$_2$ target cell recently 
developed for Hall~A is $\varnothing$0.063~m, 0.18~m long and has a wall thickness of 
approximately 0.2~mm.  The cell is machined from a high-strength aluminum alloy, AL7075-T6, 
and has a maximum allowed pressure of about 100~psia. It is expected that minor modifications 
to the cryogenic target's piping systems will also be required to satisfy the increased volume 
of condensed hydrogen.

The proposed system is expected to work equally well with liquid deuterium, which condenses at 
a slightly higher temperature than hydrogen (23.3~K versus 20.3~K at atmospheric pressure). The 
expansion ratio of LD$_2$ is 13~\% higher, which implies a storage pressure of about 60~psia. 
Therefore, the new target cell must be engineered and constructed to work with both LH$_2$ and 
LD$_2$. 

\section{Project Planning}
\label{sec:project_planning}

\begin{table}[ht]
\begin{center}
\begin{tabular}{|l|l|}
\hline \hline
\textbf{ Group} & \textbf{Members} \\
\hline
CPS & \textbf{Sean~Dobbs}, Donal~Day, Pavel~Degtyarenko,  
    Gabriel~Niculescu, \\
    & Tim~Whitlatch, Todd~Satogata, \textbf{Igor~Strakovsky},\\ 
    & Bogdan~Wojtsekhowski, Eugene~Chudakov \\
KPT & \textbf{Igor~Strakovsky}, Pavel~Degtyarenko, Tim~Whitlatch, \\
    & Eugene~Chudakov \\
KFM & \textbf{Mikhail~Bashkanov}, Nick~Zachariou, Dan~Watts, 
    Tim~Whitlatch, \\
    &\textbf{Moskov~Amaryan}, Shankar Adhikari \\
Electron Beamline & \textbf{Todd~Satogata} \\
Secondary particle yields & \textbf{Igor~Strakovsky}, 
    \textbf{Ilya~Larin}, \textbf{Moskov~Amaryan} \\
Cryogenic Target & \textbf{Chris~Keith} \\
Trigger / DAQ & \textbf{Sergey~Furletov}, \textbf{Alexander~Somov} \\
\hline \hline
\end{tabular}
\end{center}
\caption{\label{tbl:klf_working_groupsA} Topical groups for the various detector systems and 
	beamline components being developed or upgraded for the KLF experiment.  Working group 
	coordinator(s) are listed in bold text.}
\end{table}
\begin{figure}[ht]
\begin{center}
\includegraphics[width=0.95\textwidth]{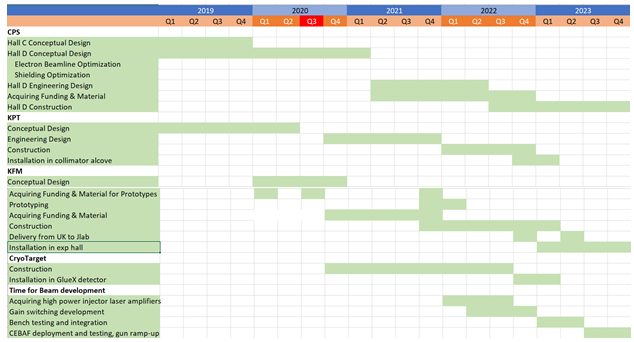}

\caption{\label{fig:klf_schedule}
 Diagram illustrating a potential timeline for the different stages of the design, 
	construction, and installation of the $K_L$ beamline.
}   
\end{center}  
\end{figure}

The KLF Collaboration represents a substantial community interested in the physics 
of strange quark hadrons, whose membership includes people currently involved in the 
physics of Hall~D and those from outside the currently established Collaborations.  
A growing core group is actively involved in the hardware and analysis activities 
required for this project.

We show the key KLF beamline and detector system projects in Tables~\ref{tbl:klf_working_groupsA}, 
along with the personnel currently involved in each project.  The main additional hardware required 
for the KLF are the new beamline components for the neutral kaon beam.  These components have been 
previously described in detail in Sec.~\ref{sec:beam}, and they are listed in 
Table~\ref{tbl:klf_working_groupsA} along with the Collaboration members currently involved  in 
their design. The CPS for Hall~D is being designed in collaboration with the CPS Collaboration, 
in order to leverage their experience with designing a CPS for Halls~A/C.  The KPT and KFM are 
smaller projects, and their designs are advanced and well-matched for the expertise of their 
working group members.

An initial high-level list of tasks required for these hardware projects and the estimated time 
to complete each of them is given in Fig.~\ref{fig:klf_schedule}.  With reasonable assumptions 
of the available resources, we expect that the new beamline elements could be built and fully 
installed by the beginning of 2024.

In addition, a growing range of members have been working on performing the detailed simulation 
and PWA studies shown in this proposal, and have been continuing to build the physics and analysis 
frameworks needed for the determination of resonance pole parameters and other physics goals 
described in this proposal.  The physics topics covered by these efforts include the primary 
hyperon and meson spectroscopy program, as well as studies of neutron-induced reactions and 
other physics possibilities with the KLF facility, as described in the body of the proposal 
and the appendices.  The range of active experimental and theoretical activities is illustrated 
by the range of talks presented at the regularly schedule KLF Collaboration 
Meetings~\cite{klfcollabmeeting}.

\section{Summary and Beam Time Request}
\label{sec:summary}

We propose to perform strange hadron spectroscopy with a secondary $K_L$ beam in the GlueX setup 
at JLab.  Precise new experimental data (both differential cross sections and recoil polarization 
of hyperons) for $K_L$p scattering with good kinematic coverage will be obtained. This will allow 
predictions from the CQM and LQCD to be checked for all families of excited $\Lambda^\ast$, 
$\Sigma^\ast$, $\Xi^\ast$, and $\Omega^{\ast}$ hyperon resonances for the first time.  In addition, 
it will permit a search for the possible existence of hybrids in the hyperon sector as predicted by 
the lattice calculations~\cite{Dudek:2012ag}.  

A complete understanding of three-quark bound states requires accurate measurements of the full 
spectra of hyperons with their spin-parity assignments, pole positions, and branching ratios.  An 
important impact of these strange hyperon spectroscopy measurements is their significance for the 
thermodynamic properties of the early universe at freeze-out, which is one of the main physics 
topics at heavy-ion colliders.

Besides hyperon spectroscopy, the experimental data obtained in the strange meson sector in the 
reactions $K_Lp\to K^\pm \pi^\mp p$ and $K_Lp\to K_S\pi^\pm n(p)$ will provide precise and 
statistically significant data for experimental studies of the $K\pi$ system. This will allow a 
determination of quantum numbers of strange meson resonances in $S$- (including $\kappa(800)$), 
$P$-, $D$-, and higher-wave states.  It will also allow a determination of phase shifts to  
account for final-state $K\pi$ interactions. Measurements of $K\pi$ form factors will be important 
input for Dalitz-plot analyses of $D$-meson and charmless $B$ mesons with $K\pi$ in final state. 

The $K_L$ facility at JLab will be {\it unique in the world}.  The high-intensity secondary beam of 
$K_L$ ($1\times 10^4~K_L$/s) would be produced in EM interactions using the high-intensity and 
high-duty-factor CEBAF electron beam with very low neutron contamination as was done at SLAC in 
the 1970s~\cite{Yamartino:1974sm}; but now, with three orders of magnitude higher intensity.  The 
possibility to perform similar studies with charged kaon beams is under discussion at J-PARC with 
intensities similar to those proposed for the $K_L$ beam at JLab. If these proposals are approved, 
the experimental data from J-PARC will be complementary to those of the proposed $K_L$ measurements.
\begin{table}[ht]

\centering \protect\caption{Expected statistics for differential cross sections of different 
  reactions with LH$_2$ and below W = 3.0~GeV for 100~days of beam time.} \vspace{2mm}
{%
\begin{tabular}{|cc|}
\hline
Reaction                     & Statistics \\
                             & (events) \\
\hline
$K_Lp\rightarrow K_Sp$       &      2.7M \\
$K_Lp\rightarrow\pi^+\Lambda$&      7M \\
$K_Lp\rightarrow K^+\Xi^0$   &      2M \\
$K_Lp\rightarrow K^+n$       &     60M \\
$K_Lp\rightarrow K^-\pi^+p$  &      7M \\
\hline
\end{tabular}} \label{tab:sum}
\end{table}

In Table~\ref{tab:sum}, we present the expected statistics for 100~days of running with a LH$_2$ 
target in the GlueX setup at JLab.  The expected statistics for the 5 major reactions are very large. 
There are however, two words of cautions at this stage. These numbers correspond to an inclusive 
reaction reconstruction, which is enough to identify the resonance, but might not be enough to 
uncover its nature. The need for exclusive reconstruction is essential to extract polarization 
observables was highlighted in Sections~\ref{sec:hyperon_sims} and Ref.~\cite{HypSim}. It further 
decrease the expected statistics, e.g., from 2M to 200k events in the $K\Xi$ case. These statistics, 
however, would allow a precise measurement of the double-differential polarization observables with 
statistical uncertainties on the order of 10~\%.  Secondly, kaon flux has a maximum around W = 3~GeV, 
which decreases rapidly towards high/low $W$'s.  Thus, the 100~days of beam time on the LH$_2$ are 
essential to maximize the discovery potential of the $K_L$ Facility and uncover the densely populated 
hyperon states at low-$W$. 

There are no data on ``neutron" targets and, for this reason, it is hard to make a realistic estimate 
of the statistics for $K_Ln$ reactions. If we assume similar statistics as on a proton target, the 
full program will be completed after running 100~days with LH$_2$ and 100~days with LD$_2$ targets.
\begin{figure}[ht]
    \begin{center}
    \includegraphics[angle=0,width=0.47\textwidth]{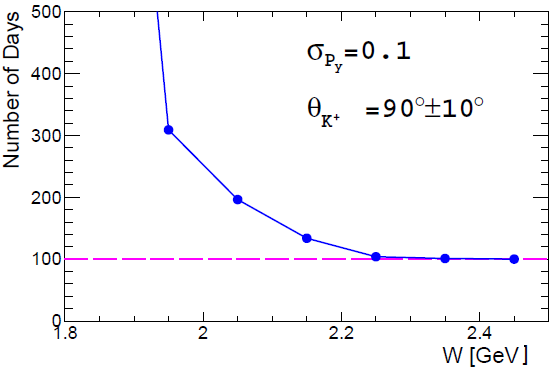}
    \includegraphics[angle=0,width=0.47\textwidth]{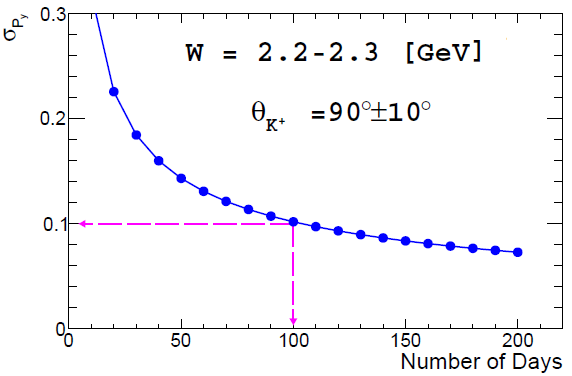}

    \centerline{\parbox{0.70\textwidth}{
    \caption[] {\protect\small 
        Required a beam time of the experiment for the $K_Lp\rightarrow K^+\Xi$ reaction.
        \underline{Left}: To reach 10~\% polarization uncertainty as a function of $W$.
        \underline{Right}: Reachable polarization uncertainty at W = 2.2~GeV and 
	$\theta_{K+}=90^\circ$.} \label{fig:DaysPy} } }
    \end{center}
\end{figure}

\subsection{Expected Statistical Accuracy:}
A coupled-channel PWA is the most direct and least model-dependent way to extract resonance 
properties. However, as shown in Appendix.~\ref{sec:A3}, it requires knowledge of both the 
differential and polarization observables at the same CM\ energy. In order to ensure that the 
duration of the experiment would be adequate to extract all observables with sufficient accuracy, 
dedicated studies were performed. One can determine the recoil polarization utilizing large 
self-analysing powers of hyperon decays. In this case, the errors on the polarization measurement 
are essentially of statistical nature, hence one can infer desired accuracy in the polarization 
measurement to a required beam time of experiment in a straightforward way. From theoretical 
perspective, the polarization error on the order of 0.1 looks essential in getting an unambiguous 
PWA solution (see Sec.~\ref{sec:Maxim}). Polarization errors larger than 0.5 would have no influence 
on convergence of the PWA fit, hence will be discarded. This tight theoretical constrains impose 
strict requirement for the duration of experiment to collect sufficient statistics in each channel. 
Fig.~\ref{fig:DaysPy} shows the expected error in measurement of polarization observable as a 
function of CM energies (left) and experiment duration (right) for the key reaction $K_Lp\to 
K^+\Xi^0$. The expected error is a complex three-fold function of kaon flux (Fig.~\ref{fig:DaysPy} 
(left)) with maximum at W = 3~GeV), cross section (Ref.~\cite{Sharov:2011xq}) and detector 
acceptance. In case of $K_Lp\rightarrow K^+\Xi^0$ reaction, it lead to a maximum statistics 
reachable in the range of 2.2 < W < 2.7~GeV.

\begin{figure}[ht]
    \begin{center}
    \includegraphics[angle=0,width=0.43\textwidth]{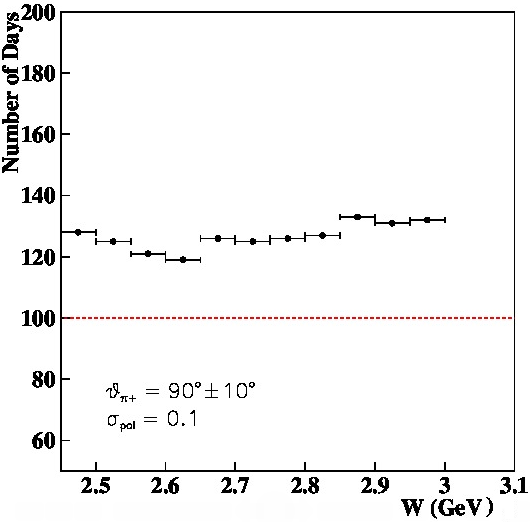}
    \includegraphics[angle=0,width=0.43\textwidth]{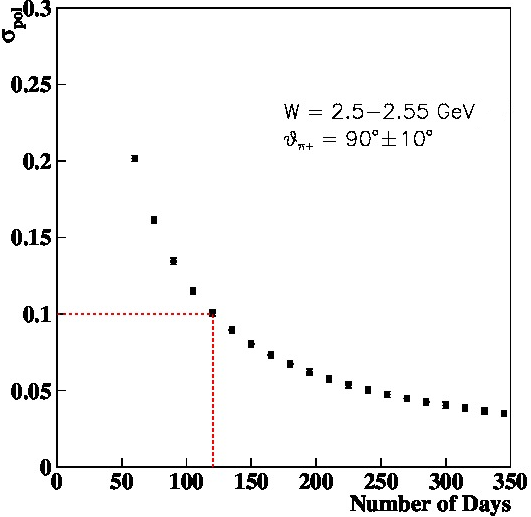}

    \centerline{\parbox{0.70\textwidth}{
    \caption[] {\protect\small 
        Required a beam time of the experiment for the $K_Lp\to
	    \pi^+\Lambda$ reaction.
        \underline{Left}: To reach 10~\% polarization uncertainty
        as a function of $W$.
        \underline{Right}: Reachable polarization uncertainty at
        W = 2.525~GeV and $\theta_{\pi^+} = 90^\circ$.} 
        \label{fig:DaysPy1} } }
    \end{center}
\end{figure}

A similar study can be performed for the another reaction channel, $K_Lp\to\pi^+\Lambda$, see  
Fig.~\ref{fig:DaysPy1}.

This reaction requires finer binning to disentangle various $\Lambda-\Sigma$ mixing effects leading 
to a similar experiment duration as in $K\Xi$ case, despite larger production cross sections.

\textbf{To summarize}: All channels we have considered so far require about 100~days beamtime for a 
nominal flux of $1\times10^4~K_L/sec$ to exhibit the beauty of strangeness physics in details and 
maximize the discovery potential of the KL Facility. Expected statistics for differential cross 
sections of different  reactions with LH$_2$ and below W = 3.0~GeV for 100~days of beam time, for 
instance, is given in Table~\ref{tab:sum}.

As  it was already mentioned in the Section~\ref{sec:kaon_spectroscopyA}, running KLF for 100~days 
on hydrogen target will increase the world statistics by about two orders of magnitude, while also 
providing a data in completely unmeasured region of $M(K\pi)<0.8$~GeV, which will result in 
significant improvement in determination of the pole position and the width of the $\kappa$ 
scalar meson. 

\subsection{Expected Systematic Uncertainties:}
Systematic uncertainties with $K_L$ beam will be reaction and kinematics dependent. The total 
systematic errors include three major sources: 
detector related, induced by the reconstruction algorithms, and overall flux estimation. The first 
two sources can be linked to the current GlueX program. Indeed after several years of running our 
understanding of the GlueX detector performance is in quite advanced state. The $K_L$ program will 
utilize this knowledge. Hence, we expect the detector related systematical errors to be of the 
similar size to that of the photon program and below $\approx$3~\%~\cite{AlGhoul:2017nbp}. The 
only source of uncertainty which cannot be estimated from ongoing GlueX program is $K_L$ flux 
related ambiguity. A dedicated $K_L$ Flux Monitor will be able to  provide a flux determination 
with an accuracy better than 5~\%. 

Additional source of systematical uncertainties in the determination of resonance parameters is 
related to theoretical extraction of ``true" observables from experimental data. Theoretical 
uncertainty is two-folded: the first part related to the absence of experimental data, e.g., the 
absence of polarization observables, or non-existence of measurements on neutron target; the second 
one is associated with particular theoretical methods employed to extract resonance parameters. We, 
in KLF, will attack this problem from both directions: precise experimental measurements will cover 
gaps in existing database, while comprehensive theoretical analysis of competing theoretical groups 
(SAID, BnGa, GW-J\"ulich, MAID, or Kent State) with uncorrelated method-dependent systematics will 
eliminate the second source of uncertainties.

In the case of strange meson spectroscopy, non-$K\pi$ scattering backgrounds are likely to play a 
role leading to systematic errors. One source of the background could be the higher mass baryon 
resonances in the case of $K^{\pm}\pi^{\mp}p$ final states. In the case when the final system is 
$K^-\pi^0\Delta^{++}$,  one source of uncertainties is due to the background under the $\pi^0$ 
peak, which according to the current GlueX measurement~\cite{AlGhoul:2017nbp} is estimated to be 
on the order of less than $1\%$. Another source of systematic uncertainties is stemming from the 
$K_L$ beam flux normalization systematics, which will be on the order of $5\%$. In the case of 
$\pi^-\Delta^{++} (K_L)$ an additional uncertainty may come from the $\Lambda \pi^+ (\pi^0)$ 
background, which should be vetoed by selecting events, where invariant mass of $p\pi^-$ lies 
above ground state $\Lambda(1116)$. Overall for the reactions with $\Delta^{++}$, we expect 
systematic uncertainties to be on the order $\sim 5\%$, while in the neutral pion exchange 
reactions systematic errors may be higher.

Systematic uncertainties for the $\kappa$ pole position determination from $K\pi$ scattering 
have already been taken into account in the estimation of the precision for its mass and width 
with KLF data in the summary Appendix~\ref{sec:A4} of the $K\pi$ scattering.


\newpage
\appendix
\etocsettocdepth.toc{section}
\section{Appendices}

\etocignoretoctocdepth 
\etocsettocstyle{\section*{\contentsname}}{}
\localtableofcontents


\newpage
\subsection{Interest of the RHIC/LHC Community in Excited Hyperon Measurements}
\label{sec:A1}

\begin{figure}[ht]
\centering
{
    \includegraphics[width=0.9\textwidth,keepaspectratio]{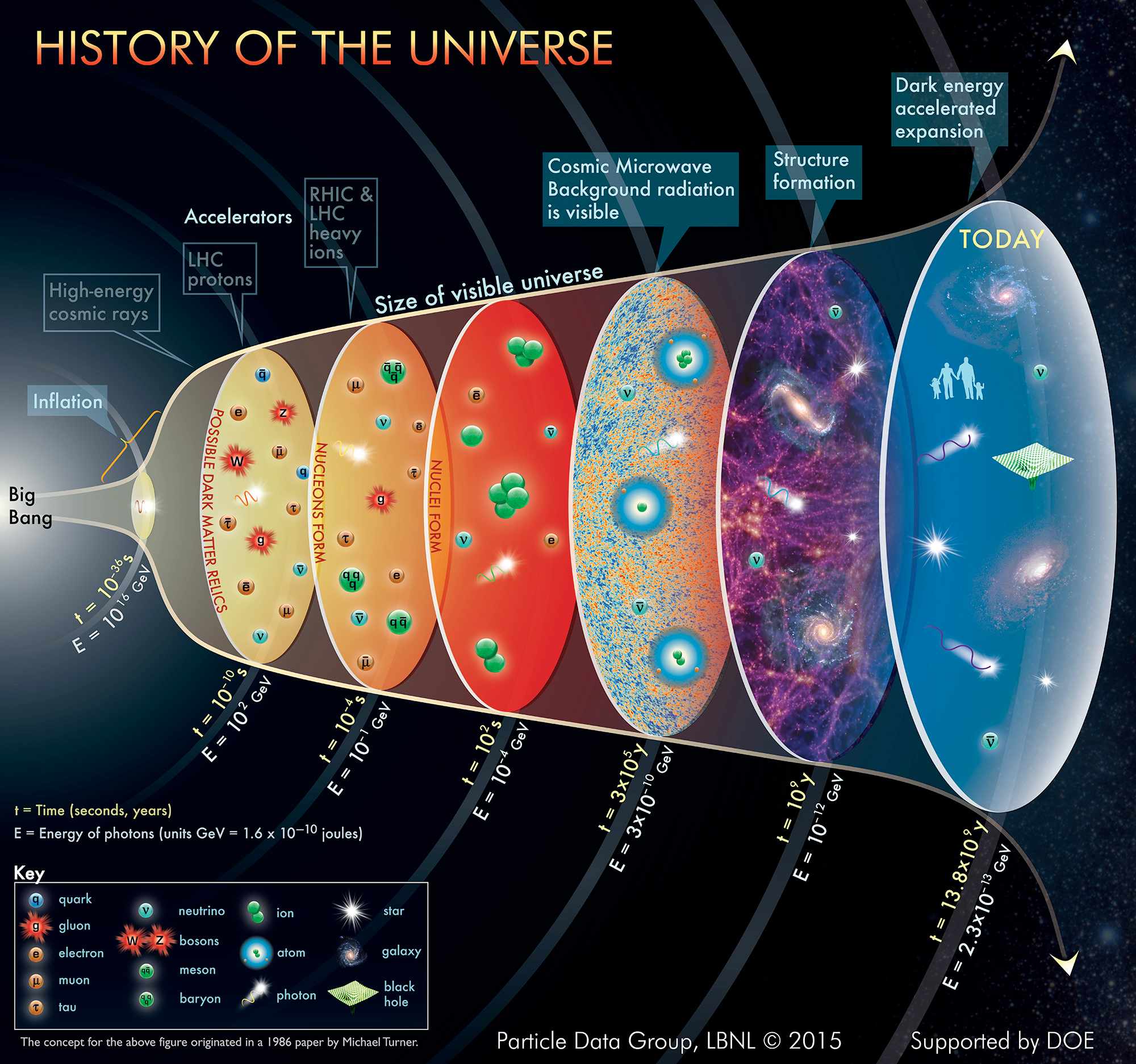} }

\centerline{\parbox{0.70\textwidth}{
    \caption[] {\protect\small
    KLF Project will provide a valuable missing input needed to shed a light 
	on thermodynamic properties of the Early Universe around 1~$\mu$s after 
	the Big Bang. }
    \label{fig:figzz} } }
\end{figure}

At temperatures on the order of the pion mass strongly interacting matter 
undergoes a transition (rapid crossover) from the confined phase with 
hadronic degrees of freedom to a deconfined phase with partonic degrees of 
freedom, Quark-Gluon Plasma (QGP). A reverse process, hadronization has 
taken place shortly after the Big Bang when the matter in the Universe 
started cooling down and underwent a chain of transitions, as illustrated 
in Fig.~\ref{fig:figzz}. The properties of the strongly interacting matter 
under extreme temperatures and densities and the transition to QGP are under 
intense study at the Relativistic Heavy-Ion Collider (RHIC) at BNL and the 
Large Hadron Collider (LHC) at CERN. To relate experimentally measured 
particle yields to theoretically predicted thermodynamic observables, a 
detailed understanding of the hadronization process of light and strange 
degrees of freedom is required.
\begin{figure}[h!]
\centering
{\includegraphics[width=0.65\textwidth,keepaspectratio]{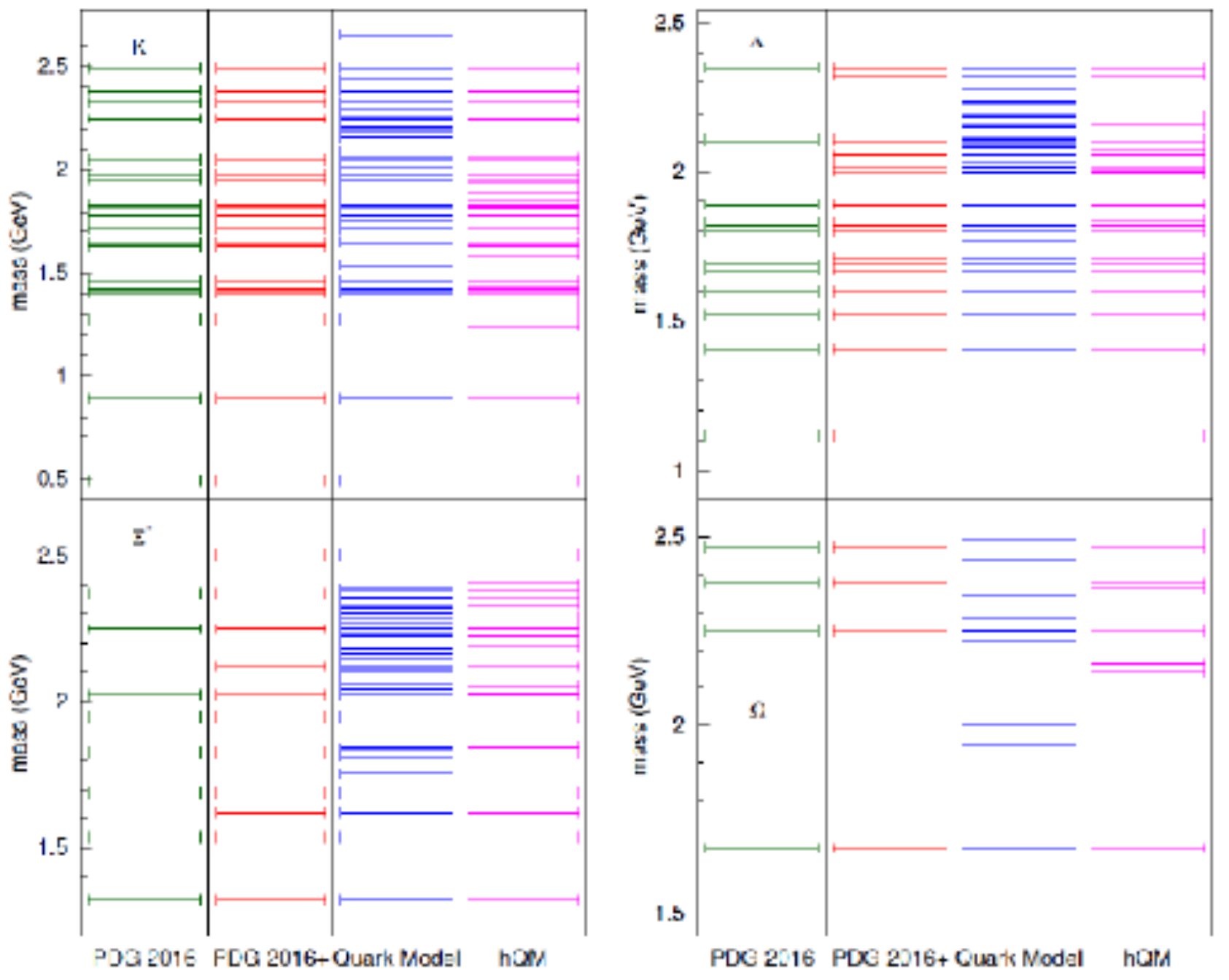} }
{\includegraphics[width=0.41\textwidth,keepaspectratio]{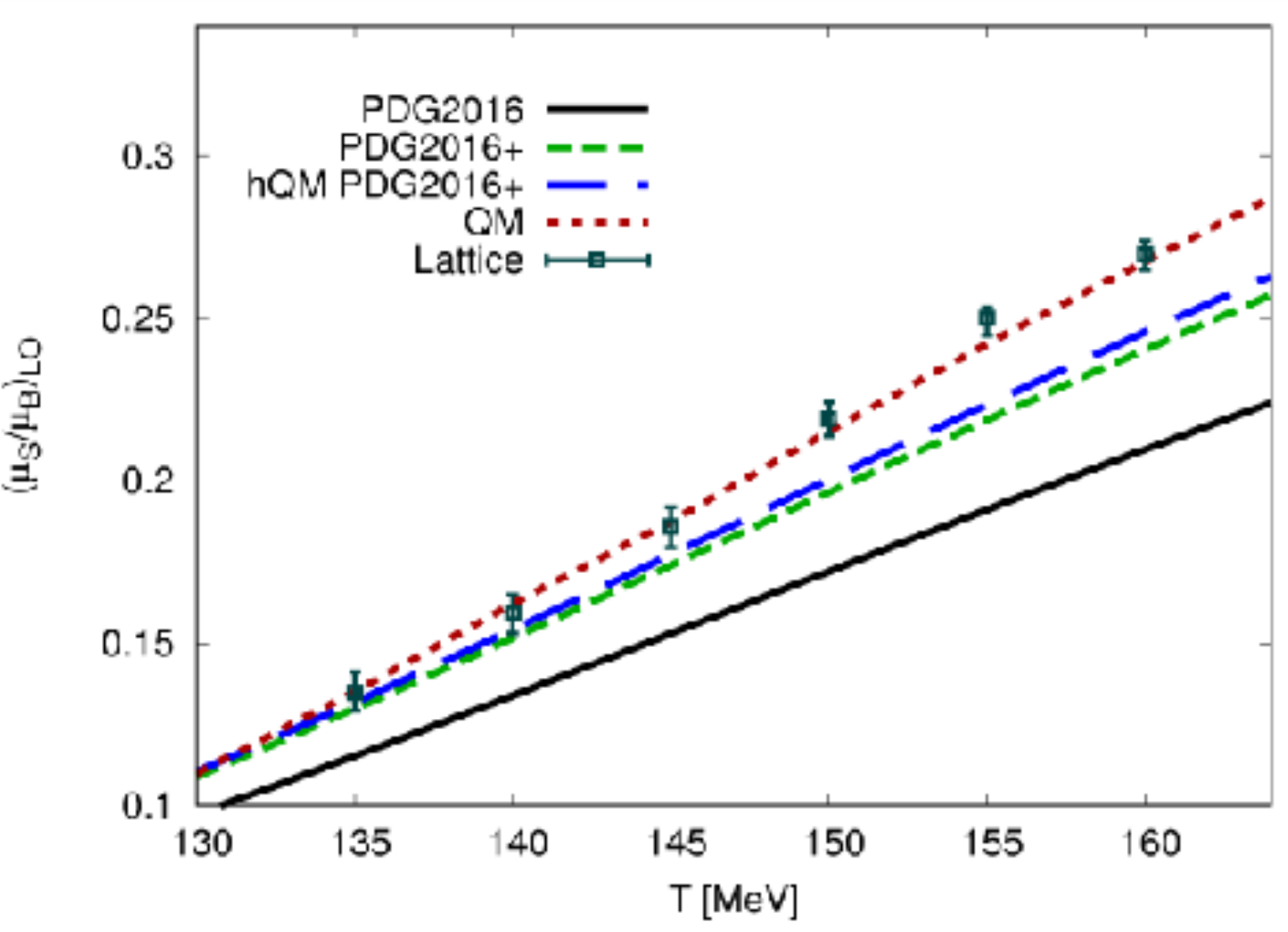} }
{\includegraphics[width=0.43\textwidth,keepaspectratio]{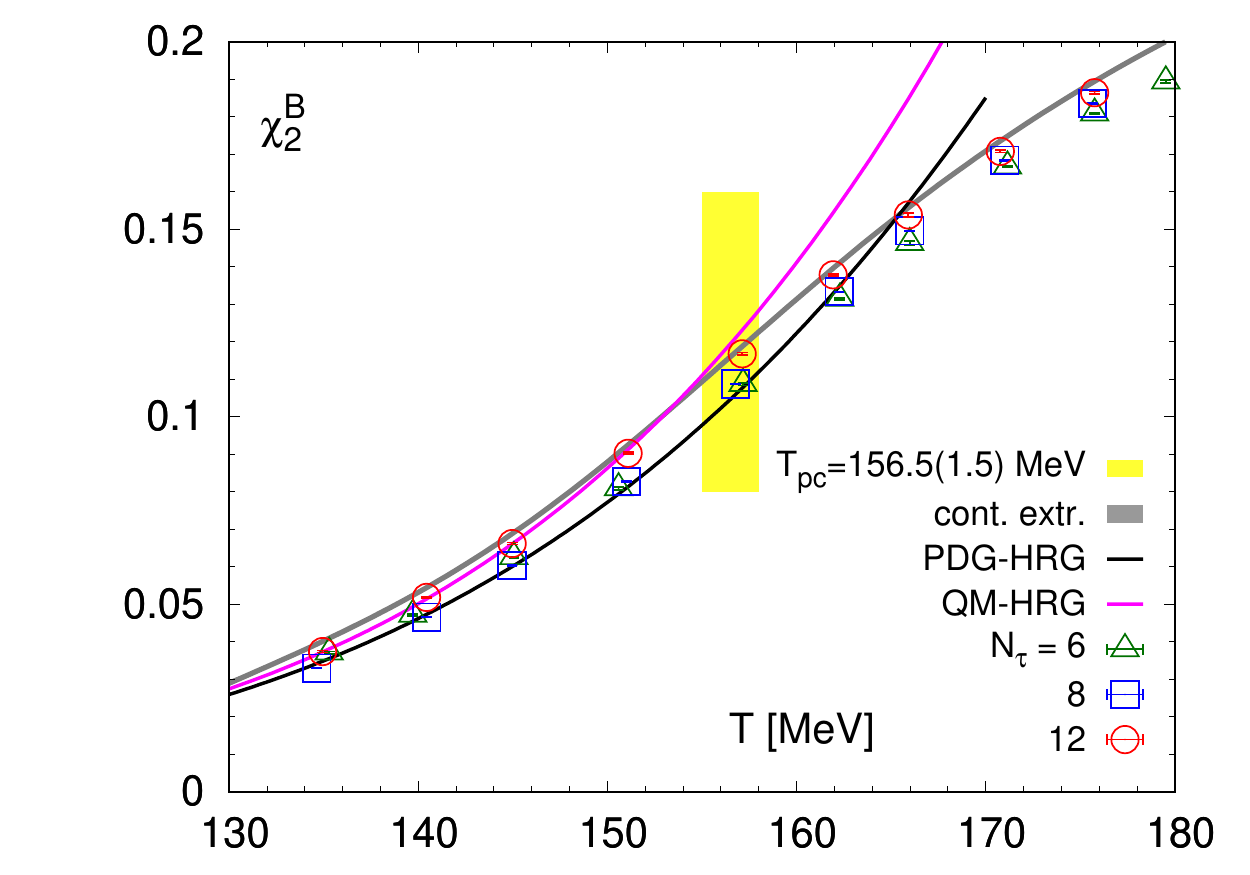} }
\centerline{\parbox{0.70\textwidth}{
 \caption[] {\protect\small
    \underline{Top}: Comparison of predicted and measured excited strange 
	hadronic states in PDG2018, PDG2018+ (including one star states), 
	QM, and hQM.
    \underline{Bottom left}: Lattice QCD calculations of the temperature 
	dependence of the leading order susceptibility ratio ($\mu_s$/$\mu_B$) 
	compared to results from HRG model calculations with varying number 
	of hadronic states.
    \underline{Bottom right}: Lattice QCD calculations of the temperature 
	dependence of the baryon number susceptibility $\chi_B^2$ compared 
	to results from HRG model calculations with varying number of 
	hadronic states.}
    \label{fig:figxx} } }
\end{figure}

The relativistic heavy-ion community at the RHIC and LHC has recently embarked 
on specific analyses to address the issue of strangeness hadronization. LQCD 
calculations in the QCD crossover transition region between a deconfined phase 
of quark and gluons and a hadronic resonance gas have revealed a potentially 
interesting sub-structure related to the hadronization process.  Studies of 
flavor-dependent susceptibilities, which can be equated to experimental 
measurements of conserved quantum-number fluctuations, seem to indicate a 
slight flavor hierarchy in the three-quark sector (u,d,s) in thermalized 
systems. Specifically, the ratios of higher-order susceptibilities in the 
strange sector show a higher transition temperature than in the light 
sector~\cite{Bellwied:2013cta}. Recently, original estimates of the 
pseudo-critical temperature~\cite{Borsanyi:2010bp,Bazavov:2014pvz} have been 
significantly improved placing the transition at 
$T_c = (156.5\pm1.5)$~MeV~\cite{Bazavov:2018mes}, 
$T_c = (158.0\pm0.6)$~MeV~\cite{Borsanyi:2020fev}. Ref.~\cite{Borsanyi:2020fev} 
has also estimated the width of the transition to be around $\Delta T = 15$~MeV. 
The difference of the specific susceptibilities calculated in 
Ref.~\cite{Bellwied:2013cta} is around 18~MeV and well outside their individual 
uncertainties. At the same time the pseudo-critical temperature associated with 
strangeness observables is certainly not consistent with the most recent 
estimates of the pseudo-critical temperatures based on the chiral condensate 
and related observables. This warrants a more detailed analysis of the 
strangeness sector, both in terms of the partonic and hadronic degrees of 
freedom.

This difference seems to be confirmed by statistical thermal-model calculations 
that try to describe the yields of emitted hadrons from a QGP based on a common 
chemical freeze-out temperature. Although the yields measured by ALICE at the 
LHC in 2.76~TeV PbPb collisions can be described by a common temperature of 
156$\pm$2~MeV, with a reasonable $\chi^2$, the fit improves markedly if one 
allows the light quark baryons to have a lower temperature than the strange 
quark baryons~\cite{Floris:2014pta}. A similar result has been found when the 
thermal fluctuations of particle yields as measured by the STAR 
Collaboration~\cite{Adamczyk:2013dal,
Adamczyk:2014fia}, which can be related to the light quark dominated 
susceptibilities of the electric charge and the baryon number on the 
lattice, have been compared to statistical model 
calculations~\cite{Alba:2014eba}.

If one assumes that strange and light quarks indeed prefer different freeze-out 
temperatures, then the question arises how this could impact the hadronization 
mechanism and abundance of specific hadronic species. In other words, is the 
production of strange particles, in particular excited resonant states, 
enhanced in a particular temperature range in the crossover region? Strange 
ground-state particle production shows evidence of enhancement, but the most 
likely scenario is that the increased strange quark abundance will populate 
excited states; therefore, the emphasis of any future experimental program 
trying to understand hadron production is shifting towards strange baryonic 
resonance production. Furthermore, recent LHC measurements in small systems, 
down to elementary proton-proton collisions, have revealed that even in these 
small systems there is evidence for deconfinement, if the achieved energy 
density, documented by the measured charged particle multiplicity is large 
enough~\cite{ALICE:2017jyt}. Therefore, future measurements of elementary 
collisions in the K-Long Facility experiment at JLab might well provide the 
necessary link to future analysis of strange resonance enhancements in 
heavy-ion collisions at RHIC and the LHC and a deeper understanding of the 
hadronization process.

This statement is also supported by comparisons between the
aforementioned LQCD calculations and model predictions based on a
non-interacting hadronic resonance gas.  The Hadron Resonance Gas
(HRG) model~\cite{Dashen:1969ep,Venugopalan:1992hy,Karsch:2003zq,
Tawfik:2004sw} yields a good description of most thermodynamic quantities in 
the hadronic phase up to the pseudo-critical temperature. The idea that 
strongly interacting matter in the ground state can be described in terms of 
a non-interacting gas of hadrons and resonances, which effectively mimics the 
interactions of hadrons by simply increasing the number of possible resonant 
states exponentially as a function of temperature, was proposed early on by 
Hagedorn~\cite{Hagedorn:1976ef}. The only input to the model is the hadronic 
spectrum: usually it includes all well-known hadrons in the {\it Review of 
Particle Physics}, namely the ones rated with at least two stars.  Recently, 
it has been noticed that some more differential observables present a 
discrepancy between lattice and HRG model results. The inclusion of 
not-yet-detected states, such as the ones predicted by the original 
Quark Model (QM)~\cite{Capstick:1986bm,Ebert:2009ub} has been proposed to 
improve the agreement~\cite{Majumder:2010ik,Bazavov:2014xya}. A systematic 
study based on a breakdown of contributions to the thermodynamic pressure 
given by particles grouped according to their quantum numbers (in particular 
baryon number and strangeness) enables us to infer in which hadron sector 
more states are needed compared to the well-known ones from the 
PDG~\cite{Alba:2017mqu}. In case of a flavor hierarchy in the transition 
region, one would expect the number of strange resonances to increase, due 
to a higher freeze-out temperature, compared to the number of light-quark 
resonances. Figure~\ref{fig:figxx} shows the effect of different strange 
hadron input spectra to the HRG model in comparison to LQCD.  
Figure~\ref{fig:figxx} (top) shows the number of states in 
PDG2018~\cite{Tanabashi:2018oca}, PDG2018+ (including one star states), 
the standard QM, and a Quark Model with enhanced quark interactions in the 
hadron (hyper-central model hQM~\cite{Giannini:2015zia}).  
Fig.~\ref{fig:figxx} (bottom left) shows a comparison of the HRG results 
to a leading-order LQCD calculation of $\mu_{s}$/$\mu_{B}$; i.e., the ratio 
to strange to baryon number susceptibility~\cite{Alba:2017mqu}. Recent 
extensions of the HRG model, for instance, accommodating repulsive 
mean-field~\cite{Huovinen:2018ziu} also indicate that inclusion of the 
predicted by the Quark Model states improves the agreement between HRG 
and LQCD results.

An interesting conclusion that arises from these studies is that the 
improvement in the listing of strange resonances between
PDG2008~\cite{Amsler:2008zzb} and PDG2018~\cite{Tanabashi:2018oca} 
definitely brought the HRG calculations closer to the LQCD data. By 
looking at details in the remaining discrepancy, which is in part 
remedied by including one-star rated resonances in PDG2018, it seems that 
the effect is more carried by singly strange resonances rather than 
multi-strange resonances, also in light of comparisons to quark models 
that include di-quark structures~\cite{Santopinto:2014opa} or enhanced 
quark interactions in the baryon (hypercentral models~\cite{Giannini:2015zia}). 
This is good news for the experiments since the $\Lambda$ and $\Sigma$ 
resonances below 2~GeV are well within reach of the KLF experiment and, 
to a lesser significance, the RHIC/LHC experiments. In this context it is 
also important to point out that the use of both hydrogen and deuterium 
targets in KLF is crucial since it will enable the measurement of charged 
and neutral hyperons. A complete spectrum of singly strange hyperon states 
is necessary to make a solid comparison to first-principle calculations.

The possible effect of missing states on the QCD thermodynamics is also 
quite pronounced in the baryon sector. Fig.~\ref{fig:figxx} (bottom right) 
shows the fluctuations of the baryon number computed in 
LQCD~\cite{Bazavov:2020bjn} and compared to the HRG model. While the HRG 
model with the PDG spectrum undershoots the continuum-extrapolated lattice 
data, adding the Quark Model predicted states brings the HRG result into 
closer agreement with LQCD.

\textbf{To summarize}: Any comparisons between experimentally verified 
strange quark-model states from YSTAR and LQCD will shed light on a 
multitude of interesting questions relating to hadronization in the 
non-perturbative regime, exotic particle production, the interaction 
between quarks in baryons and a possible flavor hierarchy in the creation 
of confined matter. 

\subsection{Strangeness Physics within Neutron Stars}
\label{sec:A2}

One of the main goal of nuclear physics is to obtain a comprehensive picture of the strong 
interaction, which can be accessed by introducing the strangeness degree of freedom in the, 
now well-understood, nucleon-nucleon ($NN$) interaction. The $NN$ interaction has a long 
history of detailed studies, and currently phenomenological approaches can describe observed 
phenomena with high accuracy~\cite{Machleidt:2001a}. On the other hand, the interaction 
between Hyperons and Nucleons ($YN$) is very poorly constrained, mainly due to difficulties 
associated with performing high-precision scattering experiments involving short-lived 
hyperon beams (currently only $\sim$1300 events from bubble chamber experiments exist on 
the interaction between hyperons and protons). This interaction, however, provides the key 
in understanding properties and phenomena associated with Neutron stars~\cite{Vidana:2013a,
Vidana:2016a}. 
\begin{figure}[ht]
\centering
    \includegraphics[width=0.45\textwidth,keepaspectratio]{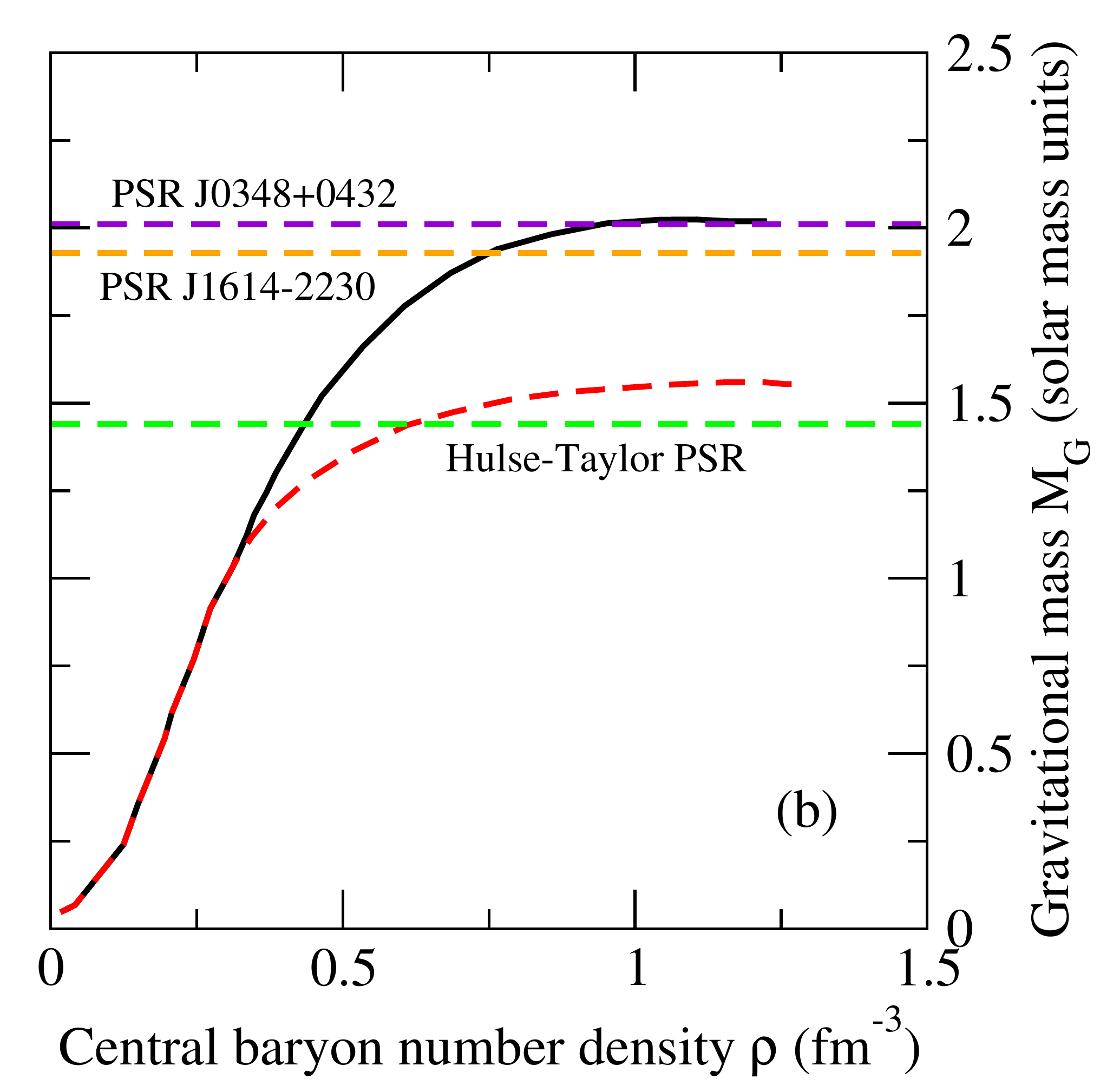}  
 \centerline{\parbox{0.70\textwidth}{
    \caption[] {\protect\small
    Effect of hyperons on EoS of neutron stars using a generic model with (black solid line) 
	and without (red dashed line) hyperons. The horizontal lines shows the observational 
	mass of the Hulse-Taylor~\protect\cite{Hulse:1975} pulsar and the observed PSR 
	J1614-2230~\protect\cite{Demorest:2010} and PSR J0348+0432~\protect\cite{Antoniadis:2013}. 
	Figure from Ref.~\protect\cite{Vidana:2018}.} \label{Fig:HyperonsinNS} } }
\end{figure}

Hyperons are expected to appear within the core of Neutron stars at densities about 2-3 times 
the nuclear saturation density. In fact, contrary to terrestrial conditions, where hyperons 
are unstable and decay into nucleons via the weak interaction in a matter of nanoseconds, the 
conditions in neutron stars can make the inverse process more energetically favourable; with 
their subsequent decay blocked by the Pauli principle. Because of this,  the existence of 
hyperons in Neutron stars has been a very hot topic in the past years as it  is predicted to 
have a very strong effect on the Equation of State of the neutron star (see 
Fig.~\ref{Fig:HyperonsinNS})~\cite{Vidana:2013a}.

Obtaining a detailed understanding of the YN interaction will allow us to address the so called 
``hyperon puzzle," which reflects how theoretical approaches cannot reconcile the predicted role 
of hyperons within neutron stars using our current poor understanding of the interaction between 
hyperons and nucleons~\cite{Vidana:2016a}.

\begin{figure}[ht]
\centering
    \includegraphics[width=0.95\textwidth,keepaspectratio]{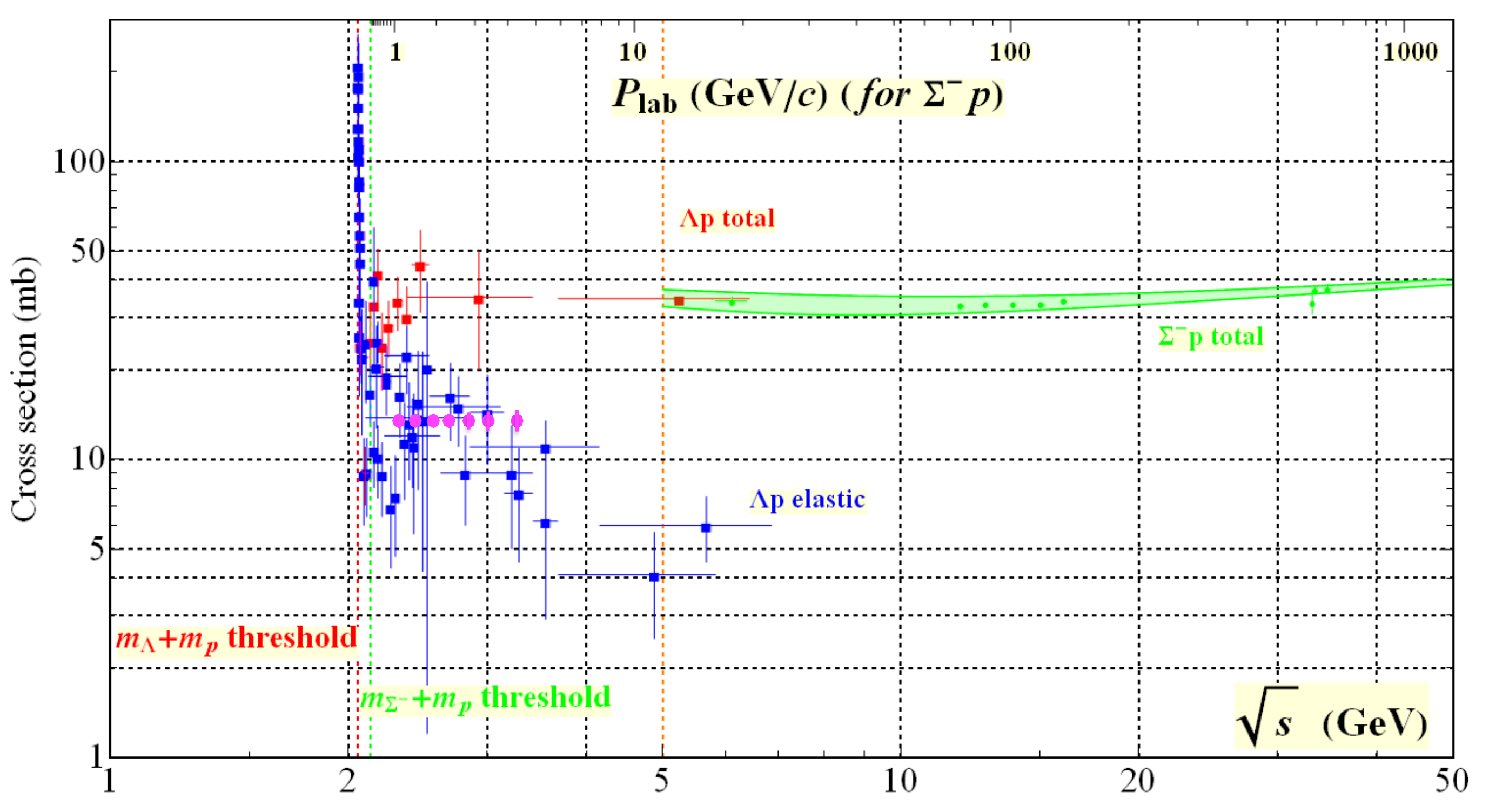}  
 \centerline{\parbox{0.70\textwidth}{
    \caption[] {\protect\small
    Total cross section for hyperon-nucleon scattering. Magenta points indicate the projected 
	measurement and statistical uncertainties from KLF and how these compare to existing 
	world data. } \label{Fig:HyperonCrossSection} } }
\end{figure}

With the recent advancement of detector and accelerator technologies we can obtain a more direct 
access on the $YN$ interaction by studying final state interaction in exclusive hyperon 
photoproduction. This approach is complementary to the alternative hypernuclear 
studies~\cite{Beane:2013a}, which are important in obtaining the many-body effects but fail in 
accessing the bare $YN$ interaction. For KLF, these studies will focus on obtaining a data set 
in which a hyperon beam, produced on a nucleon and tagged by the detection of the pion 
($K_Lp\to\pi^+\Lambda/\Sigma$) or kaon ($K_Lp\to K^+\Xi$), rescatters with a secondary nucleon 
within the target cell. Large acceptance detectors, like the GlueX detector, allows a full 
reconstruction of the event by the detection of the hyperons decay products.  The KLF, which 
facilitates a copious production of hyperons, will provide us with unprecedented statistics to 
study the $YN$ interaction. Utilising the self-analyticity of hyperons allows us to obtain 
further stringent constraints on the underlying dynamics and address the ``hyperon puzzle". 
Experiments with deuterium target can also be used to study the isospin dependence of the $YN$ 
interaction for the first time. Such approaches have been successfully applied to data from 
CLAS~\cite{Nick:g13proposal,Nick:EFB2019} with results being prepared for publication. The 
KLF facility, will allow a much more precise measurement as it will provide $\sim40$ times 
higher statistics and a significant increase in kinematical coverage. 
Figure~\ref{Fig:HyperonCrossSection} shows the projected KLF cross section measurement uncertainties 
(magenta points) and how these compare to existing world data.

\textbf{To summarize}: Providing direct experimental constraints on the Equation of State 
of neutron stars is important for the following reason: while non-perturbative ab initio 
lattice QCD calculations are very successful in the region of the phase diagram with high 
temperatures and low densities, the region of high densities and almost zero temperature, 
relevant for neutron stars, is completely out of reach for non-perturbative LQCD calculations. 
As a result, a theoretically sound Equation of State with fully controlled systematics for 
strongly interacting matter at high density is not presently available. 

\subsection{Hyperon Spectroscopy}
\label{sec:A3}

The present experimental knowledge of the spectra of strange hyperons remains remarkably incomplete. 
For example, only the lowest negative-parity doublet and the positive parity singlet of the $\Lambda$ 
hyperon are well established. In the case of the $\Sigma$ and $\Xi$ hyperons, only the lowest decuplet 
resonance states $\Sigma(1385)$ and $\Xi(1530)$ are well established. In this section, we discuss 
this context for the proposed hyperon spectroscopy program and a few highlights of proposed 
measurements which will address the outstanding questions in the field.  

\subsubsection{Previous Measurements for Hyperons:}
\label{sec:A3.1}
\begin{figure}[ht]
\centering
{
    \includegraphics[width=0.23\textwidth,keepaspectratio,angle=90]{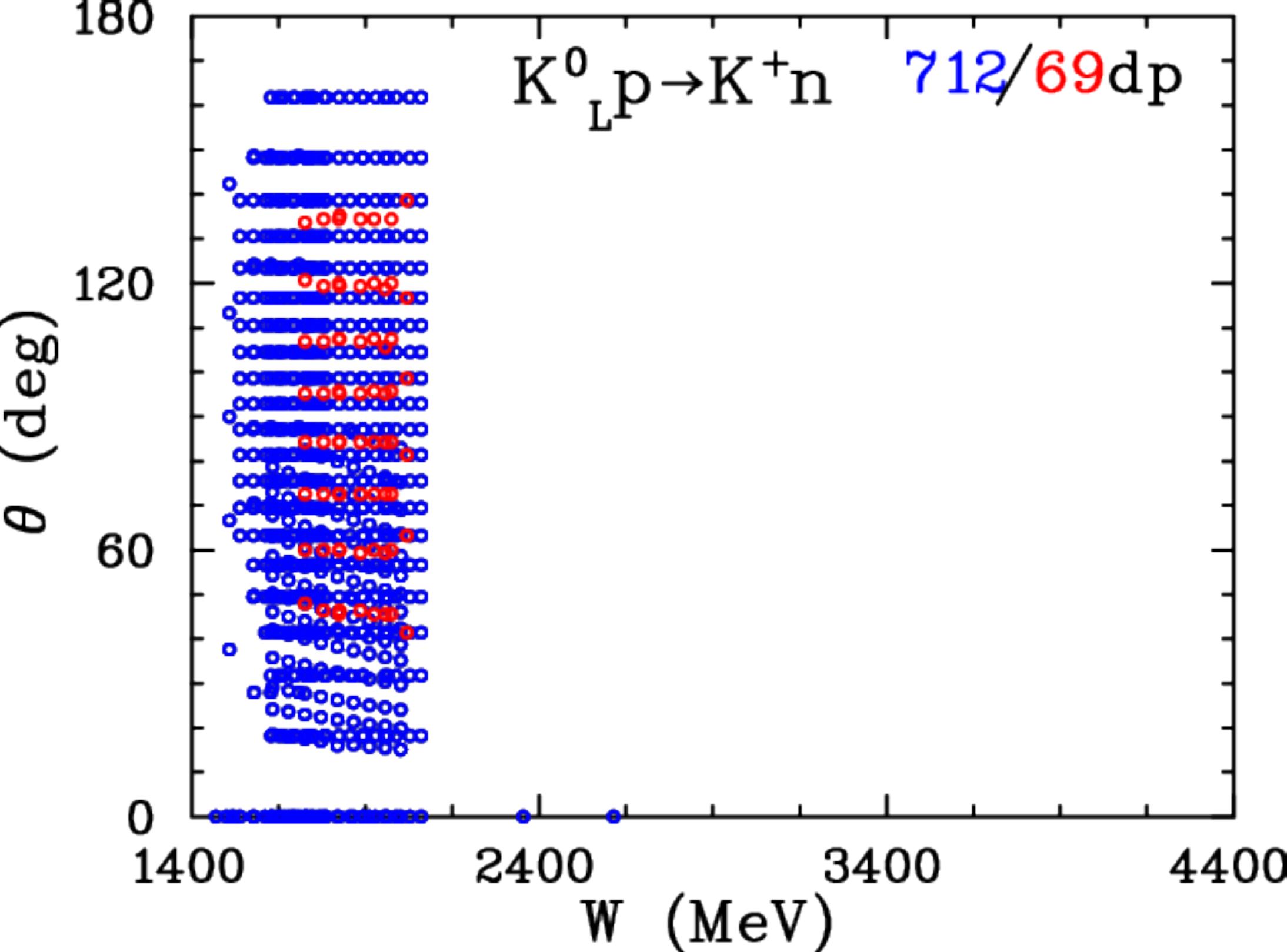} }
{
    \includegraphics[width=0.23\textwidth,keepaspectratio,angle=90]{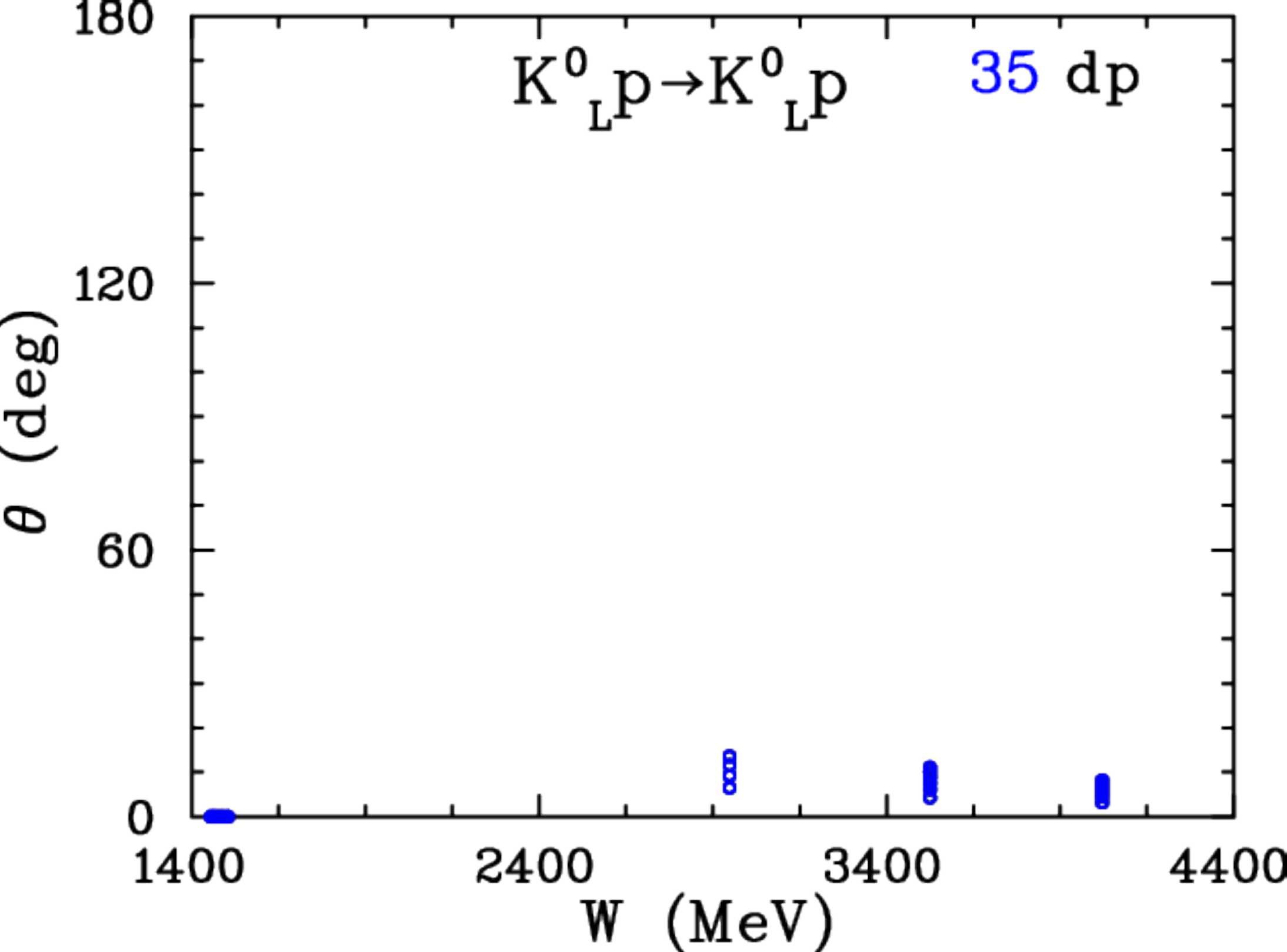} }
{
    \includegraphics[width=0.23\textwidth,keepaspectratio,angle=90]{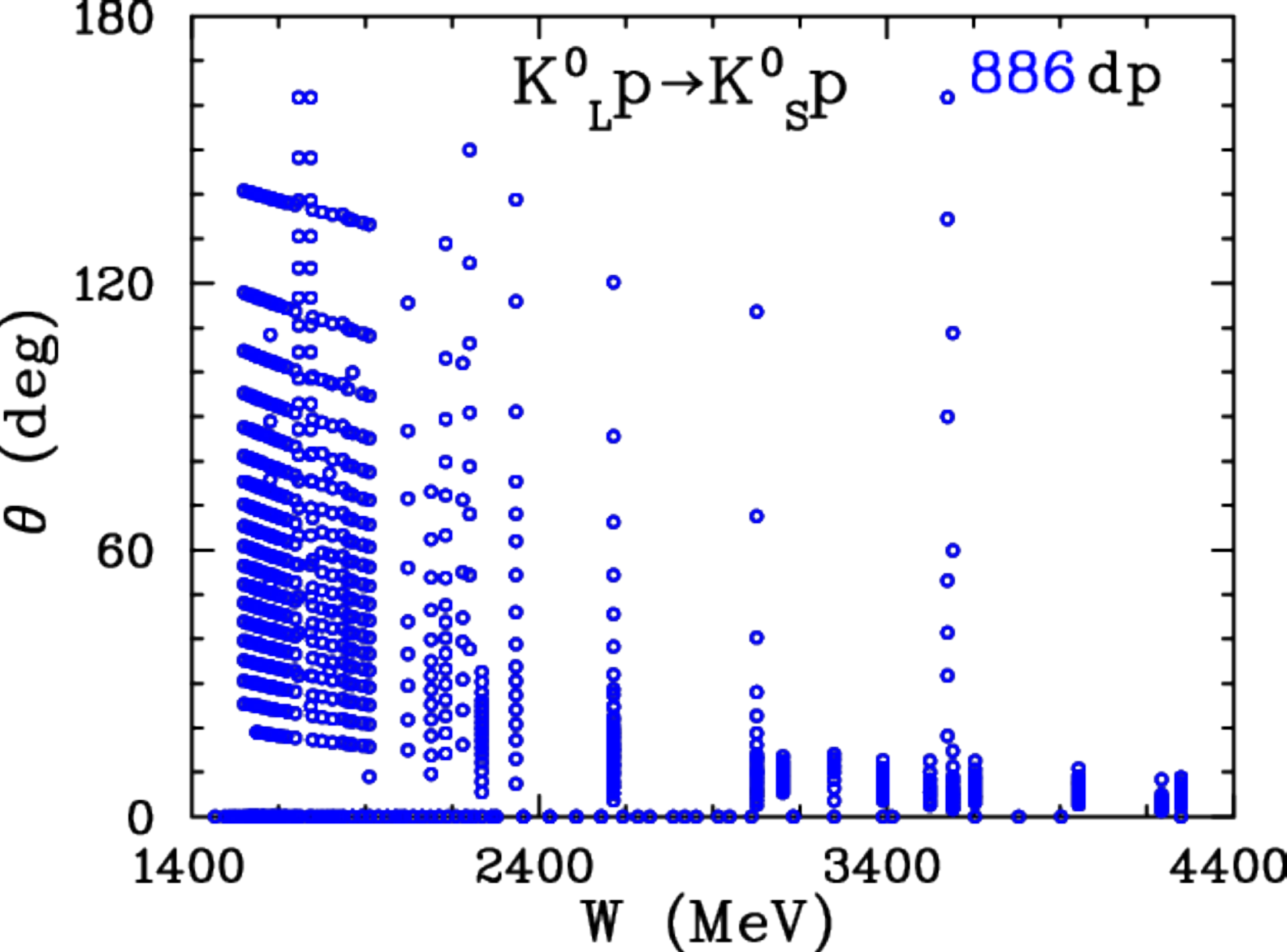} }
{
    \includegraphics[width=0.23\textwidth,keepaspectratio,angle=90]{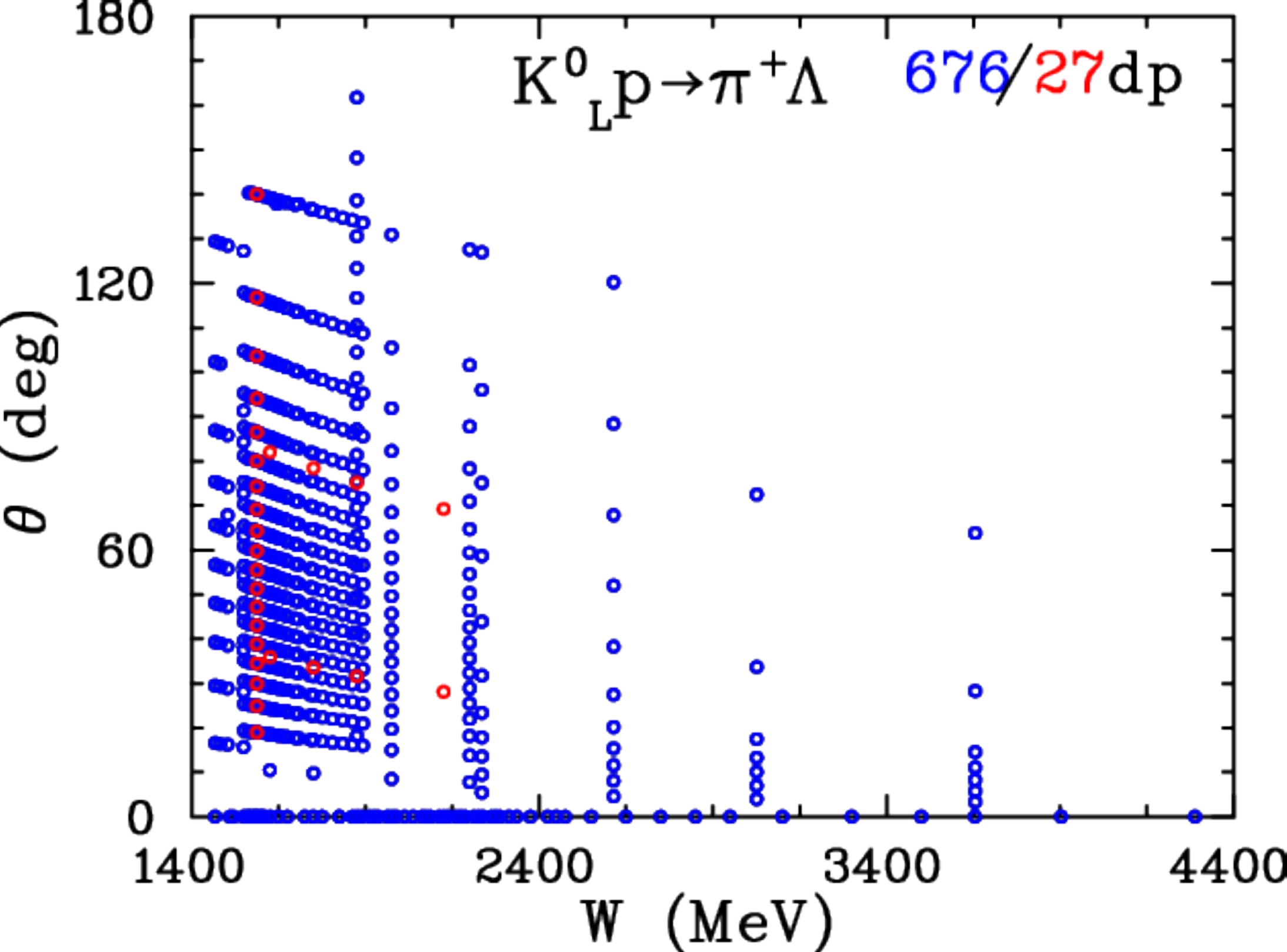} }
{
    \includegraphics[width=0.23\textwidth,keepaspectratio,angle=90]{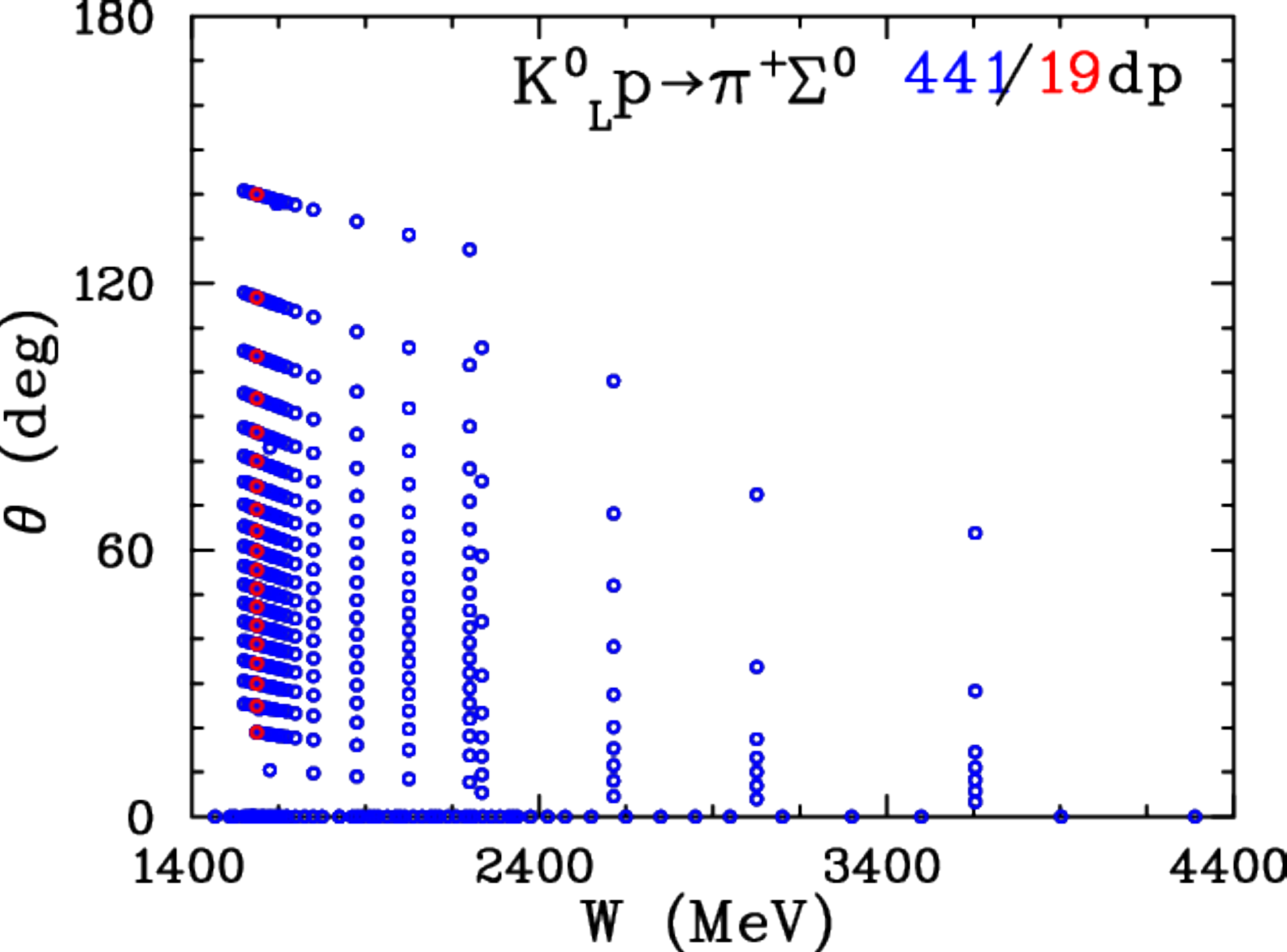} }
{
    \includegraphics[width=0.23\textwidth,keepaspectratio,angle=90]{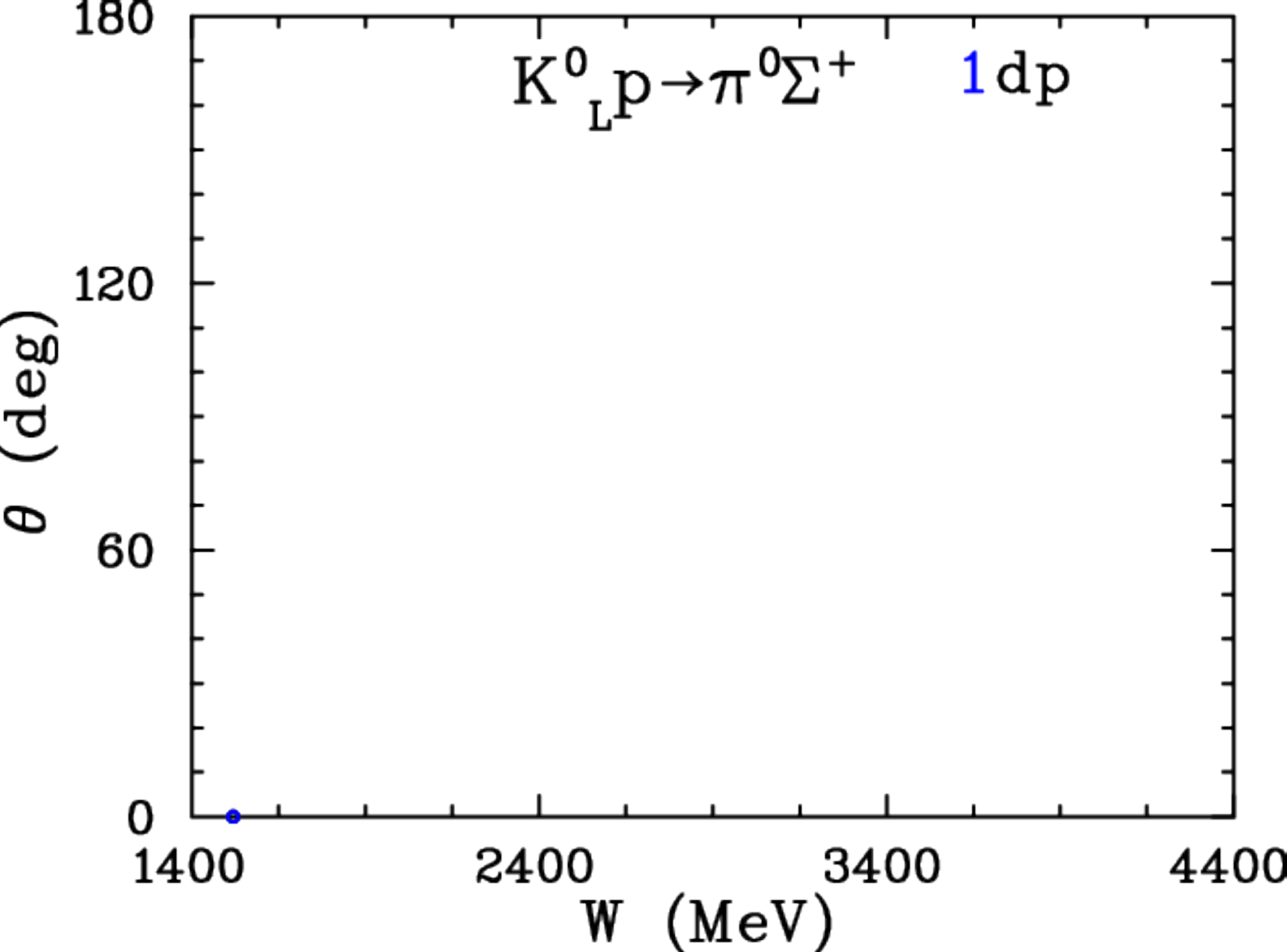} }

    \centerline{\parbox{0.80\textwidth}{
     \caption[] {\protect\small
    Experimental data available for $K_Lp\rightarrow K^+n$,
        $K_Lp\rightarrow K_Lp$, $K_Lp\rightarrow K_Sp$, 
	    $K_Lp\rightarrow\pi^+\Lambda$,
        $K_Lp\rightarrow\pi^+\Sigma^0$, and 
	    $K_Lp\rightarrow\pi^0\Sigma^+$ as a
        function of CM\ energy $W$~\protect\cite{Durham}. The
        number of data points (dp) is given in the upper
        righthand side of each subplot [blue (red) shows the
        amount of unpolarized (polarized) observables]. Total
        cross sections are plotted at zero degrees.}
        \label{fig:data} } }
\end{figure}

While a formally complete experiment requires the measurement, at each energy, $W$, and angle, $\theta$, of at least three independent observables, the current database for $K_Lp\rightarrow \pi Y$ and $KY$ is populated mainly by unpolarized cross sections. Figure~\ref{fig:data} illustrates this quite clearly.

The initial studies of the KLF program at GlueX will likely focus on two-body and quasi-two-body processes: elastic $K_Lp\rightarrow K_Sp$ and charge-exchange $K_Lp\rightarrow K^+n$ reactions, then two-body reactions producing $S = -1$ ($S = -2$) hyperons as $K_Lp\to\pi^+\Lambda$, $K_Lp\rightarrow\pi^+\Sigma^0$, and $K_Lp\to\pi^0\Sigma^+$ ($K_Lp\rightarrow K^+\Xi^0$). Most of the previous measurements induced by a $K_L$ beam, were collected for W = 1454~MeV and with some data up to W = 5054~MeV. Experiments were performed between 1961 and 1982 with mostly hydrogen bubble chambers at ANL, BNL, CERN, DESY, KEK, LRL, NIMROD, NINA, PPA, and SLAC. Note that some of data were taken at EM facilities at NINA~\cite{Albrow:1970pd} (a short overview about NINA experiments is given by Albrow recently~\cite{Albrow:2016asa}) and SLAC~\cite{Brody:1969mx}. The goal of the Manchester University group that worked at the Daresbury 5-GeV electron synchrotron NINA was CP-violation, which was a hot topic back to the mid 1960s. The main physics topics that the SLAC group addressed were studies of the systematics for particle/anti-particle processes through the intrinsic properties of the K-longs.

The first paper that discussed the possibility of creating a practical neutral kaon beam at an electron synchrotron through photoproduction was an optimistic prediction for SLAC by Drell and Jacob in 1965~\cite{Drell:1964gc}.  With significant developments in technology, high-quality EM facilities, such as JLab~\cite{LoI}, are now able to realize a complete hyperon spectroscopy program.

The overall systematics of previous $K_L$p experiments varies between 15~\% and 35~\%, and the energy binning is much broader than hyperon widths. The previous number of $K_L$-induced measurements (2426 $d\sigma/d\Omega$, 348 $\sigma^{tot}$, and 115 $P$ observables)~\cite{Durham} was very limited. Additionally, we are not aware of any measurements on a ``neutron" target.

Our knowledge about the non-strange sector is more advanced vs.\ the strange one~\cite{Tanabashi:2018oca}.  For the non-strange case, for instance, phenomenology has access to 51k data of $\pi N\to\pi N$ and 39k data of $\gamma N\rightarrow\pi N$ below W = 2.5~GeV~\cite{SAIDweb}. 
\subsubsection{Partial-Wave Analysis for Hyperons:}
Here, we \underline{summarize} some of the physics issues involved with $K_L$ scattering processes. Following Ref.~\cite{Hohler:1984ux}, the differential cross section and polarization for $K_Lp$ scattering are given by
\begin{equation}
        \frac{d\sigma}{d\Omega} = \lambdabar^2 (|f|^2 + |g|^2) \qquad {\rm and} \qquad P\frac{d\sigma}{d\Omega} = 2\lambdabar^2 {\rm Im}
        (fg^\ast),
\end{equation}
where $\lambdabar = \hbar/k$, with $k$ the magnitude of CM momentum for the incoming meson.  Here $f = f(W,\theta)$ and $g = g(W,\theta)$ are the usual spin-non-flip and spin-flip amplitudes at CM energy $W$ and meson CM production angle $\theta$. In terms of partial waves, $f$ and $g$ can be expanded as
\begin{equation}
        f(W,\theta) = \sum_{l=0}^\infty [(l+1)T_{l+}
        + lT_{l-}]P_l(\cos\theta),
\end{equation}
\begin{equation}
        g(W,\theta) = \sum_{l=1}^\infty [T_{l+} - T_{l-}]P_l^1
        (\cos\theta),
\end{equation}
where $l$ is the initial orbital angular momentum, $P_l(\cos\theta)$ is a Legendre polynomial, and $P_l^1(\cos\theta)$ is an associated Legendre function. The total angular momentum for the amplitude $T_{l+}$ is $J=l+\frac{1}{2}$, while that for the amplitude $T_{l-}$ is $J=l-\frac{1}{2}$. For hadronic scattering reactions, we may ignore small CP-violating terms and write
\begin{equation}
        K_L = \frac{1}{\sqrt{2}} (K^0 - \overline{K^0}) \qquad {\rm and} \qquad
        K_S = \frac{1}{\sqrt{2}} (K^0 + \overline{K^0}).
\end{equation}

We may generally have both $I=0$ and $I=1$ amplitudes for $KN$ and $\bar KN$ scattering, so that the amplitudes $T_{l\pm}$ can be expanded in terms of isospin amplitudes as
\begin{equation}
        T_{l\pm} = C_0 T^0_{l\pm} + C_1 T^1_{l\pm},
\end{equation}
where $T_{l\pm}^I$ are partial-wave amplitudes with isospin $I$ and total angular momentum $J = l \pm \frac{1}{2}$, with $C_I$ the appropriate isospin Clebsch-Gordan coefficients.

We plan to do a coupled-channel PWA with new KLF data in combination with available new J-PARC $K^-$ measurements when they will come.  Then the best fit will allow to determine model-independent (data-driven) partial-wave amplitudes and associated resonance parameters (pole positions, residues, BW parameters, etc.) as the GWU/SAID group does, for instance, for the analysis of $\pi N$-elastic, charge-exchange, and $\pi^-p\to\eta n$ data~\cite{Arndt:2006bf}.

In the following sections, we outline some of the outstanding questions in the hyperon spectrum, overview the available data on $KN$ scattering and introduce theoretical predictions for the unexplored domain of measurements with ``neutron" targets.

\subsubsection{The $\Lambda(1405)1/2^- - \Lambda(1520)3/2^-$ Doublet:}
In the quark model, the $\Lambda(1405)1/2^- - \Lambda(1520)3/2^-$ doublet is a flavor singlet of three quarks (\textit{uds}). Dynamical versions of this model, with two-body interactions between the quarks can describe the low mean energy of this multiplet, but not the 115~MeV splitting between them. This has led to suggestions that there may even be two different 1/2$^-$ states $-$ one dynamical low $\overline{K}N$ resonance at 1405~MeV, and an unresolved higher state close to 1520~MeV~\cite{Liu:2016wxq}. A two pole structure of $\Lambda(1405)$ was indeed found in Ref.~\cite{Oller:2000fj}. The narrow pole lies slightly below $\bar KN$ threshold, and is fixed by the scattering data rather well, see Ref.~\cite{Cieply:2016jby} for the comparison of different coupled-channel approaches. However, the position of the second pole is determined less precisely, and may lie much further below $\bar KN$ threshold and deeper in the complex plane. Recent photoproduction data on $\pi\Sigma$ by CLAS~\cite{Moriya:2013eb} may be used to reduce the theoretical ambiguity on this (second) pole of $\Lambda(1405)$~\cite{Mai:2014xna} and a single pole structure is not ruled out~\cite{Anisovich:2019exw}. Modern lattice QCD (LQCD) calculations also support the view that its structure is a $\bar KN$ state~\cite{Kamleh:2016dwe,Molina:2015uqp}. In Skyrme's topological soliton model for the baryons, the low-lying $\Lambda(1405)$ state also appears naturally as a mainly 5-quark state~\cite{Scoccola:1988wa,
Callan:1987xt}. Lattice calculations based on the sequential Bayesian do, however, indicate that the multiplet may have a mainly 3-quark structure~\cite{Liu:2016rwa}.

In the case of those lowest energy flavor-singlet $1/2^-$ - $3/2^-$ parity doublets in the strange, charm and bottom hyperon spectra: 
$\Lambda(1405) - \Lambda(1520)$, $\Lambda_c(2595) - \Lambda_c(2625)$, 
$\Xi_c(2790) - \Xi_c(2815)$, and $\Lambda_b(5912) - \Lambda_b(5920)$~\cite{Tanabashi:2018oca} the ratio between the splittings in these three doublets are 14.4:3.7:3.1:1.0. These ratios agree qualitatively and within 30\% with the corresponding inverse ratios of the $K, D,$ and $B$ meson masses: 10.7:2.8:1.0. As these resonances all contain one light quark pair the latter is what one should expect from the gradual approach to heavy-quark symmetry with increasing meson (or constituent quark) mass if the quark structure of these three multiplets is similar.  As described in Sec.~\ref{sec:Maxim}, data on neutron targets described in this proposal have the potential to provide key insights into settling the nature of the $\Lambda(1405)$.

\subsubsection{Low-Lying Positive-Parity Resonances:}
In the spectra of the nucleon and the $\Lambda$ and $\Sigma$ hyperons, the lowest positive-parity resonances all lie below the lowest negative-parity multiplets except for the flavor singlet doublet $\Lambda(1405)1/2^- -\Lambda(1520)3/2^-$.  This reversal of normal ordering cannot be achieved in the constituent quark model with purely color-spin dependent quark interactions. These low-lying positive-parity resonances are the $N(1440)$, $\Lambda(1600)$, and $\Sigma(1660)$1/2$^+$ states. Their low masses do however appear naturally, if the interactions between the quarks are flavor dependent~\cite{Glozman:1995fu}. Present day LQCD calculations have not yet converged on whether these low-lying states can be described as having a mainly three-quark structure~\cite{Liu:2014jua,
Liu:2016rwa,Liu:2020mSun}.

In the spectrum of the $\Xi$, the $\Xi(1690)$, or possibly the recently discovered $\Xi(1620)$~\cite{Sumihama:2018moz} may be such a 1/2$^+$ state as well, although the quantum numbers of those states are yet to be determined.

In the corresponding decuplet spectra, a similar low-lying positive-parity state has so far only been definitely identified in the $\Delta(1232)$ spectrum: namely, the $\Delta(1600)3/2^+$. The $\Sigma(1840)3/2^+$ resonance very likely represents the corresponding positive-parity $\Sigma^\ast$ state. It should be important to identify the corresponding $3/2^+$ state in the spectrum of the $\Xi^\ast$.

\subsubsection{Negative-Parity Hyperon Resonances:}
In the spectrum of the nucleon, two well-separated groups of negative-parity resonances appear above the 1/2$^+$ state $N(1440)$. This lowest energy group consists of the $N(1535)1/2^-$ and the $N(1520)3/2^-$ resonances.  There is a direct correspondence in the $\Lambda(1670)1/2^-$ and the $\Lambda(1690)3/2^-$ resonances. There is also a repeat of this group in the spectrum of the $\Sigma$ hyperon in the two resonances $\Sigma(1620)1/2^-$ (tentative) and $\Sigma(1670)3/2^-$.

The $N(1535)$ resonance has a large (32--52~\%) decay branch to $\eta N$, even though its energy lies very close to the $\eta N$ threshold. This pattern repeats in the case of the $\Lambda(1670)$, which also has a substantial (10--25~\%) decay branch to the corresponding $\eta\Lambda$ state. As the still uncertain $\Sigma(1620)1/2^-$ resonance is located almost exactly at the threshold for $\eta\Sigma$, there is naturally no signal for an $\eta\Sigma$ decay from it.

In the spectrum of the $\Xi$ hyperon, none of the hitherto determined negative-parity multiplets is complete. The state $\Xi(1820)3/2^-$ may be the analog in the $\Xi$ spectrum of the states $N(1520)$, $\Lambda(1670)$, and $\Sigma(1670)$. It should be important to identify the lowest $1/2^-$ resonance in the $\Xi$ spectrum. If that resonance lacks an $\eta$ decay branch, it would demonstrate that the $\eta$ decay of the $1/2^-$ resonances in the spectra of the nucleon, $\Lambda$ and $\Sigma$ involves two quarks.

It should also be important to determine whether the uncertain ``bumps" referred to in the Particle Data Tables labelled $\Sigma(1480)$, $\Sigma(1560)$, and $\Xi(1620)$ represent true resonances~\cite{Tanabashi:2018oca}.

About 120~MeV above the $1/2^--3/2^-$ pair of nucleon resonances $N(1535)$ and $N(1520)$, the nucleon spectrum has three negative-parity resonances close in energy to one another. This multiplet is formed of the $N(1650)1/2^-$, $N(1700)3/2^-$, and $N(1675)5/2^-$ resonances.

The analogs in the spectrum of the $\Lambda$ of the first and last of these nucleon resonances are the $\Lambda(1800)1/2^-$ and the $\Lambda(1830)5/2^-$ resonances. The missing 3/2$^-$ state in this $\Lambda$ resonance multiplet has not yet been identified.
A common feature of all the 1/2$^-$ resonances in these multiplets is their substantial $\eta$ decay branch.

\subsubsection{Excited $S = -2$ and $S = -3$ Baryons:}
$SU(3)$ flavor symmetry allows as many $S = -2$ baryon resonances as there are $N$ and $\Delta$ resonances combined ($\sim$27); however, until now only three states, $\Xi(1322)1/2^+$, $\Xi(1530)3/2^+$, and $\Xi(1820)3/2^-$, have their quantum numbers assigned and only a few more states have been observed~\cite{Tanabashi:2018oca}.  For the discovery of excited cascade baryons, we envision a PWA similar to the $S = -1$ sector but more complicated as one is dealing with a three-body final state. 

The experimental situation with $\Omega^{-\ast}$s is even worse than for the $\Xi^\ast$ case -- there are very few data for excited states~\cite{Hassall:1981fs}. The main reason for such a scarce dataset is the very low cross section for their indirect production with pion or photon beams.

A major effort in LQCD calculations involves the determination of inelastic and multi-hadron scattering amplitudes, and the first calculation to study an inelastic channel was recently
performed~\cite{Wilson:2014cna,Dudek:2014qha}.  For lattice calculations involving baryons that contain one or more strange quarks an advantage is that the number of open decay channels is generally smaller than for baryons comprised only of the light $u$ and $d$ quarks.

\subsubsection{Heavy Quark Symmetry and the Hyperons:}
Heavy quark symmetry~\cite{Isgur:1991wq} provides a powerful tool for comparing the structure of hyperons with heavy (charm and bottom) flavor quarks to those with strange quarks. This symmetry follows from the fact that the strength of quark spin-orbit couplings scale with the inverse of the constituent mass. In the case of hyperons with light and heavy quarks this implies that the heavy quark spin decouples from those of the light quarks. Heavy quark symmetry suggests, that the ratio of the sizes of such spin-orbit splittings in the corresponding multiplets in the spectra of the strange, charm and beauty hyperons should approximately correspond to the ratio of the inverses of the corresponding constituent quark (or approximately) meson ($K$, $D$, $B$) masses. Where the spin-orbit splittings conform to this scaling law the implication is that the structure of the corresponding hyperon resonances in the different flavor sectors are similar.

The approach to this symmetry can used as a phenomenological tool to compare the spin-orbit splittings between the $\Xi$ hyperons in the different flavor sectors. Hitherto such comparable splittings are only known for the lowest negative parity doublets in the strange, charm and beauty hyperon spectra, with two light-flavor and only one single heavy quark.  The recent observation of a rich spectrum of narrow 
$\Xi_{cc}$~\cite{Aaij:2017ueg,Aaij:2018gfl}, $\Xi_{b}$~\cite{Aaij:2018yqz}, and $\Omega_c$~\cite{Aaij:2017nav} states has yielded significant theoretical developments in heavy quark spectroscopy.  The determination of the mass spectrum and quantum numbers of the $S = -2$ Cascade resonances beyond the $\Xi(1530)$ at KLF will be essential to test these developments to the strange quark sector and compare with recent and forthcoming LQCD calculations~\cite{Edwards:2012fx}.  This is \textit{a fortiori} the case for the spectrum of the $\Omega$ hyperons.

\subsubsection{$KN$ and $\bar KN$ Final States:}  
\begin{figure}[ht]
\centering

{
    \includegraphics[width=0.32\textwidth,keepaspectratio,angle=90]{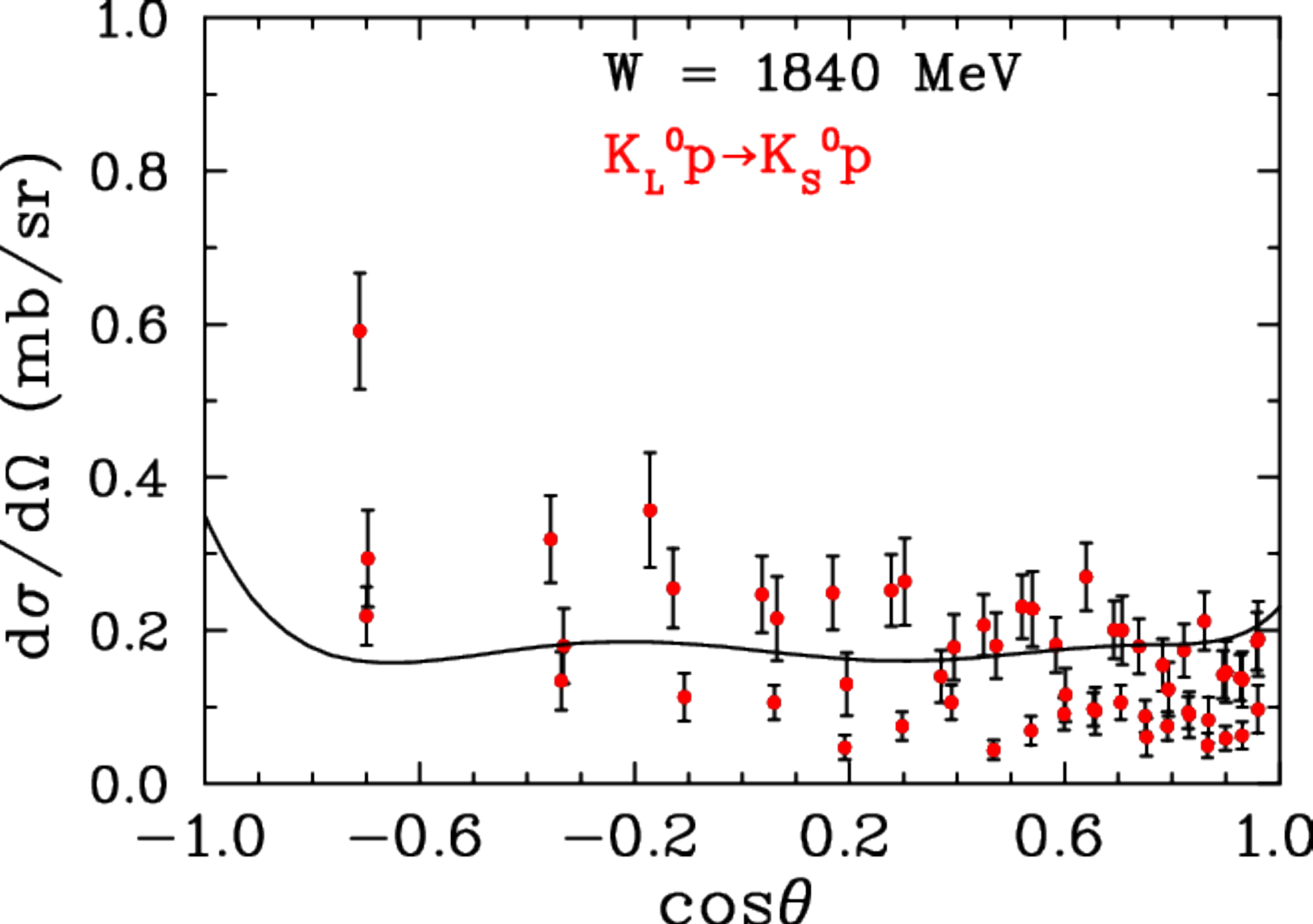} }
{
    \includegraphics[width=0.32\textwidth,keepaspectratio,angle=90]{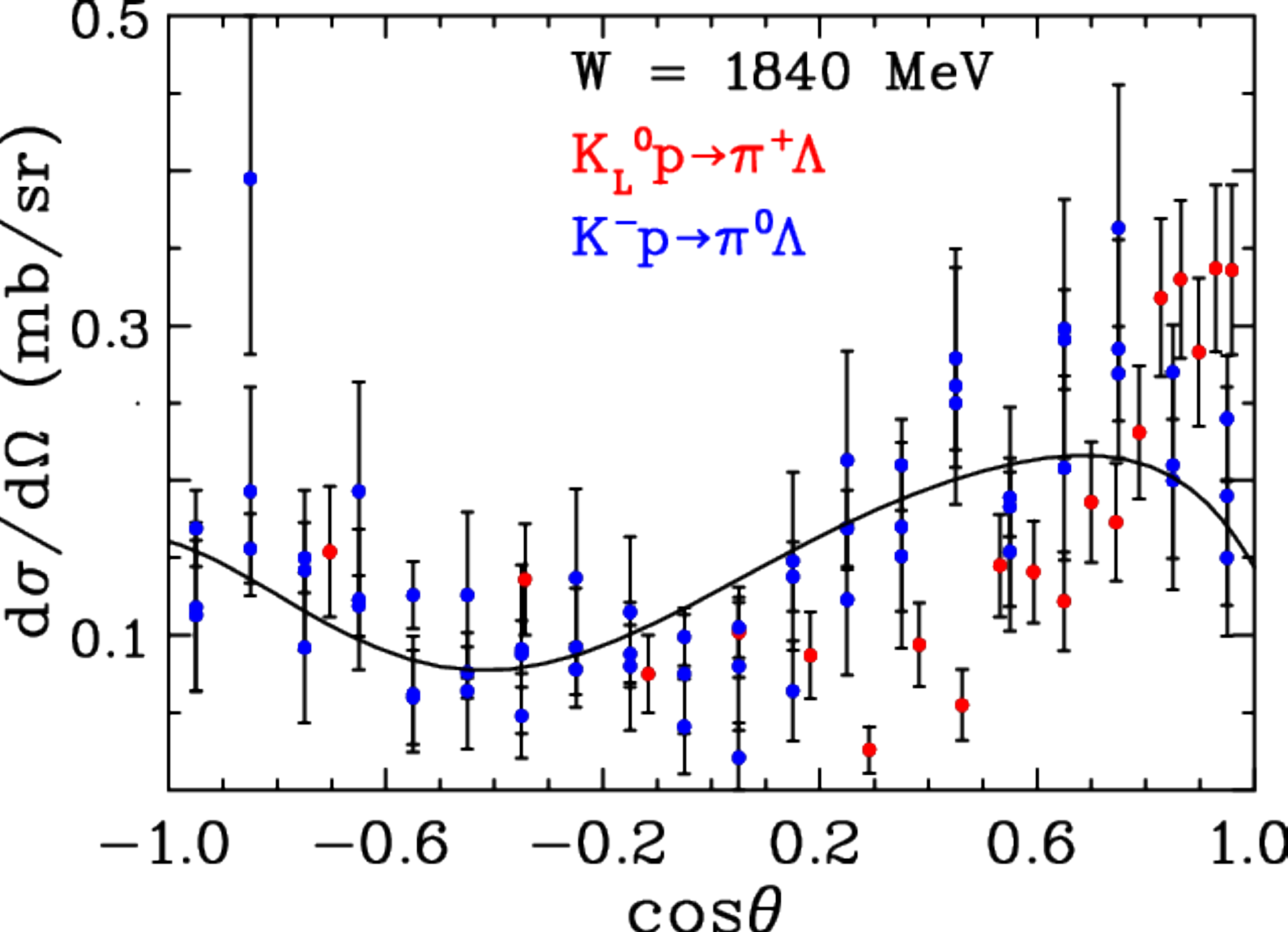} }
{
    \includegraphics[width=0.32\textwidth,keepaspectratio,angle=90]{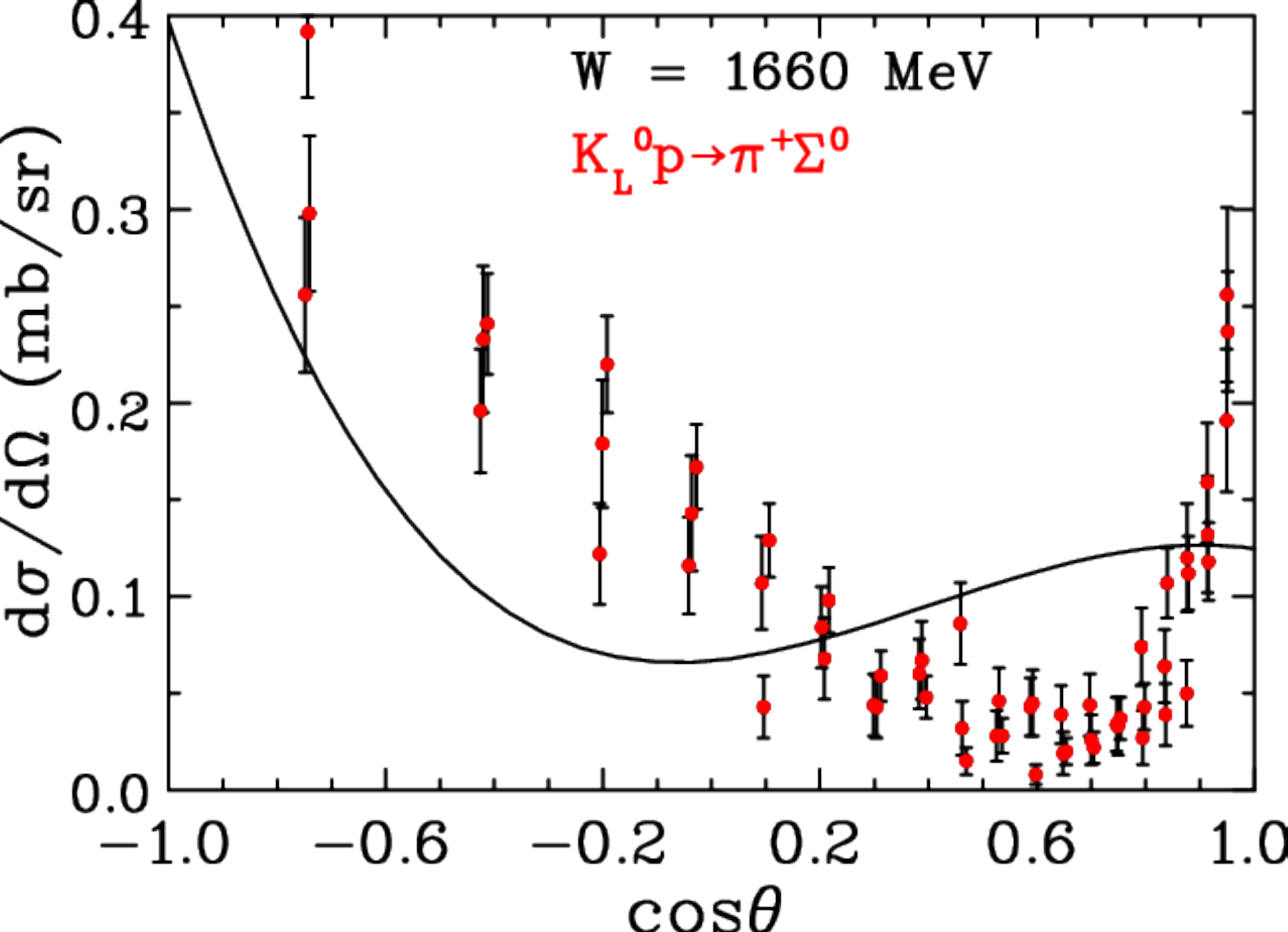} }

\centerline{\parbox{0.80\textwidth}{
 \caption[] {\protect\small 
        Selected differential cross section data for $K_Lp\to K_Sp$ (left), $K_Lp\to\pi^+\Lambda$ (middle) at W = 1840~MeV and $K_Lp \to\pi^0\Sigma^+$ at W = 1660~MeV (right) from Ref.~\protect\cite{Manley:2016blh}.  The plotted data from previously published experimental data are those data points within 20~MeV of the kaon CM\ energy indicated on each panel~\protect\cite{SAIDweb}. Plotted uncertainties are statistical only. The curves are predictions using amplitudes from a recent PWA~\protect\cite{Zhang:2013cua,Zhang:2013sva}, combined with $KN\to KN$ amplitudes from the SAID database~\protect\cite{SAIDweb}. }
        \label{fig:KLp_KSp} } }
\end{figure}

A fair amount of data are available for the reaction, $K^+n\to K^0p$, measured on a deuterium target.  Figure~\ref{fig:KLp_KSp} shows a sample of available differential cross sections for $K_Lp\to K_Sp$ compared with predictions determined from a recent PWA of $\bar KN\to\bar KN$ data~\cite{Zhang:2013cua,Zhang:2013sva}, combined with $KN\rightarrow KN$ amplitudes from the SAID database~\cite{SAIDweb}.  The predictions at lower and higher energies tend to agree less well with the data.
\begin{figure}[ht]
\centering
\floatbox[{\capbeside\thisfloatsetup{capbesideposition={right,center},capbesidewidth=7cm}}]{figure}[\FBwidth]
    {\caption{Comparison of selected polarization data for $K^-p\to\pi^0\Lambda$ and
    $K_Lp\to\pi^+\Lambda$ at W = 1880~MeV, from Ref.~\protect\cite{Manley:2016blh}.  The plotted data from previously published experimental data are those data points within 20~MeV of the kaon CM\ energy indicated on each panel~\protect\cite{SAIDweb}. The curves are from a recent PWA of $K^-p\to\pi^0\Lambda$ data~\protect\cite{Zhang:2013cua,Zhang:2013sva}.}
    \label{fig:KLp_piLambda_P}}
    {\includegraphics[width=0.4\textwidth,keepaspectratio,angle=90]{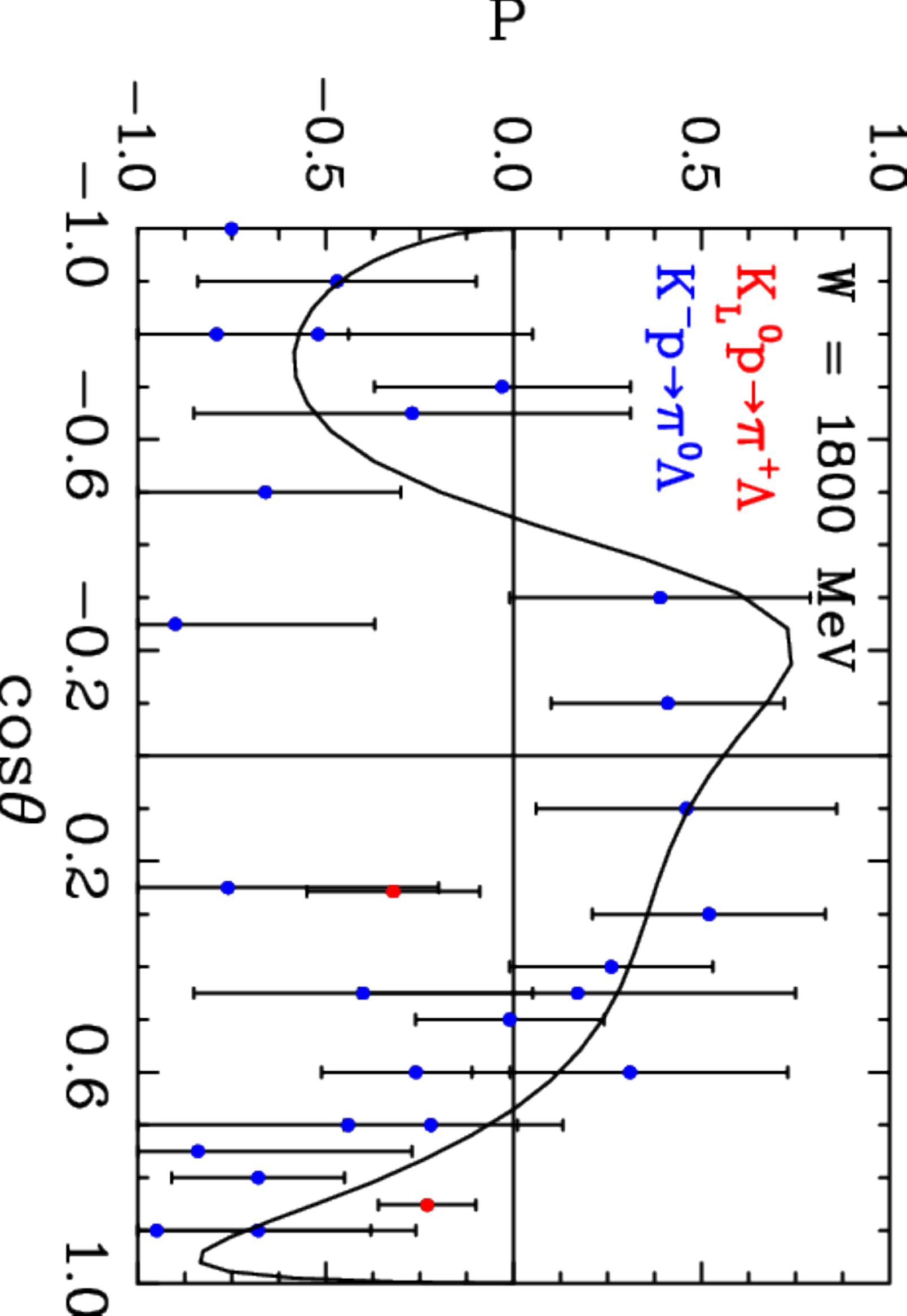}}
\end{figure}

\subsubsection{$\pi\Lambda$ Final States:}
The $K^-p\rightarrow\pi^0\Lambda$ and $K_Lp\rightarrow\pi^+\Lambda$ amplitudes imply that observables for these reactions measured at the same energy should be the same except for small differences due to the isospin-violating mass differences in the hadrons. No differential cross section data for $K^-p\rightarrow\pi^0\Lambda$ are available at CM energies W < 1540~MeV, although data for $K_Lp\to\pi^+\Lambda$ are available at such energies.  At 1540~MeV and higher energies, differential cross section and polarization data for the two reactions are in fair agreement, as shown in Figs.~\ref{fig:KLp_KSp} and \ref{fig:KLp_piLambda_P}. It should be stressed that polarized measurements are tolerable for any PWA solutions (Fig.~\ref{fig:KLp_piLambda_P}).

\subsubsection{$\pi\Sigma$ Final States:}
Figure~\ref{fig:KLp_KSp} shows a comparison of differential cross section data for $K^-p$ and $K_Lp$ reactions leading to $\pi\Sigma$ final states at W = 1660~MeV (or P$_{\rm lab}$ = 716~MeV/$c$).  The curves are based on energy-dependent isospin amplitudes from a recent PWA~\cite{Zhang:2013cua,Zhang:2013sva}. No differential cross section data are available for $K_Lp\to\pi^0\Sigma^+$. As this example shows, the quality of the $K_Lp$ data is comparable to that for the $K^-p$ data.  It would, therefore, be advantageous to combine the $K_Lp$ data in a new coupled-channel PWA with available $K^-p$ data. Note that the reactions $K_Lp\to\pi^+\Sigma^0$ and $K_Lp\rightarrow\pi^0\Sigma^+$ are isospin selective (only $I = 1$ amplitudes are involved) whereas the reactions $K^-p\to\pi^-\Sigma^+$ and $K^-p\to\pi^+\Sigma^-$ are not.  New measurements with a $K_L$ beam would lead to a better understanding of $\Sigma^\ast$ states and would help constrain the amplitudes for $K^-p$ scattering to $\pi\Sigma$ final states.

\subsubsection{$K\Xi$ Final States:}
The threshold for $K^-p$ and $K_Lp$ reactions leading to $K\Xi$ final states is fairly high (W$_{\rm thresh}$ = 1816~MeV). There are no differential cross section data available for $K_Lp\rightarrow K^+\Xi^0$ and very few (none recent) for $K^-p\to K^0\Xi^0$ or $K^-p\rightarrow K^+\Xi^-$.  Measurements for these reactions would be very helpful, especially for comparing with predictions from dynamical coupled-channel (DCC) models~\cite{Kamano:2014zba,Kamano:2015hxa} and other effective Lagrangian approaches~\cite{Jackson:2015dva}. The {\it Review of Particle Physics}~\cite{Tanabashi:2018oca} lists only two states with branching ratios (BR) to $K\Xi$, namely, $\Lambda(2100)7/2^-$ (BR $<$ 3~\%) and $\Sigma(2030)7/2^+$ (BR $<$ 2~\%). A recent theoretical prediction of the total cross section for the $K^{+}\Xi^{0}$ process was provided in Ref.~\cite{Feijoo:2018den}.

\subsubsection{Theory for ``Neutron" Target Measurements:}
\label{sec:Maxim}
The so-called coupled-channel Chiral Unitary approaches (UChPT) implement unitarity exactly via a re-summation of a chiral potential to a certain chiral order. They successfully describe all available anti-kaon-nucleon scattering data and predict the mass and width of the sub-threshold resonance in the Isospin $ I= 0$ channel, the $\Lambda(1405)1/2^-$. Furthermore, such models lead to the prediction of the second pole in the complex energy plane with the same quantum numbers as $\Lambda(1405)1/2^-$. This is usually referred to as the two-pole structure of the $\Lambda(1405)1/2^-$, see the current review by the Particle Data Group~\cite{Tanabashi:2018oca} for more details.
\begin{figure}[ht]
\centering
\floatbox[{\capbeside\thisfloatsetup{capbesideposition={right,center},capbesidewidth=7cm}}]{figure}[\FBwidth]
{\caption{Pole positions of $\Lambda(1405)$ in chiral unitary 
        approaches -
        $KM$ from Ref.~\protect\cite{Ikeda:2012au},
        $B$ from Ref.~\protect\cite{Mai:2014xna},
        $M$ from Ref.~\protect\cite{Guo:2012vv}, and
        $P$ from Ref.~\protect\cite{Cieply:2011nq} as
        compared in Ref.~\protect\cite{Cieply:2016jby}.
        Each symbol represents the position of the first (black)
        and second (red) pole in each model.}\label{fig:maxim3}}
{\includegraphics[width=0.5\textwidth,keepaspectratio]{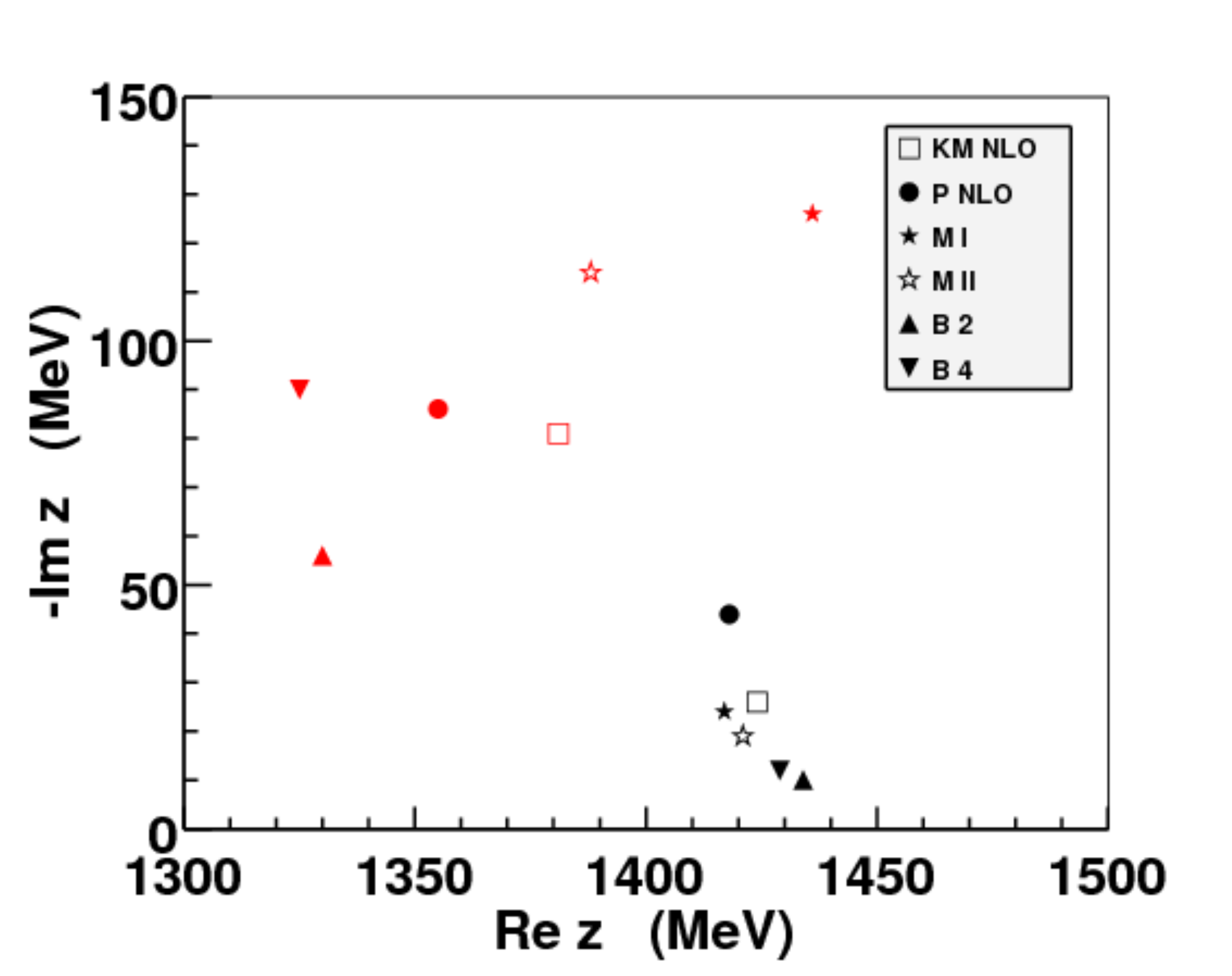}}
\end{figure}

In the most recent formulation, such UChPT approaches rely on a chiral amplitude for meson-baryon scattering up to next-to-leading chiral order. The unitarity constraint is imposed via the Bethe-Salpeter equation either in the full off-shell formulation~\cite{Mai:2012dt,Bruns:2010sv} or in the so-called on-shell approximation~\cite{Mai:2014xna,Ikeda:2012au}, while the off-shell effects are rather small~\cite{Mai:2012dt}. Among the on-shell approaches~\cite{Mai:2014xna,Ikeda:2012au,Guo:2012vv,Cieply:2011nq} a quantitative comparison was performed in Ref.~\cite{Cieply:2016jby}, which shown that using the available  experimental data, the models predict very different behavior of the scattering amplitude on and off the real energy-axis. This systematic uncertainty becomes evident, when comparing the pole positions of the $\Lambda(1405)1/2^-$ in these models, see Fig.~\ref{fig:maxim3}. The position of the narrow (first) pole seems to be constrained  rather well, while the predictions for the position broad (second) pole cover a very wide region of the complex energy-plane. This uncertainty is present even within models of the same type. This ambiguity can be traced back to the fact that the experimental data used to fix the parameters of the models is rather old and imprecise.
\begin{figure}[ht]
\centering
{
{\footnotesize
        $\bf~~~~~~~~~~~
        K_Ln\rightarrow K^-p ~~~~~~~~
        K_Ln\rightarrow {\bar K}^0n ~~~~~~~~~
        K_Ln\rightarrow\pi^0\Lambda ~~~~~~~~~~~~
        K_Ln\rightarrow\pi^0\Sigma^0~~~~~~~
        K_Ln\rightarrow\pi^-\Sigma^+~~~~~~~
        K_Ln\rightarrow\pi^+\Sigma^-~~$}\\
    \includegraphics[width=0.992\textwidth,keepaspectratio, trim=0 1.4cm 0.4cm 0.6cm, clip]{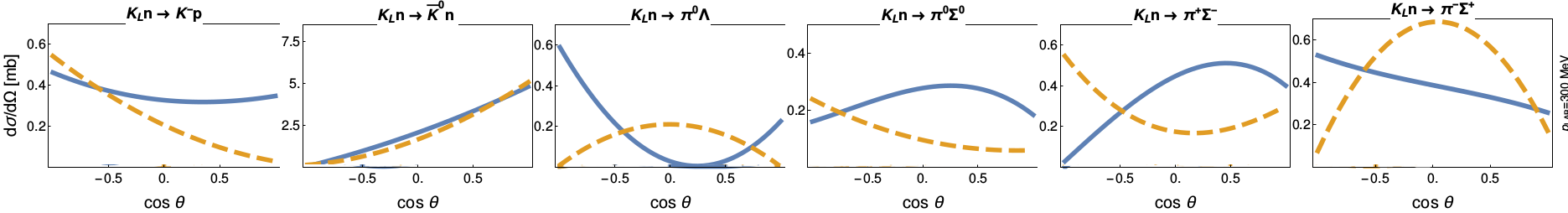}
\\[0.1cm]
    \includegraphics[width=\textwidth,keepaspectratio, trim=0 0.1cm 0.4cm 0.7cm, clip]{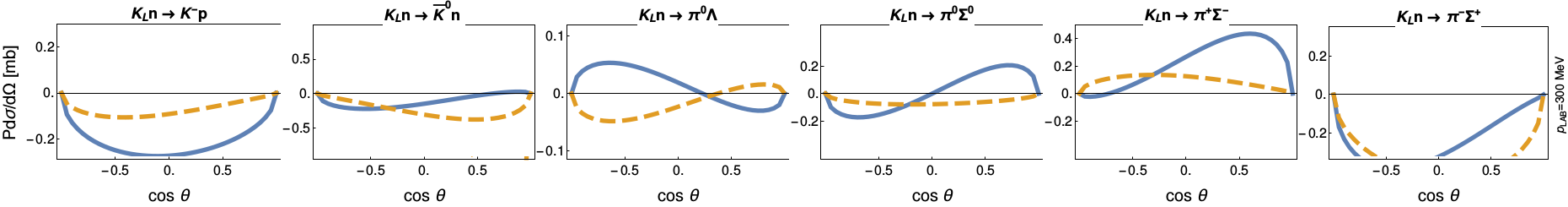} }
\vspace{-0.6cm}~\\
    \caption{Theoretical predictions for $d\sigma/d\Omega$ (top) and 
	$Pd\sigma/d\Omega$ (bottom) as a function of CM\
    $\cos$ of a meson production angle for kaon lab-momentum of 
    300~MeV/$c$ of initial neutral kaon beam.  Orange dashed and
    blue solid lines show predictions within Model-B2 and Model-B4, 
    respectively. 
}\label{fig:maxim1}
\end{figure}
\begin{figure}[ht]
\centering
\floatbox[{\capbeside\thisfloatsetup{capbesideposition={right,center},capbesidewidth=6cm}}]{figure}[\FBwidth]
{   
    \caption{Predictions for $d\sigma/d\Omega$ and $P d\sigma/d\Omega$ as a function of CM\ $\cos$ of a meson production angle, $\theta$, for the reaction $K_{L}n\to K^{+}\Xi^{-}$. Each column is associated with kaon laboratory momentum of 300 and 700~MeV/$c$ of initial neutral kaon beam. Orange (dashed) and blue lines show predictions within Model-B2 and Model-B4, respectively.}\label{fig:maxim4}}
	{\includegraphics[width=0.6\textwidth,keepaspectratio,trim=0 0 8cm 0, clip]{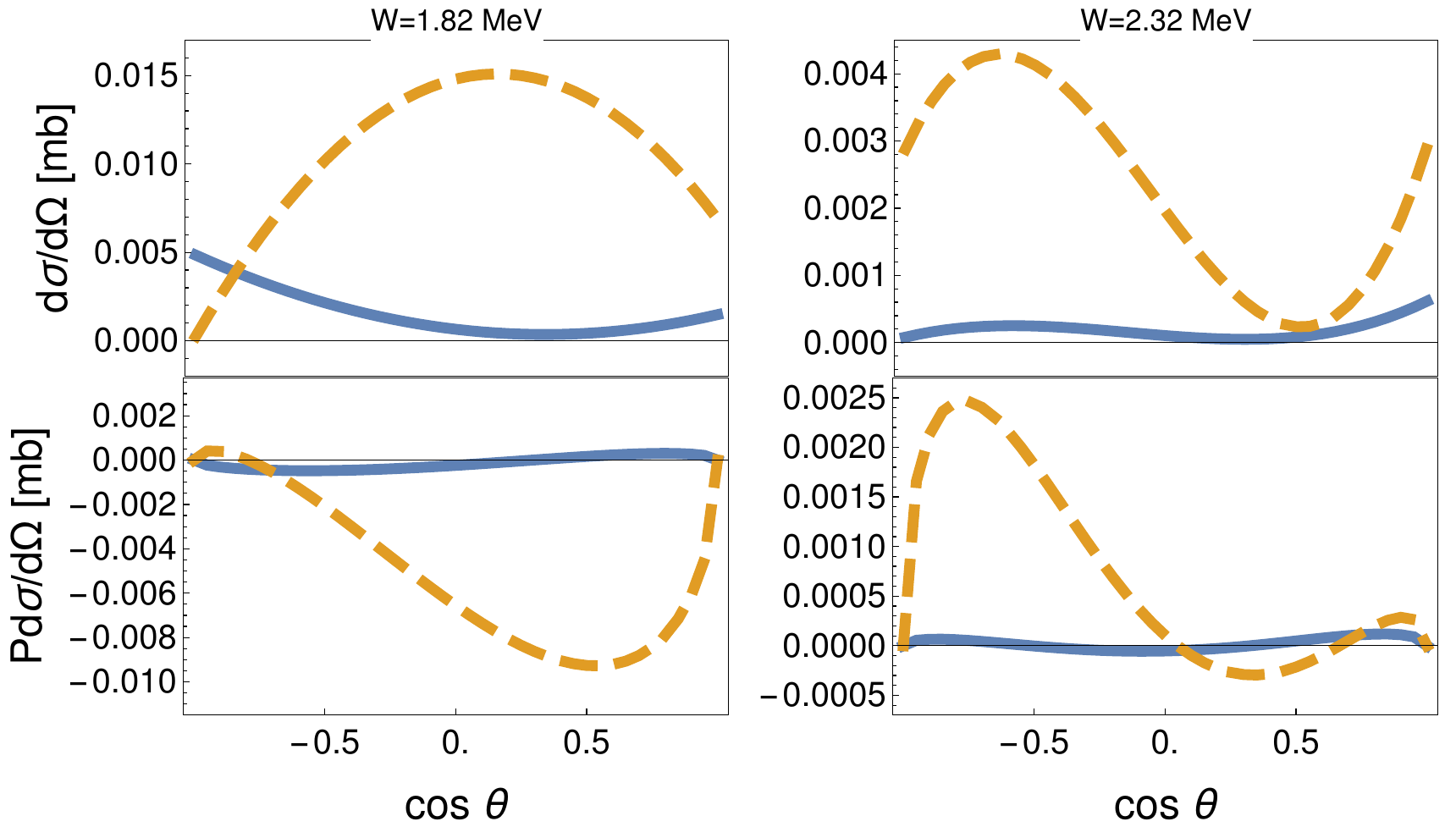}}
\end{figure}

The $K_L$ beam scattered off a neutron target, while measuring the strangeness $S = -1$ of the final meson-baryon states, will become a strongly desired new source of experimental data to pinpoint the properties of the anti-kaon-nucleon scattering amplitude. To make this statement more quantitative, we compare predictions of both solutions of the model from Ref.~\cite{Mai:2014xna}. These solutions agree with all scattering, threshold as well as the photoproduction data for the $\Sigma\pi$ line shapes by the CLAS Collaboration~\cite{Moriya:2013eb}. The predicted differential cross sections ($d\sigma/d\Omega$) as well as polarized ones ($P\cdot d\sigma/d\Omega$) for the $K_Ln$ scattering with the final states $K^-p$, $\bar K^0n$, $\pi^0\Lambda$, $\pi^{0/+/-}\Sigma^{0/-/+}$ are presented in Fig.~\ref{fig:maxim1}.  There is no evident agreement between both prediction of these observables in the energy range aimed to study in the proposed $K_L$ experiment. This is very encouraging since the actual data can sort out one (or both) solutions as unphysical, which was not possible using the present experimental data.  As for the $K\Xi$ final states being measured at KLF, both solutions of the here presented model can be used for a theoretical estimate. The reason of being able to do so is that $K^+\Xi^-$ and $K^0\Xi^0$ channels are part of the channel space of ground state octet mesons-baryon channels dynamically implemented into the present model. The result of such a prediction is depicted and addressed further in Fig.~\ref{fig:maxim4}.

\textbf{To Summarize}: The pole positions have been determined for several $\Lambda^\ast$s and $\Sigma^\ast$s but the information about the pole positions have not been determined for $\Xi$ or $\Omega$ hyperons~\cite{Tanabashi:2018oca}.  Our plan is to do a coupled-channel PWA with new KLF data in combination with available and new J-PARC $K^-p$ measurements when they will be available. Then the best fit will allow the determination of data-driven (model independent) partial-wave amplitudes and associated resonance parameters (pole positions, residues, BW parameters, and so on). See Appendix~\ref{sec:A3} for a more detailed discussion. Additionally, PWAs with new KLF data will allow a search for ``missing" hyperons via looking for new poles in complex plane positions. It will provide a new benchmark for comparisons with QCD-inspired models and LQCD calculations.                          
\subsubsection{Partial-Wave Analysis for Hyperons:}
In spite of their model dependence, partial-wave BW parameters have for quite some time been the preferred connection between experiment and QCD in hadronic spectroscopy. More recently, however, pole parameters (e.g., pole positions and residues) have justifiably become the preferred connection, and this fact has also been recognized by the Particle Data Group in recent editions of the {\it Review of Particle Physics}~\cite{Tanabashi:2018oca}. Therefore, the extraction of pole parameters from experimental data becomes a procedure of utmost importance.

Extraction of pole parameters is usually performed in two ways: \\
(a) in an energy-dependent way (ED) or \\
(b) in an energy-independent procedure through SES PWAs. \\
In an ED procedure, one measures as many observables as possible to be close to the complete set and then fits the observables with parameters of a well-founded theoretical model that describes the reaction in question. Continuity in energy is enforced by the features of the theoretical model. In a SE procedure, one again measures as many observables as possible but attempts to extract partial waves by fitting energy-binned data independently, therefore, reducing the theoretical input. A discrete set of partial waves is obtained, and the issues of achieving continuity in energy have recently been extensively discussed either by introducing the constraints in analyticity~\cite{Osmanovic:2017fwe} or through angle- and energy-dependent phase ambiguity~\cite{Wunderlich:2017dby,Svarc2018}.

In energy-dependent models, pole parameters have been extracted in various ways. The most natural way is the analytic continuation of theoretical model solutions into the complex-energy plane. Simpler single-channel pole extraction methods have been developed such as the speed plot~\cite{Hohler:1994rt}, time delay~\cite{Kelkar:2008na}, the N/D method~\cite{Chew:1960iv}, regularization procedures~\cite{Ceci:2006zw}, and Pad\'e approximants~\cite{Masjuan:2014psa}, but their success has been limited. In single-energy analyses, the situation is even worse: until recently no adequate method has been available for the extraction of pole parameters. All single-channel methods involve first- or higher-order derivatives, so partial-wave data had to be either interpolated or fitted with an unknown function, and that introduced additional and, very often, uncontrolled model dependencies.

That situation has been recently overcome when a new Laurent+Pietarinen (L+P) method applicable to both, ED and SES models, has been introduced~\cite{Svarc:2013laa,
Svarc:2014zja,Svarc:2014sqa,Svarc:2014aga,Svarc:2015usk}. The driving concept behind the single-channel (and later multichannel) L+P approach was to replace solving an elaborate theoretical model and analytically continuing its solution into the full complex-energy plane, with an approximation actualized by  local power-series representation of partial-wave amplitudes having well-defined analytic properties on the real energy axis, and fitting it to the given input. In such a way, the global complexity of a model
is replaced by a much simpler, and almost model-independent expansion, limited to the regions near the real energy axis. And this is sufficient to obtain poles and their residues. This procedure gives the simplest function with known analytic structure that fits the data. Formally, the introduced L+P method is based on the Mittag-Leffler
expansion~\footnote{Mittag-Leffler expansion~\cite{Mittag-Leffler} is the generalization of a Laurent expansion to a more-than-one pole situation. For simplicity, we call it Laurent expansion.} of partial-wave amplitudes near the real-energy axis, where we represented
the regular background term by a conformal-mapping-generated, fastly converging power series called a Pietarinen
expansion~\footnote{This type of conformal mapping expansion
was introduced by Ciulli and Fisher~\cite{Ciulli,CiulliFisher}. It was described in details and also used in pion-nucleon scattering by
Pietarinen~\cite{Pietarinen:1972nk,Pietarinen:1975se}. The procedure was named Pietarinen expansion by H\"{o}hler in
Ref.~\cite{Hohler:1984ux}.}. In practice, the regular background part is usually fitted with three Pietarinen expansion series. Each of them approximates the most general function which describes the background, and has a branch point at $x_{bp}$, while all free parameters are then
fitted to the chosen channel input. The first Pietarinen expansion with branch-point $x_P$ which is restricted to an unphysical energy range represents all left-hand cut contributions. The next two Pietarinen expansions describe background in the physical range, and the used branch points $x_Q$ and $x_R$ are defined by the analytic properties of the analyzed partial wave. A second branch point is usually fixed to the elastic channel branch point describing threshold
effects, and the third one is either fixed to the dominant channel threshold value, or let free.

Thus, solely on the basis of general physical assumptions about analytic properties of the fitted process like number of poles and number and location of conformal mapping branch points, the pole parameters in the complex energy plane are obtained. In such a way, the simplest analytic function with a set of poles and branch points which is fitting the input is actually constructed. This method is equally applicable to both theoretical and experimental input~\footnote{Observe that fitting partial wave data originating from experiment as energy independent analysis is even more favorable.}.

The transition amplitude of the multichannel L+P model is parametrized as
\begin{eqnarray} \label{Eq:MCL+P}
        T^a(W)&=&\hspace{-1mm}\sum _{j=1}^{{N}_{pole}} \frac{g^{a}_{j} }
        {W_j-W} \hspace{-0.5mm}
        +\hspace{-0.5mm}\sum_{i=1}^3\sum_{k_i=0}^{K_i^{a}} c^{a}_{k_i}
        \left(\frac{\alpha^a_i\hspace{-1mm}-\hspace{-1mm}\sqrt{x^a_i-W}}
        {\alpha^a_i\hspace{-1mm}+\hspace{-1mm}\sqrt{x^a_i - W }}
        \right)^{k_i}, \nonumber\\[-1ex]
\end{eqnarray}
where $a$ is a channel index, $W_j$ are pole positions in the complex $W$ (energy) plane, $g^{a}_i$ coupling constants. The $x^{a}_i$ define the branch points, $c^{a}_{k_i}$, and $\alpha^{a}_i$ are real coefficients. $K^a_i, \,i=1,2,3$\, are Pietarinen coefficients in channel $a$. The first part represents the poles and the second term three branch points. The first branch point is chosen at a negative energy (determined by the fit), the second is fixed at the dominant production threshold, and the third branch point is adjusted to the analytic properties of fitted partial wave.

To enable the fitting, a reduced discrepancy function $D_{dp}$ is defined as
\begin{eqnarray} \label{eq:Laurent-Pietarinen}
        D_{dp} &=&\sum _{a}^{all}D_{dp}^a;\qquad  \nonumber \\
        D_{dp}^a &=&  \frac{1}{2 \, N_{W}^a - N_{par}^a} \times
        \sum_{i=1}^{N_{W}^a}
        \left\{ \left[ \frac{{\rm Re} \,T^{a}(W^{(i)})-{\rm Re} \,
        T^{a,exp}(W^{(i)})}{ Err_{i,a}^{\rm Re}} \right]^2 \right.\nonumber\\
        &&\left.
        +\quad\ \,\left[ \frac{{\rm Im} \, T^{a}(W^{(i)})-{\rm Im} \,
        T^{a,exp}(W^{(i)})}{ Err_{i,a}^{\rm Im}} \right]^2 \right\}
        + {\cal P}^a , \nonumber
\end{eqnarray}
where
\begin{eqnarray}
        {\cal P}^{a} &=& \lambda^a_{k_1} \sum _{k_1=1}^{K^a} (c^a_{k_1})^2 \,
        {k_1}^3 +  \lambda_{k_2}^a \sum _{k_2=1}^{L^a} (c^a_{k_2})^2 \,
        {k_2}^3 + \nonumber
        + \lambda_{k_3}^a \sum _{m=1}^{M^a} (c^a_{k_3})^2 \, {k_3}^3 \nonumber
\end{eqnarray}
is the Pietarinen penalty function, which ensures fast and optimal convergence. $N_{W}^a$ is the number of energies in channel $a$, $N_{par}^a$ the number of fit parameters in channel $a$, $\lambda_c^a, \lambda_d^a, \lambda_e^a$ are Pietarinen weighting factors, $Err_{i,a}^{\rm Re, \, Im} \ldots$  errors of the real and imaginary part, and $c_{k_1}^a, c_{k_2}^a, c_{k_3}^a$ real coupling constants.
\begin{center}
\begin{figure}[ht]
\centering
{
    \includegraphics[width=1\textwidth,keepaspectratio]{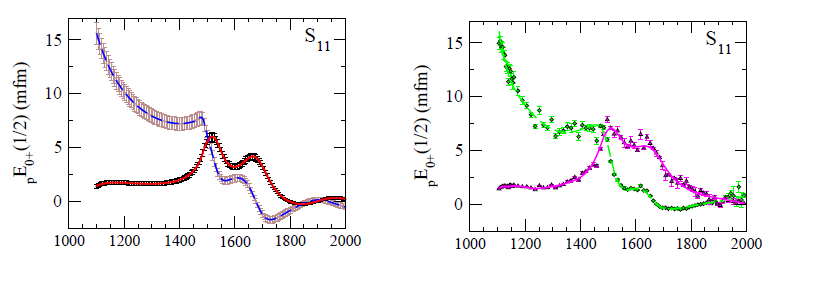} }

	\centerline{\parbox{0.70\textwidth}{
    \caption[] {\protect\small  
    L+P fit to CM12 GW/SAID pion photoproduction $_p E_{0+}$ ED and SESs~\protect\cite{Workman:2012jf}.} 
	\label{Fig:L+PCM12} } }
\end{figure}
\end{center}

In order to obtain reliable answers in the L+P model we have to build knowledge about the analytic structure of the fitted partial wave into the fitting procedure. Because we are looking for poles, we only have to define which branch points to include. Their analytic form will be determined by the number of Pietarinen coefficients.  As we have only three branch points at our disposal we expect that the first branch-point will describe all subthreshold and left-hand cut processes, the second one is usually fixed to the dominant channel opening, and the third one is to represent background contributions of all channel openings in the physical range. So, in addition to choosing the number of relevant poles, our anticipation of the analytic structure of the observed partial wave is of great importance for the stability of the fit.

The L+P model has been successfully applied to both theoretical models and discreet partial-wave data. As an example, in Fig.~\ref{Fig:L+PCM12}, we give the achieved quality of the fit for the CM12 GW/SAID pion photoproduction amplitudes~\cite{Workman:2012jf}.                
\subsubsection{Strange Hadrons from the Lattice:}
Our knowledge of the excited-state spectrum of QCD through the solution of 
the theory on a Euclidean-space lattice has undergone tremendous advances 
over the past several year.  What we characterize as excited states are 
resonances that are unstable under the strong interaction, and their 
properties are encapsulated in momentum-dependent scattering amplitudes.

The methodology for obtaining momentum-dependent phase shifts for elastic 
scattering from the shifts in energy levels on a Euclidean lattice at 
finite volume was provided many years ago~\cite{Luscher:1990ux} and 
extended to systems in motion~\cite{Rummukainen:1995vs}, but its 
implementation for QCD remained computationally elusive until recently. 
A combination of theoretical, algorithmic, and computational advances 
has changed this situation dramatically, notably in the case of mesons. 
There have been several lattice calculations of the momentum-dependent 
phase shift of the $\rho$ mesons~\cite{Aoki:2007rd,Feng:2010es,Dudek:2012xn,
Guo:2016zos,Alexandrou:2017mpi,Bulava:2016mks,Lang:2011mn}. This has now 
been extended to $K\pi$ scattering~\cite{Brett:2018jqw} in both $P$- and 
$S$-wave.
\begin{figure}[ht]
 \includegraphics[width=0.95\textwidth]{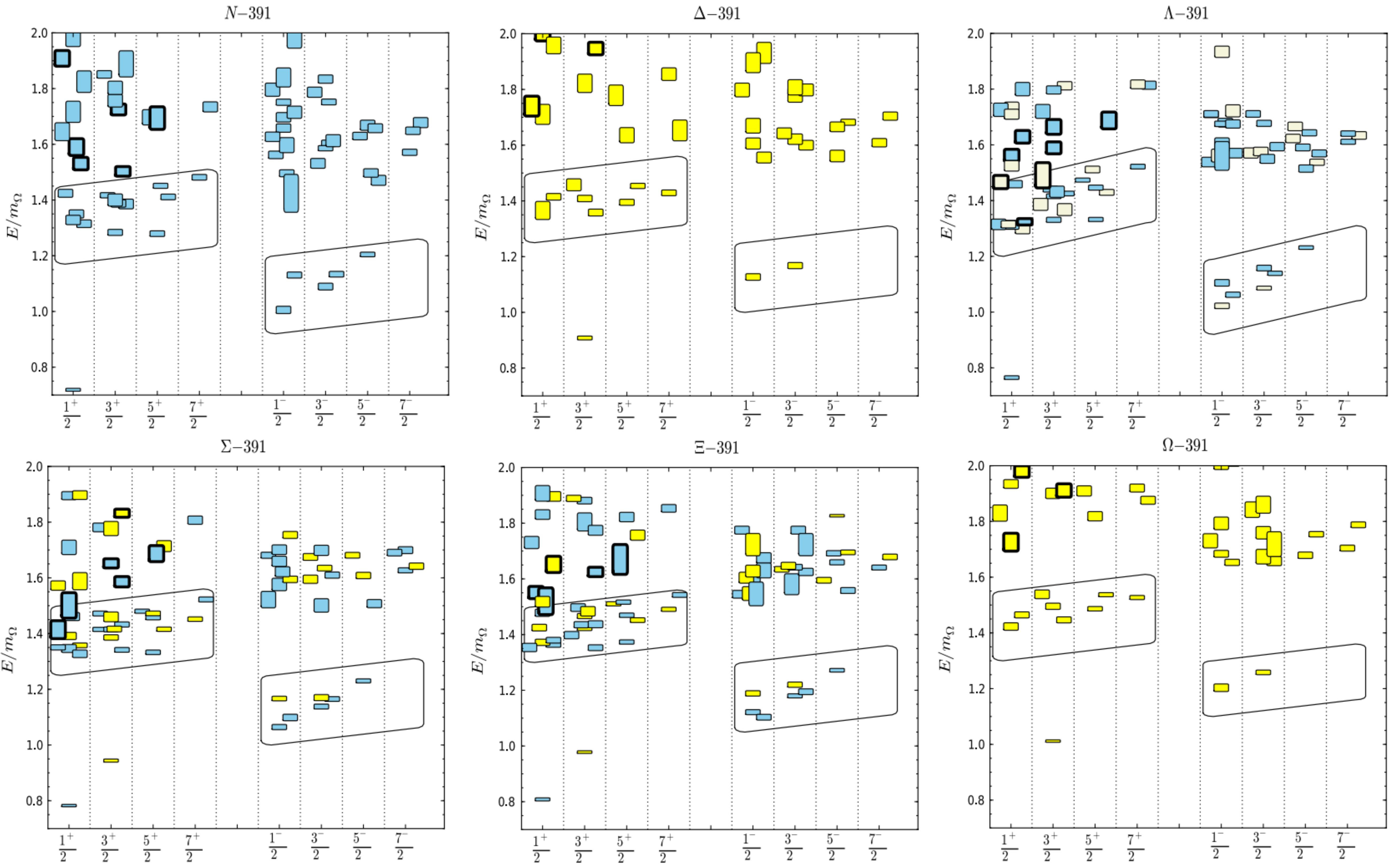}

    	\centerline{\parbox{0.80\textwidth}{
        \caption[] {\protect\small 
        Results for baryon excited states using an ensemble with $m_\pi = 
	391$~MeV are shown versus $J^P$~\protect\cite{Edwards:2012fx}. 
	Colors are used to display the flavor symmetry of dominant operators 
	as follows: blue for \textbf{$8_F$} in $N$, $\Lambda$, $\Sigma$, and
        $\Xi$; beige for \textbf{$1_F$} for $\Lambda$; yellow for 
	\textbf{$10_F$} in $\Delta$, $\Sigma$, $\Xi$, and $\Omega$. The 
	lowest bands of positive- and negative-parity states are highlighted 
	within slanted boxes.  Hybrid states, in which the gluons play a 
	substantive role, are shown for positive parity by symbols with 
	thick borders.} \label{fig:lqcd3} } }
\end{figure}

The formulation to extract amplitude information has been extended to the 
coupled-channel \\
case~\cite{Guo:2012hv,Briceno:2012yi,Meissner:2014dea,Liu:2005kr,
Lage:2009zv,Morningstar:2017spu,Doring:2011vk,Doring:2011nd,Doring:2012eu}, 
and applied to the case of the coupled $K\bar K -
\pi\pi$~\cite{Wilson:2015dqa} system describing the $\rho$ resonance to the 
$\eta K - \eta\pi$ system~\cite{Wilson:2014cna,Dudek:2014qha}, and to the 
emblematic isoscalar sector~\cite{Briceno:2017qmb,Briceno:2016mjc}.  Most 
recently, a calculation of coupled isoscalar $\pi\pi$, $K\bar{K}$, and 
$\eta\eta$ scattering for both $S$ and $D$ wave has been 
performed\cite{Briceno:2017qmb}, revealing a qualitative interpretation of 
$\sigma$, $f_0$, and $f_2$ mesons similar to that seen in experiment.  
Collectively, these papers provide a comprehensive picture of SU(3) nonets 
both in the tensor and scalar sectors, albeit at unphysically large pion mass.

The application to baryons is far more limited, but nonetheless, important 
insights have been gained.  In an approach in which the excited-state hadrons 
are treated as stable particles, a spectrum of baryons at least as rich as 
that of the quark model is revealed~\cite{Edwards:2011jj,
Engel:2013ig}, and evidence has been presented for ``hybrid" baryon states, 
beyond those of the quark model, in which gluon degrees of freedom are 
essential~\cite{Dudek:2012ag}.  Notably, this picture extends to the 
spectrum of $\Lambda$, $\Sigma$, $\Xi$, and $\Omega$ states where the 
counting of states reflects $SU(6) \times O(3)$ symmetry, and the presence 
of hybrids is common across the spectrum.  In Fig.~\ref{fig:lqcd3}, baryon 
spectra from~\cite{Edwards:2012fx} are presented in units of $\Omega$ mass 
from LQCD calculations with ensemble $m_\pi = 391$~MeV (not yet at physical 
$m_\pi$).

The calculations for the baryon sector are incomplete, in that the 
momentum-dependent scattering amplitudes characterizing multi-hadron 
states have not been extracted.  In comparison with the calculations 
for mesons cited above, the challenges are more computational than 
theoretical or conceptual.  Nonetheless, the first direct calculation of 
the $I = 3/2$ $N \pi$ system in $P$-wave has now been 
performed~\cite{Andersen:2017una}, revealing a BW description of the 
amplitude commensurate with a phenomenological description of the $\Delta$ 
resonance.  Thus we can be confident that the progress made in the meson 
sector will be reflected for the case of baryons in the coming years.  
Indeed, many of the algorithmic and computational challenges, notably the 
need to perform calculations at physical quark masses, and the increasing 
complexity of the Wick contractions as the number of hadrons, and therefore 
quarks, is increased, are being addressed with the recently launched 
Exascale Computing Project of the DOE and NNSA, whose application to 
lattice QCD is described in Ref.~\cite{Brower:2017vth}. 


\subsection{$\pi K$ Scattering Amplitudes and Strange Meson Resonances}
\label{sec:A4}
It is instructive to compare the spectrum of the kaons to the corresponding spectra 
of the $D$ and $D_s$ mesons (see Ref.~\cite{Godfrey:1998pd}). With exception of the 
still uncertain spin $0$ state $\kappa$ or $K_0^\ast(700)$ the known parts of the 
kaon, the $D$ and $D_s$ mesons are qualitatively very similar, but with somewhat 
different orderings~\cite{Tanabashi:2018oca}.

The established part of the strange meson spectrum~\cite{Godfrey:1998pd,
Tanabashi:2018oca} begins with the $0^-$ ground state, followed by the $1^-$ 
$K^\ast(892)$ vector meson meson state, which is followed by the two $1^+$ states 
$K_1(1270)$, $K_1(1400)$, and then the recurrence $K^\ast(1410)$ of the $1^-$ vector 
meson state and the scalar meson state $K_0^\ast(1430)$.  The spectrum of the $D$ 
meson differs only in that the corresponding scalar meson state $D_0^\ast(2400)$ 
slightly below, rather than slightly above the two $1^+$ states $D_1(2420)$ and 
$D_1(2430)$.

This comparison of the spectra of the $K$, $D$ and $D_s$ mesons reveals the importance 
of settling the existence of the $\kappa$ or $K_0^\ast(700)$, as its existence would 
settle the existence of a light scalar nonet below 1~GeV.  This can be compared to the 
corresponding low-lying scalar meson states in the spectra of the charm and charm-strange 
mesons. In all these spectra, the first recurrence of that low scalar meson is well 
established by the states $K_0^\ast(1430)$, $D_0^\ast(2400)$, and $D_{s0}^\ast(2317)$.  
Given the very large width of the non-strange scalar meson $f_0(500)$ (or $\sigma$), it 
may be expected that the $\kappa$ and the lowest charm strange, charm and charm-strange 
mesons will have similar large widths and threshold effects.  Moreover, establishing 
firmly the existence of the $\kappa/K_0^\ast(700)$, with similar characteristics to the 
$\sigma/f_0(500)$, would also kill the glueball interpretations of the 
latter~\cite{Narison:2000dh} or the dilatonic interpretation~\cite{Crewther:2013vea}.

\subsubsection{Previous Measurements for Strange Mesons:} 
\label{sec:A4.1}
Most of experimental data on $K\pi$ elastic scattering are obtained from $KN$ scattering 
data under the assumption that they occur in reactions with one pion exchange. First 
measurements of $K\pi$ scattering in the reaction $K^-d\to p_spK^-\pi^-$ produced in 
deuterium bubble chamber at 5.5~GeV/$c$ were performed at Argonne ~\cite{Cho:1970fb}, 
then in the reaction $K^-d\to p_spK^-\pi^-$ in the deuterium bubble chamber at 3~GeV/$c$ 
were performed at Saclay~\cite{Bakker:1970wg}, in the reaction $K^{\pm}N$ in hydrogen 
bubble chamber in the momentum range 2.0 to 12.0~GeV/$c$ were produced at 
CERN~\cite{Linglin:1973ci}, in the reaction $K^-p\to K^-\pi^-\Delta^{++}$ at 
4.25~GeV/$c$ in hydrogen bubble chamber were produced at CERN as 
well~\cite{Jongejans:1973pn}. Finally, high statistics measurements for reactions 
$K^{\pm}p\to K^{\pm}\pi^+n$ and $K^{\pm}p\to K^{\pm}\pi^-\Delta^{++} $ at 13~GeV/$c$ 
using spectrometers were performed at SLAC~\cite{Estabrooks:1977xe} and in the reaction 
$K^-p\to K^-\pi^+n$ at 11~GeV/$c$ were performed with the LASS 
Spectrometer~\cite{Aston:1987ir}. 

\subsubsection{Strange Exotics:} 
Two important motivations for new measurements of $\pi K$ scattering amplitudes, is the 
attention received by Chiral Perturbation Theory~\cite{Bernard:1990kx,Bernard:1990kw,
Bernard:1991zc,Bijnens:2004bu}, resonance and unitarized models~\cite{Dobado:1992ha,
Jamin:2000wn,GomezNicola:2001as,Nebreda:2010wv,Guo:2011pa,vanBeveren:1986ea}, and the need 
to confirm the existence of the exotic $\kappa$ meson (or $K^\ast_0(700)$) in the $I=1/2$ 
$S$-wave. This state would be the strange counterpart of the $\sigma$ (or $f_0(500)$) 
meson which is now rather well established from $\pi\pi$ scattering (see the 
review~\cite{Pelaez:2015qba}).

For spectroscopy, the relevance of this state, which according to the Review of Particle 
Physics~\cite{Tanabashi:2018oca} still ``needs confirmation'', is twofold: \\
First, establishing firmly its existence will settle the longstanding debate on whether 
there is a low-lying scalar nonet, with the $\sigma/f_0(500)$, $f_0(980)$, and $a_0(980)$ 
as partners. \\
But, second, because there is mounting evidence that such nonet cannot be generated solely
from the quark model as quark-antiquark states, but emerge only when considering properly
unitarized meson-meson interactions~\cite{Pelaez:2004xp,vanBeveren:1986ea,Oller:1997ng,
Oller:1998zr,Close:2002zu,Pelaez:2003dy} or tetraquark/molecule configurations~\cite{Jaffe:1976ig,
Barnes:1985cy,Black:1998zc}. \\

For Chiral Perturbation Theory, the interest is on the low energy parameters, particularly 
the scalar scattering lengths. Below, we discuss the existing tension between dispersive 
analyses of experimental data~\cite{Buettiker:2003pp,Pelaez:2016tgi}, theoretical 
predictions from Chiral Perturbation Theory~\cite{Bernard:1990kx,Bijnens:2004bu}, and 
lattice calculations~\cite{Beane:2006gj,Flynn:2007ki,Fu:2011wc,Sasaki:2013vxa}. One of 
the main difficulties for extracting reliable values form experiment is that the existing 
$\pi K$ data starts at 750~MeV, and one needs an extrapolation down to the threshold at 
$\sim$635~MeV. Thus, the new KLF input at low energies, together with the general 
improvement in statistics, will settle this issue.

At this point, it is worth noting the decisive role that the precise low-energy data from 
the NA48/2 experiment~\cite{Batley:2010zza} played for the revision of the $\sigma/f_0(500)$ 
in the PDG2018.  In this regard, improved measurements of the $S$-wave $\pi K$ phase-shifts 
at low energy ($E \lapprox 1$~GeV) would be highly desirable in order to play a similar role 
for the $\kappa/K_0^\ast(700)$.  
Precision measurements of the $S$-wave phase-shifts would allow application of the Pad\'e 
approximant method for determining the positions of the resonances (see, 
e.g.,~\cite{Masjuan:2014psa}. This method has been recently applied to $\pi K$ scattering, 
and a $\kappa$ pole has been found in Ref.~\cite{Pelaez:2016klv} using as an input the fit 
to data constrained with Forward Dispersion Relations obtained in Ref.~\cite{Pelaez:2016tgi}.

Alternatively, the most rigorous way to determine this resonance pole is using Roy-Steiner 
(RS) type equations~\cite{Roy:1971tc,Steiner:1971ms}. These equations rely on the first 
principles like analyticity, crossing as well as data. They provide a suitable framework 
for performing extrapolations in the low energy region, E < 1~GeV, of the $S$ and $P$ 
partial waves given sufficiently precise inputs at higher energies, essentially in the 
range 1 -- 2~GeV. Extrapolations to complex values of the energies can be performed with 
the same accuracy as on the real axis. Unlike the Pad{\'e} approximant approach, the 
extrapolation of the $I = 1/2$ $S$-wave from the RS equations requires inputs from other 
partial waves as well since the equations form a coupled system. Based on the existing 
data set, an estimate of the $\kappa$ pole position from the RS equations was performed 
in Ref.~\cite{DescotesGenon:2006uk}. Note that no input on $\pi K$ scattering in the 
scalar partial waves below 1~GeV was used for this estimate. Using this RS equations 
with the data produced in KLF would produce an actual experimental and rigorous 
determination of the $\kappa$ pole.

In the $P$-wave, finally, the studies by the LASS 
Collaboration~\cite{Aston:1986jb,Aston:1987ir} have identified besides the well known 
$K^\ast(892)$ a new meson, the $K^\ast(1410)$. This meson has an unexpectedly low mass 
as it appears to be essentially degenerate with the non-strange $\rho(1450)$ or 
$\omega(1420)$ vector mesons. Its properties are not very precisely known at present.

\subsubsection{Status of $\pi K$ Scattering Measurements:}
\begin{figure}[ht]
\centering
{   
    \caption{\small 
    Illustration of the contribution from one-pion exchange, which is dominant at small 
	momentum transfer, to the production amplitude. 
    \underline{Left}: $K_Lp\to K^-\pi^0 \Delta^{++}$ or 
    \underline{Right}: $K_Lp\to K_L\pi^-\Delta^{++}$.} \label{fig:OPE}}
{\includegraphics[width=0.65\textwidth,keepaspectratio, trim=0 0 0 0, clip]{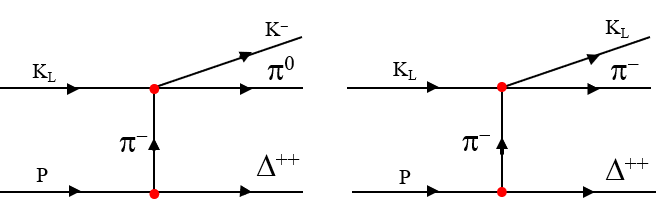}}
\end{figure}

The traditional method for measuring $\pi K\to\pi K$ amplitudes is from production experiments like
\begin{equation}
        Kp\to K\pi p,\quad Kp\to K\pi n,\quad Kp\to K\pi\Delta 
        \label{eq:production}
\end{equation}
focusing on the region of small momentum transfers $|t| < 0.1 - 0.2$~GeV$^2$, which is accessible 
with kaon beams of a few GeV. In this region, the amplitude is dominated by the one pion exchange 
(OPE) contribution, see Fig.~\ref{fig:OPE}. Depending on the particular production process 
Eq.~\eqref{eq:production}, these will carry different Clebsch-Gordan coefficients for combinations 
of the $I = 1/2$ and $I = 3/2$ $K\pi$ scattering amplitudes. For example, the reaction  $K_L p\to 
K^-\pi^0\Delta^{++}$ (Fig.~\ref{fig:OPE} (left)) is proportional to $(1/3)\cdot(-T^{1/2}+T^{3/2})$ 
while reaction $K_Lp\to K_L\pi^-\Delta^{++}$ (Fig.~\ref{fig:OPE} (right)) is proportional to 
$(1/3)\cdot(T^{1/2}+2T^{3/2})$, where $T^{1/2}$ and $T^{3/2}$ correspond to the $K \pi$ scattering 
isospin $I = 1/2$ and $I = 3/2$ amplitudes, respectively. These two independent measurements in 
principle allow a separation of isospin $1/2$ and $3/2$ amplitudes. A similar method (OPE 
approximation) was used for the extraction of $\pi\pi\to\pi\pi$ elastic scattering amplitudes, 
further details can be found in the book~\cite{Martin:1976mb} or in the recent review by 
Pel\'aez~\cite{Pelaez:2015qba}. The two experiments performed at SLAC~\cite{Estabrooks:1977xe,
Aston:1987ir} have the largest statistics and provide the best determinations of the $\pi K$ 
scattering amplitudes at present. They cover the energy ranges $0.73\le E\le 1.85$~GeV 
(Ref.~\cite{Estabrooks:1977xe}) and $0.83\le E\le 2.52$~GeV (Ref.~\cite{Aston:1987ir}), 
respectively.  References to earlier works can be found in the review~\cite{Lang:1978fk}.
-
\begin{figure}[ht]
\centering
{
    \includegraphics[width=7.7cm, height=5.0cm]{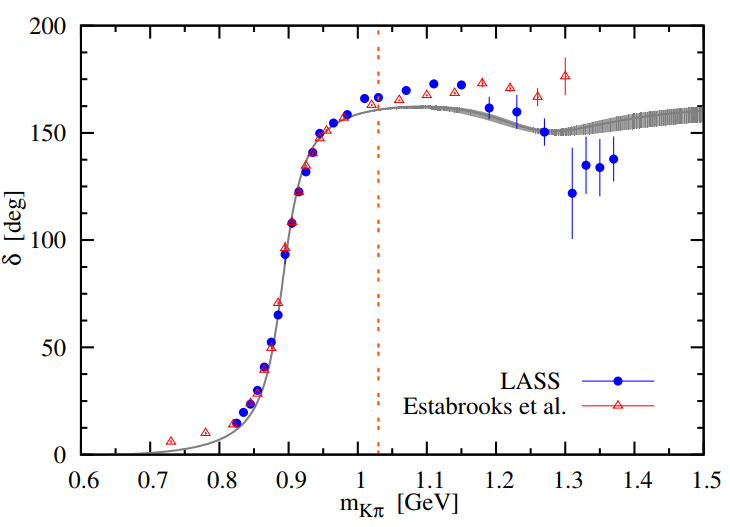}
    \includegraphics[width=8.7cm, height=5.3cm]{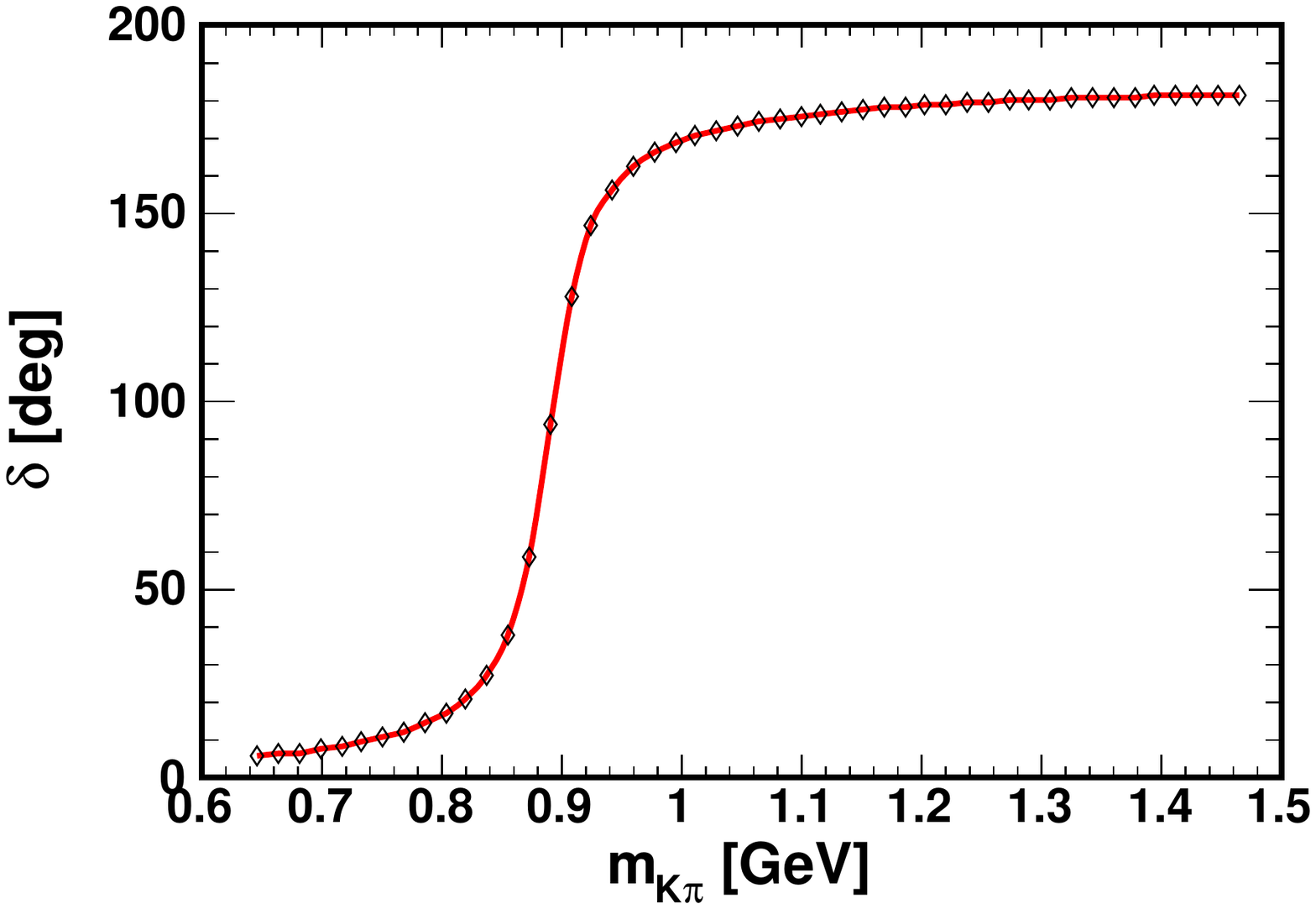}
}
\centerline{\parbox{0.80\textwidth}{
    \caption[] {\protect\small 
    Phase of the $\pi K$ vector form factor.
    \underline{Left}: Experimental data from Estabrooks \textit{et al.}~\protect\cite{Estabrooks:1977xe} 
	and LASS Collaboration~\protect\cite{Aston:1987ir}. The opening first inelastic $K^{\ast}\pi$ 
	channel is indicated by the dashed vertical line. The gray band represents the extrema from 
	the fits of Table~3 in the paper by Boito \textit{et al.}~\protect\cite{Boito:2010me} 
	constrained by the data on $\tau \to K\pi\nu_{\tau}$ decay from Belle 
	experiment~\protect\cite{Epifanov:2007rf} with additional constrains from a compilation of 
	$K_{l3}$ decay analyses~\cite{Antonelli:2010yf}.
    \underline{Right}: 100~days of running with KLF, simulated with the $K^{\ast-}(892)$ (see text 
	for details).}  \label{fig:boitoPphase} } } 
\end{figure}

A completely different approach to measuring the $\pi K$ phase-shifts makes use of the Watson's 
theorem for weak decay form factors. In this manner, the phase-shift difference $\delta_S-\delta_P$ 
was determined by analyzing the $D^+\to K^-\pi^+e^+\nu$ by the \babar\
Collaboration~\cite{delAmoSanchez:2010fd}. The results are in agreement with the LASS 
determination but more statistics are needed before one reaches a comparable precision. 
Similarly, from the measurement of the energy distribution in the decay $\tau^-\to K_S\pi^-\nu$ 
by the Belle Collaboration~\cite{Epifanov:2007rf} the $P$-wave phase has been 
determined~\cite{Boito:2010me}, relying on the analyticity properties of the form factor. Their 
result is shown in Fig.~\ref{fig:boitoPphase} (left). While Fig.~\ref{fig:boitoPphase} (right) 
shows the expected results for 100~days of running with KLF, simulated with $K^{\ast-}(892)$ only. 
As one can see the proposed measurement will dramatically increase statistical precision of the 
world data and extend the measurement  to the very low mass elastic scattering, as well as high 
mass inelastic scattering regions.

Since Watson's theorem is valid in the energy region of elastic scattering, these alternative phase 
determinations provide important information on the effective onset of inelastic scattering in the 
various partial waves. This figure also shows that the determination of the phase shift in the region 
of the $K^\ast(1410)$ resonance is not very precise and could be improved.

The same form factors which appear in the $\tau\to K\pi\nu$ decays are also involved in the $K_{l3}$ 
decay amplitudes: $K\to\pi e\nu$, $K\to\pi\mu\nu$. A series of new $K_{l3}$ experiments were undertaken 
recently in order to improve the determination of $V_{us}$ (see Ref.~\cite{Antonelli:2010yf}). As shown 
in Ref.~\cite{Antonelli:2010yf}, an optimal analysis of the $K_{l3}$ data is achieved by using a 
description of the two form factors involved based on phase dispersive representations rather than 
phenomenological polynomial or pole forms as done previously.The $\pi K$ scattering also plays an 
important role in a number of three-body decays, like $D\to K\pi\pi$. Recently, a method was 
developed~\cite{Niecknig:2015ija} which allows to compute the effect of the three-body re-scattering 
in terms of the known two-body $\pi\pi$ and $\pi K$ $T$-matrices.  This could be useful for identifying 
small $CP$ violating effects in the charm sector.

\subsubsection{Theory:}
\begin{figure}[ht]
\centering
{   
    \caption{\small 
    \underline{Left}: Results for the $S$-wave phase shifts extrapolated below 1~GeV based on the 
	Roy-Steiner dispersive equations (from Ref.~\protect\cite{Buettiker:2003pp}) compared with the 
	experimental data from Ref.~\protect\cite{Estabrooks:1977xe}. 
    \underline{Right}: Expected phase shift statistics  for $S$-wave isospin $I = 1/2$ (upper half) 
	and $I = 3/2$ (lower half) states simulated for 100~days of running with KLF. In  both panels, 
	phase  shifts are plotted as a function of invariant mass of $K\pi$ system.}
    \label{fig:royphaseS}
}
\includegraphics[width=7cm,  height=5.3cm]{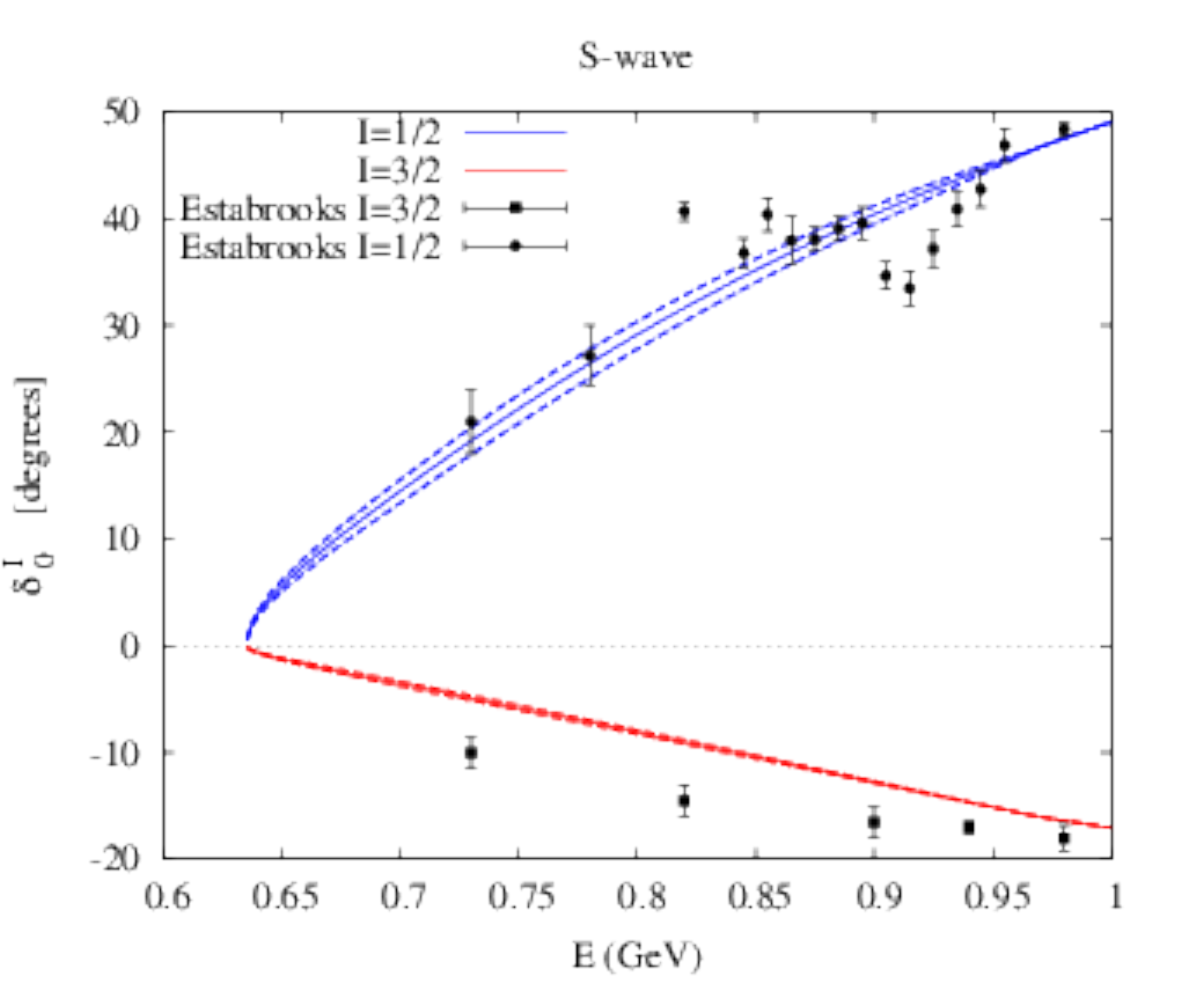}
\includegraphics[width=7cm,  height=5cm]{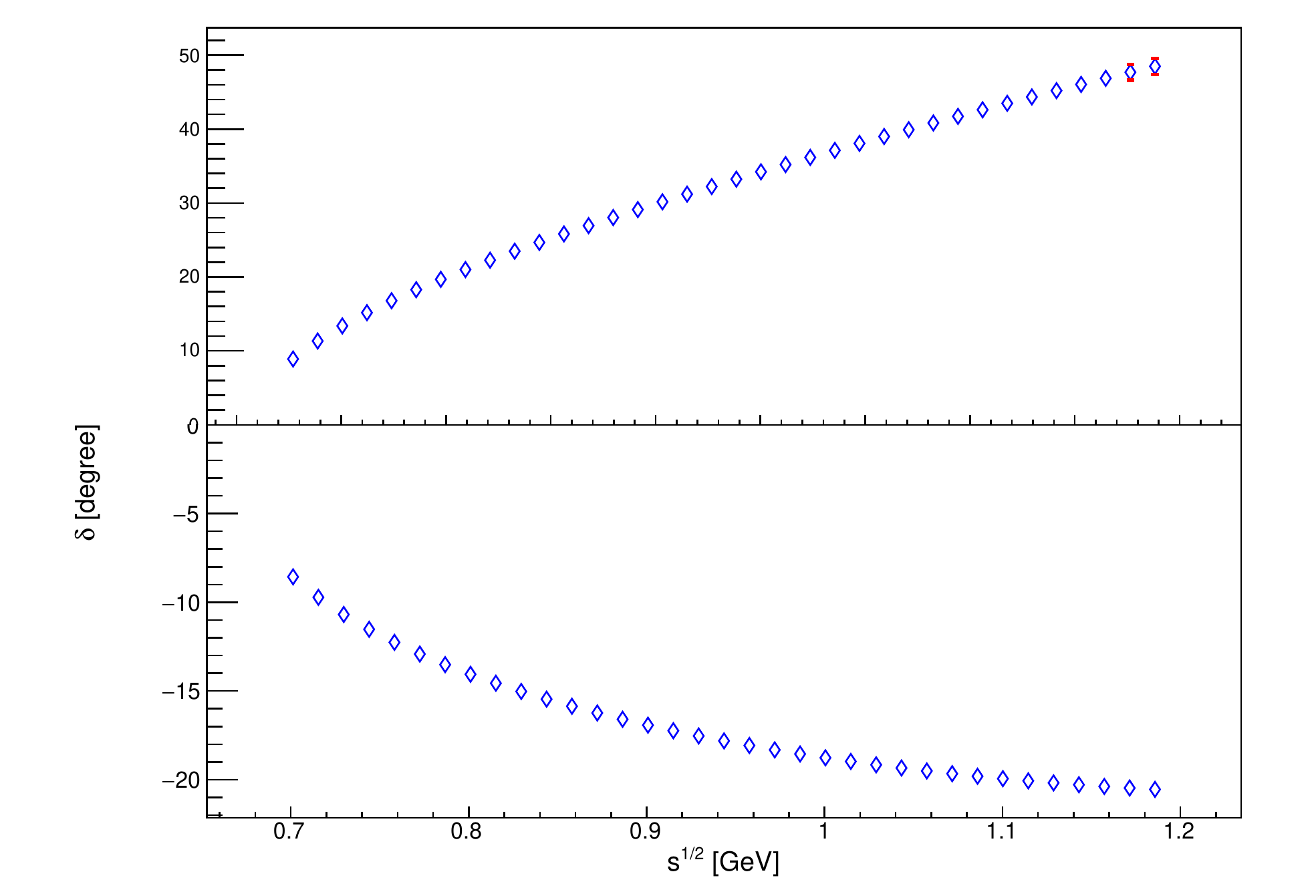}
\end{figure}

Up to  now the only experimental data on s-wave for both  isospin  states  $I = 1/2$ and $I = 3/2$ from 
the same experiment were obtained at SLAC by Estaabrooks \textit{et al.}~\protect\cite{Estabrooks:1977xe}. 
This limits precision to  extract the mass and the width of the $\kappa$.  On  the left panel of  
Fig.~\ref{fig:royphaseS}, the SLAC data are presented with the theoretical extrapolation of the 
Roy-Steiner dispersive equations from Ref.~\protect\cite{Buettiker:2003pp}). On the  right  panel of 
Fig.~\ref{fig:royphaseS}, the $S$-wave phase shifts are presented with statistics based on 100~days of 
running with KLF.  As one can  see  there is dramatic improvement not only in statistical uncertainties, 
but also in the number of points reaching to the much lower invariant masses of $K\pi$  system. The  
simulated data are obtained from the reactions $K_Lp\to K^-\pi^0\Delta^{++}$ and $K_Lp\to 
(K_L)\pi^-\Delta^{++}$, where the $K_L$ in the latter case is reconstructed via missing mass technique.

More details about MC simulations of $K\pi$ scattering can be found in Ref.~\cite{Amaryan:2020zz}.

Pions and kaons are QCD pseudo-Goldstone bosons, therefore the $\pi K$ amplitudes at low energy can 
be expressed as a chiral expansion. The NLO calculation was performed in Ref.~\cite{Bernard:1990kx} 
who predict the following results for the scattering lengths,
\begin{equation}
        a_0^{1/2} =  0.19\pm 0.02,\quad
        a_0^{3/2} = -0.05\pm 0.02
\end{equation}
(in units of $m_\pi^{-1}$). Verifying these predictions would provide an important check of the 
three-flavor chiral expansion. Based on experimental phase-shift measurements this is possible, in 
principle, using dispersion relations for extrapolating down to the threshold. The Roy-Steiner 
equations provide a suitable framework for that. This is illustrated in Fig.~\ref{fig:royphaseS} 
which shows the extrapolated results for the $S$-waves in the region $E\le 1$~GeV, based on 
experimental inputs from Refs.~\cite{Estabrooks:1977xe,Aston:1987ir} in the region $E > 1$~GeV. It 
is clear that the availability of more precise data in the range $E\le 1$~GeV would greatly strengthen 
the efficiency of this method. We note that a direct experimental estimate of the scattering length 
difference was performed recently~\cite{Adeva:2017oco} based on the lifetime of the $\pi^+K^-$ atom 
at the DIRAC experiment at CERN. Unfortunately, the experimental errors are still too large and do 
not provide really precise information about the pion-kaon scattering lengths.

Alternatively, scattering phase shifts can be computed in lattice QCD using L{\"u}scher's 
method~\cite{Luscher:1990ux}. Results for $\pi K$ phase shifts were first obtained in 
Refs.~\cite{Lang:2012sv,Prelovsek:2013ela} and in Ref.~\cite{Wilson:2014cna}. In this last work, 
the influence of one inelastic scattering two-body channel is accounted for and m$_\pi$ = 391~MeV. 
Very recently, results for m$_\pi$ = 230~MeV have been presented~\cite{Brett:2018jqw}. Once physical 
values for $m_\pi$ are reached, these lattice QCD results can be compared directly to experimental 
measurements of the $\pi K$ phase shifts which provides a direct probe of the quality of the numerical 
QCD solution.

\textbf{To Summarize}: 
\label{sec:A4_sum}
As it is discussed above, there are many aspects of $\pi K$ scattering that require improvement on 
the existing measurements. First of all,  it is  the quest to establish existence or  non-existence 
of scalar $\kappa$ meson either to complete  scalar meson nonet or to find an alternative way to 
explain well established non-strange $\sigma$, $a_0$, and $f_0$  meson family. Besides there are 
some fundamental questions that need to  be clarified. In particular, currently there is a sizable 
tension between the values of scattering lengths obtained from dispersive analyses of 
data~\cite{Buettiker:2003pp,Pelaez:2016tgi}, on one side, and the predictions from Chiral 
Perturbation Theory~\cite{Bernard:1990kx,Bijnens:2004bu} and lattice calculations~\cite{Beane:2006gj,
Flynn:2007ki,Fu:2011wc,Sasaki:2013vxa}, on the other side.  The values of the threshold parameters are 
related to two important questions. On the one hand, for phenomenology,  establishing the convergence 
and reliability of SU(3) Chiral Perturbation Theory. On the other hand, for the foundations of QCD, 
the size of the strange versus the non-strange chiral condensate, i.e., the detailed pattern of the 
QCD spontaneous chiral symmetry breaking is very important.

As previously noticed and as shown in Fig.~\ref{fig:royphaseS}, the existing $\pi K$ data starts at 
750~MeV, and one needs an extrapolation down to the threshold at $\sim$635~MeV. Hence, the new KLF 
data at low energies, together with the general improvement in statistics, will be determinant to 
resolve this tension.


\subsection{Current Hadronic Projects} 
\label{sec:A5}
Past measurements involving kaon scattering measurements were made at a 
variety of laboratories, mainly in the 1960s and 1980s when experimental 
techniques were far inferior to the standards of today (short summary is 
given in Appendices~\ref{sec:A3.1} and \ref{sec:A4.1}.  It is important 
to recognize that current projects are largely complementary to the 
proposed JLab KL hadron beam facility. We \underline{summarize} the 
status of the J-PARC, Belle, \babar\, PANDA, COMPASS, and LHCb efforts 
here.

\subsubsection{J-PARC, Japan:} 
While J-PARC has a whole program of charged strange particle and 
hypernuclear reactions~\cite{Ohnishi:2019cif}, the photon beam at KLF 
allows unique access to other channels. J-PARC provides separated 
secondary beam lines up to 2~GeV/$c$ (Table~\ref{tab:jparc}). The 
operation of the Hadron Experimental Facility resumed in April of 2015 
following a two-year suspension to renovate the facility after the 
accident that occurred in May 2013~\cite{JPARC_accident}. The primary 
beam intensity is currently 50~kW, and can be upgraded to 85~kW. This 
will correspond to $\sim$10$^9$~ppp (particles per pulse) for pion beam 
intensity and to $\sim$10$^6$~ppp for negative kaon beam flux. The 
$K/\pi$ ratio is expected to be close to 10, which is realized with 
double-stage electrostatic separators. One of the main problems in the 
$K/\pi$ separation is a high duty-factor of the J-PARC Complex.
\begin{table}[ht]
\centering \protect\caption{J-PARC beamlines in the Hadron Experimental 
	Facility from Ref.~\protect\cite{Fujioka:2017gzp}. Top part of
	Table gives information about beamlines in the present hall,
	while bottom part information is about new beamlines in the 
	extended area.}
\vspace{2mm}
{%
\begin{tabular}{|c||c|c|c|c|}
\hline
Beamline & Particle & Momentum & Number of particles & Characteristics \\
         &          &  (GeV/$c$) & per spill           & \\
\hline
K1.8     & $K^\pm$, $\pi^\pm$ & $<$2.0      & 10$^6$ $K^-$  & separated \\
K1.8BR   & $K^\pm$, $\pi^\pm$ & $<$1.1      & 10$^5$ $K^-$  & separated \\
KL       & $K_L$              & 2.1 in ave. & 10$^7$ $K_L$  & to 16$^\circ$ \\
High-p   & p                  &             & 10$^{10}$ p   & primary protons \\
         & $\pi^\pm$          & $<$31       & 10$^7$ $\pi$  & \\
\hline
K1.1     & $K^\pm$, $\pi^\pm$ & $<$1.2      & 10$^6$ $K^-$  & separated \\
         &                    & 0.7$\sim$0.8&               & lower momentum \\
         &                    &             &               & [K1.1BR] \\
HIHR     & $\pi^\pm$          & $<$2.0      & $2.8\times10^8$ $\pi^-$  & separated \\
         &                    &             &               & $\times10$ better $\Delta p/p$ \\ 
K10      & $K^\pm$, $\pi^\pm$, $\bar{p}$    & $<$10         & 10$^7$ K$^-$ & separated \\
new KL   & $K_L$              & 5.2 in ave. & 10$^8$ $K_L$  & to 5$^\circ$ \\
         &                    &             &               & n/$K_L$ optimized \\
\hline
\end{tabular}} \label{tab:jparc}
\end{table}

With $K^-$ beams, currently there is no proposal specific for $S = -1$ 
hyperons, but the cascades will be studied in the early stage of 
E50~\cite{E50}, hopefully in a few years. The beam momentum bite, 
$\Delta p/p$, is strongly depending on the configuration of the beam 
line spectrometer, but one can determine beam momentum with the 
resolution of $\Delta p/p\sim10^{-3}$ or $10^{-4}$.  One can think 
that the systematic study for $S = -1$ hyperons even with charged kaons 
is desirable and J-PARC folks think that such a study is definitely 
needed but currently there is no room to accept a new proposal to require 
a long beamline. J-PARC is focusing on hypernuclei physics~\cite{Naruki2015}.  

There is no $K_L$ beamline for hyperon physics at J-PARC. It is 100~\% 
dedicated to the study of CP-violation. The momentum is spread out from 
1 to 4~GeV/$c$, there is no concept of $\Delta p/p$ since the beam cannot 
be focused with EM devices.

\subsubsection{Belle, Japan:}
At $B$ Factories large samples of charmed baryons are produced in the decays 
of $B$ mesons as well as from the $e^+ e^-\to c\bar{c}$ 
continuum~\cite{Bevan:2014iga}. The masses, widths and branching fractions 
of many ground state singly-charmed baryons and their excitations have been 
measured at Belle~\cite{Mizuk:2004yu}, like $\Xi_c(2645)^{0,+}$, 
$\Xi_c(2790)^{0,+}$, $\Xi_c(2815)^{0,+}$, $\Xi_c(2980)^{0,+}$, 
$\Xi_c'$~\cite{Lesiak:2004zy,Lesiak:2008wz,Yelton:2016fqw}; 
$\Sigma_{c}(2455)^{0,++}$, and \\ $\Sigma_{c}(2520)^{0,++}$~\cite{Lee:2014htd}. 

Many charmed baryons and their decay modes have been discovered at 
Belle~\cite{Abe:2001mb} or confirmed after LHCb recently, for example: 
$\Xi_c(3055)^+$~\cite{Kato:2013ynr}; $\Lambda^{+}_{c} \rightarrow p K^{+} 
\pi^{-}$~\cite{Yang:2015ytm}; $\Xi_c(3055)^0$, \\ 
$\Xi_c(3055)^+/\Xi_c(3080)^+\to\Lambda D^+$~\cite{Kato:2016hca}; 
$\Omega_c(3000)$, $\Omega_c(3050)$, $\Omega_c(3066)$, 
$\Omega_c(3090)$~\cite{Yelton:2017qxg}; $\Xi_c^0\to\Xi^-\pi^+$, $\Xi_c^0\to 
\Lambda K^-\pi^+$, $\Xi_c^0\to pK^-K^-\pi^+$~\cite{Li:2018qak}; 
$\Xi_c^+\to\Xi^-\pi^+\pi^+$, $\Xi_c^+\to p K^-\pi^+$~\cite{Li:2019atu}. 
Fifty times larger data sample expected from Belle~II will open new era in 
the precision studies of charmed baryons in the coming 
decade~\cite{Kou:2018nap}. 

\begin{figure}[ht]
\centering
{\includegraphics[width=0.5\textwidth,keepaspectratio]{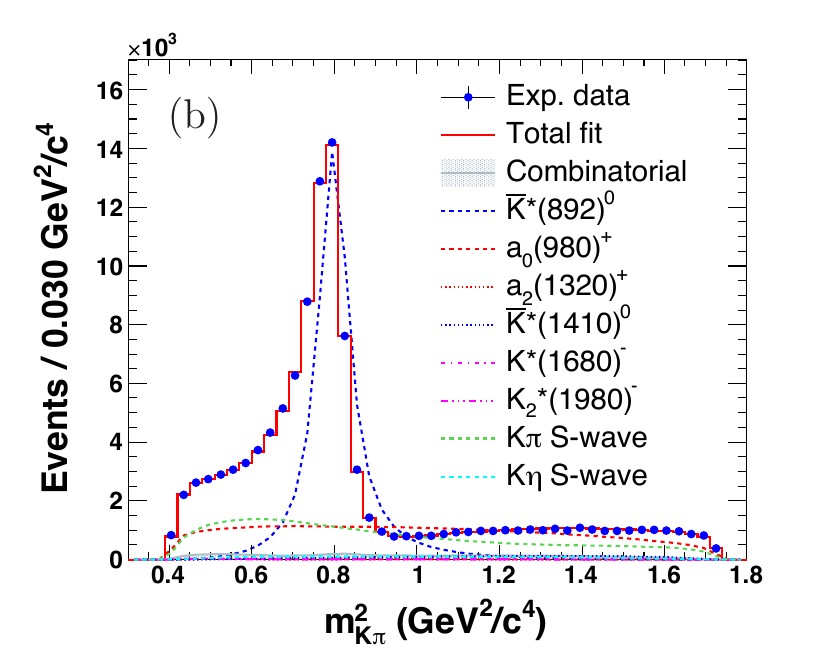} }

\centerline{\parbox{0.60\textwidth}{
 \caption[] {\protect\small 
   Squared $K^-\pi^+$ invariant mass distribution from Dalitz plot analysis 
of $D^0\to K^-\pi^+\eta$ decay at Belle.}
	\label{fig:Belle-Kpi} } }
\end{figure}

Recently, the Belle Collaboration posted an article in arXiv where it studies 
$D^0\to K-\pi^+\eta$ decay~\cite{Chen:2020sog}. In this analysis, it is 
clearly shown that along with the contribution from higher mass $K^\ast$'s 
the $S$-wave below $K^\ast(892)$ is needed to get a reasonable fit to the 
experimental squared invariant mass distribution of $K^-\pi^+$ from the 
threshold up to 1.8~GeV$^2$ (see Fig.~\ref{fig:Belle-Kpi}).
Actually, predictions for a sizeable $S$-wave $\bar K \pi$ contribution in 
addition to $\bar K^\ast$ excitation have already been done in the related 
reaction $\bar B^0\to J/\psi \pi^+ K^-$ reaction (see Fig.~6 of 
Ref.~\cite{Bayar:2014qha}).

\subsubsection{\babar\, USA:} 
Charmonium decays can be used to obtain new information on light meson 
spectroscopy. In $e^+e^-$ interactions, samples of charmonium decays can 
be obtained using different processes, in particular \etac decays can be 
produced in two-photon interactions. In this process, events are selected 
in which the $e^+$ and $e^-$  beam particles are scattered at small angles 
and remain undetected. Two-photon events are isolated from background 
requiring the conservation of the transverse momentum with respect to the 
beam axis. The study of the $\eta_c$ three-body hadronic decays is found 
to proceed almost entirely through the intermediate production of scalar 
meson resonances.

\babar\ experiment has performed Dalitz plot analyses of $\eta_c \to 
K^+K^-\eta$ ($\eta\to\gamma\gamma$ and $\eta\to \pip \pim \piz$), $\eta_c 
\to K^+K^-\piz$, and $\eta_c \to \KS \Kpm \pimp$. These three \etac decays 
have provided new information on the properties of the strange scalar mesons.   

\begin{figure}[ht]
\centering
{
\includegraphics[width=0.32\textwidth,keepaspectratio]{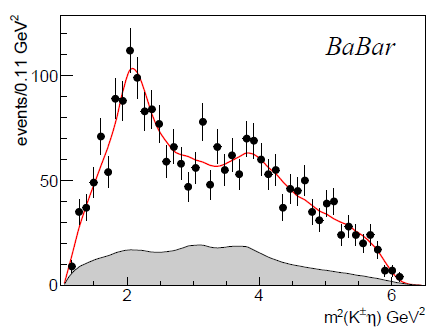}
\includegraphics[width=0.32\textwidth,keepaspectratio]{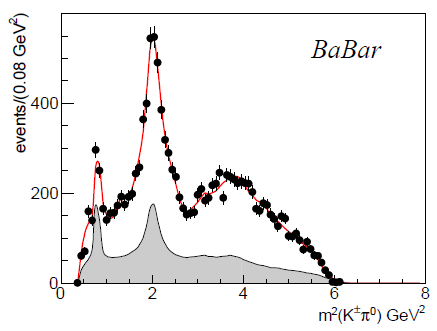}
\includegraphics[width=0.32\textwidth,keepaspectratio]{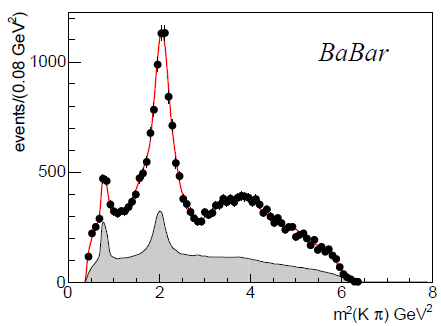}
}

\centerline{\parbox{0.70\textwidth}{
\caption[] {\protect\small
    \underline{Left}: The $m^2(\Kpm \eta)$ from $\eta_c \to K^+K^-\eta$ 
	Dalitz plot projection from \babar. The superimposed curve results 
	from the Dalitz plot analysis. The shaded region show the background 
	estimate obtained by interpolating the results of the Dalitz plot 
	analyses of the sideband regions.
    \underline{Middle}: The $m^2(\Kpm \piz)$ from $\eta_c \to K^+K^-\piz$ 
	Dalitz plot projection from \babar. The superimposed curve results 
	from the Dalitz plot analysis. The shaded region show the background 
	estimate obtained by interpolating the results of the Dalitz plot 
	analyses of the sideband regions.
    \underline{Right}: The $m^2(K \pi)$ from $\eta_c \to \KS\Kpm\pimp$ 
	Dalitz plot projection from \babar. The superimposed curve results 
	from the Dalitz plot analysis. The shaded region show the background 
	estimate obtained by interpolating the results of the Dalitz plot 
	analyses of the sideband regions.
    }
	\label{fig:fig1_babar} } }
\end{figure}

The isobar model Dalitz plot analysis of the $\eta_c \to K^+K^-\eta$ has 
provided the unexpected observation of $K^\ast_0(1430)\to K 
\eta$~\cite{Lees:2014iua}. The Dalitz plot projection on the $m^2(\Kpm 
\eta)$ is shown in Fig.~\ref{fig:fig1_babar} (left), togheter with the 
fit projection. A clear signal of $K^\ast_0(1430)$ can be observed.
\begin{figure}[ht]
\centering
{
\includegraphics[width=0.7\textwidth,keepaspectratio]{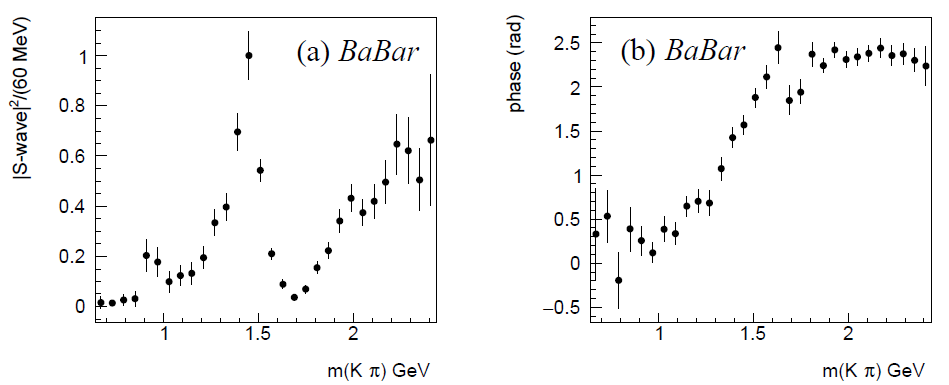} 
}

\centerline{\parbox{0.70\textwidth}{
 \caption[] {\protect\small
    (a) Squared $K\pi$ $\mathcal{S}$-wave and (b) phase 
	from the QMI analysis of $\eta_c\to K^+K^-\piz$, 
	and $\eta_c\to \KS \Kpm \pimp$ decays.}
	\label{fig:fig4_babar} } }
\end{figure}

The corresponding $m^2(\Kpm \piz)$ projection from the $\eta_c 
\to K^+K^-\piz$ Dalitz plot analysis is shown in 
Fig.~\ref{fig:fig1_babar} (middle). The decay is dominated by the 
$K^\ast_0(1430) K$ final state.

In the Dalitz plot analysis of  $\eta_c \to\Kp\Km\piz$, a likelihood 
scan allows to obtain the best-fit parameters for the $K^{\ast}_0(1430)$:
\begin{equation}
\begin{split}
	m(K^{\ast}_0(1430)) = (1438 \pm 8 \pm 4) \ \mev \\
	\Gamma(K^{\ast}_0(1430)) = (210 \pm 20 \pm 12) \ \mev.
\end{split}
\end{equation}

The observation of the $K^\ast_0(1430)$ in both the $K \eta$ and $K\piz$ 
decay modes permits a measurement of the corresponding branching ratio:
\begin{equation} 
	\frac{BR(K^\ast_0(1430) \to \eta K)}{BR(K^\ast_0(1430) 
	\to \pi K)} = 0.092 \pm 0.025^{+0.010}_{-0.025}.
\end{equation}

\babar\ has also performed a Quasi Model Independent Partial Wave Analysis 
(QMI)~\cite{Aitala:2005yh} of $\eta_c \to K^+K^-\piz$, and $\eta_c \to \KS 
\Kpm \pimp$ decays~\cite{Lees:2015zzr}. In the QMI, the $K \pi$ mass 
spectrum is divided into 30 equally spaced mass intervals 60 \mev wide 
and for each bin two new free parameters are added to the fit, the amplitude 
and the phase of the $K \pi$ $\mathcal{S}$-wave. The corresponding Dalitz 
plot projection on $m^2(K\pi)$ from $\eta_c \to \KS \Kpm \pimp$ is shown in 
Fig.~\ref{fig:fig1_babar} (right) and is dominated by the $K^\ast_0(1430)$ 
resonance.

The resulting $K \pi$ $\mathcal{S}$-wave squared amplitude and phase, 
averaged over the two \etac decay modes, is shown in Fig.~\ref{fig:fig4_babar}. 
The $K \pi$ $\mathcal{S}$-wave is dominated by the $K^\ast_0(1430)$ resonance, 
with little evidence of $\kappa(700)$. The phase motion shows the expected 
behaviour for the $K^\ast_0(1430)$ resonance.

\subsubsection{PANDA, Germany:}
The PANDA experiment~\cite{Lutz:2009ff} will measure annihilation reactions 
of antiprotons with nucleons and nuclei in order to provide complementary 
and in part uniquely decisive information on a wide range of QCD aspects. 
The scientific scope of PANDA is ordered into several pillars: hadron 
spectroscopy, properties of hadrons in matter, nucleon structure and 
hypernuclei. Antiprotons are produced with a primary proton beam, collected 
and phase-space cooled in the Collector Ring (CR), and then transferred to 
the High Energy Storage Ring (HESR) where they are stacked, further phase-space 
cooled, and then directed onto an internal target located at the center of the 
$\overline{P}$ANDA detector. The facility will start with a luminosity of 
$10^{31}$~cm$^2$/s and a momentum resolution of $\Delta p / p = 10^{-4}$, 
and later improve to $2\times 10^{32}$ and $4\times 10^{-5}$, respectively.  
The large cross section into baryon-antibaryon final states (e.g., 
$\sim$1~$\mu$b for $\Xi\overline{\Xi}$ or $0.1~\mu$b for 
$\Omega\overline{\Omega}$) make spectroscopic studies of excited multi-strange 
hyperons a very compelling part of the initial program of PANDA, which is 
expected to commence by 2025~\cite{Ritman16}.

\subsubsection{COMPASS, CERN:}
In 2008 and 2009, the COMPASS experiment at CERN used a negatively charged 
hadron beam with a momentum of 190\,GeV/$c$ to study the light meson 
spectrum. This secondary beam from the CERN SPS accelerator was composed of 
97\% pions, which produced a world-leading data set of diffractively produced 
three-pion final states~\cite{Adolph:2015tqa}. An admixture of about 2.6\% of 
kaons in the beam was tagged by Cherenkov detectors in the beam line, which 
allows the COMPASS collaboration to investigate the spectrum of strange mesons 
in various final states. The number of signal events in the sample of Kaon 
diffraction into $K^-\pi^-\pi^+$, for instance, surpasses any previous 
measurement by a factor of 4~\cite{Wallner:2020}. An elaborate partial-wave 
decomposition revealed signals in the mass region of well-known states, such 
as $K_1$(1270) and $K_1$(1400). In addition, potential signals from excited 
states, such as $K_1$(1650), were observed.

As a continuation of these effort, the newly formed COMPASS++/AMBER 
Collaboration has included a kaon spectroscopy program  in its Letter of 
Intent~\cite{Adams:2018}. By installing a radio-frequency (RF) separation 
stage in the beam line, the kaon contribution in the beam can be considerably 
enhanced, aiming at a 20-times larger data set compared to what has been 
measured so far. The energy of the kaon beam would be somewhere between 
50~GeV and 75~GeV, where diffractive production via Pomeron exchange is 
the dominant process and kaon beam diffraction can be well separated from 
target excitation. The charged kaon beam can also be used to extend the 
ongoing $\chi\text{PT}$ investigations~\cite{Adolph:2014kgj} into the 
strangeness sector, e.g., to measure the polarizability of the kaon.

Since the program with RF-separated kaon beams will require a considerable 
commitment by CERN to upgrade the existing beam line infrastructure, it was 
not included in the SPSC proposal for COMPASS++/AMBER Phase-1~\cite{Adams:2019}. 
The full project described in the Letter of Intent is expected to stretch across 
the next 10 to 15~years. 

\subsubsection{LHCb, CERN:} 
The LHCb experiment is designed to study the properties and decays of heavy 
flavored hadrons produced in proton-proton collisions at the LHC. The 
collisions production recorded between 2010 and 2016 provided the world's 
largest data sample of beauty and charm hadrons. Various spectroscopy studies 
have been conducted in LHCb collaborations for such states. Recently, 
important results in exotics hadron search have been obtained, such as the 
discovery of the first pentaquark states~\cite{Aaij:2015tga} and the 
existence confirmation of the exotic meson $Z_c(4430)$~\cite{Aaij:2014jqa}, 
these results have increased the interest in spectroscopy of heavy hadrons. 
Moreover, LHCb collaboration published several spectroscopy studies of 
unexplored charmed baryons, for example the observation of new $\Omega_c$ 
states~\cite{Aaij:2017nav} and the observation of the 
$\Xi_{cc}$~\cite{Aaij:2013voa}. Regarding the spectroscopy studies of bottom 
baryons, LHCb collaboration published many interesting achievements in this 
topic, among them the recently observed $\Omega_b$ states via strong decays 
to $\Xi_b^0K^-$~\cite{Aaij:2020cex}, observed in Tevateron a decade ago. In 
the mesonic sector, divers spectroscopy studies has been performed, for 
example, in the charmonium sector~\cite{Aaij:2019evc} and the search for 
the excited $D_s$ and $B_s$ mesons~\cite{Aaij:2012pc,Aaij:2014xza,
Aaij:2015sqa,Aaij:2018rol}.


\subsection{Additional Physics Potential with a $K_L$ Beam}
\label{sec:A6}
There are two particles in the reaction $K_Lp\to\pi Y$ and $KY$ that can carry 
polarization: the target and recoil baryons. Hence, there are two possible 
double-polarization experiments: target/recoil. The total number of observables 
is three. The formalism and definitions of observables commonly used to 
describe the reaction $K_Lp\to KY$ is given in Appendix~\ref{sec:A3}. 
Although one cannot easily measure recoil polarization with GlueX, the 
self-analyzing decay of hyperons makes this possible. Double-polarization 
experiments, using, e.g., a polarized target like FROST~\cite{Keith:2016rfh}, 
will however be left for future proposal(s).

As stated in the summary of Mini-Proceedings of the Workshop on Excited 
Hyperons in QCD Thermodynamics at Freeze-Out (YSTAR2016)~\cite{sum} (see 
Appendix~\ref{sec:A1}): a very interesting further opportunity for the KL 
Facility is to investigate KL reactions on complex nuclei. By requiring an 
appropriate beam momentum along with a fast forward-going pion, events can 
be identified in which a hyperon is produced with low relative momentum 
with respect to the target nucleus, or even hypernuclear bound states. 
Baryons with strangeness embedded in the nuclear environment, hypernuclei 
or hyperatoms, are the only available tool to approach the many-body aspect 
of the three-flavor strong interaction. Furthermore, appropriate events with 
a forward-going $K^+$ could deposit a doubly-strange hyperon into the 
nucleus, enabling searches for and studies of double-$\Lambda$ hypernuclei.

Similarly, the scattering of kaons from nuclear targets could be a favorable 
method to measure the matter form factor (and, therefore, neutron skin) of 
heavy nuclei, with different and potentially smaller systematics than other 
probes. The character of the neutron skin, therefore, has a wide impact and 
the potential to give important new information on neutron star structure 
and cooling mechanisms~\cite{Steiner:2004fi,Horowitz:2000xj,Xu:2009vi,
Steiner:2010fz,ToddRutel:2005fa}, searches for physics beyond the standard 
model~\cite{Wen:2009av,Pollock:1999ec}, the nature of 3-body forces in 
nuclei~\cite{Tsang:2012se,Hebeler:2010jx}, collective nuclear
excitations~\cite{Centelles:2008vu,Carbone:2010az,Chen:2010qx,Tamii:2011pv} 
and flows in heavy-ion collisions~\cite{Li:2008gp,Tsang:2008fd}. Theoretical 
developments and investigations will be required to underpin such a program, 
but science impact of such measurements would be high. 

The high flux $K_L$ beam allows a first measurement of a $K_L$ beta-decay, 
$K^0_L\to {K^+}{e^-}{\bar \nu_e}$~\cite{Shishov:2019zyn}, having about 700 
in-flight beta-decay events produced during 200~days of the beam with ability 
to measure about 20-30 of these rare decay events. A fairly simple dedicated 
detector system might be necessary for achieving decent detection efficiency 
for this extremely rare decay branch, $BR\sim 4\times 1
0^{-9}$~\cite{Shishov:2019zyn}.

Further potential exists to search for---or exclude---possible exotic 
baryonic states that cannot easily be described by the usual 
three-valence-quark structure. Recent results from LHCb provide tantalizing 
hints for the existence of so-called pentaquarks that include a charm valence 
quark~\cite{Aaij:2015tga}; however, the interpretation of those results is 
under discussion. In contrast, elastic scattering of $K_L$ with a hydrogen 
target gives unambiguous information on the potential existence of such 
states in the strange sector. With the given flux of $K_L$ at the proposed 
facility, a clear proof of existence or proof of absence will be obtained 
within the integrated luminosity required for the excited hyperon spectroscopy 
program that forms the basis of this proposal. 

The physics potential connected with studies of CP-violating decays of the 
$K_L$ is very appealing; however, that topic is not currently the focus of 
this proposal, since a detailed comparison with the competition from existing 
and upcoming experiments is needed in order to identify the most attractive 
measurements that could be done at the proposed KL Facility at JLab.


\end{document}